\documentclass[11pt,a4paper,openany]{book}

\usepackage{amsmath,bm,amsfonts,amssymb,pifont}

\usepackage{mathtools}
\usepackage[english]{babel}
\usepackage{tikz}
\usetikzlibrary{calc,shapes.geometric}
\usepackage{textgreek}
\newcommand{\Rho}{\mathrm{P}}   

\catcode`\;=12  
\catcode`\:=12
\usepackage[T1]{fontenc}
\usepackage[utf8]{inputenc}
\usepackage{kpfonts}

\usepackage{geometry}
\usepackage{bm}
\usepackage{multirow}
\usepackage{graphicx, xcolor, float, array}
\usepackage{colortbl}
\usepackage{tikz}
\usetikzlibrary{arrows.meta}

\usepackage{siunitx}

\DeclareGraphicsExtensions{.pdf}
\renewcommand{\arraystretch}{1.2} 
\usepackage{booktabs,tabularx,ragged2e,caption,subcaption,titling,textcomp}
\usepackage{longtable, needspace}
\usepackage{fancyhdr,titlesec,nameref,mdframed,parskip,enumitem}

\newcolumntype{Y}{>{\RaggedRight\arraybackslash}X}

\usepackage{needspace}
\usepackage{placeins}

\newcommand{\AppendixRunningTitle}{}

\fancypagestyle{appendixfancy}{%
  \fancyhf{}%
  \fancyhead[C]{\itshape\AppendixRunningTitle}%
  \fancyfoot[C]{\thepage}%
}

\fancypagestyle{appendixplain}{%
  \fancyhf{}%
  \fancyhead[C]{\itshape\AppendixRunningTitle}%
  \fancyfoot[C]{\thepage}%
}

\newcommand{\BeginTextAppendixLayout}{%
  \raggedbottom
  \allowdisplaybreaks[1]%
  \setlength{\parskip}{0pt plus 0.5pt}%
  \setlength{\abovedisplayskip}{7pt plus 2pt minus 3pt}%
  \setlength{\belowdisplayskip}{7pt plus 2pt minus 3pt}%
  \setlength{\abovedisplayshortskip}{4pt plus 1pt minus 1pt}%
  \setlength{\belowdisplayshortskip}{4pt plus 1pt minus 1pt}%
  \setlength{\textfloatsep}{8pt plus 2pt minus 3pt}%
  \setlength{\floatsep}{8pt plus 2pt minus 3pt}%
  \setlength{\intextsep}{8pt plus 2pt minus 3pt}%
  \setlength{\skip\footins}{6pt plus 2pt minus 2pt}%
  \interfootnotelinepenalty=10000%
  \pagestyle{appendixfancy}%
  \titlespacing*{\section}{0pt}{2.2ex plus 0.8ex minus 0.3ex}{0.8ex plus 0.2ex minus 0.2ex}%
  \titlespacing*{\subsection}{0pt}{1.8ex plus 0.6ex minus 0.2ex}{0.5ex plus 0.2ex minus 0.1ex}%
  \titlespacing*{\subsubsection}{0pt}{1.4ex plus 0.5ex minus 0.2ex}{0.4ex plus 0.2ex minus 0.1ex}%
}

\newcommand{\EndTextAppendixLayout}{\par}

\newcommand{\SetAppendixHyperPrefix}[1]{%
  \renewcommand{\theHchapter}{#1.chapter}%
  \renewcommand{\theHsection}{#1.section.\arabic{section}}%
  \renewcommand{\theHsubsection}{#1.section.\arabic{section}.\arabic{subsection}}%
  \renewcommand{\theHsubsubsection}{#1.section.\arabic{section}.\arabic{subsection}.\arabic{subsubsection}}%
  \renewcommand{\theHfigure}{#1.figure.\arabic{figure}}%
  \renewcommand{\theHtable}{#1.table.\arabic{table}}%
  \renewcommand{\theHequation}{#1.equation.\arabic{equation}}%
}

\newcommand{\SetAppendixNumbering}[1]{%
  \renewcommand{\thesection}{#1.\arabic{section}}%
  \renewcommand{\thesubsection}{#1.\arabic{section}.\arabic{subsection}}%
  \renewcommand{\thesubsubsection}{#1.\arabic{section}.\arabic{subsection}.\arabic{subsubsection}}%
  \renewcommand{\thefigure}{#1\arabic{figure}}%
  \renewcommand{\thetable}{#1\arabic{table}}%
  \renewcommand{\theequation}{#1\arabic{equation}}%
  \setcounter{section}{0}%
  \setcounter{subsection}{0}%
  \setcounter{subsubsection}{0}%
  \setcounter{figure}{0}%
  \setcounter{table}{0}%
  \setcounter{equation}{0}%
}

\newcommand{\ManualAppendixChapter}[3]{%
  \clearpage
  \refstepcounter{chapter}%
  \gdef\AppendixRunningTitle{APPENDIX #1: #2}%
  \addcontentsline{toc}{chapter}{APPENDIX #1: #2}%
  \markboth{APPENDIX #1: #2}{}%
  \thispagestyle{appendixplain}%
  {\centering\normalfont\huge\bfseries APPENDIX #1:\par}%
  \vspace{0.25\baselineskip}%
  {\centering\normalfont\Large\bfseries #2\par}%
  \vspace{0.75\baselineskip}%
  \label{#3}%
}

\makeatletter
\newcommand{\BeginAppendixLabelPrefix}[1]{%
  \def\AppendixLocalPrefix{#1:}%
  \let\AppendixOrigLabel\label
  \let\AppendixOrigRef\ref
  \renewcommand{\label}[1]{\AppendixOrigLabel{\AppendixLocalPrefix##1}}%
  \renewcommand{\ref}[1]{%
    \@ifundefined{r@\AppendixLocalPrefix##1}%
      {\AppendixOrigRef{##1}}%
      {\AppendixOrigRef{\AppendixLocalPrefix##1}}%
  }%
}
\makeatother

\makeatletter
\newcommand{\BeginAppendixStarredSectionTOCAnchors}{%
  \let\AppendixOrigAddContentsLine\addcontentsline
  \renewcommand{\addcontentsline}[3]{%
    \def\AppendixTOCLevel{##2}%
    \def\AppendixSectionLevel{section}%
    \ifx\AppendixTOCLevel\AppendixSectionLevel
      \phantomsection
    \fi
    \AppendixOrigAddContentsLine{##1}{##2}{##3}%
  }%
}
\makeatother

\usepackage{changepage}
\usepackage[toc,acronym,nonumberlist]{glossaries}
\usepackage{hyperref}

\hypersetup{
  colorlinks=true,
  linkcolor=blue,
  urlcolor=blue,
  citecolor=blue
}

\usepackage{chngcntr}
\counterwithout*{footnote}{chapter}

\newcommand{\ManualNumberedChapter}[3]{%
  \clearpage
  \setcounter{chapter}{\numexpr#1-1\relax}%
  \refstepcounter{chapter}%
  \chapter*{#1 #2}%
  \addcontentsline{toc}{chapter}{#1 #2}%
  \markboth{#1 #2}{}%
  \setcounter{section}{0}%
  \label{#3}%
}

\newenvironment{abstract}
   {\begin{center}\bfseries Abstract\end{center}\normalsize}
   {\par}
\newenvironment{tenptabstract}
  {\par\begingroup\fontsize{10.2pt}{12pt}\selectfont}
  {\par\endgroup}
\newenvironment{acknowledgement}
   {\begin{center}\bfseries \large Acknowledgements\end{center}\normalsize}
   {\par}  
\newenvironment{quotation_it}
   {\begin{quote}\itshape}
   {\end{quote}}

\tikzset{legend stub/.style={line width=1.2pt, baseline=-0.5ex}}

\makeatletter
\begingroup
  \catcode`\;=12   
  \catcode`\:=12   
  \gdef\linestub#1{%
    \tikz[legend stub,#1]\draw (0,0)--(.7,0);%
  }%
\endgroup
\makeatother
 
\newcommand{\legendlinemark}[3]{%
  \tikz[baseline=-0.5ex]{%
    \draw[line width=0.8pt,#1,color=#2] (0,0)--(1.0,0);
    \node at (0.5,0) [draw,color=#2,fill=#2,#3,inner sep=0.8pt,minimum size=4pt] {};
  }%
}
  
\definecolor{azureblue}{HTML}{0073E6}
\definecolor{teal}{HTML}{008B59}
\definecolor{signalred}{HTML}{FF3B30}
\definecolor{vividviolet}{HTML}{A600FF}

\newcommand{\citequote}[1]{\textit{#1}}

\newcommand{\ProcessGroup}[1]{%
  \noalign{\Needspace{4\baselineskip}}
  \multicolumn{3}{@{}l}{\textbf{#1}}\\[0.0ex]
  \midrule[0.15pt]
}

{%
   \fancyhf{}
   \fancyhead[LE]{\thepage\hfill\MakeUppercase{\leftmark}}
   \fancyhead[RO]{\thepage\hfill\nouppercase{\rightmark}}
   \fancyfoot[CE]{\thepage}

   \renewcommand{\sectionmark}[1]{\markright{\thesection.\ #1}}
}

\newcolumntype{Y}{>{\centering\arraybackslash}m{3.5cm}}

\makeglossaries
\newglossaryentry{quality}{
    name={quality (in the field of health)},
    description={Quality (in the field of health) is defined by the National Academy of Medicine as the degree to which health services for individuals and populations increase the likelihood of desired health outcomes and are consistent with current professional knowledge \cite{CMS2026QualityMeasurementImprovement}}} 
    
\newglossaryentry{qualityassurance}{
    name={quality assurance},
    description={([…]; General) Quality assurance (QA) is the system that provides confidence in the safety, reliability and accuracy of a medical device or process. A QA system is based on a set of procedures that allow the monitoring of the medical device or process, such that if a fault occurred it would be quickly noticed, and the equipment removed from service/process suspended for review. A QA system should encompass the life-cycle of a medical device/process, instilling quality principles at specification, tendering, acceptance, commissioning, in-service and decommissioning stages. A mature QA system should inform the specification and tendering process, just as acceptance and commissioning should inform the in-service QA.\\ ~ \\
    QA systems generally categorize faults relative to their importance, with patient-safety foremost, followed by lower-level faults relating to equipment reliability and/or data-accuracy. Patient safety is important to equipment vendors, and safety issues should have been exhaustively tested by the manu\-facturer. Nevertheless, safety tests form part of routine QA procedures; these tests are designed to check patient-critical systems which may not be encountered during routine clinical operation. Some equipment faults may be spotted first by clinical staff, and so fault-logs also form part of the QA system.\\ ~\\
    Most QA activity is involved in the assessment of accuracy and reliability of equipment and processes, to help verify  that the data produced by a medical device or process is accurate. The tests undertaken to determine accuracy and reliability are defined under quality control (QC) procedures, which form part of the overall QA system. Process optimization is another aspect of QA that aims to reduce errors in clinical procedures. This is of particular importance in areas such as diagnostic imaging, where a variety of clinical and technical factors can invalidate the usefulness of a scan. As well as the economic impact of poorly optimized procedures, there are additional risks associated with ionizing radiation that necessitate a ‘right-first-time’ ethos. Thus, process review, QC results and fault-logs are all aspects that contribute to a responsive QA system \cite{tabakov2021encyclopaedia}}
}

\newglossaryentry{qualitycontrol}{
    name={quality control},
    description={(General) Quality control (QC) refers to the processes by which the data-accuracy and safety of a medical device is assessed. Quality control describes any set of tests that should be carried out on an equipment to determine whether it is fit for purpose.\\    
    Generally, a minimum set of tests that should be carried out on a piece of equipment is defined by a professional body or a national regulatory authority. These will set-out required tests and the acceptable performance for each test. Acceptable performance may be defined in terms of absolute accuracy, or a tolerance of variation from a previous measurement (e.g., at commissioning).\\ ~ \\
    QC then concerns itself with monitoring of data-accuracy over the in-service use of the medical device, and informs the QA system, which encompasses the broader clinical application of the device, and all potential-related sources of error, over the lifecycle of the device \cite{tabakov2021encyclopaedia}}
}
 
\newglossaryentry{qualityimprovement}{
    name={quality improvement},
    description={Quality improvement is the framework used to systematically improve care. Quality improvement seeks to standardize processes and structure to reduce variation, achieve predictable results, and improve outcomes for patients, healthcare systems, and organizations. Structure includes things like technology, culture, leadership, and physical capital; process includes knowledge capital (e.g., standard operating procedures) or human capital (e.g., education and training) \cite{CMS2026QualityMeasurementImprovement}}
} 

\newglossaryentry{qualitymanagement}{
    name={quality management},
    description={Quality management puts in place a plan which should support an organization, product or service to consistently function well. It has four main components: quality planning, quality assurance, quality control and quality improvement. Quality management is focused not only on product and service quality, but also on the means to achieve it. Quality management, therefore, uses quality assurance and control of processes as well as products to achieve more consistent quality. Quality control is also part of quality management \cite{wikipedia_quality_management}}
}

\newglossaryentry{sociotechnicalsystems}{
  name={socio-technical systems},
  description={
  From a systems-theory perspective, radiotherapy systems in general, and external-beam radiotherapy with PBS proton therapy in particular, are complex socio-technical systems. A socio-technical system is an organized set of people, technologies, processes, information flows, and institutional rules that are structured to achieve a specific clinical or operational objective \cite{Trist1951Longwall,Carayon2014HumanFactorsHealthcare}. In contrast to purely technical systems, which can often be modeled as deterministic, decomposable, or at least technically controllable, socio-technical systems are usually complex because their behavior depends not only on technical components, but also on human decisions, communication, organizational context, adaptation, and feedback. \\~\\
  A socio-technical system can be analytically divided into a technical subsystem and a social subsystem. The technical subsystem comprises hardware, software, devices, data structures, interfaces, and physical processes; the social subsystem comprises people, roles, responsibilities, procedures, communication pathways, training, safety culture, and organizational governance. These two subsystems are analytically distinguishable but operationally interdependent: they cannot be optimized independently without affecting the behavior, safety, and performance of the overall system. Their interaction gives rise to dependencies between human communication, human--machine interaction, workflow design, automation, technical reliability, and institutional decision-making. \\~\\
  Socio-technical systems are open and dynamic systems. Their behavior changes with technology, clinical practice, staffing, institutional learning, software updates, data pathways, and evolving quality-management requirements. Therefore, clinical, human-factors, safety, and organizational objectives must be considered explicitly in system design, validation, operation, and risk management. \\~\\
  Among healthcare-related socio-technical systems, radiation oncology in general, and proton or particle beam therapy in particular, require special care because small deviations in imaging, planning, data transfer, treatment delivery, or human decision-making can translate into clinically relevant dose, range, or localization errors. PBS proton therapy may be considered especially complex within EBRT because it combines conventional radiotherapy workflow dependencies with proton-specific sensitivities related to CT-to-stopping-power conversion, range uncertainty, robust optimization, spot and energy-layer delivery, motion/interplay, beam modifiers, treatment-record fidelity, log-file observability, and adaptive anatomical change}
}

\newcommand{\glsQM}{\glslink{qualitymanagement}{QM}}

\newcommand{\glsQC}{\glslink{qualitycontrol}{QC}}

\newcommand{\glsQA}{\glslink{qualityassurance}{QA}}

\newcommand{\glsQI}{\glslink{qualityimprovement}{QI}}

\begin{document}

\title{PTCOG Treatment Efficiency Subcommittee \\Risk Assessment Report \\on Patient-Specific Quality Assurance}
\author{~\\
  ~\\
  Frank Emert\textsuperscript{1},
  Christina Vallhagen Dahlgren\textsuperscript{2},
  Chin-Cheng Chen\textsuperscript{3},\\
  Heng Li\textsuperscript{4},
  Jay Flanz\textsuperscript{5},
  Daniel Robertson\textsuperscript{6} ~\\
  ~\\  
  ~\\
}
\date{\normalsize accepted as PTCOG White Paper\\[0.5em] June 10, 2026}

\begin{minipage}[h]{\textwidth}
  \vspace{1cm}
  \maketitle
  \vspace{2cm}
  \begin{flushleft}
  \begin{quotation}
    \normalsize
      \textsuperscript{1} Center for Proton Therapy, Paul Scherrer Institute, Villigen PSI, Switzerland.\\ 
      \hspace{0.2cm} \href{mailto:frank.emert@psi.ch}{frank.emert@psi.ch}\\[0.3em]
      \textsuperscript{2} The Skandion Clinic, Uppsala, Sweden.\\[0.3em]
      \textsuperscript{3} St. Jude Children’s Research Hospital, Memphis, Tennessee, USA.\\[0.3em]
      \textsuperscript{4} Johns Hopkins Medicine, Washington DC, USA.\\[0.3em]
      \textsuperscript{5} Harvard Medical School, Cambridge MA, USA; Retired.\\[0.3em]
      \textsuperscript{6} Department of Radiation Oncology, Mayo Clinic, Phoenix, Arizona, USA.\\ 
      \hspace{0.2cm} \href{mailto:Robertson.Daniel@mayo.edu}{Robertson.Daniel@mayo.edu}\\
  \end{quotation}

  \vspace{2.5cm}

  \normalsize
  \textcopyright~2026 PTCOG Treatment Efficiency Subcommittee and authors. All rights reserved.\\[0.3em]
  This preprint is distributed via arXiv under the arXiv.org perpetual, non-exclusive license to distribute this article. The certified final version will be hosted on the official PTCOG website and will receive a DOI assigned by Elsevier; its long-term accessibility and citation will be governed by PTCOG and the associated DOI registration, not by journal publication.
  \end{flushleft}
\end{minipage}
\pagebreak
\begin{abstract}
\begin{tenptabstract}

Patient-specific quality assurance (PSQA) in pencil beam scanning proton therapy (PBS-PT) is often discussed as a technical verification problem: a treatment plan is measured, recalculated, or checked against delivery records, and the result is judged acceptable or unacceptable. This report takes a broader view. It treats PSQA as a workflow-embedded risk-control strategy and asks a more fundamental question: how do different PSQA methods reshape the same clinical risk landscape? Specifically, it frames PSQA as a clinical decision and governance problem: which combination of verification layers provides independent, timely, and workflow-appropriate control of the dominant patient-specific risks?

Motivated by the original PTCOG Treatment Efficiency Subcommittee initiative on treatment log records for PSQA, we developed a process-driven Failure Mode and Effects Analysis (pFMEA) framework for a generic PBS-PT workflow. Forty-four validated PSQA-relevant failure modes were assigned to 20 process steps and evaluated in an expert-scored, semi-quantitative pFMEA under a common no-PSQA baseline and three alternative PSQA ap\-proach\-es: measurement-based PSQA, log file-based PSQA, and independent secondary dose calculation. To compare these pathways consistently, we introduced a staged mathematical formalism se\-pa\-rating preparatory Data-stage effects, method-specific Full-stage verification, cumulative end-state effects, and the derived Data-to-Cum bridge that quantifies the additional verification benefit on the common baseline scale. Risk changes were analyzed at the failure-mode, process-step, workflow-region, and full-workflow levels using baseline-risk-weighted aggregation.

Within this expert-scored, baseline-anchored pFMEA model, log file-based PSQA showed the largest cumulative work\-flow-level risk-score reduction (-40.3\%), followed by measurement-based PSQA (-34.7\%) and independent secondary dose calculation (-8.5\%). This model-derived ranking should not be interpreted as a winner-takes-all result, a universal replacement hierarchy, or a probability-calibrated estimate of clinical risk. Rather, log file-based PSQA was strongest in delivery-, recording-, and machine-state-related workflow regions; measurement-based PSQA provided broad compensatory mitigation across planning, preparation, translation, setup, and delivery-adjacent steps; and independent secondary dose calculation was highly selective but essential for image-, model-, and calculation-sensitive failure modes. Root-cause analysis further confirmed the socio-technical nature of PSQA risk: 33 of 44 validated failure modes (75.0\%) were human-, procedure-, communication-, or human-system-interaction-related, whereas the dominant effects remained clinically meaningful dose-, plan-, and delivery-related consequences.

The report therefore argues against replacing one PSQA method by another on purely technical grounds. Instead, it supports a risk-informed hybrid architecture in which each method is assigned to the work\-flow regions where its risk-control signature is strongest. The proposed framework is semi-quantitative, not probabilistic; its value lies in transparent, baseline-anchored, stage-resolved comparison. It provides a practical and mathematically traceable structure for institutions seeking to evaluate, implement, or evolve measurement-based, log file-based, and calculation-based PSQA in proton therapy.

In practical terms, pre-treatment log file-based PSQA may offer major efficiency advantages over conventional measure\-ment-based PSQA, may require less dedicated measurement equipment, and can allow treatment fields to be tested under delivery conditions that more closely resemble the actual patient treatment. When combined with routine or daily log file-based dose evaluation, it may provide greater overall risk-score reduction within the present model than measurement-based PSQA or independent secondary dose calculation alone. This does not make measurement-based PSQA obsolete: measurements remain indispensable for failure modes that are not observable in machine log data, and they provide substantial complementary risk reduction, especially when new techniques, treatment sites, delivery configurations, or commissioning states are introduced and require appropriate physical validation.

A fully independent secondary dose calculation remains valuable regardless of whether a center primarily uses log file-based or measurement-based PSQA. For contemporary PBS-PT workflows, at least one of the primary or secondary dose calculations should ideally be an independently commissioned three-dimensional calculation in patient geometry, preferably using an accurate Monte Carlo algorithm. Consequently, log file-based PSQA should not be used as a primary replacement strategy unless the log data, treatment records, and upstream data-transfer pathways on which their interpretation depends are themselves explicitly validated, monitored, and governed within the institutional QA program.

The resulting decision message is therefore conditional and local: PSQA in PBS proton therapy should be designed as a baseline-anchored, stage-resolved, risk-informed hybrid architecture tailored to local workflow maturity, va\-li\-da\-ted implementation conditions, and the clinical consequences of undetected failure modes.
\end{tenptabstract}
\end{abstract}

\newpage
\begin{acknowledgement}
We gratefully acknowledge the vision and leadership of Niek Schreuder and Chris Beltran, past co-chairs of the PTCOG Treatment Efficiency Subcommittee and organizers of the excellent series of Treatment Efficiency Workshops held in Knoxville, TN, culminating in the 2018 workshop on log file-based PSQA.  This white paper is a direct outgrowth of those meetings, and we express our gratitude for their efforts to gather physicists from around the world to address these important topics.

We would also like to thank the many participants in these workshops whose ideas and discussion helped to lay the groundwork for this white paper.
\end{acknowledgement}

\tableofcontents

\setcounter{chapter}{0}
\chapter*{0 PROLOGUE}
\addcontentsline{toc}{chapter}{0 PROLOGUE}
\markboth{0 PROLOGUE}{}
\label{ch:prologue}
\section{Risk Analysis in Radiation Oncology}

\subsection{Foreword}
Within the PTCOG Treatment Efficiency Subcommittee, patient-specific quality assurance (PSQA) has been the subject of various workshops and numerous discussions for many years, including discussions about its optimization as well as its effective and efficient integration into state-of-the-art particle therapy treatments.

The following prologue is divided into three sections. The first outlines the theoretical problem, provides definitions\footnote{Please refer to the glossary for selected definitions and explanations of quality terminology, which are marked when they first appear in the prologue.}, introduces a uniform, contemporary terminology for \gls{quality} and risk management (RM), and briefly summarizes the methodological basis for data collection, analysis, and evaluation in this report. The second prologue section provides a review of key literature and reference documents, focusing on the two Treatment Efficiency Subcommittee white papers presented in May 2016 \cite{flanz2016white}\cite{flanz2016ptcog}. Specifically, their summary provides the methodological framework, sets the theoretical background, and describes the practical tools to analyze, evaluate, and further develop key aspects of safety, risk, and \gls{qualitymanagement} in the field of particle therapy. The third section briefly explains concepts and terminology employed in the workflow and process description of particle therapy as they appear in this paper.
\vspace{-0.5cm}
\begin{quotation_it}
\noindent
\definecolor{orange}{rgb}{1,0.5,0}
  \begin{mdframed}[linewidth=1pt, linecolor=red]
  \textbf{This prologue is recommended for readers desiring a description of the methodological and theoretical foundations of process-based Failure Modes and Effects Analysis (pFMEA), in combination with a review of the relevant literature.}
  \end{mdframed}
  \begin{mdframed}[linewidth=1pt, linecolor=orange]
  \textbf{Strictly speaking, studying the prologue is not necessary to understand the results of this work, but it does provide context and background information that may enhance the reader's perspective of PSQA.}
  \end{mdframed}
  \begin{mdframed}[linewidth=1pt, linecolor=green]
  \textbf{However, without this foundation, the nuances and implications of the PSQA definitions, in particular, could be missed. Therefore, readers are encouraged to review the Prologue to gain a deeper understanding of the underlying principles and concepts that shape the contents of this report.}
  \end{mdframed}
\end{quotation_it}

\subsection{Problem Statement}
\label{subsec:prologue_problemstatement}
A major goal of radiation treatment is to deliver a dose distribution which is within the targeted error tolerance of a prescribed treatment. To mitigate potential errors, one can evaluate the relevant processes and equipment in order to understand and reduce the risks associated with their application.

The considerations that follow attempt to outline the theoretical-methodological basis of the process-oriented, prospective, Failure Mode and Effect Analysis-based risk assessment of patient-specific quality assurance (PSQA-pFMEA) in proton or particle therapy (PT) using pencil beam scanning (PBS). The aim is to conceptualize and define the organizational framework and functionalities that should be in place. Some of the safety results have been identified and cataloged in existing international standards (ICRU, AAPM, IAEA, ISO, etc.), but a first principles analysis has generally been lacking, as well as the ability to account for different treatment processes and equipment. Essential interrelationships are explained below, but a complete, scientific discussion of all correlations would go beyond the scope of this paper. For this purpose, reference is made to the relevant literature.

It helps to understand how conventional and particle-operated external beam radiotherapy (EBRT) systems have developed over time. This includes both their technological capabilities and their organizational structures. The cultures and management concepts influence how these systems function and are operated. In this context, quality management systems play an important role, as they are nowadays closely linked to the functioning and integration of risk management concepts.

These quality management phases shaped important discussions in the scientific field of the EBRT domain during the last decade and have concrete consequences for the practical implementation of the PSQA-pFMEA described in this report. Process-based modeling has only relatively recently been applied to medical systems, and the concept of using this methodology for system-specific assessment continues to evolve. These processes do not always produce the same results as the proposals used in traditionally more technical or \gls{qualitycontrol}-oriented standards. This assessment will therefore cover both the physical/technical control level and the treatment planning system\cite{icru2007report}\footnote{In a methodically strict sense, it would be necessary at this point to describe all interactive software systems that are integrated into the patient workflow processes. However, for feasibility reasons and by focusing on PSQA, for which the treatment planning system (TPS) plays the central role, it is omitted. This can be done without the findings and results losing any of their relevance and significance. The reasons for this are that 
\begin{itemize}
\item the intrinsic, systemic uncertainty in a semi-quantitative process-based PSQA-pFMEA in PBS-PT is very likely to be significantly higher compared to the influence arising from the software systems not considered.
\item  the processes in proton therapy centers apparently have a significantly greater influence on the outcome of any risk assessment than the complete coverage of purely technical software components.
\end{itemize}
The latter argument is supported  by the fact that numerous studies, inter alia by IAEA, have concluded that at least 2/3 of reportable EBRT incidents have human error as their main cause and thus as the most significant failure mode. Since human error in any QM system of an EBRT institution is assigned exclusively to its process components, it must be concluded that the safety and quality in patient irradiation is influenced more by the quality of dynamic, socio-technical factors of the corresponding organization than by the reliability of its technical-physical equipment and its software control. Therefore, in our specific case it is also methodologically justifiable to ignore software systems that are of secondary meaning for PSQA.}. 

\subsection{Quality and Risk Management}
\label{subsec:prologue_QM+RM}

The \gls{quality} of treatment regimens as well as patient and personnel safety are of paramount importance in external beam radiotherapy. Therefore, quality management (\glsQM{}) has an essential steering function for the way EBRT systems are developed, implemented, organized, and operated. Quality control (\glsQC{}) has traditionally focused on the technical and product-related devices, so it enhances the fault-free functioning of technical subsystems and components. Generally, these are medical products, i.e., medical and physics equipment, related technical devices, information technology (IT) hardware and software systems, etc., including the operating and application conditions in each phase of their life cycles.

In contrast, \gls{qualityassurance} (\glsQA{}) in EBRT has a much broader scope, going far beyond technical QC. This becomes immediately obvious when EBRT systems are understood as \gls{sociotechnicalsystems} in the sense of their definition. In this context, QA effectively points to a set of instructions to (i) help remember the QC-relevant tasks, (ii) help guide the smooth, planned and trouble-free flow of human functions in the sequence and control of the operational business, and (iii) provide a framework for the complex interplay of all components, processes and interactions of the technical with the human-social subsystem in all their mutual dependencies and effects.  \gls{qualityimprovement} (\glsQI{}) in healthcare, has been defined as
\begin{quotation_it}
the combined and unceasing efforts of everyone — healthcare professionals, patients and their families, researchers, payers, planners and educators—to make the changes that will lead to better patient outcomes (health), better system performance (care) and better professional development \cite{paul2007quality}.
\end{quotation_it}

If one focuses specifically on EBRT systems, i.e. on new, innovative irradiation devices, techniques and their process-controlled application, it becomes immediately clear that there is no QI without preceding risk assessment and risk management:
\begin{quotation_it}
  \begin{center}
    \vspace{0.1cm}
    \boxed{\textbf{RM drives QI!}}
  \end{center}
\end{quotation_it}
In general, risk management comprises the sub-areas of risk identification, risk assessment, risk analysis, risk mitigation and risk evaluation. Risk identification, as the first step, is thereby associated with the perception of the level of risk by the individuals involved and can therefore be subjective. In order to identify the appropriate risk level, all relevant role representatives of the system should participate in all the risk management steps identified in the paragraph.

Following risk identification, there's a natural progression into assessing and analyzing these risks. Here, the focus shifts to understanding the nature and potential impact of each risk. This phase is not just about listing possible issues but delving into their likelihood and consequences. It's a critical step that shapes the entire risk management strategy, helping prioritize efforts towards the most significant risks.

The development of targeted strategies to mitigate the identified risks constitutes risk mitigation. Risk evaluation immediately follows, in which the effectiveness of risk mitigation strategies is appraised. If the risk under consideration has been sufficiently reduced, the assessment for that particular risk ends. (Note that some risks may be linked and it would require all the linked ones to be handled before completing any one risk.) If the risk mitigation is deemed insufficient, further risk mitigation considerations are made. These considerations are part of an iterative approach to risk management. Considering the changed risk, the process then starts again with risk identification, revealing its iterative nature.

\section{Literature on Risk Analysis for Particle Therapy}

\subsection{ICRU Reports 50 and 78}
The various well-known ICRU reports address the prescribing, recording, and reporting of the respective EBRT modalities. These reports provide the information necessary to standardize techniques and technical procedures and harmonize clinical descriptions and prescriptions. Thus, the radiobiological, physical, technical, treatment planning, and clinical aspects of different EBRT modalities are described in a largely uniform terminology to provide new users with the basic background that will enable them to understand, implement, and apply these techniques.

The only mention of a quality term in the ICRU reports to date appears as the following in Appendix 1 of Report 50 \cite{jones1994icru} on Photon Beam Therapy as follows:
\begin{quotation_it}
\samepage
\textbf{Quality Control}\\
(i) Check and confirm systems.\\
(ii) Verification imaging (when during treatment, frequency, and acceptability). \\
(iii) In-vivo dosimetry (type of measurements, frequency, and acceptability).
\end{quotation_it}

ICRU Report 78 \cite{icru2007report} specifically addresses proton beam therapy. This report focusses on the technical subsystem as opposed to the socio-technical aspects. For example, the report states in the executive summary section on quality assurance that while 
\begin{quotation_it}
\samepage
a rigorous quality assurance (QA) program is required to ensure reproducible, accurate, and safe fulfillment of treatment prescriptions [...,] QA checks are often technology- and equipment-specific and focus principally on various aspects of dose delivery, patient positioning, and treatment planning as well as on radiation protection.
\end{quotation_it}

\subsection{PTCOG Safety Group Report on Aspects of Safety in Particle Therapy}
The \textit{PTCOG Safety Group Report on Aspects of Safety in Particle Therapy} \cite{flanz2016ptcog} highlights the importance of safety in radiation therapy in general and particle therapy in particular. It is emphasized that a particle therapy facility must undertake to ensure safe and accurate treatments and create a culture of safety. In this regard, the responsibility for establishing procedures and treatment parameters lies with the radiation therapy department and equipment manufacturers. The incidents reported in various public articles question what responsibilities and processes are necessary for safety. Reference should be made to the implementation of quality management systems based on ISO 9000 \cite{iso1992standards} standards and additional country-specific standards for radiation therapy.

Due to the diversity and complexity of radiation therapy, the potential risks associated with particle therapy devices, and the importance of analyzing risks and interfaces, the main goal of this report is to provide information and tools for self-assessment and implementation of safety measures. In doing so, it is important to create a safety culture and establish proactive quality management methods, focusing on the analysis of safety issues, alternative approaches, and general guidelines for improving safety in radiation therapy. Definitions, hazard analyses, risk assessments, corrective actions, patient care workflows, and quality assurance fundamentals can be applied in this context, while also obtaining information from other agencies and resources such as the FDA's Adverse Event Reporting System and IAEA reports on radiotherapy errors.

Some areas of this Safety Group Report have been updated, corrected and enhanced in Chapter 15 of the book Particle Therapy Technology for Safe Treatment \cite{flanz2022particle}. This book details how to identify potential system technical errors and how to incorporate them in risk management.

\subsubsection{Hazards and Risks}
In general, hazards and risk mitigation measures should be discussed in the context of processes that enhance safety. By definition, hazards are circumstances that can lead to unplanned or undesired events. Equipment, people, or a combination of both can interact with that hazard causing an accident. It is important to protect a system from hazards and to integrate mitigation measures using prevention, detection, and response to reduce accidents. Ensuring 100\% success for a complex system may not be achievable. In fact, the frequency of detection and time to respond are critical factors in radiation therapy environments, since errors must be detected, and the system should react to mitigate the error. This raises the fundamental question of whether errors can generally be prevented or whether the focus should be on detection and response. In this context, there is also the question of whether and which tasks should be performed by humans or by machines. Given the differences between conventional radiotherapy and proton beam therapy (particle beam therapy facilities have generally been designed to have more data which is logged in detail), various methods of risk mitigation can be discussed in terms of error analysis and human intervention, including fail-safe design, quality controls, and teaching personnel through training.

With this in mind, risk and criticality is introduced according to international standards such as IEC 601-1-4 \cite{iec1995medical} and ISO 13489-1 \cite{iso2023safety}, where the term 'risk' is defined as the combination of the probable frequency of occurrence of a hazard and the severity of the damage caused by an associated accident. Hierarchies of severity and frequency levels categorize risks using various ordinal scales that rank severities and different probabilities of an event occurring. The number of levels can vary depending on specific circumstances, and they can be subdivided in different ways. When categorizing and quantifying risks, the calculation of the risk potential number (RPN) is often used as a combination of severity, probability, and other factors such as detectability and preventability. The RPN determines the criticality of a risk and guides the selection of appropriate mitigation actions. RPNs can be calculated using a variety of approaches, including RPN tree or risk graph, brute force calculation, and binary combination. Various concepts and methods exist for defining and evaluating risks and criticality in safety analyses, which provide a basis for further hazard and failure analyses.

\subsubsection{Risk Assessment Methods}
Starting with Heinrich's "Domino Accident Model" and Reason's "Swiss Cheese Model", which consider multiple sources of error, various assessment models and methods can be distinguished that are used in accident analysis and risk assessment. In addition, newer models such as the Functional Resonance Accident Model (FRAM) and the System Theoretic Accident Model and Process (STAMP) are used, which emphasize the importance of considering organizational factors, social culture, and the environment when analyzing accidents. The need for sophisticated analytical models increases as systems become more complex. Three commonly used tools for system safety analysis are fault tree analysis (FTA), event tree analysis (ETA), and failure mode and effects analysis (FMEA), which can be used to identify hazards, their causes, and potential corrective actions. Regardless of the method chosen, it is important to include appropriate data, ask the right questions, and choose the right level of detail for the analysis. A safety culture that promotes robust reporting can contribute significantly to improving any approach. Reviewing risk mitigation measures and defining a safe state are essential aspects of the analysis. A brief review of the history of risk assessment in healthcare, including the evolution of hazard analysis and failure analysis in various industries, can outline how FMEA is used in healthcare, particularly radiation oncology, and recognized by organizations such as the FDA.

Hazard analysis itself is a systematic process aimed at identifying potential hazards that may arise from the workflow and operation of a system. A hazard analysis presents a top-down approach. Hazard analyses include selecting a hazard, identifying the affected subsystem, envisioning how the subsystem may cause the hazard, evaluating the significance of the hazard, creating a mitigation plan, and determining how the effectiveness of the mitigation will be verified. The analysis typically focuses on identifying hazards at a higher level, such as the subsystem level, rather than component or subprocess level details. Hazard analysis inputs include the identification of hazards and the subsystems that can contribute to those hazards, along with their associated responsibilities and requirements. Hazard tables serve as visual representations to support hazard analysis. While it may not be possible to identify every potential hazard or failure condition, a comprehensive analysis can be achieved by considering relevant hazard types outlined in standards such as ISO 14971 \cite{iso2019medical} and using techniques like FMEA. In addition, the concept of root cause analysis (RCA) provides a tool to help identify the underlying causes of hazards. RCA is a detailed analysis aimed at understanding the root causes of problems and is often used in healthcare settings.

In general, FMEA represents a method used to identify failure modes that can lead to hazards and to prioritize them for safety improvements. It takes a bottom-up approach by examining each step or component of a process and imagining what can fail and what effects can lead to hazards. The FMEA process starts by creating a list of failure modes in a particular component or workflow. The significance of each failure mode is evaluated based on the factors probability, severity, and detectability. The goal is to prioritize safety improvements by addressing the most critical hazards. Brainstorming sessions involving representatives from different disciplines are often helpful in performing the analysis and fostering a culture of safety. Once the failure modes are identified, their impacts and hazards are evaluated. This evaluation can be done using different scales, but it is important to use what makes sense for the specific environment and be consistent within the system. Risk mitigations are then proposed, and their effectiveness is evaluated. FMEA provides a link between the bottom-up approach of examining failure modes and the top-down approach of hazard analysis. The iterative nature of FMEA, where risk mitigation measures lead to a reanalysis and a lower risk level, allows it to be applied to various systems, improving the understanding of potential failure modes and prioritizing safety improvements. It is important for clinics to perform FMEA analysis and be consistent with the assessment details. Additionally, FMEA can facilitate the exchange of information and best practices between different clinics.

\subsubsection{Risk Mitigation}
Mitigation of the risks identified in FMEA is paramount to risk reduction, and a dearth of literature on this topic can be identified, particularly in the field of radiation oncology and patient safety. Different strategies can be used to mitigate risk, such as introducing specific measures to reduce the probability of risk or providing general guidance based on the different levels of risk. The first method is to introduce risk reduction measures for specific, typical errors or hazards and continuously evaluate their effectiveness on risk reduction. The second method establishes an overarching risk mitigation strategy based on the hazard category, using different types of mitigation methods depending on the level of risk. For example, there are three conceivable approaches to mitigating risk: making the risk impossible, making the error visible, and providing resilience. Protocol- and communication-based mitigation measures, general quality assurance, and facility safety design, among others, play a key role. Intensive staff training, effective communication structures and regular meetings are prerequisites to promote proper information flow and a safety culture within a given system such as a particle therapy center.

The importance of safety extends across the entire patient treatment workflow and goes far beyond technical safety considerations for the function of the radiotherapy device, so that the entire multidisciplinary team and all systems involved must be included. To systematically identify and mitigate potential errors, it is recommended to visualize the key steps in the treatment process graphically as a map or flowchart, including consultation, simulation, treatment planning, physics quality assurance, and patient irradiation. As part of the consultation process, comprehensive information gathering, consideration of various factors in decision-making and the importance of effective communication between different disciplines are imperative.

\subsubsection{Quality Assurance}
Quality assurance (QA) itself is an essential part of the clinical workflow and involves verifying the correctness of processes and equipment that must meet established quality requirements. It is important to integrate quality assurance into the overall quality management system (QMS) and to include QA checks throughout the workflow. Machine QA focuses on verifying that the proton therapy system is producing the radiation parameters required for treatment planning, while clinical QA confirms that the dose distribution in the treatment plan matches the actual dose delivered. Both machine and clinical quality assurance are essential to the accuracy and safety of particle therapy.

Radiation QA protocols must be in place that include system acceptance testing, clinical commissioning, and ongoing quality control. Quality assurance is in itself an error prevention strategy and helps to identify and respond to potential errors in the system. Specifically, dosimetric radiation parameters can be identified as critical aspects of QA. These parameters include absolute dose, relative dose distribution, and absolute position of the dose distribution. For scatter and scanning beam delivery methods, specific beam parameters result that must be measured as part of QA and their transformation into the relevant clinical beam parameters must be determined. It is important to distinguish between the measurement of clinical beam parameters with external instruments and the measurement of the device condition itself, which determines the beam parameters. The frequency of QA measurements depends on the potential risk to patient safety if a particular activity or control is not performed correctly.

\subsubsection{Incident Learning Systems}
Improving the safety and quality of particle beam therapy could be aided by a shared database, which, while not an exhaustive list of potential failures in the field because technology and procedures change rapidly, underscores the importance of sharing information about potential risks among clinical users. The concept of sharing experiences between clinicians can be compared to practices in the commercial nuclear power industry, where the Institute of Nuclear Power Operators (INPO) sends daily alerts to power plants about identified conditions that could be relevant to other plants. Similarly, a common database of potential risk scenarios would be beneficial to the safety of clinical operations. This database should serve as a clearinghouse for potential risk areas and provide for anonymity.

In addition, the establishment of an institutional system for reporting critical events without blame, similar to critical incident reporting systems (CIRS), is proposed. These systems, referred to as Radiation Oncology Incident Learning Systems (RO-ILS), can contribute to further risk reduction and quality management. For example, the distinction between actual accident reports and hypothetical risk scenarios could be discussed within the database, while maintaining patient privacy. The structure of such a database should reflect the AAPM white paper on accident reporting databases. A common process map is envisaged as a generic template for conducting FMEA assessments. Quality control and monitoring of incoming data is important, as is the need for an anonymous means of communication with contributors to clarify descriptions or address other issues.

\subsection{Particle Therapy Efficiency: Aspects of Quality Assurance}
The second white paper, \textit{Particle Therapy Efficiency: Aspects of Quality Assurance} \cite{flanz2016white} discusses the potential for improving the efficiency of quality assurance processes in particle therapy facilities. Several causes contribute to the inefficiency of current QA practices, including the use of procedures similar to those used in photon therapy, measures deemed necessary for regulatory and legal purposes, and processes that are dictated by available instrumentation, available personnel, and machine performance. According to the authors, these processes can be made more efficient without compromising safety and accuracy.

In doing so, the development of a quality assurance program should be appropriately configured for the system being evaluated. When possible, established methods such as risk analysis and lean principles should be followed, and regulatory rules must be followed. It is important to balance rules with flexibility and there should be room for proven new methodologies, such as log-based QA. However, the prevailing approach relies on techniques recommended by organizations such as AAPM and ASTRO, or their working groups, and often interpreted as mandatory. It is hoped that the use of risk analysis methods, such as through application of the AAPM TG-100 \cite{huq2016taskgroup100} methodology, or reference \cite{flanz2022particle}, and the flexibility to enhance the spirit of the guidelines where appropriate, will lead to more efficient QA practices.

It should be noted that the quality terms used in this section are only partially consistent with the definitions used in the previous section of this report. For example, quality assurance follows the stand-alone definition (from the Oxford Dictionary) as \citequote{the maintenance of a desired level of quality in a service or product, especially by attention to each stage of the delivery or production process.} Subsequently, quality assurance is essentially interpreted as maintaining the existing level of quality, and quality improvement is implicitly understood as efficiency improvement in terms of the PTCOG Treatment Efficiency subcommittee’s group title only. Nevertheless, the overall methodological concept is in agreement with our approach.

\subsubsection{Efficiency in Quality Assurance}
Several prerequisites exist for QA in particle therapy, with a focus here on treatment planning and beam delivery. Critical elements of beam delivery include clinical commissioning, oncology information system data, correspondence between imaging and radiation delivery, and radiation delivery modality. Each of these elements and their associated activities can be considered potential sources of inefficiency and should be examined to identify areas for improvement.

Potential problems that impede optimization of quality assurance programs include outdated processes, differences between facilities, lack of stakeholder acceptance, information silos, and billing problems. There are also associated costs, such as investments in optimized instruments, training staff, and measurement time. Educating stakeholders and promoting acceptance of optimized approaches are essential for successful implementation.

Contributions to improving QA efficiency could include a methodology for analyzing and prioritizing critical parameters, assessing the severity and likelihood of errors, calculating risk, and determining appropriate mitigation factors. For example, defined, standardized frequencies and intervals for measurements (real-time, daily, monthly, yearly, and/or one-time) could implement measurements as potential mitigation strategies. It will be necessary to streamline measurement steps and analysis requirements, automate processes, and eliminate redundant tasks.

The time required for QA activities in particle therapy is, at this time, significantly higher compared to photon therapy, and a wider range of equipment will continue to be used for measurements in particle therapy. For optimization, multiple measurements could be combined into one session and instruments could be integrated to improve efficiency.

A prominent example of how aspects of quality assurance in particle therapy can be reconsidered and made more efficient is patient-specific QA. By incorporating a log file-based QA approach (in combination with an independent Monte Carlo dose calculation), the traditional measurement-based QA method could be optimized or made more efficient. Overall, the existing challenges and potential solutions for enhanced QA efficiency in particle therapy depend on the application of risk analysis, optimization, and collaboration among all stakeholders.

\subsubsection{AAPM TG-100}
In the preface of their book \citequote{Quality and Safety in Radiotherapy} \cite{thomadsen2013quality}, the authors (mostly representatives of the AAPM Task Group 100) outline, that
\begin{quotation_it}
  radiotherapy became one of the first medical disciplines to establish standardized quality assurance. In light of the increasing complexity of radiotherapy treatments and technology and the diversity of methods used by facilities for various clinical procedures, it is no longer practical to rely solely on generic, prescriptive lists of comprehensive quality assurance (QA) steps to ensure quality and safety for patients.
\end{quotation_it}

Their methodological approach, which is described in the AAPM TG 100 report and represents a process-based FMEA-based risk assessment of IMRT, essentially reflects how 
\begin{quotation_it}
  quality and safety […] techniques consider not only QA for equipment, but also for procedures as a whole in the context of the facility.
\end{quotation_it}

\subsubsection{AAPM TG-224}
In analogy, the 2019 AAPM Task Group 224 \cite{arjomandy2019taskgroup224} report on comprehensive proton therapy machine quality assurance states, 
\begin{quotation_it}
  [a] robust QA program considers the potential for system failure and performs the necessary steps to measure system performance to detect these potential failures. To help the reader understand the rationale for quality assurance, this report briefly discusses the methodology and parameter requirements that help ensure consistent proton beam system performance.
\end{quotation_it}

Furthermore, in the following section on the methodology of quality assurance, it is noted that 
\begin{quotation_it}
  quality assurance has three main branches: (a) general equipment functionality, including dosimetry, imaging, and mechanical QA; (b) patient-specific QA; and (c) treatment planning system (TPS) QA [.., but] this report focuses on general equipment functionality QA or machine QA. Patient-specific and TPS QA are not addressed in detail.
\end{quotation_it}
The presented reference \cite{aapm2015summer} for the QA branches on PSQA and TPS QA missing in the AAPM TG-224 report, deals in general with the proton therapy workflow and contains in a section on hazard analysis a good compilation of important facts and practice-relevant correlations regarding risk management, FMEA and related methods, which are supported with additional literature references. There one also finds practice-oriented representations of the relationship between a PBS-PT workflow and QA.

\subsubsection{ICRU Report 93}
Also published in 2019, ICRU Report 93 \cite{jakel2016icru93} on light ion therapy states in the introduction to its QA chapter that quality assurance in this regard follows essentially the same guidelines as the other EBRT modalities. At the same time, the AAPM TG-100 report is explicitly referenced, \citequote{which describes modern quality management applied to radiation oncology.} Furthermore, it is also specifically mentioned, \citequote{PTCOG has published a white paper, which gives advice for prioritizing QA measures and how to make QA procedures more efficient.}

Almost full agreement with the overall methodological approach presented in the previous section, incorporating its main components of quality and risk management, is evident in ICRU Report 93, when the section on recommendations for light ion therapy of the QA chapter, from which has already been quoted, states:
\begin{quotation_it}
\nopagebreak
It should be stressed that QA is always part of a comprehensive quality management system. This includes additional tasks like the development of standard operating procedures and user manuals, definition of maintenance plans, training and further education of the operating staff, management of human and financial resources as well as risk analysis and other, more general issues of safety and radiation protection. […] The Particle Therapy Cooperative group (PTCOG) has installed a Task Group 2, which issued a report on ‘Risk Analysis of Charged Particle Therapy Facilities’ (Flanz et al., 2016b). In this report systematic methods are described for assessing and improving the safety of an ion-beam irradiation system. An example of the application of a specific method to analyze failure modes and resulting consequences, the so-called Failure Mode and Effects Analysis (FMEA), to particle therapy is given by Cantone et al. (2013).
\end{quotation_it}

\section{Workflow and Process Description for PBS-PT Treatments}

\subsection{Workflow Models and Granularity}
\label{subsec:workflow_granularity}
Generally, a workflow or process representation is characterized by either
\begin{itemize}
    \item a simplified, descriptive and usually hierarchically ordered list or collection of specifications of what should happen where, when, how, why and by whom, or
    \item a graphically represented modeling of a part of reality to be depicted, e.g., a socio-technical system, and if necessary, its temporal development, which is mostly realized by a special (programming) language with its own vocabulary and grammar. Examples are flowcharts, event-driven process chains, UML activity diagrams (UML = Unified Modeling Language), Business Process Modeling (and) Notation (BPMN) and others.
\end{itemize}

The relationship between the total number of process units and subunits modeled, i.e., the degree of "model granularity", and the purpose and scope of the model is critical to the practical modeling of structures. For example, the set of all flowcharts in detailed work instructions of all roles in a radiotherapy (RTTs, dosimetrists, operators, medical physicists, radio-oncologists, etc.) may be suitable to provide a complete static system description, but it is completely unsuitable for a rough system overview to represent higher-level functions and interfaces.

The modeling of the temporal system behavior will not be further analyzed and discussed here, but there may be time ordered effects that affect the failure modes (e.g. software race conditions). This also applies to decision-related feedback loops within the system (adaptive RT/PT approaches).

All technical medical devices, EBRT units (LINACs, PBS PT systems, etc.) and especially all IT systems involved, in short, all components used for radiation treatment of a cancer patient, can (and some must legally) be quasi completely "modeled through". Some aspects of the combined technical and social subsystem (in systems theory terminology), i.e., the human-human, human-machine interactions can be more complex (e.g., it is possible that rules may not be followed). The model granularity depends on the functionality of the model.

Regarding the terminology for the granularity of the modeling, the following is agreed upon, which essentially also corresponds to the use of terms in the literature considered.

\begin{itemize}
    \item \textbf{Workflow}\\ The entire process (map) of a patient treatment from the initial contact to the therapy facility to the downstream follow-up.
    \item \textbf{Process}\\ The workflow consists of a list of functional (high-level) processes. Classically, these include the simulation, delineation, treatment planning, treatment delivery and verification phases.
    \item \textbf{Process step}\\ A process contains steps that have a functional meaning in the model.
    \item \textbf{Task}\\ The process steps in the model contain those action-bound tasks necessary to fulfill the function. These include human interactions and communications as well as simple man-machine interactions (e.g., reading a parameter from the screen, input of a date into a software system, etc.). This level of task is neglected in this report. They matter in methods such as Hierarchical Task Analysis (HTA), for example, when performing a human reliability assessment.
\end{itemize}

\subsection{Process Feedback and Adaptive Therapy}

Starting in 1993 with ICRU Report 50 \cite{jones1994icru}  for photon beam therapy the corresponding workflow was introduced in a rather sequential order. Nevertheless, the report already states:

\begin{quotation_it}
There should be a continuous feed-back between all the different steps. A difficulty at a given point may question all the decisions made at previous steps.
\end{quotation_it}

This implies a feedback loop, an early indication of the adaptive nature of any EBRT workflow.

The first paper defining the term \citequote{adaptive radiation therapy} (carrying the same title) was - to the best of our knowledge - published by D. Yan et al. in 1997 \cite{yan1997adaptive}. The workflow figures visible in the ICRU Reports 50 \cite{jones1994icru}, 78 \cite{icru2007report} and 93 \cite{jakel2016icru93}, illustrate the evolution of the workflow concept from a simple process sequence to nested, feedback organizational units of increasing complexity.

\chapter*{1 INTRODUCTION}
\setcounter{chapter}{1}
\addcontentsline{toc}{chapter}{1 INTRODUCTION}
\markboth{1 INTRODUCTION}{}
\setcounter{section}{0}
\label{ch:introduction}
{%
  \renewcommand{\sectionmark}[1]{\markright{\thesection.\ #1}}%
  Radiation therapy using scanned particle beams requires the coordination of computerized treatment planning systems, advanced accelerator and beam delivery systems, as well as data transfer, translation, and record and verify systems. Verifying the safety and accuracy of particle therapy treatments is a complex endeavor, encompassing the objectives of (i) checking that beam delivery parameters are within permissible tolerances during treatment fractions, (ii) ensuring end-to-end data consistency along the transfer through the system components involved, (iii) reviewing the agreement of generated machine control information to deliver planned dose distributions with recorded protocol data for treatment documentation, as well as (iv) proactive control and adjustment of the patient dose if a tendency for deviation from the physician's treatment prescription appears.

In addition to rigorous commissioning and regular quality assurance of treatment planning and delivery systems, individual patient treatments are typically verified via a dose measurement in a phantom \cite{lomax2004treatment,zhu2011patient,furukawa2013patient,mackin2014spot}. This process of measurement-based patient-specific quality assurance (PSQA\textsubscript{meas}) provides an end-to-end test of the entire system (for some dose locations) and gives an additional level of verification that the treatment plan can be physically delivered at those locations. However, PSQA\textsubscript{meas} can be resource-intensive, impacting the cost of treatment, the number of patients who can be treated, and the speed at which new or modified treatment plans can progress from planning to treatment delivery. Furthermore, some have questioned the value of PSQA using phantom measurements, citing a low ability of measurement-based PSQA to detect reported incidents \cite{ford2012quality} and the insensitivity of the popular Gamma test to relevant dosimetric errors \cite{nelms2011per,carrasco2012dvh}.

In recent years several groups have investigated the use of independent dose calculation software \cite{li2013use,mackin2013improving,meier2015independent,hernandez2019automation,deng2020technical,meijers2019log,yamada2021validation} and treatment log file analysis \cite{matter2018alternatives,winterhalter2019log,johnson2019highly,ates2023development} to decrease the required quantity of PSQA measurements and increase the ability to detect clinically relevant treatment delivery errors. The aims of the PTCOG Treatment Efficiency Subcommittee working group on log file-based PSQA are:

\begin{itemize}
  \item to objectively evaluate the viability, benefits, and challenges of using treatment logs for PSQA
  \item to compare the relative efficacy of measurement-based PSQA, log file-based PSQA, and independent secondary dose calculation\footnote{Note that at this point the term \citequote{'independent dose calculation'} is extended to \citequote{'independent secondary dose calculation'}. The reason for this further essential differentiation compared to the 'usual' (short) terminology used in the references is context-specific for this report and will be introduced and explained in detail in the next sections.}, or combinations thereof.
  \item to investigate the requirements for safe and effective implementation of log file-based PSQA
  \item to identify technical requirements for machine logs to facilitate their use in PSQA
\end{itemize}

In this chapter, we will review the purposes for PSQA and evaluate the benefit provided by measurement-based and log file-based approaches as well as application of an independent secondary dose calculation. In the remainder of the report, we evaluate the impact of these risk mitigation approaches in the context of a risk assessment for scanning particle beam treatment delivery. We then give a summary of recommendations for implementation of log file-based PSQA, and we conclude with a few remarks about these topics as they relate to adaptive radiotherapy. 
  \section{Historical Perspective of PSQA}
\label{sec:intro_psqahistory}

Before intensity-modulated radiation therapy (IMRT) was widespread, AAPM TG-40, \textit{Comprehensive QA for Radiation Oncology} \cite{kutcher1994comprehensive}, recommended that 
\begin{quotation_it}
an independent calculation of the dose at one point in the plan, preferably at the isocenter or at a point near the center of the tumor
\end{quotation_it} 
should be performed for every patient plan prior to treatment \cite{kutcher1994comprehensive}.

This \textit{hand calculation} could be performed with little difficulty using standard techniques and lookup tables. It provided quality assurance of the individual patient treatment plan, checking that the delivered dose (at one or several locations) from the plan matched the physician’s treatment intent. When in vivo dosimetry (e.g., with diodes) was introduced, it was recommended for patient QA by national authorities like those in Sweden and Norway and by ESTRO \cite{huyskens2001practical}. However, it is now mostly replaced by independent dose calculations and pre-treatment PSQA for IMRT and VMAT.

The advent of IMRT made existing hand calculation methods impractical or impossible \cite{alber2008guidelines}\cite{low2011dosimetry}. Additionally, the increased complexity of treatment planning, data transfer, and treatment delivery made it difficult to verify treatment accuracy through a manual inspection of the machine delivery parameters. Consequently, pre-treatment patient-specific dosimetric measurements became the standard for IMRT treatment verification. Despite the universal adoption of phantom measurement-based IMRT QA, it has been shown that this method is not especially good at identifying relevant errors \cite{kry2014institutional,jin2015correlation,heilemann2013sensitivity,crowe2016technical,nelms2013evaluating,kim2014sensitivity,stasi2012pretreatment,allred2021method}. The AAPM TG-218 report gives guidance on methods and tolerances for measurement-based PSQA for IMRT in an effort to improve its utility and consistency \cite{miften2018tolerance}.

More recently, commercial software for independent IMRT dose calculations has become widespread, including the use of machine log files to verify treatment delivery or recalculate delivered dose. Similar tools are beginning to be available for intensity modulated proton therapy (IMPT). However, despite the availability of these tools, there has been a general hesitance to give up patient-specific measurements. There may be financial, cultural, administrative, and legal reasons for the persistence of measurement-based IMRT QA, but technical reasons exist as well.

A recent photon-focused review by Decabooter et al. provides an instructive historical complement to this discussion by tracing how PSQA evolved in parallel with the increasing technical complexity of external-beam MV photon radiotherapy \cite{decabooter2026psqa}. In particular, Figure~1 of that review is informative because it condenses, in a single technique--technology--PSQA timeline, the progression from 3D conformal radiotherapy through IMRT, VMAT, modern SRS/SBRT, and online adaptive radiotherapy toward future risk-based, automated, and clinically interpretable PSQA.

Although this representation is intentionally photon-centered and therefore cannot be transferred directly to PBS proton therapy, its organizing logic is directly relevant to the present report. It illustrates that PSQA measures do not evolve independently of treatment technology; rather, they change as treatment concepts, dose-calculation methods, delivery control, imaging, data pathways, and workflow complexity change. Table~\ref{tab:proton_psqa_timeline} adopts this organizing principle for proton therapy and reformulates it around proton-specific delivery concepts, range-sensitive technology, imaging, data-transfer and dose-calculation requirements, and risk-control measures.

Accordingly, the historical evolution of PSQA in proton therapy can be summarized as a transition from hardware- and phantom-centered verification in broad-beam and early scanning delivery, through measur\-ement-, calculation-, and data-transfer-centered PSQA in PBS/IMPT, toward risk-informed, log file-supported, delivered-dose-oriented, and adaptive PSQA. The past and present stages summarized in Table~\ref{tab:proton_psqa_timeline} provide the technical basis for the comparative risk assessment developed in this report, whereas the future-oriented entries anticipate implementation topics revisited later in Sections~\ref{sec:discussion_relation_to_rm_psqa_context}, \ref{sec:log_adoption_recommend} and~\ref{sec:outlook_apt_complex}.

\begin{table}[h!]
\begingroup

\centering
\footnotesize
\renewcommand{\arraystretch}{1.15}
\setlength{\tabcolsep}{4.5pt}

\captionsetup{
  font=small
}

\caption{Proton-adapted overview of treatment concepts, treatment technology, and (PS)QA/risk-control measures across major stages of proton therapy evolution. The table is intended as a compact text-equivalent timeline to complete the historical PSQA perspective presented above.}
\label{tab:proton_psqa_timeline}

\begin{tabularx}{\textwidth}{>{\bfseries\RaggedRight\arraybackslash}p{0.19\textwidth}YYYY}
\toprule
\textbf{Time / Characteristic} & \textbf{Past: broad-beam and early scanning} & \textbf{Present: PBS/IMPT and robust planning} & \textbf{Future: data-driven and adaptive PT} \\
\midrule

\textbf{Treatment concept}
&
Passive scattering; double scattering; uniform scanning; early conformal proton therapy.
&
PBS / spot scanning; SFUD / MFO; IMPT; robustly optimized proton therapy; motion-aware planning.
&
Log file-supported IMPT; hypofractionated and complex proton therapy; adaptive proton therapy; online or near-real-time workflows. \\

\addlinespace[0.3em]

\textbf{Treatment technology}
&
SOBP and range modulation; apertures and compensators; range shifters; beam-specific hardware; delivery-mode-specific machine QA.
&
Energy-layer sequencing; spot position, spot size, and MU control; scanning dynamics; gantry, couch, and nozzle states; robust optimization; 4D evaluation; analytical-to-MC dose calculation; TPS--OIS--TCS transfer.
&
Structured treatment records; DICOM RT Ion Plan and treatment-record objects; automated spot-level checks; GPU / fast MC; daily imaging; log file dose reconstruction; in-vivo range or dose signals; AI-assisted anomaly detection. \\

\addlinespace[0.3em]

\textbf{(PS)QA and risk control}
&
Patient-specific hardware verification; range and output checks; phantom measurements; point and planar dose checks; end-to-end tests; independent MU or dose sanity checks.
&
Measurement-based PSQA; 2D / 3D detector arrays; film and scintillator systems; $\gamma$-index and dose comparison; independent dose calculation; MC recalculation; data-transfer verification; commissioning and class-solution validation.
&
Risk-informed hybrid PSQA; log file-based PSQA and log file QA; MC-based independent secondary dose calculation; selective measurement; delivered-dose reconstruction; daily or fraction-wise evaluation; adaptive workflow QA; DVH- and structure-based criteria; living pFMEA feedback. \\

\bottomrule
\end{tabularx}

\vspace{0.35em}
\begin{minipage}{0.98\textwidth}
\footnotesize
\emph{Abbreviations:} PBS = pencil beam scanning; IMPT = intensity-modulated proton therapy; PT = proton therapy; SFUD = single-field uniform dose; MFO = multi-field optimization; SOBP = spread-out Bragg peak; MU = monitor units; MC = Monte Carlo; TPS = treatment planning system; OIS = oncology information system; TCS = treatment control system; QA = quality assurance; PSQA = patient-specific quality assurance; DVH = dose-volume histogram; pFMEA = process-driven Failure Mode and Effects Analysis.
\end{minipage}

\arrayrulecolor{black}
\endgroup
\end{table} 
  \section{Measurement-Based PSQA and the Value Provided}
\label{sec:intro_psqameas}

Measurement-based PSQA – whether for photon or proton therapy – does a number of things including (i) providing an independent check of the dose calculation from the treatment planning system, (ii) verification of the accuracy of data transfer from the treatment planning system to the treatment delivery system and deliverability on the treatment machine, and (iii) aiding in the continuous quality assurance of the treatment planning system and the treatment delivery system. In summary, measurement-based PSQA provides a patient-specific end-to-end test of the entire planning and delivery process.

\subsection{Dose Calculation Verification}
\label{subsec:dose_calc_verification}
Some older reports and guidelines from AAPM \cite{low2011dosimetry,ezzell2003guidance,palta2003intensity,gibbons2014monitor} and ESTRO \cite{alber2008guidelines} indicate that patient-specific meas\-ure\-ment-based IMRT QA can theoretically be replaced by the combination of an independent dose calculation, verification of correct data transfer, and enhanced machine QA. These sources also agree that extensive measurement-based validation of the system (including commissioning) is essential before stepping away from performing measurements on every patient. Some weaknesses of measurement-based IMRT QA are also given, including cases where an independent dose calculation may catch errors that a measurement would miss.

Recently, AAPM Task Group 219 \cite{zhu2021report} provided updated guidance about independent calculation-based dose verification for IMRT. Their report emphasized the value of both measurement-based PSQA as well as independent dose calculation. They noted that
\begin{quotation_it}
secondary calculations cannot catch hardware delivery errors, such as MLC leaf motor slippage, etc. For this reason, independent MU calculation for IMRT is unlikely to completely replace measurement-based methods for patient-specific QA.
\end{quotation_it}

TG-219 recommended that 
\begin{quotation_it}
secondary dose/MU calculation should be performed for every IMRT/VMAT plan, at least in 1D but preferably in 2D/3D, regardless of the method of measurement-based verification utilized.
\end{quotation_it}
While TG-219 steps back from the suggestion of earlier reports that PSQA measurements could be replaced by a combination of \citequote{independent/secondary}\footnote{It should be explicitly noted at this point that TG-219 uses the attributes 'independent' and 'secondary' as synonyms in the sense of an uninfluenced alternative. As indicated before, the importance of this terminological differentiation will become clear soon.} dose calculations and enhanced machine QA, it does suggest that \citequote{independent/secondary} dose calculations may be used to decrease the frequency of patient-specific dose measurements. However, it warns that independent/secondary dose calculations are insufficient to replace measurement-based PSQA due to their inability to detect treatment delivery errors caused by dose miscalibration and hardware faults. 

\subsection{Data Transfer Accuracy}

Several sources describe the importance of verifying data transfer accuracy from the treatment planning system all the way to the treatment control system \cite{icru2007report,alber2008guidelines,ezzell2003guidance,palta2003intensity,hartford2009american}. The AAPM guidance document on IMRT makes the interesting point that 
\begin{quotation_it}
the logic of [a] ‘phantom plan’ methodology is that it verifies the correct transcription of IMRT delivery parameters, leaf sequence, and MU calculation. \cite{ezzell2003guidance}
\end{quotation_it}
In the AAPM Task group 201 report on quality management of data transfer for EBRT, the authors state that with the evolving treatment complexity
\begin{quotation_it}
we (mainly the medical physicists) will need to do and know more about ensuring the quality of data transfers.\cite{siochi2021report}
\end{quotation_it}

There are many methods of verifying accurate data transfer. While (most) medical physicists are not information technology professionals, they can confirm the presence of data transfer validation and advocate as needed for provisioning or improvement of these systems.

\subsection{Plan Deliverability}

Verifying deliverability is important for safety as well as logistical reasons. AAPM Task Group 120 notes that a limitation of independent dose calculation methods (without PSQA measurements) is that
\begin{quotation_it}
the deliverability of the IMRT plan is not validated on the actual treatment device. For this reason, patient‐specific validation is often conducted using direct dose distribution measurements in homogeneous solid media. \cite{low2011dosimetry}
\end{quotation_it}
That being said, standard PSQA does not always check at the prescribed treatment angle and couch position, or all the beams. In these cases, the verification of plan deliverability is arguably incomplete.

\subsection{Treatment Planning System QA}

The ESTRO IMRT guidebook states: 
\begin{quotation_it}
Even if later on independent dose calculations are chosen as the preferred method of IMRT verification, at the start of the clinical implementation of a new IMRT technique it is strongly recommended to perform measurements of the 3D dose distribution delivered to a phantom and to compare these with planned dose distributions. \cite{alber2008guidelines} 
\end{quotation_it}

This process serves as continued commissioning of the treatment planning system by evaluating multiple real patient dose calculations. Additionally, the ESTRO booklet states that early in the process, a large number of individual patient fields should be measured in order to identify systematic errors that might be missed by composite dose measurements.

The report of the 2003 AAPM Summer School on IMRT includes an account of the IMRT QA approach of a major academic medical center, which – after thorough initial validation and an extensive period of patient-specific measurements – performs patient-specific measurements on a very limited basis. However, patient-specific measurements are performed at this center for
\begin{quotation_it}
new treatment sites, new MLC, or new software\cite{palta2003intensity}.
\end{quotation_it}

At least one proton treatment center represented in the working group has reported that deviations in patient-specific measurements have identified limitations in the treatment planning system commissioning, which led to revisions in the beam model. This may occur long after initial commissioning of the system as practice patterns, treatment sites, and planning approaches evolve. As discussed in the ESTRO IMRT booklet, thorough measurements of dose distributions are a valuable tool when implementing a new \textit{class solution}, treatment site, or planning approach.

\subsection{Treatment Delivery System QA}
\label{subsec:TDS_QA}

Most reports agree that comprehensive QA of the treatment delivery system is an essential component of overall IMRT QA, whether patient dose verifications are measurement-based or calculation-based. In most cases, validation of machine performance is discussed separately from patient-specific plan verification. However, when moving from measurement-based PSQA to calculation or log file-based PSQA, it is important to reconsider the frequency and extent of delivery system QA.

Centers may rely on patient-specific QA measurements as a supplementary quality assurance test on the treatment delivery system. But even these measurements are more of a ‘spot-check’ at some subset of the treatment parameters. Machine QA may be used to enhance the parameter phase space. If these measurements are removed or decreased through a transition to calculation or log file-based treatment verification, the frequency of some machine QA tests may need to be increased, and a more comprehensive set of tests with the appropriate tools may be needed for all components of the treatment delivery system.

\subsection{End-to-end Testing}
\label{subsec:end_to_end_testing}
The value of \textit{patient-specific end-to-end testing} is highlighted in the ACR practice parameter \cite{hartford2009american} and discussed in most other relevant reports. A common theme is that all failure modes are not known for new and complex systems. While the ACR document is firm in stating the necessity of this end-to-end testing via dosimetric measurement in a phantom for every patient, \cite{hartford2009american} other sources suggest that it may be acceptable to relax this requirement after extensive systemic validation and with appropriate QA of the individual aspects of the treatment process \cite{alber2008guidelines,ezzell2003guidance,palta2003intensity}.

They also note that patient-specific measurements fall short in some ways from being a complete end-to-end test, such as in the case where fields are all delivered at a single gantry angle, instead of the planned treatment angle. Some sources suggest possible workflows that exclude patient-specific measurements entirely \cite{alber2008guidelines,palta2003intensity}, while some examples are given of centers that perform patient-specific measurements on a subset of patients, while verifying most patient plans through a combination of independent dose or MU calculations, testing of data transfer accuracy, and comprehensive machine QA \cite{icru2007report,palta2003intensity}.

An important point of consideration for scanning beam proton therapy is the differences in the equipment, information technology, and software among different proton centers. There are several major manufacturers of proton therapy equipment, and there is considerable variability even between centers built by the same vendor. To some extent, nearly every proton therapy system is a custom solution.

Proton therapy delivery systems have not reached the relatively steady state of technology that is currently found in the photon IMRT world. This state of affairs suggests that the field as a whole has not left the \textit{early adoption} phase where additional caution is advisable. That being said, individual centers may reach a point, through extensive validation, that they no longer consider individual measurements to be necessary for every patient, as in the case of the Paul Scherrer Institute \cite{trnkova2016factors}.

  \section{Log File-Based PSQA}
\label{sec:intro_psqalog}

\subsection{Photon Therapy}
Several institutions have developed treatment log analysis systems for photon IMRT. Some of these efforts are directed towards IMRT QA \cite{litzenberg2002verification,agnew2012implementation,rangaraj2013catching,sun2013initial,tyagi2012real}, while others prioritize machine QA\cite{stell2004extensive,agnew2014monitoring} and verifying delivery accuracy \cite{li2003validation}. Commercial log file analysis solutions have also been developed and integrated into clinical practice \cite{mcdonald2017validation}. Some of these systems verify delivery parameters but do not actually recalculate dose \cite{agnew2012implementation}, while others use analytical or Monte Carlo methods as a dose calculation second check \cite{mcdonald2017validation,sun2012evaluation}.

Log file-based IMRT QA can identify machine setting and feedback problems that would not be identified by measurement-based IMRT QA \cite{agnew2012implementation,rangaraj2013catching,li2003validation,sun2012evaluation}. If it replaces measurement-based IMRT QA, log file-based PSQA can improve operational efficiency and save time and money \cite{litzenberg2002verification,agnew2012implementation,rangaraj2013catching}. However, others have found that a move from measurement-based to log file-based IMRT QA will require significant additional machine QA in order to maintain treatment safety and quality \cite{agnew2014monitoring}.

Debate continues about whether measurement-based IMRT QA can be replaced with log file and calculation-based approaches \cite{smith2011necessary,childress2015parallel,siochi2013patient,pawlicki2008moving,kruse2013comment}. Some groups have developed log-based IMRT QA but did not abandon phantom measurements \cite{sun2013initial,tyagi2012real}, while others have exhibited a willingness to abandon phantom-based measurements in favor of log file-based IMRT QA \cite{rangaraj2013catching,mcdonald2017validation}.

In general, the photon PSQA literature provides an important historical and conceptual basis for the present analysis, particularly regarding the limitations of routine pretreatment phantom measurement, the need for risk-based PSQA selection, and the transition toward log-file, calculation-based, image-based, and automated verification. However, PBS proton therapy introduces additional modality-specific risks related to range, stopping-power conversion, spot delivery, energy-layer sequencing, beam modifiers, motion/interplay, and vendor-specific data pathways. Photon PSQA concepts must therefore be translated at the level of risk-control principles rather than copied at the level of devices or action levels.

\subsection{Particle Therapy}
Particle Therapy systems operate on a system of detailed machine settings and feedback, which facilitates production of delivery log files. For years this ability remained unused and not applied to QA. However, in recent years, log file analysis for patient and machine QA has begun to be explored in proton centers.

Some approaches rely on analysis of treatment delivery parameters without a recalculation of the dose, \cite{li2013use,ates2023development,zhu2015towards} or on recalculation of the beam fluence. Log files have also been used to recalculate dose using analytical methods in water \cite{mackin2014spot} and in patient CT scans \cite{matter2018alternatives,cohilis2022development}, and also with Monte Carlo methods \cite{winterhalter2019log,johnson2019highly,jeon2023monte,marmitt2020platform,kry2019independent}. Comparative studies have found log file-based PSQA to be more efficient and more effective at catching relevant errors than measurement-based approaches \cite{matter2018alternatives,meijers2020feasibility,Wolter2026SensitivityEfficiency}.

Much work has also gone into the development and use of dose recalculation systems for treatment verification based on treatment log files \cite{toscano2019impact}. These systems have seen considerable use in some clinics, providing information on treatment delivery accuracy \cite{ates2023development,scandurra2016assessing}, including the accuracy of dose delivery in the context of motion mitigation technologies \cite{yamada2021validation}.

  \section{Independent Dose Calculation (IDC)}
\label{sec:intro_IDC}

\subsection{Terminology}
\label{subsec:intro_terminology}
As indicated previously, in the cited and analyzed references of this PSQA report, the term \citequote{Independent Dose Calculation} (IDC) uses the attribute \citequote{'independent'} as a synonym for \citequote{'secondary'}, which is particularly evident in TG-291 through the combined notation \citequote{'independent/secondary'}.
 
In the context of IDCs \citequote{'independent/secondary'} as such primarily refers to a mathematically different algorithm for dose calculation compared to its corresponding TPS implementation. Thereby, the risk-reducing potential of the alternative algorithm is all the more pronounced the lower its (mathematical) \citequote{similarity} to the TPS calculation algorithm appears. Additionally, it often operates at a different level of dose calculation accuracy. A representative example is ray tracing\footnote{Ray tracing in its simplest form serves as an initial (adaptable) model calculating the linear proton propagation in (homogeneous) media under energy deposition.}, which is supplemented by more precise simulation methods on a Monte Carlo basis \cite{aitkenhead2020automated,botnariuc2024evaluation,chang2020standardized,holmes2024fast}.

\subsection{From IDCs to Independent Secondary Dose Calculations (ISDCs)}
In principle, IDC usage increases the overall quality level of treatment planning and verification in terms of reliability, but it may or may not be interpreted as a truly independent part of a dedicated PSQA process in a treatment workflow. For example, a recalculation with a different algorithm within the primary TPS may serve as a fast, easy-to-use safety check, but its workflow is too similar to the original TPS dose calculation to serve as an independent, reliable, and fully-accepted PSQA substitute. A similar example is an additionally available TPS dose calculation that uses the same calculation algorithm but is implemented in a different programming language, representing a low level of calculation independence \cite{meier2015independent}. Such an IDC, while not sufficiently independent to replace PSQA, may provide value as a complementary safety check, as it may be sensitive to failure modes that are more difficult to detect via measurement- or log file-based PSQA.

If the desire is to partially or completely replace other PSQA methods, a more sophisticated form of IDC is needed. Ideally, this would be a separate, stand-alone system, including the elements of inter-system data transfer. Therefore, in a process-oriented context, such a 'fully-developed' IDC has to be considered as an integral part of the patient workflow and serves as a second ('secondary') implementation of an independent dose calculation (ISDC)\footnote{The authors are aware that the chosen paired terminology 'independent, secondary' DC or 'secondary, independent' DC (ISDC) may sound somewhat confusing. Nevertheless, this compact version of a term was preferred over a longer, possibly clearer, but cumbersome variant due to its already existing use (see \ref{subsec:intro_terminology} \nameref{subsec:intro_terminology}). In each case, the meaning is always an 
\begin{quotation_it}
  \begin{enumerate}
    \item independent implementation of a second dose calculation, which is 
    \item based on a calculation algorithm that is independent of the TPS, whereby 
    \item the corresponding software component is (ideally) represented by a separate PSQA process step or a complete PSQA application and 
    \item can be clearly identified as such.
  \end{enumerate}
\end{quotation_it}
In the course of this report, their application is usually clearly outlined and distinguished in the given context.}
As such, an ISDC serves as an individual, inter-connected system component such as the TPS, TDS, OIS, etc., which is additionally included into the workflow-related data transfer.

Historically, ISDCs started clinically in conventional EBRT modalities and in the meantime became common for photon-based IMRT QA \cite{mcdonald2017validation,sjoelin2018clinical,clemente2015dosimetric}, where they have demonstrated improved results in detecting poor treatment plans compared to traditional measurement-based IMRT-QA \cite{kry2019independent}. Roughly at the same time, but with significantly less spread and different, site-dependent deployment concepts, the first IDC prototype applications for independent dose calculations in proton therapy were also developed. These applications were essentially in-house developments that used both independent analytical calculation algorithms and Monte Carlo methods \cite{mackin2013improving,hernandez2019automation,magro2022dosimetric} and discussed the first workflow integration of an IDC \cite{meier2015independent}.  Through the targeted clinical use and the corresponding integration into the clinical workflow, a quasi-seamless transition from IDC to ISDC arose – depending on the institution – which, with the availability of commercial systems also in proton therapy, could be considered complete.

Monte Carlo-based independent dose calculations (MC-IDCs) have demonstrated particular promise for detecting dose differences caused by shortcomings in analytical dose calculation algorithms, whereby some systems have been directly implemented clinically as early representatives of MC-ISDCs \cite{deng2020technical,marmitt2020platform,aitkenhead2020automated}. Modern computing techniques like GPUs have been employed to speed up almost every implementation of Monte Carlo dose calculation applications \cite{beltran2016clinical,gajewski2021commissioning,shan2022virtual} and some institutions have found that ISDCs can decrease or even eliminate the need for the treatment planning aspects of patient-specific measurements \cite{meijers2020feasibility,winterhalter2018validating,grevillot2021gate}.

  \section{Establishing a General Definition of Patient-Specific QA}
\label{sec:general_psqa_definition}

\subsection{Motivation and Objectives}
As outlined in the prologue, modern radiation therapy – in particular with scanned particle beams – should be understood as a complex socio-technical system whose safety and quality cannot be ensured solely by directing, monitoring, and verifying technical parameters by means of quality control (QC). In fact, a holistic approach is needed that also integrates the principles of quality management (QM) and risk management (RM) using dynamic, process-based system views, so that technical quality controlling develops into overall quality assurance (cf.~Subsec.~\ref{subsec:workflow_granularity}).

In the previous sections of the introduction, the historical development of patient-specific quality assurance was outlined, whereby the initial focus was on measurement-based verification methods (PSQA\textsubscript{meas}) for intensity-modulated photon (IMRT) and proton (IMPT) treatments. Over time, alternative or supplementary PSQA concepts emerged, such as using log files to verify treatment planning and independent secondary dose calculations, which were designed to potentially reduce effort, increase efficiency and improve the detection of clinically relevant errors.

The features of measurement-based PSQA were discussed on the basis of the various objectives that any PSQA must cover (cf.~Subsec.~\ref{subsec:dose_calc_verification}--\ref{subsec:end_to_end_testing}). Historically, a primarily quality-control oriented perspective was chosen, focusing on monitoring and verification of machine and calculation parameters in the context of a single patient treatment, because the results should (ideally) be generally valid, i.e. independent of the individual workflow of a particle center.

With such a perspective, which seems to be limited to pure quality control, PSQA tends to overlook the broader system-level factors, such as workflow management and human-machine interactions, with software integrity and data transfer, as well as the increasing complexity of treatment delivery technology. As already mentioned, PSQA must be embedded in the larger framework of total quality assurance as such to truly achieve patient safety and treatment quality, whereby RM principles and prospective analysis such as failure modes and effects analysis (FMEA) must be incorporated.

Therefore, the motivation for overcoming this one-sided perspective on QC is twofold:
\begin{enumerate}[noitemsep]
\item The traditionally QC-based conceptualization of PSQA methods (measurement-based, log-file-based, independent dose calculations) is to be aligned with the comprehensive QM/RM approach defined in the prologue.
\item It must be made clear that the choice of a PSQA strategy or a combination thereof should be primarily specific to the risk-reduction potential in order to maximize patient safety at the highest level of complexity and uncertainty. While increases in efficiency and effectiveness are important, they are subordinate to safety.
\end{enumerate}

\subsection{Proposed Definition}
\label{subsec:psqa_definition}
\begin{quotation_it}
\textbf{Patient-Specific Quality Assurance (PSQA)} in particle therapy refers to a systematic, workflow-oriented and risk-based process designed to ensure that the planned radiation dose is delivered to the patient as envisaged and within clinically acceptable tolerances. It involves both controlling technical machine and physical (dose) calculation parameters (QC) and is an integral part of an overarching quality management system. PSQA encompasses the evaluation and control of all relevant socio-technical interactions – machine function, software behavior, data integrity, human procedures and communication processes – within a defined treatment process. By embedding prospective risk assessments (e.g. FMEA) and continuous improvement mechanisms, PSQA achieves its fundamental objective: maintaining or improving patient safety and treatment efficiency through the dynamic interaction of QC, QM and RM measures.
\end{quotation_it}

\subsection{Further Relevance}

This comprehensive definition of PSQA reflects the framework for the application of the workflow-based FMEA approach that was introduced in theory in the prologue. As such, it is used to evaluate and apply the various PSQA methods described in the introduction with regard to their common objectives of safety and quality. By not considering PSQA as just machine- or calculation-oriented testing, the subsequent FMEA approach evaluates the potential of each PSQA strategy to identify, prioritize, and mitigate risks throughout the site-specific treatment workflow of each particle center.

This holistic evaluation ensures that decisions to implement, modify or replace PSQA methods (e.g., switching from measurement-based PSQA to log file analysis or integrating independent dose calculations) are guided by their ability to reduce risk and improve overall safety. Thus, the above definition provides the conceptual and methodical basis for the next steps in this work, where FMEA-based investigations will influence the selection and refinement of PSQA practices.

}

\chapter*{2 PSQA RISK ASSESSMENT}
\setcounter{chapter}{2}
\addcontentsline{toc}{chapter}{2 PSQA RISK ASSESSMENT}
\markboth{2 PSQA RISK ASSESSMENT}{}
\setcounter{section}{0}
\label{ch:psqa_risk_assessment}
{%
  \renewcommand{\sectionmark}[1]{\markright{\thesection.\ #1}}%
  \section{Development of the Present PSQA Report}
\label{sec:development_of_psqa_report}

\subsection{Motivation}
The PTCOG Treatment Efficiency Subcommittee (TxEff SC) has a single mission: "Fighting cancer effectively and efficiently." Thereby, the effectiveness, efficiency and safety of particle therapy workflows is improved by the use of innovative approaches to quality assurance and risk management. At the center of this mission is the systematic evaluation of patient-specific quality assurance methods. In this context, the focus is on their ability to optimize resource utilization and at the same time ensure the highest standards of patient safety. By adopting a comprehensive, workflow-oriented perspective, the TxEff SC promotes the integration of advanced PSQA methods into clinical practice. Ultimately, this should improve both process safety, treatment efficiency, and patient outcomes.

\subsection{Historical Background}
After the PTCOG TxEff SC had held several meetings on QA-related topics in the mid-2010s, medical physicists representing proton therapy centers from around the world met again in Knoxville, TN, in December 2018 for another workshop on log file-based patient-specific quality assurance (PSQA\textsubscript{log}). As such, the workshop participants formed three groups to investigate the risks and impacts of using PSQA\textsubscript{log} in treatment planning, data transfer between systems, and treatment delivery separately. In each of these areas, risks were identified based on participants' clinical experience and technical expertise, and the extent of mitigation by PSQA\textsubscript{meas} and PSQA\textsubscript{log} was evaluated for each of these risks.

Following this workshop, a smaller group of TxEff SC members organized a dedicated working group. Since 2020, this group has expanded on the original target corridor and supplemented the initial results of the workshop(s) by 
\begin{enumerate}
    \item analyzing and comparing the results with their own site-specific work, experiences and environments related to log file-based PSQA, 
    \item investigating the analysis and comparison of log protocols with other established PSQA concepts and practices, and 
    \item utilizing formalized methods of modern risk assessment and analysis as the basis for their work, as such
    \item designing a generic proton therapy workflow as the collective core of their work processes and implement its PSQA components as standardized processes using an appropriate FMEA (Failure Modes and Effects Analysis) approach.
\end{enumerate}

\subsection{Objective}
The main objective of any PSQA is to ensure that the planned conditions are met at each stage of treatment and, in particular, to verify that these can be ensured before the patient is irradiated. The aim is to adapt the available technology to the state of development at present and use it to the best possible effect. In this sense, each individual PSQA application strives to fulfil the mission of the PTCOG TxEff SC, namely to aim for the most effective, i.e. the safest, most efficient and most innovative treatment and to use the radiation technology that proves to be optimal, taking into account resource optimization aspects.

  \section{PSQA-related Risks in a Proton Therapy Workflow}

\subsection{Workflow-related Considerations and Constraints}
The implementation of patient-specific quality assurance in a proton therapy workflow requires a comprehensive understanding of the intricate interdependencies between process steps, technological components, and human factors. In particular, pencil beam scanning proton therapy represents a highly dynamic socio-technical system, where the successful delivery of safe and effective treatment depends on the seamless integration of treatment planning, data transfer, and treatment delivery processes. 

The PTCOG TxEff SC working group recognized two critical considerations and constraints in de\-ve\-lo\-ping a structured approach to PSQA risk assessment within the context of PBS-PT workflows:
\begin{enumerate}
    \item \textbf{Technological heterogeneity and institutional variability:} 
    Technological advancements have recently widened the range of QA tools available for diverse EBRT modalities, including IMPT. Nevertheless, QA requirements depend on the specific treatment system, the imaging and planning software, and the available QA equipment. One core purpose of a risk assessment approach is to offer structured methods for evaluating transitions from one clinical modality to another. Conducting a thorough, site-specific risk assessment of QA in the PBS-PT workflow enables a detailed comparison of alternative PSQA methods. While a general framework can illustrate overarching principles, the greatest benefit will be achieved by tailoring the analysis to each individual clinic.

    \item \textbf{Workflow complexity and practical feasibility:} 
    The dynamic nature of PBS-PT workflows — characterized by iterative feedback loops (e.g., adaptive planning) and multi-step processes — calls for thorough risk assessment. However, conducting a comprehensive, process-driven Failure Modes and Effects Analysis (pFMEA) for every step of the entire proton therapy workflow (from the initial physician consultation to the end of treatment) was deemed impractical for this study. Given the primary goal of determining the relative effectiveness of measurement-based and log file-based PSQA for risk reduction, both individually and in combination with an independent secondary dose calculation (ISDC), the working group decided to bypass process steps unlikely to be significantly affected by a particular PSQA method. Instead, the focus was placed on those workflow stages most directly impacted by PSQA practices, starting after the virtual CT simulation phase and extending through treatment delivery. This targeted approach allows for sufficient analytical depth and clinical relevance without introducing unnecessary complexity.
\end{enumerate}

To systematically address these challenges, the working group emphasized the importance of workflow granularity, as defined in Chapter~\ref{subsec:workflow_granularity}, as a foundational element for FMEA application. Workflow granularity refers to the appropriate resolution of process modeling, ensuring that each process step is detailed enough to capture critical interactions without unnecessary complexity. This concept is particularly relevant in PBS-PT, where both high-level processes (e.g., treatment planning) and granular sub-processes (e.g., data transfer between planning and delivery systems) must be evaluated for potential failure modes.

By adopting a granularity-driven perspective, the working group developed workflow models tailored to the evaluation of PSQA-related risks. These models are designed to:
\begin{itemize}
    \samepage
    \item Identify critical process steps where specific PSQA methods (e.g., measurement-based, log file-based, ISDC) can provide meaningful risk mitigation.
    \item Facilitate the comparative analysis of PSQA methods across distinct stages of the workflow, thereby quantifying their relative efficacy.
    \item Support iterative feedback and adaptive planning mechanisms, ensuring that evolving clinical and technological requirements are incorporated into the QA process.
\end{itemize}

A representative PBS-PT workflow that encompasses all relevant process steps including all PSQA alternatives as different tracks of the associated process is described in Figure \ref{fig:PSQA_all} of Appendix A. This model serves as the structural basis for the FMEA analyses that are subsequently performed, in which the interactions of socio-technical factors, human-machine interactions and system dynamics are systematically evaluated. 

The various PBS-PT workflow models are described in Appendix A and explained there at the beginning in terms of their various functions. In a methodological sense, Figures \ref{fig:PSQA_all} and \ref{fig:PSQA_noQA} are of particular meaning, whereby Figures \ref{fig:PSQA_calc}, \ref{fig:PSQA_meas} and \ref{fig:PSQA_log} are also presented for the sake of completeness. 

Figure or Workflow 
\begin{itemize}
    \item[A.1] represents the generic core workflow, as it describes the virtually identical processes in all participating proton centers, and it juxtaposes all three PSQA methods as different tracks in parallel (\textbf{PSQA\textsubscript{all}} view). For each track a separate FMEA is performed.
    \item[A.2] yields the generic core workflow with an explicit listing of all process steps and the identification of associated, possible failure modes that could have a potentially risky influence on patient treatment in the absence of any PSQA (\textbf{PSQA\textsubscript{noQA}} view)\footnote{At this point (and for the following sections), it is particularly important to bear in mind that there is a one-to-many relationship between process steps and failure modes, i.e. for each process step, one or more failure modes may exist that are considered PSQA-relevant. Ultimately, all failure modes that are \textbf{not} PSQA-monitored or -protected, summarized at the process step level, provide the process step-dependent and failure mode-specific risks as reference or baseline values for each PSQA measure determined by the corresponding FMEA and all PSQA\textsubscript{noQA} risk values.}.
    \item[A.3] represents the generic core workflow with independent secondary dose calculation as a separate PSQA method (\textbf{PSQA\textsubscript{calc}} perspective).
    \item[A.4] shows the generic core workflow with measurement-based PSQA method (\textbf{PSQA\textsubscript{meas}} perspective).
    \item[A.5] displays the generic core workflow with log file-based PSQA method (\textbf{PSQA\textsubscript{log}} perspective).
\end{itemize}
\vspace{-0.1cm}
\subsection{Methodology for the Application of a Process-driven FMEA}
It is assumed that the reader is familiar with the general concept of an FMEA and, if necessary, is aware of various application techniques under variable conditions of use, which is in the present case a process-driven FMEA (pFMEA)\footnote{The pFMEA methodology is employed as a structured approach to identify potential failure modes within each process step, determine their effects, assess their causes, and prioritize corrective actions. This semi-quantitative method involves scoring each failure mode based on occurrence (O), severity (S), and detectability (D), with scores ranging from 1 to 10. The Risk Priority Number (RPN), calculated as the product of these scores (O*S*D), provides a quantitative measure of risk, ranging from 1 to 1000.}. At this point, reference is made again to the frequently cited and used AAPM TG-100 report, including supplementary documents, as an introductory source of information (see references in the prologue). In this context, three essential aspects have to be emphasized:
\begin{enumerate}
  \item It should be explicitly noted that the specific application of FMEA described by AAPM TG-100 (in parti\-cular its use in conjunction with a Fault Tree Analysis (FTA)) cannot be considered as a universally valid, representative and ideally suited approach to risk minimization in every EBRT context.
  \item The FMEA method described and applied in this PSQA report differs significantly from the one described by AAPM TG-100, with the motivation for the present application being related to both the parti\-cular PSQA conditions in PBS-PT and the TxEff SC working group's views on fundamental methodological issues related to how best to implement a risk assessment.
  \item This has three implications:
  \begin{enumerate}
    \item Due to resource constraints and the complexity of a PBS-PT workflow (which is much more extensive than in the IMRT-FMEA use case of TG-100), that includes various hardware, software and PSQA techniques, no complementary FTA application was carried out. In practice, such an approach would not have been feasible because an FTA requires (relatively) accurate data and information about a given system under (relatively) constant operating conditions to be carried out, which, given the limitations discussed in the previous section, is not the case here.
    \item The AAPM TG-100 report describes a process-driven FMEA approach, which is essentially a single-step pFMEA. This means that the workflow is only run through once and a single risk assessment is performed. The TG-100 procedure does not include iterative assessments or baseline assessments as reference conditions\footnote{TG-100 does recommend a post-implementation ``testing of effectiveness'' (i.e., an ``evaluation run'') of the QM interventions derived from the initial pFMEA/FTA---typically via monitoring and review of incident-reporting data---and that the QM program be periodically updated.\cite{huq2016taskgroup100} However, this step is framed as program validation and continuous improvement and is not formulated as a repeated, baseline-referenced pFMEA re-scoring of the workflow; within the terminological framework of the present work, TG-100 is therefore treated as a single-run pFMEA rather than an iterative pFMEA procedure.}.
    \item The process-driven FMEA (pFMEA) used here iteratively evaluates the effects of PSQA measures on risk mitigation, whereby a baseline assessment is used as a reference that assumes that no PSQA measures are present or active. This approach allows for multiple iterations of the workflow, whereby each phase introduces specific PSQA measures and their risk assessment in stages, enabling a more comprehensive, dynamic and realistic risk assessment.
  \end{enumerate} 
\end{enumerate}
\vspace{-0.2cm}
\subsection{PSQA-specific Assumptions and Risk Identification}
As stated earlier, the effort for risk assessments had to be kept at a reasonable level. To make the differences in the various PSQA-FMEA assessments clearer, only those failure modes in their respective process steps were assessed that were expected to contribute to a risk change and to show different risk reduction by various PSQA methods (for details see Figure \ref{fig:PSQA_noQA}).

As such, some assumptions had to be made in order to establish a meaningful and well-defined comparison between the clinical impact of the various PSQA methods. The prerequisites for measurement-based PSQA, log file-based PSQA and independent secondary dose calculation are defined for the purposes of this study in the following subsections.

In each subsection, the substantially critical elements required for a risk assessment of the corresponding PSQA method are identified. Finally, the prerequisites with which the individual PSQA procedures contribute to the resulting risk assessments are presented in tabular form and compared using relevant influencing factors.

\begin{table}[h!]
\centering
\begin{tabularx}{\textwidth}{>{\raggedright\arraybackslash}m{3.5cm}|Y|Y|Y}
\hline
& \textbf{Measurement-based PSQA} & \textbf{Log File-based PSQA} & \textbf{Independent Secondary Dose or MU Calculation} \\
\hline
\textbf{Additional hardware} & 2D ionization array (Film, Gel) & None & None \\
\hline
\textbf{Additional software} & Dose comparison, Redo plan for QA & Dose calculation and comparison & Dose calculation \\
\hline
\textbf{Room occupancy} & Inflexible & Flexible & None \\
\hline
\textbf{Treatment parameters} & Amended & Amended or identical & Amended or identical \\
\hline
\textbf{Beam modifier} & Measured & Ignored or simulated & Simulated \\
\hline
\textbf{Phantom/CT} & Homogeneous water/phantom only & Homogeneous digital phantom or patient CT scan & Homogeneous digital phantom or patient CT scan \\
\hline
\textbf{Point dose} & Yes & Yes & Yes (mostly used) \\
\hline
\textbf{2D Dose} & Yes & Yes & Yes \\
\hline
\textbf{3D Dose} & Rarely, and with very low resolution & Yes & Yes \\
\hline
\textbf{Pre-treatment QA} & Yes & Yes & Yes \\
\hline
\textbf{Daily and cumulative dose reconstruction} & No & Yes & No \\
\hline
\end{tabularx}
\caption{\small Comparison of the assumed risk mitigation settings (including some options) for different PSQA methods and independent dose or MU calculation}
\label{tab:psqa-comparison}
\end{table}

\subsubsection{Measurement-based PSQA}
Measurement-based PSQA requires 
\begin{enumerate}[noitemsep]
  \item detectors for 2D or 3D dose measurements, 
  \item treatment room occupancy, 
  \item proton beam delivery and measurement, and 
  \item data analysis software for documentation. 
\end{enumerate}
The 2D ionization chamber array is widely used for PSQA, which provides the measured planar dose distribution in real time\cite{mackin2014spot}\cite{mackin2013improving}. The uses of radiochromic film and gel for PSQA were reported but are not considered practical in the clinic due to time-consuming post-processing\cite{chan2017patient}\cite{vandecasteele2013evaluation}.

The dose distribution can only be measured and compared with a TPS recalculation of the dose in homogeneous water or plastic phantoms with appropriate correction of water-equivalent thickness in a PSQA plan generated for this specific purpose. The gantry angle and snout position with range shifter in this PSQA plan may need to be adjusted. Often only vertical or horizontal gantry angles (0° or 90°) can be used in measurements, and the snout position may need to be pulled back to avoid collisions with the water tank. Beam modifiers such as range shifter and aperture should be used in these measurements.

The 2D planar dose distributions at one or more depths inside the treatment volume are measured field by field\cite{arjomandy2019taskgroup224}. Neither the total target dose nor dose sparing for organs-at-risk is compared with the calculation due to the modified beam parameters in the QA plan. The resolution of measured dose distributions may also be limited by the detector volume and spacing on the 2D array, as well as the depths selected for measurements. The 2D plane doses at depths beyond the $R_{80}$ are rarely reported in measurements. Measurement-based PSQA tends to have a higher gamma analysis passing rate when measured at depths with low dose gradients. The measurement-based PSQA can be performed in the patient-specific treatment rooms after hours, or a fixed-beam room with matched beam data, which would either extend or share the beam time.
\subsubsection{Log File-based PSQA}
Unlike the measurement-based approach, the log file-based PSQA does not require an additional QA plan with modified beam parameters. The field can be delivered at the prescribed gantry angle without any external dosimeters just as it would be done during treatment. The log file can be imported to either the primary treatment planning system or to an independent software for forward dose calculation\cite{winterhalter2019log}. The MU chamber used to collect the log files is upstream of any of the beam modifiers.

The dosimetric effect caused by the existence of beam modifiers including the air gap can be only simulated in log file-based PSQA. The 3D dose distribution can be reconstructed in homogeneous water or inhomogeneous patient planned CT and then compared with the calculations. The 2D or 3D gamma evaluation can be used to present the similarity of the dose distribution in homogeneous water or phantom, as well as the inhomogeneous patient planned CT. Dose-volume histograms can also be compared, which presents the clinical significance for both target volumes and organs-at-risk\cite{li2013use}.

\subsubsection{Independent Secondary Dose or MU Calculation}
The AAPM Task Group 219 recently reported the recommendations on the clinical implementation of independent secondary dose or MU calculations as part of the verification for photon IMRT\cite{zhu2021report}. The independent secondary dose or MU calculation should be and has been a part of the patient-specific QA regardless of the use of photon or proton beams. Historically there have been few commercial software options for independent dose or MU calculations for proton beams. This has begun to change, and currently independent (secondary) dose or MU calculation are performed using a mixture of in-house developed software, open-source codes, or a different dose calculation algorithm in the same treatment planning system\cite{ma2014gpu}\cite{wan2015fast}\cite{lin2017benchmarking}. The independent (secondary) dose or MU calculation can be performed using a homogeneous phantom or heterogeneous patient CT voxels with amended or identical treatment parameters.

\subsubsection{Summary}
The characteristics and their assumed risk mitigation settings - including a few options - for measurement-based and log file-based PSQAs, as well as for independent secondary dose or MU calculations are listed in Table \ref{tab:psqa-comparison}. 

  \section{PSQA-Related FMEA Procedures}
\label{sec:psqa_fmea_procedures}
\subsection{PSQA-Dependent pFMEAs}
The working group of the PTCOG Tx Eff SC concluded that it would be appropriate to carry out a separate process-based analysis of the failure modes and effects (pFMEA) for each proton PSQA approach. Thereby, – as previously described – an independent modeling of the entire PBS-PT workflow was carried out first, and was then limited to those process steps and their failure modes that influence each form of PSQA. A baseline scoring, i.e. an FMEA without considering any PSQA-relevant measures, was subsequently carried out. The risk assessment performed for this “naked” PBS-PT workflow provided the PSQA-independent, quasi workflow-inherent risk estimates for each process step, which were used as basic risk reference values.

In this way, it was possible to compare the further PSQA-specific FMEAs or risk assessments and their potential for risk reduction both in absolute terms against a baseline and relatively among each other. At the same time, this approach revealed both the process step-dependent and failure mode-specific influence of each PSQA method on its individual risk contribution and allowed the determination of the progression or development of risk contributions within a single PSQA-specific PBS-PT workflow.

As previously described, the expert panel developed the following FMEA-relevant workflow diagrams of patient treatment:
\begin{itemize}[noitemsep]
  \item without any PSQA (PSQA\textsubscript{noQA}), 
  \item with measurement-based PSQA (PSQA\textsubscript{meas}), 
  \item with log file-based PSQA (PSQA\textsubscript{log}), and 
  \item with an independent secondary dose calculation (PSQA\textsubscript{calc}).
\end{itemize}

\subsection{Classification of Related Process Steps between PSQA Methods}
\label{subsec:psqa_classification}
\begin{center}
\begin{figure}[h!]
    \centering
    \includegraphics[clip, trim=0.56cm 5cm 0.56cm 5cm, width=\textwidth]{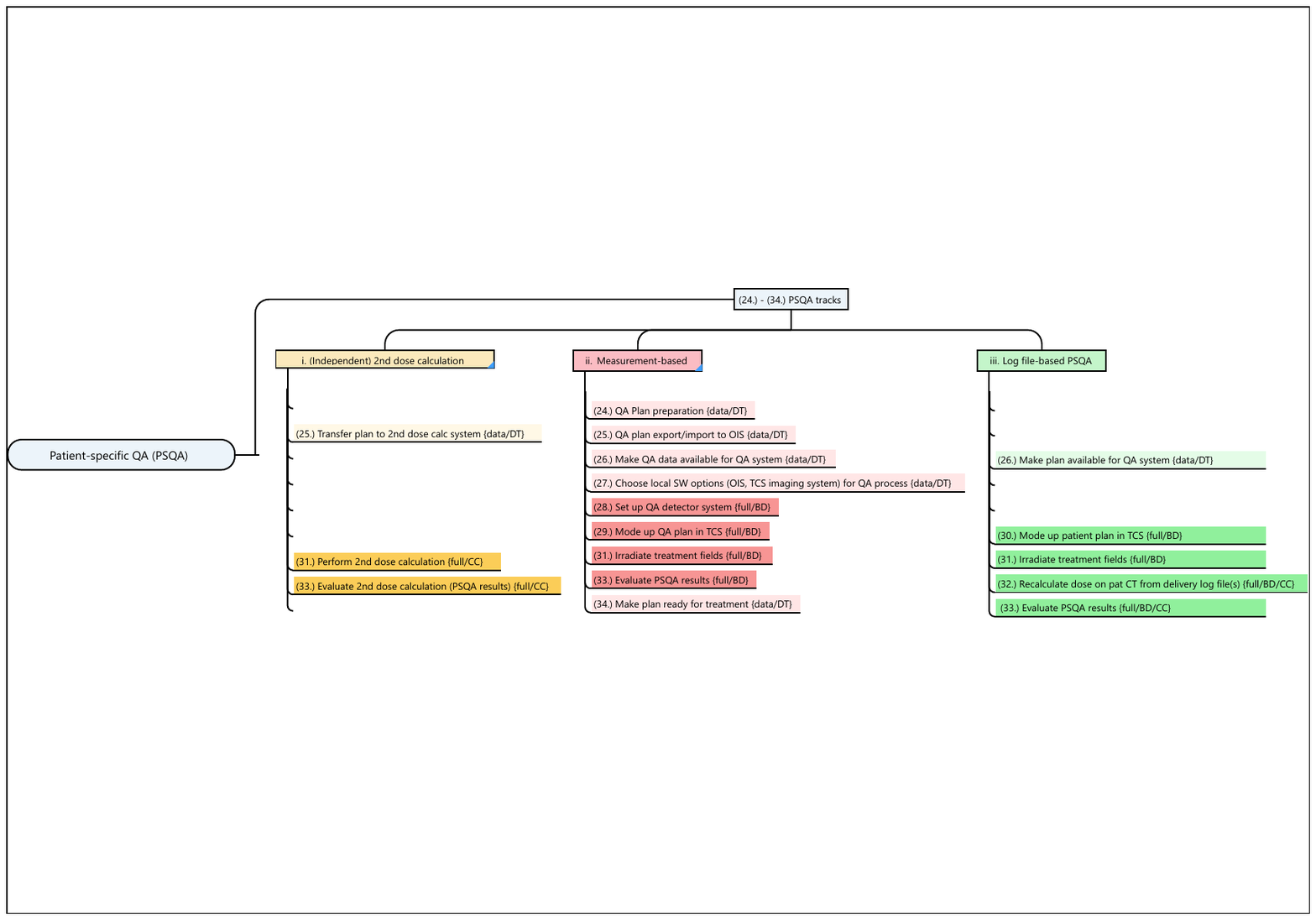}
    \caption{\small Overview of process steps associated with different PSQA methods in a PBS-PT workflow. Each track demonstrates the steps involved in Independent Secondary Dose Calculation (yellow), Measurement-Based PSQA (red), and Log File-Based PSQA (green), emphasizing their unique process step granularity. Lighter colors refer to data manipulation steps, while darker colors indicate steps included in the full QA risk evaluation.}
    \label{fig:PSQA-tracks}
\end{figure}

\begin{table}[hb!]
    \centering
    \begin{tabular}{|>{\centering\arraybackslash}p{1.6cm}|
                    >{\centering\arraybackslash}p{3.5cm}|
                    >{\centering\arraybackslash}p{3.5cm}|
                    >{\centering\arraybackslash}p{3.5cm}|
                    >{\centering\arraybackslash}p{1.8cm}|}
        \hline
        \textbf{Process Step Number} & \textbf{PSQA\textsubscript{calc}} & \textbf{PSQA\textsubscript{meas}} & \textbf{PSQA\textsubscript{log}} & \textbf{PSQA Category} \\
        \hline \hline
        (24) & --- & QA Plan preparation & --- & \{data/DT\} \\
        \hline
        (25) & Transfer plan to 2\textsuperscript{nd} dose calculation system & QA plan export/import to OIS & --- & \{data/DT\} \\
        \hline
        (26) & --- & Make QA data available for QA system & Make plan available for QA system & \{data/DT\}\\
        \hline
        (27) & --- & Choose local SW options (OIS, TCS imaging system) for QA process & --- & \{data/DT\} \\
        \hline
        (28) & --- & Set up QA detector system & --- & \{full/BD\} \\
        \hline
        (29)/(30) & --- & Mode up QA plan in TCS & Mode up QA plan in TCS & \{full/BD\} \\
        \hline
        (31) & Perform 2\textsuperscript{nd} dose calculation & Irradiate treatment fields & Irradiate treatment fields & \{full/CC/BD\} \\
        \hline
        (32) & --- & --- & Recalculate dose on patient CT from delivery log file(s) & \{full/BD\} \\
        \hline
        (33) & Evaluate 2\textsuperscript{nd} dose calculation (PSQA\textsubscript{calc} result) & Evaluate PSQA\textsubscript{meas} result & Evaluate PSQA\textsubscript{log} result & \{full/CC/BD\} \\
        \hline
        (34) & --- & Make plan ready for treatment & --- & \{data/DT\} \\
        \hline
    \end{tabular}
    \caption{\small Classification of process steps associated with different PSQA methods in the PBS-PT workflow.}
    \label{tab:psqa-steps}
\end{table}
\end{center}
The analysis of the different PSQA methods in the context of PBS-PT workflows (ISDC, measurement- and log file-based PSQA) reveals distinct sets of process steps associated within each method. Each approach is modeled based on the principle of appropriate process step granularity, ensuring a clear representation of their individual workflow components. Figure~\ref{fig:PSQA-tracks} provides a comprehensive overview of these tracks and their classification, while Table \ref{tab:psqa-steps} lists the explicit annotations of the PSQA process graphic in a structured way for full readability and traceability.

What does 'classification' mean in this context? If one were to strictly follow the previous explanations on the iterative execution of the FMEAs after the baseline has been determined, each PSQA method to be analyzed with each of its process steps would have to lead to a separate, complete iteration of the risk assessment with regard to the entire PBS-PT workflow. This would be necessary because – strictly speaking – the introduction of a measure that influences or mitigates risk at each process step would influence the prerequisites for risk estimation across the entire workflow, which would lead to a total of 17 additional iteration cycles or risk assessments (3 for PSQA\textsubscript{calc}, 9 for PSQA\textsubscript{meas}, 5 for PSQA\textsubscript{log}).

As can easily be imagined, this effort was also beyond the capabilities of the working group. In addition, such processing of the PSQA-FMEAs along all of their process steps would have made a reasonably realistic comparison of their mutual efficacy on risk mitigation extremely difficult to nearly impossible. This was due to the fact that a reasonable quantitative risk accumulation of all estimated process step influences per PSQA technique would have had to be carried out, which seemed practically unfeasible. Even in the best case, this would have led to absolutely unrealistic distortions in the comparison of the risk reduction potentials of the PSQA methods.

As such, the problem was analyzed by first identifying similarities between all PSQA methods and their sets of process steps. It was found that each set of process steps could be roughly divided into two categories of measures, whereby the first group of measures belongs to the respective PSQA data transfer ($DT = \text{Data Transfer}$) and largely describes the initial phase of the corresponding PSQA application. 

After the data transfer has been completed, the second category or group of process steps completes the PSQA application by performing PSQA-relevant irradiation steps or measurements ($BD$ = Beam Delivery) and/or (subsequently) (dose) calculations ($CC = \text{Completed Calculation}$) including the respective final PSQA evaluation. The only exception to this classification is process step (34) for PSQA\textsubscript{meas}, which is best understood as a plan manipulation, but which – assuming a comparable risk influence – has been assigned to the $DT$ category.

From these considerations arise the indices of the risk estimates used in the further course of this report, i.e. the RPN values, whereby $DT$ (or the term $data$) designates the (initial) process category Data Transfer, while $BD$ and $CC$ refer to the second, final  process category with 'Beam Delivery' or 'Completed Calculation'. They conclude the respective PSQA application and are therefore also referred to by the term 'full'. Thus, the following applies to the indices: $DT \equiv data$ and $DT+BD \mapsto BD \equiv full$ and $DT+CC \mapsto CC \equiv full$. In addition, the light color coding in Figure \ref{fig:PSQA-tracks} indicates $DT$ or $data$ process steps, while the dark color coding represents the corresponding $BD$, $CC$ or $full$ process steps.

In summary, this classification supports a systematic understanding of the roles and contributions of each PSQA method together with their grouped process steps within the PBS-PT workflow. The integration of the different PSQA tracks ensures structured and modular workflows, enabling effective comparison and risk evaluation of PSQA methodologies. Detailed diagrams illustrating the specific workflows are provided in the Appendix A (see Figures~\ref{fig:PSQA_calc} to \ref{fig:PSQA_log}).

\subsection{Expert-Scoring Dataset and Data Structure}
\label{subsec:expert_scoring_dataset}

\begin{table}[h!]
\centering
\footnotesize
\renewcommand{\arraystretch}{1.18}
\setlength{\tabcolsep}{4.5pt}

\caption{\small Structure of the expert-scoring data set used for the comparative pFMEA analysis. The table separates data collection, scoring, transformation, aggregation, and uncertainty-relevant layers.}
\label{tab:ses_data_structure}

\begin{tabularx}{\textwidth}{p{0.24\textwidth}p{0.35\textwidth}X}
\toprule
\textbf{Layer} & \textbf{Data element} & \textbf{Function in the present pFMEA} \\
\midrule

Expert-scoring layer
&
Five center-specific expert-scoring inputs
&
Captures institutional and technical variability across participating proton-therapy centers. The centers constitute an expert-elicitation panel, not a random statistical sample of all PBS-PT facilities. \\

\addlinespace[0.25em]

Workflow layer
&
Generic PBS-PT workflow with process, process-step, and PSQA-track structure
&
Defines the common coordinate system for all subsequent scoring. It ensures that measurement-based PSQA, log file-based PSQA, and independent secondary dose calculation are compared against the same underlying treatment workflow. \\

\addlinespace[0.25em]

Failure-mode layer
&
Process step, Risk ID, failure mode, cause, effect, notes, and PSQA-specific comments
&
Provides traceability from each scored entry back to its workflow location and clinical meaning. The final analysis set corresponds to the PSQA-relevant Risk IDs listed in Table~\ref{tab:riskid_catalogue}. \\

\addlinespace[0.25em]

Risk-state layer
&
NoQA baseline; Meas-Data; Meas-Full; Log-Data; Log-Full; Calc-Data; Calc-Full
&
Separates the common baseline state from method-specific preparatory Data-stage effects and final Full-stage verification effects. The Cum state is derived mathematically and remains anchored to NoQA. \\

\addlinespace[0.25em]

Elementary scoring layer
&
Occurrence \(O\), severity \(S\), and detectability \(D\), each scored on a 1--10 ordinal TG-100-style scale
&
Defines the elementary expert judgments from which all RPN-derived quantities are calculated. The scoring scale supports structured comparison but remains ordinal and semi-quantitative. \\

\addlinespace[0.25em]

Failure-mode RPN layer
&
\(\mathrm{RPN}=O\times S\times D\) for each failure mode and risk state
&
Transforms the elementary \(O\), \(S\), and \(D\) scores into failure-mode-specific RPN values for the NoQA baseline and each PSQA-dependent risk state. \\

\addlinespace[0.25em]

Risk-change layer
&
Failure-mode-level absolute and relative changes:
\(\Delta^{*,(pc)}_{\mathrm{abs},i,l}\) and
\(\delta^{*,(pc)}_{\mathrm{rel},i,l}\)
&
Quantifies how each PSQA method changes each failure mode at the Data, Full, and Cum levels. The Cum quantity provides the strict NoQA-anchored basis for inter-method comparison. \\

\addlinespace[0.25em]

Aggregation layer
&
Process-step metrics \(\rho\) and workflow-level metrics \(P\)
&
Aggregates failure-mode effects to process-step, process-region, and workflow levels using baseline-risk-weighted normalization. This prevents low-burden failure modes with large relative changes from dominating the global interpretation. \\

\addlinespace[0.25em]

Uncertainty and coverage layer
&
Number of contributing center scores, inter-center score dispersion, and exclusion or interpretive flags for insufficiently scored entries
&
Represents uncertainty from expert judgment, institutional heterogeneity, and limited center participation. Because \(n=5\), dispersion measures should be interpreted descriptively as scorer heterogeneity and sensitivity indicators, not as inferential confidence intervals. \\

\bottomrule
\end{tabularx}
\end{table}

The pFMEA data set used in this report was organized as a structured expert-elicitation matrix rather than as an observational incident registry or a statistically sampled multi-institutional cohort. Consistent with established radiotherapy FMEA practice, the analysis followed a process-map--to--failure-mode--to--score structure: the generic PBS-PT workflow was first decomposed into process steps, PSQA-relevant failure modes were assigned to those steps, and each failure mode was then scored using occurrence, severity, and detectability criteria adapted from the TG-100 FMEA framework \cite{huq2016taskgroup100,Cantone2013FMEAProton,Taylor2023RTQAFMEA,Kornek2024PFMEA}.

The scoring data were provided by five proton therapy centers represented in the working group. Each center-specific scoring sheet contributed expert ratings for the same failure-mode catalogue and for the same set of PSQA-dependent risk states. The elementary unit of the scoring data set was therefore not a patient, a treatment fraction, or an incident, but a failure-mode--by--risk-state entry within a defined PBS-PT workflow. For each retained failure mode, the relevant scores were assigned for the common no-PSQA baseline and for the Data- and Full-stage states of the investigated PSQA methods: measurement-based PSQA, log file-based PSQA, and independent secondary dose calculation. The cumulative effect was not scored independently; it was derived from the stage-wise values and expressed relative to the common no-PSQA baseline.

Table~\ref{tab:ses_data_structure} summarizes the resulting data structure. This structure is important for interpretation because the numerical outputs of this report are not direct measurements of absolute clinical risk. They are structured, baseline-anchored, expert-derived risk surrogates used to compare how alternative PSQA architectures modify the same workflow under a common scoring framework.

Failure modes for which fewer than a majority of participating experts could provide meaningful scores were retained in the catalogue where appropriate for transparency, but should not be treated as quantitatively equivalent to fully scored entries. Conversely, failure modes retained in the quantitative analysis should be interpreted as expert-consensus risk signatures within the defined PBS-PT workflow, not as estimates of population-level event frequency. This distinction is central to the validity of the subsequent comparisons: the report evaluates relative changes in a common expert-scored workflow model, not observed clinical incidence rates across proton centers.

  \section{Operational Framework for a Comparative Risk Assessment}
\label{sec:operational_framework}

This section presents a unified operational, partially mathematical framework for comparing patient-specific quality assurance (PSQA) strategies within particle therapy workflows. The framework builds on classical process-driven Failure Modes and Effects Analysis (pFMEA) using the Risk Priority Number (RPN) methodology.  It accommodates PSQA-dependent pFMEAs and the process step classification scheme described in section \ref{subsec:psqa_classification}. It is designed not only to quantify the risk contributions at both local (process step) and global (workflow) levels but also to enable direct, evidence-based comparisons among different PSQA methods (measurement-based, log file-based, and independent secondary dose calculation). In addition, its fundamental limitations and uncertainties are discussed within the framework of a pFMEA.

\subsection{Foundational Principles}
\label{subsec:op_fw_foundational_principles}

Risk management in radiotherapy requires that we compare alternative quality assurance strategies in a quantitative manner\footnote{In the remainder of this section and in Appendix \ref{ch:methodological_implications}, we will take a critical and in-depth look at the concept of “quantifiability” in the RM methodology presented, which will lead, among other things, to a discussion of the “semi-quantitative nature” of the PSQA-pFMEA to be introduced. In order not to anticipate these discussions, we will initially use the simplifying adjective \emph{quantitative} in the sense of a number-based analysis.}. In this framework we assume that:

\begin{enumerate}
    \item The RPN approach based on pFMEA can serve as an operational basis for a comparative, semi-quanti\-ta\-tive risk assessment within a defined evaluation framework.
    \item Failure modes can be identified for every relevant process step in the treatment workflow.
    \item PSQA methods can modify the key parameters (occurrence, severity, detectability) in ways that they can be operationally assessed through a structured expert elicitation.
    \item Differentiated risk mitigation across various PSQA modalities can provide meaningful guidance for clinical implementation.
\end{enumerate}

Although risk assessment remains partly subjective and failure modes may be interdependent, the process-oriented formalism developed here is intended to provide a transparent and internally consistent basis for a comparative PSQA assessment.

\subsection{Risk Priority Number Calculation}
\label{subsec:op_fw_RPN_calculation}

Following established pFMEA methodology (adapted for radiotherapy as in AAPM TG-100), each failure mode (FM) is characterized by three parameters:

\begin{itemize}[noitemsep]
    \item \textbf{Occurrence (O)}: the probability or frequency of the failure.
    \item \textbf{Severity (S)}: the impact or clinical consequence if the failure occurs.
    \item \textbf{Detectability (D)}: the likelihood that the failure will remain undetected.
\end{itemize}

Each parameter is scored on a scale from 1 to 10 according to defined criteria. As such, Table~\ref{tab:tg100_fmea} contains the qualitative descriptions and corresponding numerical ranks used in this study in exactly the form as envisaged in the AAPM TG-100 report \cite{huq2016taskgroup100} in order to standardize the evaluation across different errors\footnote{At this point, it should be noted that the terminology and specification of the short descriptions from the table (e.g., 'Limited toxicity or tumor underdose' or 'Wrong dose, dose distribution, location, or volume') are described in detail in \cite{huq2016taskgroup100}, without being explicitly presented here.}. This scoring system ensures consistency in evaluating risks across different PSQA workflows.

The \textbf{core baseline} Risk Priority Number (RPN) is then computed as
\begin{subequations}
  \label{eq:RPN_pair}  
  \begin{equation}
      \mathrm{RPN}^{\mathrm{noQA}}_{\mathrm{FM}} = \mathrm{O}^{\mathrm{noQA}}_{\mathrm{FM}} \times \mathrm{S}^{\mathrm{noQA}}_{\mathrm{FM}} \times \mathrm{D}^{\mathrm{noQA}}_{\mathrm{FM}} \,
      \label{eq:RPN_noQA}
  \end{equation}
  \text{and the \textbf{FM-specific}, \textbf{process category (pc)-based}, and \textbf{PSQA-dependent} RPNs are calculated accordingly by}
  \begin{equation}
      \mathrm{RPN}^{\mathrm{PSQA},\mathrm{pc}}_{\mathrm{FM}} = \mathrm{O}^{\mathrm{PSQA},\mathrm{pc}}_{\mathrm{FM}} \times \mathrm{S}^{\mathrm{PSQA},\mathrm{pc}}_{\mathrm{FM}} \times \mathrm{D}^{\mathrm{PSQA},\mathrm{pc}}_{\mathrm{FM}} \,
      \label{eq:RPN_pc}
  \end{equation}
\end{subequations}
with $\mathrm{PSQA} \in \{\mathrm{meas}, \mathrm{log}, \mathrm{calc}\}$ and $\mathrm{pc} \in \{\mathrm{data}, \mathrm{full}\}$. 

Note that the scales for \(\mathrm{O}\) and \(\mathrm{D}\) are highly non-linear. Although one could use an exponential formulation, the standard multiplicative form in Equations~\ref{eq:RPN_pair} is retained for consistency and convenience in comparing risk profiles.

\begin{table}[ht]
\centering
\footnotesize 
\renewcommand{\arraystretch}{1.5} 
\setlength{\tabcolsep}{5pt} 

\begin{tabular}{|m{1.0cm}|m{2.0cm}|m{1.7cm}|m{3.0cm}|m{2.7cm}|m{3.5cm}|}
\hline \hline
\multirow{2}{*}{\centering \textbf{Rank}} 
& \multicolumn{2}{m{3.7cm}|}{\centering \textbf{Occurrence (O)}} 
& \multicolumn{2}{m{5.7cm}|}{\centering \textbf{Severity (S)}} 
& \multicolumn{1}{m{3.5cm}|}{\centering \textbf{Detectability (D)}} \\ \cline{2-6}
& \multicolumn{1}{m{2.0cm}|}{\centering \textbf{Qualitative}} 
& \multicolumn{1}{m{1.7cm}|}{\centering \textbf{Frequency in \%}}
& \multicolumn{1}{m{3.0cm}|}{\centering \textbf{Qualitative}} 
& \multicolumn{1}{m{2.7cm}|}{\centering \textbf{Categorization}} 
& \multicolumn{1}{m{3.5cm}|}{\centering \textbf{Estimated Probability of failure going undetected in \%}} \\ \hline

\centering 1 
& \multirow{2}{*}{\parbox{2cm}{\centering Failure\hspace{1cm} unlikely}} 
& \centering 0.01 
& \centering No effects
& \centering -
& \parbox{3.5cm}{\centering 0.01} \\ \cline{1-1} \cline{3-6}
\centering 2 
& 
& \centering 0.02 
& \multirow{2}{*}{\parbox{2.7cm}{\centering Inconvenience}}
& \multirow{2}{*}{\parbox{2.7cm}{\centering Inconvenience}}
& \parbox{3.5cm}{\centering 0.2} \\ \cline{1-3} \cline{6-6}

\renewcommand{\arraystretch}{5}
\centering 3 
& \multirow{3}{*}{\parbox{2cm}{\centering Relatively\hspace{1cm} few~failures}} 
& \centering 0.05
& 
& 
& \parbox{3.5cm}{\centering 0.5} \\ \cline{1-1} \cline{3-6}
\centering 4 
& 
& \centering 0.1
& \parbox{2.7cm}{\centering Minor dosimetric error}
& \parbox{2.7cm}{\centering Suboptimal plan or treatment}
& \parbox{3.5cm}{\centering 1.0} \\ \cline{1-1} \cline{3-6}
\centering 5 
& 
& \centering <0.2
& \multirow{2}{*}{\parbox{2.7cm}{\centering Limited toxicity or tumor underdose}}
& \multirow{4}{*}{\parbox{2.7cm}{\centering Wrong dose, dose distribution, location, or volume}}
& \parbox{3.5cm}{\centering 2.0} \\ \cline{1-3} \cline{6-6}
\renewcommand{\arraystretch}{1.5}

\centering 6 
& \multirow{2}{*}{\parbox{2cm}{\centering Occasional\hspace{1cm} failures}} 
& \centering <0.5
& 
& 
& \parbox{3.5cm}{\centering 5.0} \\ \cline{1-1} \cline{3-4} \cline{6-6}
\centering 7 
& 
& \centering <1
& \multirow{2}{*}{\parbox{2.7cm}{\centering Potentially serious toxicity or tumor underdose}}
& 
& \parbox{3.5cm}{\centering 10} \\ \cline{1-3} \cline{6-6}

\centering 8 
& \multirow{2}{*}{\parbox{2cm}{\centering Repeated\hspace{1cm} failures}} 
& \centering <2
& 
& 
& \parbox{3.5cm}{\centering 15} \\ \cline{1-1} \cline{3-6}
\centering 9 
& 
& \centering <5
& \centering Possible very serious toxicity or tumor underdose
& \multirow{2}{*}{\parbox{2.7cm}{\centering Very wrong dose, dose distribution, location, or volume}}
& \parbox{3.5cm}{\centering 20} \\ \cline{1-4} \cline{6-6}

\centering 10 
& \centering Failures inevitable 
& \centering >5
& \centering Catastrophic
& 
& \parbox{3.5cm}{\centering >20} \\ \hline \hline

\end{tabular}
\caption{\small Scales and descriptions of the $O$, $S$, and $D$ values used in the AAPM TG-100 FMEA.}
\label{tab:tg100_fmea}
\end{table}

\subsection{Operational Summary of the Stage-Wise pFMEA Risk-Quantification (\textit{Data}, \textit{Full}, \textit{Cum})}
\label{subsec:op_fw_psqa_operational_summary}

This subsection provides an operational description and summary of methods how the stage-wise PSQA effects are quantified and aggregated from the failure-mode (FM) level to process-step (PS) and workflow levels. The full
mathematical specification (including the complete notation system, formal definitions, identities, and properties) is provided in the Appendix B, where the full technical derivations are provided.

\subsubsection{Core Objects and Stage Logic}
We consider a workflow represented as a set of process steps (PS) indexed by $i$, each with an associated set of failure modes (FMI) indexed by $l \in \mathrm{FMI}_i$.

For each FM $(i,l)$ we start from a \emph{baseline} risk priority number (RPN) without PSQA, $\mathrm{RPN}^{\mathrm{noQA}}_{i,l}$, and evaluate the impact of three PSQA modalities $^{*} \in \{\mathrm{Meas}, \mathrm{Log}, \mathrm{Calc}\}$ along two \emph{sequential} scoring phases:
\begin{itemize}
  \item \textbf{Data-stage} (\textit{NoQA} $\to$ \textit{Data}): captures the incremental effect of PSQA components that act on the data-handling and information-transfer layer (stage-specific outcome
  $\mathrm{RPN}^{*,\mathrm{data}}_{i,l}$), \emph{normalized to the NoQA baseline}.
  \item \textbf{Full-stage} (\textit{Data} $\to$ \textit{Full}): captures the incremental effect of the remaining (verification/delivery-related) PSQA components beyond \textit{Data}, producing $\mathrm{RPN}^{*,\mathrm{full}}_{i,l}$, \emph{normalized to the method-specific Data-stage state}.
  \item \textbf{Cumulative (Cum) effect} (\textit{NoQA} $\to$ \textit{Full}): the \textit{Full}-track net effect of Data and Full combined, expressed relative to the \emph{common} \textit{NoQA} baseline.
\end{itemize}

This explicit two-stage structure reflects the practical reality that PSQA may (i) introduce additional data-handling steps that can inflate risk for certain FMs, while (ii) later verification actions can mitigate risks. The \emph{derived, not scored} \textit{Cum} category provides the strict \textit{NoQA}-anchored net effect of the \textit{Full}-track.

\subsubsection{Failure-mode level: absolute and relative effects}
At the FM level, the PSQA impact is quantified in two complementary ways:
\begin{itemize}
  \item \textbf{Absolute effect} (in RPN units): stage-wise absolute increments are computed as differences between the appropriate stage outcome and its stage baseline. Concretely, \textit{Data} uses $\mathrm{RPN}^{\mathrm{noQA}}_{i,l}$ as baseline, \textit{Full} uses $\mathrm{RPN}^{*,\mathrm{data}}_{i,l}$ as baseline, and \textit{Cum} uses $\mathrm{RPN}^{\mathrm{noQA}}_{i,l}$ as baseline. By construction, the cumulative \emph{absolute} change equals the sum of the \textit{Data} and \textit{Full} absolute increments (additive decomposition of the \textit{Full}-track).
  \item \textbf{Relative effect} (dimensionless): stage-wise relative changes are computed as \[ \delta_{\mathrm{rel}} := \mathrm{RPN}_{\mathrm{new}}/\mathrm{RPN}_{\mathrm{baseline}} - 1\,,\] with the same stage-dependent baseline choices as above. Since TG-100--style RPNs are strictly positive, these ratios are well-defined. Interpretation is uniform:
  $\delta_{\mathrm{rel}}>0$ indicates \emph{risk inflation},
  $\delta_{\mathrm{rel}}=0$ indicates \emph{no net change}, and
  $\delta_{\mathrm{rel}}<0$ indicates \emph{risk mitigation} (with a theoretical lower bound at $-1$).
\end{itemize}

\subsubsection{Mission-critical nuance: why \textit{Cum} is the strict basis for inter-method comparisons}
The \textit{Full}-stage relative effect is \emph{intentionally} normalized to $\mathrm{RPN}^{*,\mathrm{data}}_{i,l}$, because it is meant to capture the \emph{incremental} mitigation (or inflation) beyond the \textit{Data}-stage within the same modality. This makes \textit{Full}-stage relative values \emph{meaningful within a given modality}, but \emph{not strictly comparable across modalities} when their \textit{Data}-stage baselines differ. 

A simple counterexample illustrates the point: two modalities can reach the same $\mathrm{RPN}^{*,\mathrm{full}}_{i,l}$ yet yield different \textit{Full}-stage $\delta_{\mathrm{rel}}$ values solely because $\mathrm{RPN}^{*,\mathrm{data}}_{i,l}$ differs. In contrast, the \textit{Cum} relative effect is anchored to the common \textit{NoQA} baseline and therefore supports strict inter-method comparisons of \textit{Full}-track net performance.

Operationally, \textit{Cum} also obeys the expected \emph{compounding identity} for sequential relative changes:
\begin{equation}
1+\delta^{*,(\mathrm{cum})}_{\mathrm{rel},i,l}
=
\bigl(1+\delta^{*,(\mathrm{data})}_{\mathrm{rel},i,l}\bigr)\,
\bigl(1+\delta^{*,(\mathrm{full})}_{\mathrm{rel},i,l}\bigr),
\end{equation}
which ensures that the \textit{Full}-track net effect remains invariant to the internal split between \textit{Data} and \textit{Full}.

\subsubsection{Process-step level: local metrics ($\bm{\rho_{\mathrm{abs}}}$ and $\bm{\rho_{\mathrm{rel}}}$)}
Because multiple failure modes may belong to one process step, FM-level effects are aggregated to local, step-level indicators. 

Two local metrics are used:
\begin{itemize}
  \item \textbf{Local absolute metric} $\rho_{\mathrm{abs},i}$: within each PS $i$, the local absolute effect is the \emph{sum} of the FM-level absolute increments across all $l\in \mathrm{FMI}_i$. This preserves the natural
  unit (RPN points) and directly represents the net step-level change in risk burden.
  \item \textbf{Local relative metric} $\rho_{\mathrm{rel},i}$: within each PS $i$, the local relative effect is a \emph{risk-weighted average} of the FM-level relative changes, where the weights are proportional to the \emph{baseline risk contribution} of each FM within that step.
\end{itemize}

The weighting is not cosmetic; it is the mechanism that makes the local relative metric (i) bounded and numerically stable on a per-step basis, (ii) interpretable as a \emph{step-level percent change}, and (iii) compatible with further aggregation to workflow-level metrics without semantic distortion. 

The baseline used for weights is \emph{stage-consistent}: for \textit{Data} and \textit{Cum}, weights derive from $\mathrm{RPN}^{\mathrm{noQA}}_{i,l}$; for \textit{Full}, weights derive from $\mathrm{RPN}^{*,\mathrm{data}}_{i,l}$. Equivalently — and crucial for transparency — the chosen weighting implies a closed-form identity: the local relative metric equals the corresponding local absolute metric \emph{normalized by the process step’s total baseline burden} (stage-appropriate denominator). Thus, $\rho_{\mathrm{abs}}$ and $\rho_{\mathrm{rel}}$ are not independent constructs; they are complementary representations of the same process step-level information on absolute and relative scales.

\subsubsection{Workflow level: global metrics on arbitrary PSQA comparison sets}
To support comparisons between PSQA methods at the level of sub-processes, processes or entire workflows, the framework allows aggregation over any nonempty set of process steps $\mathcal{I}$ (a ``comparison set''). Two global indicators mirror the local
ones: 
\begin{itemize}
  \item \textbf{Global absolute metric} $\mathrm{P}_{\mathrm{abs},\mathcal{I}}$: the sum of the local absolute metrics over all $i\in\mathcal{I}$ (preserving the RPN unit and representing the net burden change on   $\mathcal{I}$).
  \item \textbf{Global relative metric} $\mathrm{P}_{\mathrm{rel},\mathcal{I}}$: a baseline-burden-weighted average of the local relative metrics over $i\in\mathcal{I}$, using stage-consistent step weights (derived from the
  step burdens used already at the local level).
\end{itemize}

As at the local level, the weighting is selected so that the global relative is exactly the global absolute total normalized by the total baseline burden on $\mathcal{I}$. Importantly, the \emph{cumulative} global relative effect is explicitly anchored to the \emph{NoQA} burden (both as baseline and as the step-weight profile), which makes it the natural quantity for strict inter-PSQA comparisons of the \textit{Full}-track net effect on any $\mathcal{I}$.

\subsubsection{Practical computation pipeline} 
In practice, the framework is applied via the following deterministic steps:
\begin{enumerate}[label=(\roman*)]
    \item score $\mathrm{RPN}^{\mathrm{noQA}}_{i,l}$ for all relevant FMs,
    \item score $\mathrm{RPN}^{*,\mathrm{data}}_{i,l}$ and $\mathrm{RPN}^{*,\mathrm{full}}_{i,l}$ for each PSQA modality,
    \item compute stage-wise absolute and relative FM-level effects with the stage-appropriate baselines,
    \item verify internal identities (additivity for absolute changes; compounding for relative changes),
    \item aggregate to local metrics with stage-consistent risk weights (or equivalently via the normalized absolute identity), and
    \item aggregate to global metrics on any chosen comparison set $\mathcal{I}$.
\end{enumerate}

\subsection{Risk-Change Metrics Across Failure Modes, Process Steps, and Workflow: An Intuitive but Formal Summary}
\label{subsec:psqa_risk_metrics_summary}
The following summary provides a conceptually simplified overview of the risk-change formalism. The complete derivations and the full set of intermediate identities are listed in Appendix B, while this subsection focuses on (i) the basis anchoring, (ii) phase or stage consistency, and (iii) the rationale for the risk contribution weights, which enable a meaningful aggregation of relative changes in failure modes across process steps and up to arbitrary workflow segments.

\subsubsection{Index structure and stage-specific risk states}
We work directly on the two-level index structure ``process step $\rightarrow$ failure mode'' and complement it by a third aggregation layer acting on \emph{arbitrary collections of process steps}.
More precisely, process steps are indexed by an absolute workflow index $i$, and each step $i$ is associated with a set of failure modes
$l \in \mathrm{FMI}_i$.
For each PSQA modality $^{\ast} \in \{\mathrm{meas},\mathrm{log},\mathrm{calc}\}$, the analysis tracks the risk-priority number (RPN) of each failure mode $(i,l)$ across the three relevant risk states:
\begin{itemize}
  \item \textbf{Baseline (\textit{NoQA}):} $\mathrm{RPN}^{\mathrm{noQA}}_{i,l}$,
  \item \textbf{\textit{Data}-stage:} $\mathrm{RPN}^{\ast,\mathrm{data}}_{i,l}$, representing the RPN after applying the modality-specific \textit{Data}-stage PSQA actions,
  \item \textbf{\textit{Full}-stage:} $\mathrm{RPN}^{\ast,\mathrm{full}}_{i,l}$, representing the RPN after applying the \textit{Full}-track PSQA actions (\textit{Data} $\rightarrow$ \textit{Full}).
\end{itemize}
This explicit separation of \emph{Data} and \emph{Full} is essential because the \textit{Full}-stage is modeled as an \emph{incremental} intervention applied \emph{on top of} the \textit{Data}-stage configuration, rather than as an independent intervention directly referenced to \textit{NoQA}.

\subsubsection{Failure-mode level: relative changes and the meaning of ``cumulative''}
At the most granular level, we quantify how each failure mode $(i,l)$ changes when moving along the track $\text{\textit{noQA}} \rightarrow \text{\textit{data}} \rightarrow \text{\textit{full}}$.
To preserve stage consistency, the \textit{Full}-stage change is referenced to the \textit{Data}-stage state (not to \textit{NoQA}), while the cumulative change is referenced directly to \textit{NoQA} and therefore represents the net \textit{Full}-track effect.

Formally, we use stage-wise \emph{relative} changes
$\delta^{\ast,(pc)}_{\mathrm{rel},i,l}$ with $\mathrm{pc} \in \{(d),(f),(c)\}$ denoting \textit{Data}, \textit{Full}, and \textit{Cum}, respectively:
\begin{subequations}
\label{eq:rel_risk_change_tripel}
   \begin{equation}
       \delta^{\ast,(d)}_{\mathrm{rel},i,l} := \frac{\mathrm{RPN}^{\ast,\mathrm{data}}_{i,l}}{\mathrm{RPN}^{\mathrm{noQA}}_{i,l}} - 1,
       \label{eq:rel_risk_stage_data}
   \end{equation}
   \begin{equation}    
       \delta^{\ast,(f)}_{\mathrm{rel},i,l} := \frac{\mathrm{RPN}^{\ast,\mathrm{full}}_{i,l}}{\mathrm{RPN}^{\ast,\mathrm{data}}_{i,l}} - 1,
   \end{equation}
   \begin{equation}
   \begin{aligned}
       \delta^{\ast,(c)}_{\mathrm{rel},i,l} 
       &:= \frac{\mathrm{RPN}^{\ast,\mathrm{full}}_{i,l}}{\mathrm{RPN}^{\mathrm{noQA}}_{i,l}} - 1 \\
       &= \bigl(1+\delta^{\ast,(d)}_{\mathrm{rel},i,l}\bigr)\bigl(1+\delta^{\ast,(f)}_{\mathrm{rel},i,l}\bigr) - 1 \\
       &= \delta^{\ast,(d)}_{\mathrm{rel},i,l} + \delta^{\ast,(f)}_{\mathrm{rel},i,l}
          + \delta^{\ast,(d)}_{\mathrm{rel},i,l}\,\delta^{\ast,(f)}_{\mathrm{rel},i,l}.
       \label{eq:delta_rel_cum_factorization}
   \end{aligned}
   \end{equation}
\end{subequations}

Eq.~\ref{eq:delta_rel_cum_factorization} is particularly instructive: the cumulative net effect is \emph{not} the mere sum of the \textit{Data}- and \textit{Full}-stage relative changes. Instead, it includes an interaction term
$\delta^{\ast,(d)}_{\mathrm{rel},i,l}\delta^{\ast,(f)}_{\mathrm{rel},i,l}$, which captures the fact that the \textit{Full}-stage acts on a risk landscape that has already been altered by the \textit{Data}-stage.

Throughout, $\delta_{\mathrm{rel}}>0$ indicates net \emph{risk inflation}, whereas $\delta_{\mathrm{rel}}\le 0$ indicates \emph{risk mitigation}. Since RPNs are nonnegative under TG-100--style scoring and typically strictly positive in our setting, the natural mitigation interval is $\delta_{\mathrm{rel}}\in(-1,0]$ (a $100\%$ reduction corresponds to $\delta_{\mathrm{rel}}=-1$).

\subsubsection{From failure modes to process-step indicators: why we introduce local weights}
A process step $i$ usually comprises multiple failure modes with different baseline RPN magnitudes.
When aggregating \emph{relative} changes across these modes, a naive arithmetic mean would implicitly treat a minor failure mode (low baseline RPN) as equally important as a dominant failure mode (high baseline RPN).
This is generally undesirable in pFMEA-based decision making, because it can overemphasize large percentage shifts in low-impact modes and underemphasize modest percentage shifts in high-impact modes.

To address this, we introduce \emph{local} (failure-mode) weights that normalize each mode by its \emph{baseline risk contribution}.
For step $i$, define the relevant stage-consistent step burdens
\begin{equation}
G^{(d)}_i := \sum_{l\in \mathrm{FMI}_i}\mathrm{RPN}^{\mathrm{noQA}}_{i,l},
\qquad
G^{\ast,(f)}_i := \sum_{l\in \mathrm{FMI}_i}\mathrm{RPN}^{\ast,\mathrm{data}}_{i,l}.
\label{eq:local_fm_weights}
\end{equation}
These burdens measure the total baseline risk mass within the process step at the stage-appropriate anchor.
They induce normalized weight vectors (probability weights) on $\mathrm{FMI}_i$:
\begin{equation}
g^{(d)}_{i,l} := \frac{\mathrm{RPN}^{\mathrm{noQA}}_{i,l}}{G^{(d)}_i},
\qquad
g^{\ast,(f)}_{i,l} := \frac{\mathrm{RPN}^{\ast,\mathrm{data}}_{i,l}}{G^{\ast,(f)}_i},
\qquad
g^{(c)}_{i,l} = g^{(d)}_{i,l},
\label{norm_fm_weight_vector}
\end{equation}
so that $\sum_{l\in\mathrm{FMI}_i} g^{(\mathrm{pc})}_{i,l} = 1$ for each process category $\mathrm{pc}$.

Intuitively, $g^{(\mathrm{pc})}_{i,l}$ is the fraction of the step's baseline burden attributable to failure mode $(i,l)$ at the stage-consistent baseline. This makes the weighting a \emph{normalization by risk contribution}: each mode contributes to the aggregated relative metric in proportion to how much it contributes to baseline process step risk.\\[-2pt]

Using these weights, the \emph{local risk-weighted relative} indicators for process step $i$ are defined as weighted averages of the failure-mode relatives:
\begin{equation}
\rho^{\ast,(\mathrm{pc})}_{\mathrm{rel},i}
:= \sum_{l\in \mathrm{FMI}_i} g^{\ast,(\mathrm{pc})}_{i,l}\,\delta^{\ast,(\mathrm{pc})}_{\mathrm{rel},i,l},
\qquad \mathrm{pc}\in\{(d),(f),(c)\},
\label{eq:local_rel_weighted}
\end{equation}
with the understanding that $g^{\ast,(d)}_{i,l}=g^{(d)}_{i,l}$ and $g^{\ast,(c)}_{i,l}=g^{(c)}_{i,l}$.
The stage dependence is therefore deliberate: \textit{Data} and \textit{Cum} relatives are \emph{NoQA-anchored}, whereas \textit{Full} relatives are \emph{Data-anchored} (and thus PSQA dependent).

This weighting has several practical consequences that motivate its use beyond mere ``preference'':
\begin{itemize}
  \item \textbf{Interpretability as a fractional step-burden change.}
  The local relative $\rho^{\ast,(\mathrm{pc})}_{\mathrm{rel},i}$ equals the fractional change of the \emph{total} step burden under the corresponding stage, rather than an arbitrary average of percentages.
  \item \textbf{Numerical robustness and boundedness.}
  Because the weights have the function of a probability vector\footnote{Here, ``probability vector'' is meant in the purely mathematical sense of a non-negative, normalized weight vector: each weight is defined as a local or global RPN-burden contribution divided by the corresponding total burden, so all weights are non-negative and sum to one. They therefore distribute the aggregate burden over failure modes or process steps analogously to probabilities over outcomes, without implying that the weights are empirical probabilities.} and all denominators are strictly positive under the scoring scheme, the aggregation is stable and does not introduce artificial scale effects.
  \item \textbf{Decision relevance.}
  If a failure mode contributes only a small share to the baseline risk of the step, even a large percentage change in that mode cannot dominate the step-level conclusion --- which matches the risk-management intent of pFMEA aggregation.
\end{itemize}

\subsubsection{From local to workflow-level indicators: global weights as a second normalization layer}
Clinical decision making often requires summarizing risk effects over a larger part of the workflow (e.g., an isolated process, a subworkflow, or the complete workflow). Let $\mathcal{I}\subseteq \mathcal{W\!F}$ denote a nonempty subset of process steps (a contiguous segment in the process map). A natural global \emph{absolute} metric is obtained by summing the corresponding local absolute metrics over $i\in\mathcal{I}$, thereby preserving dimensional consistency (RPN units).

For global \emph{relative} metrics, however, the same structural issue reappears: different process steps have different baseline risk burdens. A simple average of the step-level relatives $\rho_{\mathrm{rel},i}$ would treat all steps equally, regardless of how much baseline risk they carry. Hence, we introduce a second normalization layer: \emph{global} (step) weights derived from baseline process step burdens.

Define the stage-consistent step-weight vectors on $\mathcal{I}$ by
\begin{equation}
\gamma^{(d)}_{\mathcal{I},i}
:= \frac{G^{(d)}_i}{\sum_{j\in\mathcal{I}}G^{(d)}_j},
\qquad
\gamma^{\ast,(f)}_{\mathcal{I},i}
:= \frac{G^{\ast,(f)}_i}{\sum_{j\in\mathcal{I}}G^{\ast,(f)}_j},
\qquad
\gamma^{(c)}_{\mathcal{I},i}
= \gamma^{(d)}_{\mathcal{I},i},
\qquad i\in\mathcal{I},
\end{equation}
so that each $\gamma$ functions as a probability vector\footnote{See the preceding explanatory note.} on $\mathcal{I}$.

The corresponding \emph{global risk-weighted relatives} are then obtained by weighting the local relatives across steps:
\begin{align}
\mathrm{P}^{\ast,(d)}_{\mathrm{rel},\mathcal{I}}
&:= \sum_{i\in\mathcal{I}} \gamma^{(d)}_{\mathcal{I},i}\,\rho^{\ast,(d)}_{\mathrm{rel},i},
&
\mathrm{P}^{\ast,(f)}_{\mathrm{rel},\mathcal{I}}
&:= \sum_{i\in\mathcal{I}} \gamma^{\ast,(f)}_{\mathcal{I},i}\,\rho^{\ast,(f)}_{\mathrm{rel},i},
\label{eq:global_rel_weighted}
\\
\mathrm{P}^{\ast,(c)}_{\mathrm{rel},\mathcal{I}}
&:= \sum_{i\in\mathcal{I}} \gamma^{(d)}_{\mathcal{I},i}\,\rho^{\ast,(c)}_{\mathrm{rel},i}.
\end{align}
Here $\mathrm{P}^{*}_{\mathrm{rel},\mathcal{I}}$ denotes the workflow-level (global) relative indicator (written as a capital symbol of $\rho$ to emphasize the additional aggregation layer). Most importantly, the cumulative global relative is \emph{explicitly anchored to the NoQA burden} both in its baseline and in its step-weight profile, which makes it the natural basis for strict inter-PSQA comparisons of the net \textit{Full}-track effect at the level of $\mathcal{I}$.

\subsubsection{Interpretation and what the weighting accomplishes (local and global)}
The introduction of $g_{i,l}$ and $\gamma_{\mathcal{I},i}$ is best viewed as a principled normalization strategy:
\begin{enumerate}
  \item \textbf{Local level (within a step).}
  The weights $g^{(\mathrm{pc})}_{i,l}$ normalize failure modes by their \emph{baseline contribution} to the process step burden. Therefore, $\rho_{\mathrm{rel},i}^{\ast,(\mathrm{pc})}$ summarizes ``how much the \emph{total} step risk mass changed'' at the stage-consistent anchor.
  \item \textbf{Global level (across steps).}
  The weights $\gamma^{(\mathrm{pc})}_{\mathcal{I},i}$ normalize process steps by their \emph{baseline contribution} to the total burden of the workflow segment $\mathcal{I}$. Therefore, $\mathrm{P}_{\mathrm{rel},\mathcal{I}}^{\ast,(\mathrm{pc})}$ summarizes ``how much the \emph{total} risk mass of the workflow segment changed''.
\end{enumerate}
Consequently, the global relatives are \emph{not} arbitrary averages of process-step percentages: they correspond exactly to fractional changes of the total burden in $\mathcal{I}$ relative to the same baseline workflow segment\footnote{Appendix~\ref{ch:math_framework} provides the closed-form identities that make this statement explicit at both local and global aggregation levels: at the process-step level, see Eqs.~\eqref{appB:eq:rho_rel_in_terms_of_rho_abs}--\eqref{eq:rho_abs_from_rho_rel} and the ratio-of-totals interpretation in Eq.~\eqref{appB:eq:ratio_of_totals_local_relatives}; at the workflow level, see Eqs.~\eqref{appB:eq:Rho_rel_in_terms_of_Rho_abs}--\eqref{appB:eq:Rho_abs_from_Rho_rel} and Eq.~\eqref{appB:eq:ratios_of_totals_for_global_relatives}.}.

\subsubsection{Implications for comparisons across stages and across PSQA modalities}

Within a fixed PSQA modality, the stage structure supports a clear interpretation of sequential interventions: the \textit{Data}-stage metrics quantify the relative deviation from the \textit{NoQA} baseline after applying the \textit{Data}-stage QA actions, while the \textit{Full}-stage metrics quantify the incremental effect of extending the \textit{Data}-stage configuration to the \textit{Full}-track.

For comparisons \emph{across} modalities, baseline anchoring becomes mission-critical. In particular, cumulative (\textit{NoQA}-anchored) relatives - both local $\rho^{\ast,(c)}_{\mathrm{rel},i}$ and global $\mathrm{P}^{\ast,(c)}_{\mathrm{rel},\mathcal{I}}$ - provide the most direct basis for strict inter-PSQA statements about the overall net effect of the \textit{Full}-track QA intervention, because they share a common reference state and a common weighting profile derived from the \textit{NoQA} burdens.

\subsection{Uncertainty Considerations}
\label{subsec:uncertainty_condiderations}
While the above formalism treats the elementary RPN values as definite, it is essential to acknowledge that $O$, $S$, and $D$ each (and as product) carry uncertainty — stemming from expert judgment, limited evidence, non-linearities, mathematical-statistical conditions, and/or context-dependent interpretation. In this subsection we focus on the statistical prerequisites.

\subsubsection{Statistical Prerequisites}
In order to quantitatively assess uncertainty in the PSQA-based risk evaluation, certain mathe\-matical-statisti\-cal prerequisites must be satisfied. The classical approach assumes that the risk metrics can be treated as continuous variables. 

However, in our context the key risk factors Occurrence ($O$), Severity ($S$), and Detection ($D$) are ordinal, discretized ratings rather than true continuous measurements. On an ordinal scale, arithmetic operations (like adding or averaging ratings, or treating them as ratios) are not strictly meaningful. The rankings convey only relative ordering (e.g., $S=8$ denotes higher severity than $S=4$, but not necessarily twice the severity). In other words, the “distance” between consecutive scores is not quantitatively uniform. 

This presents a limitation for the direct RPN application as product of $O$, $S$, or $D$ as factors, which do not have a consistent physical significance. 
Consequently, treating $O$, $S$, $D$ as interval-scale variables must be done with caution. Also any uncertainty estimates are at best approximations, serving as rough indicators of confidence rather than exact statistical measures. It is also noteworthy that these ratings are typically based on expert judgment; thus their uncertainty may stem from subjective variability or ambiguity, which is not always well-modeled by classical assumptions (e.g., normality).

Several methods can be employed to handle uncertainty in the risk assessment. These methods do not require strict linearity or interval-scale data, though they come at the cost of additional complexity:

\begin{itemize}[label=$\rightarrow$,leftmargin=6.3mm,itemsep=2pt]
  \item \textbf{Monte Carlo simulation:} By assigning probability distributions to the $O$, $S$, and $D$ inputs (reflecting the uncertainty or confidence in each rating) and repeatedly sampling from these distributions, one can numerically construct the distribution of the resulting risk metric. Monte Carlo methods are especially useful when the propagation of uncertainty involves non-linear combinations or discrete inputs, as is the case with the product or matrix-based risk scoring; they impose no requirement that a one-unit change in $O$ or $S$ has a fixed effect. In practice, this means generating a large number of random scenarios (e.g., thousands of trials) where $O, S, D$ values are varied according to their uncertainty ranges, and observing the spread of the calculated risk outcome. This approach provides an empirical estimate of uncertainty (e.g. confidence intervals for the risk score) without relying on the linear approximation.

  \item \textbf{Bootstrap resampling:} If historical data or multiple expert assessments are available for the risk factor scores, bootstrap techniques can be used to gauge uncertainty. The idea is to resample from the observed set of $(O,S,D)$ values or expert judgments to create “synthetic” datasets, from which many realizations of the aggregated risk measure are computed. The variability across these realizations reflects the uncertainty due to limited data or inter-expert variability. The bootstrapping method relies on only minimal assumptions regarding the underlying distributions and is well-suited for assessing the robustness of risk classification in the face of uncertainties in the sample. For example, one could resample the pool of expert ratings for a specific failure mode and observe how the resulting risk rating fluctuates, thereby obtaining a confidence estimate for the scoring.

  \item \textbf{Bayesian methods:} In a Bayesian framework, the ordinal risk scores can be treated as random variables with prior probability distributions that capture our initial belief about their likely values (and uncertainties). By incorporating evidence (for example, actual failure occurrences or additional expert input), these priors are updated to posterior distributions. A Bayesian analysis can propagate uncertainty by computing the full probability distribution of the risk outcome given the uncertainties in $O$, $S$, and $D$. This often involves constructing a probabilistic model (such as a Bayesian network) that links the risk factors to the outcome. The result is not a single “error bar” but a probabilistic characterization of risk (e.g., the probability that a certain risk level is exceeded). Bayesian approaches are powerful in that they allow the integration of diverse sources of information and explicit quantification of uncertainty; for example, radiotherapy studies have used Bayesian network models to complement qualitative risk analysis, incident analysis, and plan-review error detection, while the broader FMEA literature has shown how FMEA-derived structures or scores can be integrated into Bayesian-network-based risk ranking \cite{Reitz2013QuantitativeRiskRadiotherapyBN,He2024HFACSBNRadiotherapyIncidents,Kalendralis2023BNRadiotherapyQA}.
\end{itemize}

It should be emphasized that while Monte Carlo, bootstrap, and Bayesian techniques can provide a more comprehensive uncertainty analysis, they are not the default choice in our PSQA-based risk assessment workflow. These methods require more extensive data or computational effort, and their implementation must be carefully tailored to the context (including making additional assumptions about distributions or prior probabilities).

\subsubsection{Interpretive Scope of the Formalism}
The preceding considerations should be read on two distinct levels. At the \emph{algebraic} level, once stage-specific baseline and PSQA-transformed scores have been assigned, the transformations and burden-weighted aggregations derived in Appendix~B are exact within the chosen score system. At the \emph{measurement} level, however, the underlying inputs $O$, $S$, and $D$ remain ordered expert ratings rather than calibrated interval-scale quantities. 

\begin{quote}
    \vspace{0.1cm}
    \begin{mdframed}[linewidth=1pt]
       This framework therefore does not claim to measure absolute risk. Rather, it uses values derived from the RPN formalism as structured surrogates to compare how different PSQA strategies alter risks within the same base workflow under a common evaluation framework.
    \end{mdframed}
\end{quote}

The broader FMEA literature has long recognized that the classical RPN is only one of several possible semi-quantitative ranking devices and that both probabilistic and generalized reformulations are possible \cite{liu2013fmea_review,parrachosantanna2012ppn,kimzuo2018general}. This interpretation is also consistent with recent probabilistic work on the distribution of RPN values, which treats $O$, $S$, and $D$ as ordered categorical inputs and models the resulting RPN in a Bayesian-multinomial setting rather than as a continuous risk variable \cite{mahmoudvand2025rpn}. Such developments strengthen the case for score-congruent uncertainty analysis, but they do not eliminate the need for careful interpretation.

Within these limits, the present formalism is best viewed as a \emph{baseline-anchored, stage-consistent, semi-quantitative framework for comparative decision support}. Its claim is therefore not that it measures absolute risk, but that, under consistent scoring, it provides coherent indices of how the introduction of a given PSQA method changes the baseline burden across failure modes, process steps, and workflow segments. The corresponding methodological implications are examined critically in Appendix \ref{ch:methodological_implications}.

}

\chapter*{3 pFMEA RESULTS}
\setcounter{chapter}{3}
\addcontentsline{toc}{chapter}{3 pFMEA RESULTS}
\markboth{3 pFMEA RESULTS}{}
\setcounter{section}{0}
\label{ch:pfmea_results}

\newcommand{\mFull}{\bigl\langle\delta^{*}_{\mathrm{rel},\mathrm{Full}}\bigr\rangle}
\newcommand{\mData}{\bigl\langle\delta^{*}_{\mathrm{rel},\mathrm{Data}}\bigr\rangle}
\newcommand{\DeltaSigned}{\Delta\!\bigl\langle\delta^{*}_{\mathrm{rel}}\bigr\rangle_{\pm}}
\newcommand{\CompRatio}{\mathrm{CR}}

\renewcommand{\sectionmark}[1]{\markright{\thesection.\ #1}}
This chapter summarizes the quantitative results of the process-driven FMEA and compares the \emph{\textbf{base}} risk landscape (\ref{fig:PSQA_noQA}) with the changes caused by the three investigated patient-specific QA modalities (\ref{fig:PSQA_all}) – independent secondary dose calculation (\textbf{PSQA\textsubscript{calc}}; \ref{fig:PSQA_calc}), measurement-based verification (\textbf{PSQA\textsubscript{meas}}; \ref{fig:PSQA_meas}) and log file-based verification (\textbf{PSQA\textsubscript{log}}; \ref{fig:PSQA_log}). 

The overview starts with a catalog that assigns each risk identifier (Risk ID) to its parent process, the discrete process step, and the failure mode (\ref{sec:res_catalogue}); this table serves as a Rosetta stone for interpreting all subsequent graphs. 
\medskip

Section \ref{sec:no_psqa_risk_landscape} continues analyzing the \textbf{\emph{baseline No-PSQA}} scenario by its Risk Priority Number (RPN) distributions, providing a reference against which any risk mitigation — or, in some cases, escalation — must be estimated or judged. Leveraging the same indexation scheme used throughout the previous chapter, we mainly present radar plots that depict the geometric imprint of each PSQA technique on the 44-axis risk space.

In the following sections these qualitative impressions are quantified by (i) failure mode-specific absolute and relative risk changes $\bigl(\Delta^{*,(pc)}_{\mathrm{abs},\,i,\,l}$ \& $\delta^{*,(pc)}_{\mathrm{rel},\,i,\,l}\bigr)$ (Eqs.~\eqref{appB:eq:abs_risk_change_pair} \& \eqref{appB:eq:rel_risk_change_pair}), (ii) process step-dependent local absolute and relative risk metrics $\bigl(\rho^{*,(pc)}_{\mathrm{abs},\,i}$ \& $\rho^{*,(pc)}_{\mathrm{rel},\,i}\bigr)$ (Eqs.~\eqref{appB:eq:local_abs_metric_triple} \& \eqref{appB:eq:local_rel_metric_weighted_triple}), and (iii) process- and workflow-wide aggregates $\bigl(\Rho^{*,(pc)}_{\mathrm{abs},\,\mathcal{I}}$ \& $\Rho^{*,(pc)}_{\mathrm{rel},\,\mathcal{I}}\bigr)$ (Eqs.~\eqref{appB:eq:global_abs_metric_triple} \& \eqref{appB:eq:global_rel_metric_weighted_triple}), together with their closed-form relations between absolute and relative metrics (Eqs.~\eqref{appB:eq:Rho_rel_in_terms_of_Rho_abs} \& \eqref{appB:eq:Rho_abs_from_Rho_rel}). For the indexation see Appendix~\ref{appB:sec:notation_system}.

\paragraph{Scoring dataset and interpretation of uncertainty.}
The results presented in this chapter should be read in light of the expert-scoring data structure described in Section~\ref{subsec:expert_scoring_dataset}. The five participating proton-therapy centers provide a deliberately expert-based and technically diverse scoring panel, but not a statistical sample from which population-level uncertainty or center-to-center generalizability can be inferred. The relevant uncertainty is therefore primarily epistemic and descriptive: it reflects expert judgment, local workflow experience, institutional technology, and the limited number of centers contributing to each scored risk state. 

Where inter-center dispersion is available in the scoring data, it should be interpreted as a heterogeneity or sensitivity indicator, not as a formal confidence interval. Consequently, the most robust findings in the following sections are not primarily the exact numerical values of individual RPN-derived changes, but the sign, relative magnitude, workflow location, and method-specific pattern of risk modification across the common NoQA-anchored comparison framework.

In recognition of the limited representation and statistical power of the results presented here, a larger dataset has been collected through an online survey of additional centers ($n=11$).  While those results are beyond the scope of this report, their analysis will be provided in a subsequent publication. 
\section{Catalogue of Risk Identifiers}
\label{sec:res_catalogue}

Table~\ref{tab:riskid_catalogue} lists every \textbf{Risk-ID} used throughout this study, mapping the alphanumeric identifier to its process number and specific failure mode\footnote{The reason why the Risk-IDs are not listed in sequential alphanumeric order is due to the fact that approximately 100 ordered Risk-IDs (process steps with their associated failure modes) were recorded at the beginning of the procedural workflow description, which were then filtered out according to their PSQA relevance before the scoring process.}. The table provides the textual key for the interpretations and discussions throughout this chapter in combination with the workflow maps provided in Appendix \ref{ch:appendix_workflows}. 

\begin{center}
\footnotesize
\begin{longtable}{@{}l p{5cm} p{8cm}@{}}
  \caption{\small Mapping of Risk-IDs to process steps and failure modes}
  \label{tab:riskid_catalogue}\\
  \toprule
  \hline
  \textbf{Risk ID} & \textbf{Process Step} & \textbf{Failure Mode} \\
  \midrule
  \endfirsthead
  
  \multicolumn{3}{@{}l}{\textit{Table \thetable{} – continued}}\\
  \toprule
  \textbf{Risk ID} & \textbf{Process Step} & \textbf{Failure Mode} \\
  \midrule
  \endhead
  
  \midrule
  \multicolumn{3}{r@{}}{\textit{Continued on next page}}\\
  \endfoot
  
  \bottomrule
  \endlastfoot
  
  \ProcessGroup{Transfer Images}
  1.d & Plan CT Images Import & Incorrect HU LUT (CT calibration curve) assigned \\
  
  \ProcessGroup{Anatomy Contouring (Delineation)}
  7.c & Normal-tissue delineation & Immobilisation device outside calculation contour \\
  8.a & Density override for artificial materials and artifacts & Incorrect (or missing) density override \\
  
  \ProcessGroup{Treatment Planning}
  11.b & Treatment fields - Insertion & Wrong treatment machine \\
  11.c & Treatment fields - Insertion & Wrong treatment accessory \\
  12.a & Treatment fields - Gantry/couch angle selection  & Incorrect gantry/couch limits \\
  12.b & Treatment fields - Gantry/couch angle selection  & Incorrect couchtop/immobilisation avoidance \\
  12.c & Treatment fields - Gantry/couch angle selection  & Poor gantry angle selection \\
  16.a & Optimization & Production of undeliverable plan \\
  17.a & Forward calculation & Improper selection of calculation algorithm \\
  17.b & Forward calculation & Inappropriate calculation uncertainty settings \\
  17.d & Forward calculation & Error in forward-calculation base data \\
  17.g\footnote{This failure mode was not (explicitly) modeled in the corresponding workflow map.} & Forward calculation & Inaccurate dose calculation \\
  
  \ProcessGroup{Plan Preparation}
  20.b & Plan export/import to OIS & Plan imported into wrong treatment room \\
  20.c & Plan export/import to OIS & Data corrupted during transfer \\
  20.e & Plan export/import to OIS & TPS prescription differs from OIS prescription \\
  20.h & Plan export/import to OIS & Important treatment parameters changed or set to not intended values\footnote{For example, gantry angle, snout, position, MU scaling.} \\
  20.i & Plan export/import to OIS & Important setup/field parameters changed or set to not intended values\footnote{For example, couch and tolerance tables.} \\
  21.a & Image guidance preparation & Wrong isocenter defined for set-up imaging \\
  21.b & Image guidance preparation & Wrong couch shift entered for plan with multiple isocenters \\
  22.a & Review and approval of treatment data & No identification of parameter discrepancy during automated check\\
  22.b & Review and approval of treatment data & No identification of parameter discrepancy during manual check \\
  23.a & Treatment scheduling & Too few fractions scheduled compared to prescription \\
  
  \ProcessGroup{Final Plan Approval}
  35.a & Plan approval & Wrong plan approved \\
  35.b & Plan approval & Plan with errors approved \\
  
  \ProcessGroup{Treatment}
  36.b & Send plan from OIS to TCS & Data corrupted during transfer \\
  36.c & Send plan from OIS to TCS & Incorrect recording/resume of partial delivery \\
  36.d & Send plan from OIS to TCS & New plan uploaded or plan with reference changed \\
  37.a & Translation of plan to TCS internal settings & Change in lookup table \\
  37.b & Translation of plan to TCS internal settings & Change in translation algorithm \\
  37.c & Translation of plan to TCS internal settings & Wrong general beam-modifying device (e.g. range shifter) \\
  37.e & Translation of plan to TCS internal settings & Plan is not deliverable (TPS allows values not possible in TDS) \\
  38.a & Patient imaging and alignment & Mistake in couch correction or couch position for fields or setup \\
  39.a & Set treatment parameters & Wrong parameters uploaded\\
  40.a & Treatment accessories & Missing or incorrect patient-specific beam-modifying device \\
  42.a & Beam delivery & Unwanted object in beam path \\
  42.b & Beam delivery & Non-therapeutic radiation detected in MU chamber\footnote{For example, noise, vibration, x-rays, dark current.} \\
  42.d & Beam delivery & Beam delivery error (MU, spot position, energy, etc.) \\
  42.g & Beam delivery & Machine abort incorrectly handled \\
  42.h & Beam delivery & Machine delivery failure \\
  42.i & Beam delivery & Mistake in follow-up of interrupted fields\footnote{For example, loss of dose not compensated due to unclear presentation in OIS, etc.} \\
  43.a & Data recording and OIS logging & Treatment not recorded to OIS \\
  43.b & Data recording and OIS logging & Incorrect recording of treatment MU \\
  
  \ProcessGroup{End}
  44.a & Final steps & Delay in post-treatment verification with possible compensations not performed \\ \hline \\
  
  \ProcessGroup{Risk IDs not taken into account\footnote{These (detailed) failure modes of the corresponding process steps were removed from the pFMEA elicitation and analysis - although they appear in the workflow map - because less than half of the participating experts were able to score them due to a lack of experience.}}
  17.e & Forward calculation & Dose distribution problem not observed due to secondary dose calculation (SDC) in water \\
  17.f & Forward calculation & Corrupted data transfer to SDC system \\
  43.c & Beam delivery & Failure to write machine log \\
  43.d & Beam delivery & Erroneous data saved to machine log \\ \hline
\end{longtable}
\end{center}
\renewcommand{\arraystretch}{1}

\section{Inspection of the \textit{No-PSQA} PBS-PT Risk Landscape}
\label{sec:no_psqa_risk_landscape} 

\subsection{The \textit{No-PSQA} Radar Plot}
\label{subsec:no_psqa_radar_plot}

\subsubsection{Why Radar Plots?}
The risk model introduced in Sec.~\ref{sec:operational_framework} and formalized in Appendix~B assigns a discrete Risk Priority Number (RPN) to each of 44 Risk-IDs, distributed over 20 process steps (with 44 failure modes) of a generic pencil-beam–scanning proton-therapy (PBS-PT) workflow. 

These 44 values form a \emph{multivariate} signature that cannot be reduced to a single histogram or bar chart without sacrificing the process–step/failure-mode context that is vital for clinical interpretation. Radar (or spider) plots solve this dilemma in three ways:

\begin{enumerate}[label=(\alph*)]
  \item \textbf{Dimensional fidelity.}  
        Each Risk-ID is represented by its own radial axis; no dimension is collapsed or aggregated.
  \item \textbf{Instant comparative overlay.}  
        Multiple polygon traces — baseline vs.\ PSQA‐adjusted — can be drawn in a single glyph, enabling a one-glance assessment of risk contraction or expansion.
  \item \textbf{Process-aware ordering.}  
        By arranging the 44 axes counter-clockwise in ascending process-step order (numerical ID; midnight start), the plot preserves the sequential nature of the workflow while highlighting inter- and intra-process-step interactions (visual proximity of pipes “|” marking process-step transitions).
\end{enumerate}

\begin{figure}[h!]
  \centering
  \includegraphics[width=0.6\textwidth]{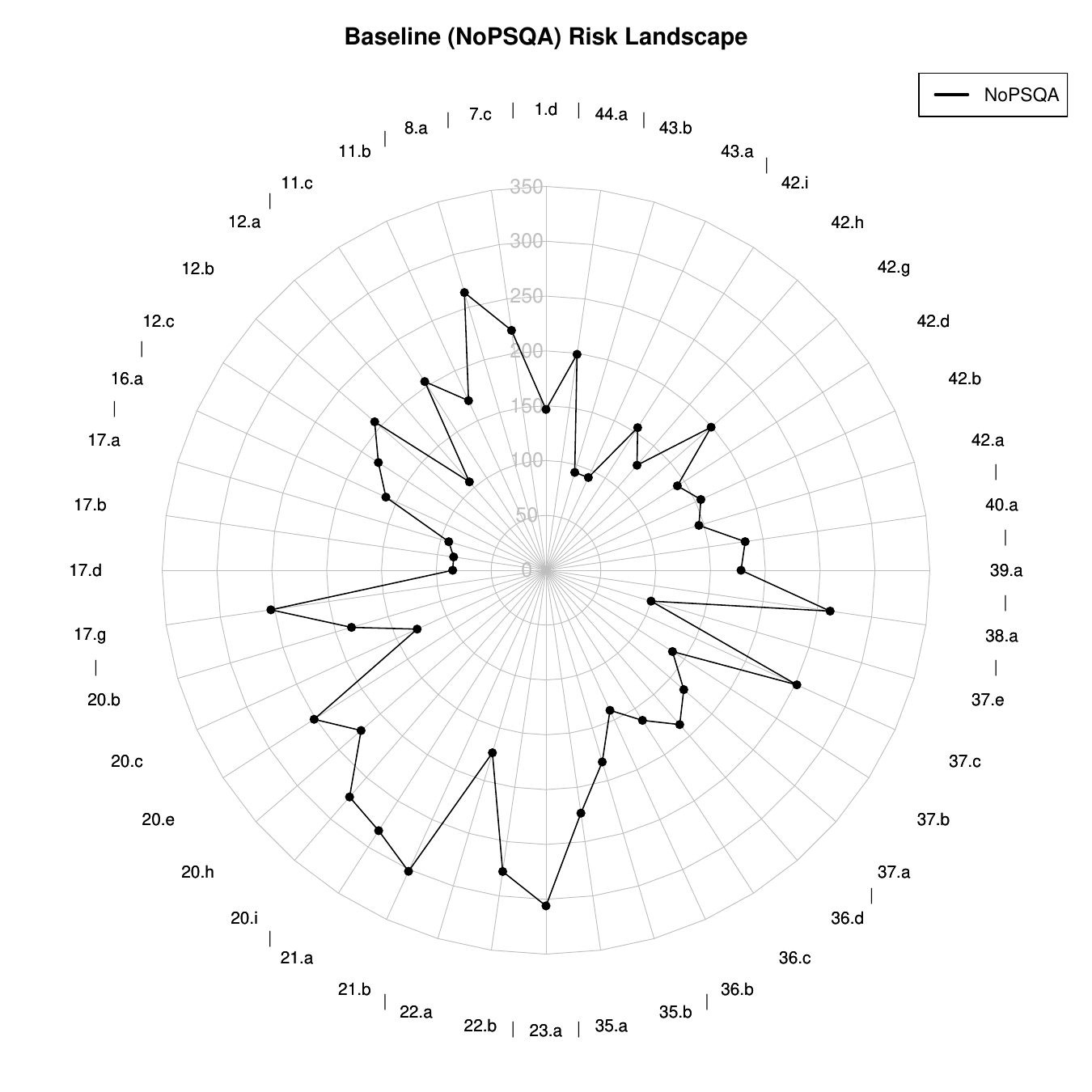}
  \caption{\small Baseline (NoPSQA) risk landscape for the PBS-PT workflow. Each radial axis corresponds to one Risk-ID (44 total); the radial spot position marks its RPN per failure mode (\#FMs = \#Risk-IDs) before patient-specific QA is applied.}
  \label{fig:no_psqa_radar_of_riskIDs_baseline}
\end{figure}
\begin{figure}[h!]
  \centering
  \includegraphics[width=0.6\textwidth,clip,trim=18mm 5mm 15mm 5mm]{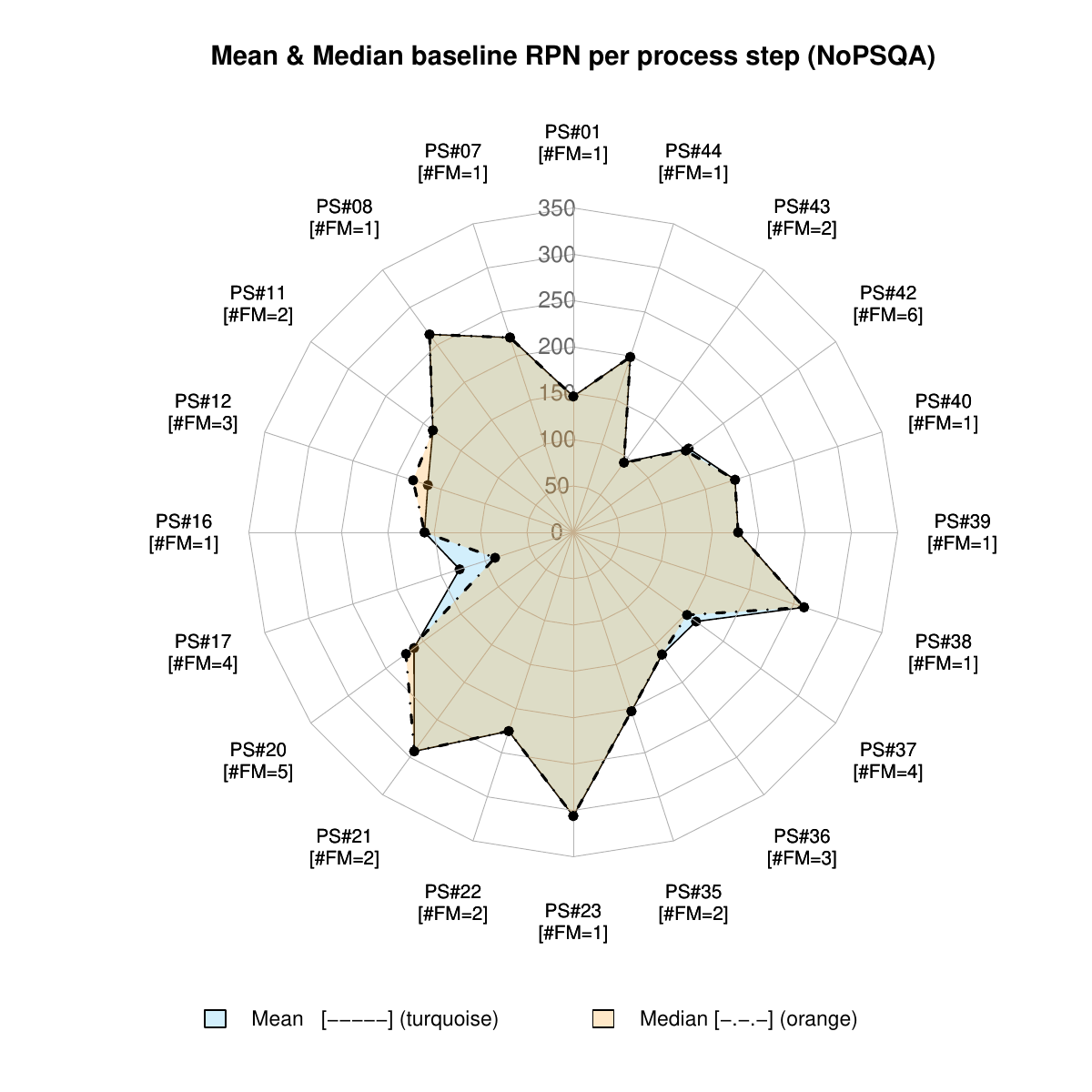}
    \caption{\small Radar representation of the \textbf{mean} (turquoise, solid outline) and \textbf{median} (orange, dashed outline) baseline RPN per process step in the \emph{No-PSQA} reference scenario. Axis labels follow the notation \texttt{PS\#\,$xx$\,[\#FM=$yy$]}, where $xx$ is the sequential process-step number and $yy$ the number of failure modes aggregated. The (almost complete) overlap of the two filled areas produces the olive shade visible in well-balanced steps.}
  \label{fig:no_psqa_radar_of_PSs_mean_median_baseline}
\end{figure}

For these reasons we adopt radar plots as the \emph{default} vehicle for subsequent visualisations of the complete workflow and its local risk behaviour.

\subsubsection{Anatomy of the Baseline Radar Plot}
Figure~\ref{fig:no_psqa_radar_of_riskIDs_baseline} depicts the \emph{NoPSQA} risk landscape — i.\,e.\ the raw pFMEA scoring before any patient-specific QA is applied.  The construction principles are as follows:

\begin{itemize}
  \item \textbf{Axes and segmentation.}  
        Forty-four equi-angular axes (each 8.2$^\circ$ between adjacent axes) span the full circle, generating 44 spider-web sectors.
  \item \textbf{Radial scale.}  
        All axes share a common linear scale from $0$ (center; in a strict sense it is $1$ on the RPN scale) to $350$ RPN units (outer rim), matching the upper bound of the baseline scores in our data set.
  \item \textbf{Polygon trace.}  
        Baseline RPNs are plotted as black dots on their respective axes and connected to form a closed polyline. Outward spikes flag process steps with intrinsically high risk; inward dents
        indicate relatively benign modes.
  \item \textbf{Labelling convention.}  
        Axis labels combine the numeric process-step ID with its associated lower-case failure-mode letter(s). Pipe symbols (‘\,\texttt{|}\,’) visually separate consecutive process steps.
\end{itemize}

This baseline glyph serves as the \emph{reference silhouette} against which all PSQA-specific polygons will be evaluated.

\subsubsection{Mean– and Median-Aggregated View of the Baseline Risk Landscape\label{subsubsec:mean_median}}
The overlaid radar chart (Fig.~\ref{fig:no_psqa_radar_of_PSs_mean_median_baseline}) shows the mean (turquoise polygon) and median (orange polygon) of the base RPNs assigned to the associated failure modes for each process step\footnote{It should be noted that the angular distances between the axes per process step (total number of PS is 20) are now (Fig.~\ref{fig:no_psqa_radar_of_PSs_mean_median_baseline}) uniformly 30°, whereas in Fig.~\ref{fig:no_psqa_radar_of_riskIDs_baseline} they are determined by the angular differences between two pipe symbols. This results in optical distortions between the two polygon lines when both figures are superimposed in the mind's eye.}. For most process steps, the two polygons almost coincide (e.g., PS\#08, PS\#38), indicating that the underlying risk ID distribution is fairly symmetrical (overlay filled in olive). 

A distinct turquoise area (e.g., PS\#17 “Forward calculation”) shows a single subgroup of failure modes in which the mean value is higher than the median. In contrast, a narrower, outwardly protruding orange region shows the opposite effect in terms of the mean and median values for PS\#12 (“Treatment fields – Selection of gantry/table angle (beam arrangement)”).

\subsection{The \textit{No-PSQA} Bar Chart}
\label{subsec:no_psqa_bar_chart}
\begin{figure}[h!]
  \centering
  \includegraphics[width=0.8\textwidth,keepaspectratio]{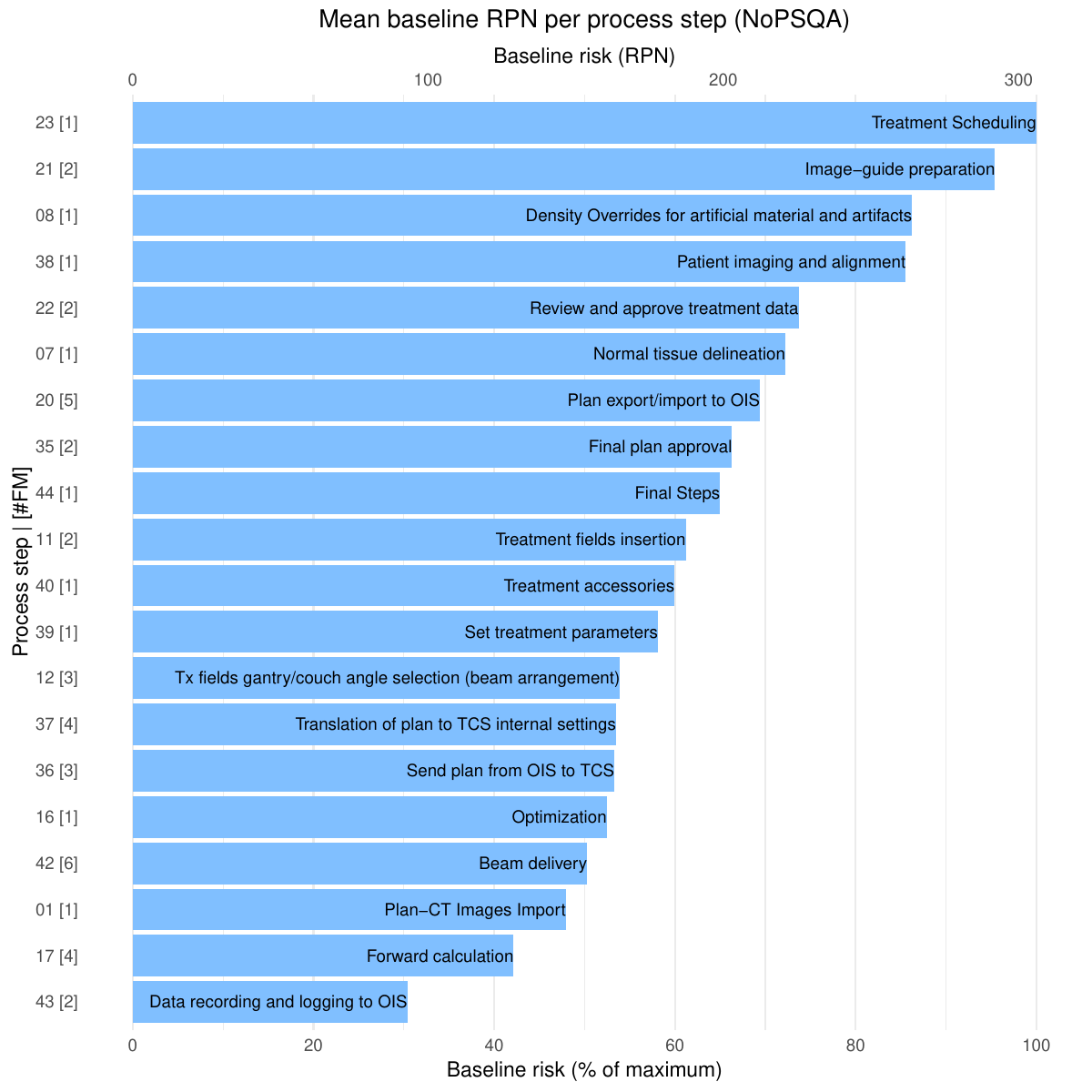}
  \caption{\small Horizontal ranking of the \emph{mean} baseline RPN per process step for the \emph{No-PSQA} case. The primary $x$-axis (bottom) gives the risk as a percentage of the maximum scored value (normalization), while the secondary axis (top) shows the corresponding absolute RPNs with a maximum of \SI{306}{RPN~units}. Labels combine the clinical descriptor with the process step ID and the number of failure modes contributing (\texttt{[\,\#FM\,$=$\,\,]}); bars are coloured in a pale blue consistent with the fill of the means in Fig.~\ref{fig:no_psqa_radar_of_PSs_mean_median_baseline}.}
  \label{fig:no_psqa_bar_of PSs_(mean)_baseline}
\end{figure}

Figure \ref{fig:no_psqa_bar_of PSs_(mean)_baseline} visualizes the aggregation of the 44 Risk-IDs into the 20 underlying process steps (shown in radar plot form in Figure  \ref{fig:no_psqa_radar_of_PSs_mean_median_baseline}) in a complementary way.

Quantifying the same information in a different visual grammar, the bar chart of Fig. \ref{fig:no_psqa_bar_of PSs_(mean)_baseline} ranks the \emph{mean} baseline RPN per process step from lowest (bottom) to highest (top).
\begin{figure}[h!]
  \centering
  \includegraphics[width=0.9\textwidth]{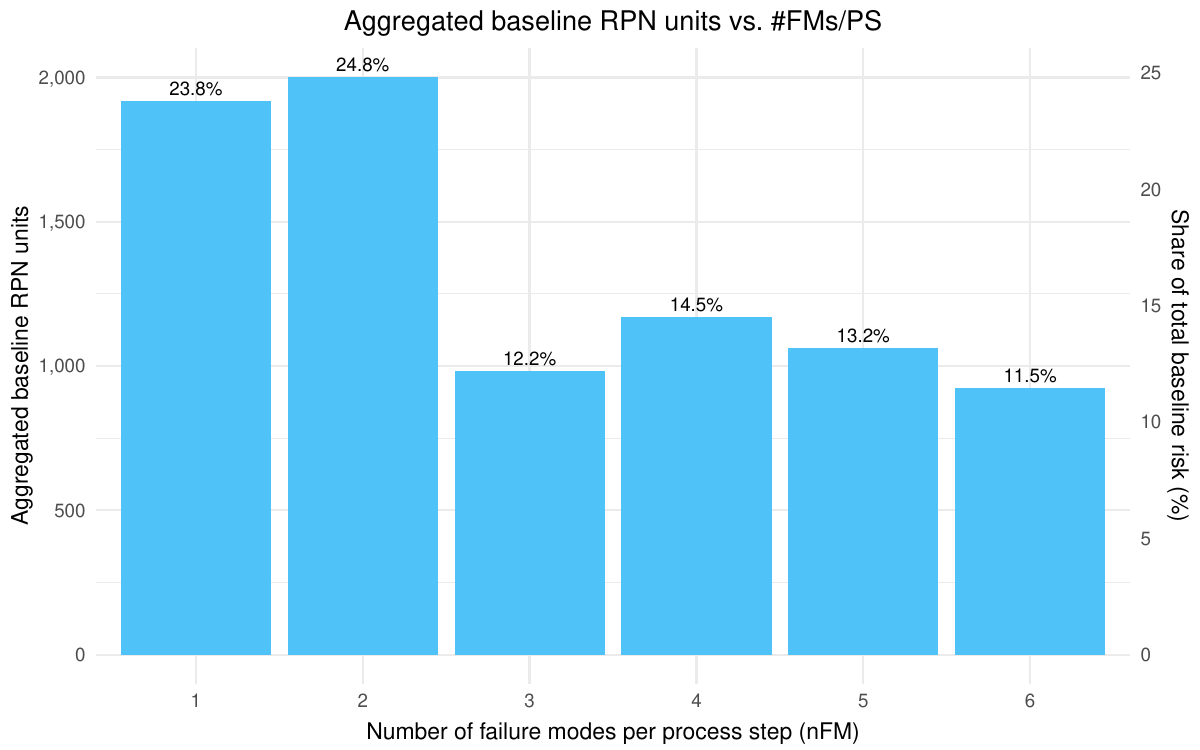}
  \caption{\small Aggregated baseline risk (RPN units, left axis) versus number of failure-modes per process step ($nFM=\#FMs$ per PS). A larger number of “simple” process steps with one–two FMs account for $\approx 49 \%$ of total risk, while the remaining $\approx 51 \%$ is distributed across a few steps containing three to six FMs, indicating comparable total risk share by higher structural complexity.}
  \label{fig:sum_of_baseline_RPNs_vs_nFM_by_PS}
\end{figure}
\vspace{-5mm}
Three observations stand out:
\begin{enumerate}[label=\alph*)]
  \item Even after \emph{averaging} on the process step level, the \emph{local} risk remains widely and asymmetrically distributed: when normalized to the maximum RPN value with about 300 RPN units (attributed to PS\#23 - \emph{Treatment Scheduling}), the bottom quarter of all process steps, i.e., only 25\%, including PS\#42 (\emph{Beam delivery}) and PS\#17 (\emph{Forward calculation}), aligns with the bottom 50\% of the maximum risk. Conversely, this means that 75\% of all process steps are within the upper half of the (normalized) risk range.
  \item On the other hand, if one argues that the number of contributing failure modes per process step (denoted $nFM_{i}$) is of substantial relevance for a process step-specific risk estimation , a look at the following histogram provides important insights.\\
      Fig. \ref{fig:sum_of_baseline_RPNs_vs_nFM_by_PS} aggregates the contribution to the total absolute workflow risk as a function of the number of failure modes per process step via their RPN sums, i.e., it determines how much “simple” process steps (with only 1–2 failure modes) contribute to the overall risk compared to “riskier” process steps ($\#[FMs] \geq 3$).\\
      While the left scale shows the sum of RPN units, the right scale – together with the percentage values on the individual bars – provides an immediate impression of how strongly the complexity of the process step's “risk structure” (based on the number of FMs) influences the absolute total risk of the workflow. Here, this measure is divided roughly equally between the “simple” ($23.8\% +24.8\%$) and “more complex” ($12.2\%+14.5\%+13.2\%+11.5\%$) process steps (PS).
  \item Finally, if one compares the bar chart from Fig. \ref{fig:no_psqa_bar_of PSs_(mean)_baseline} with the process step-specific radar plot (Fig. \ref{fig:no_psqa_radar_of_PSs_mean_median_baseline}), it can be seen in both figures that the process steps in the 20's clearly indicate the most risky workflow phase under \emph{No-PSQA} conditions. This observation can be identified as the process of \emph{Plan preparation (VII.)} with the help of Table \ref{tab:riskid_catalogue}. When subsequently transferring this result to the workflow map (\ref{fig:PSQA_noQA}), it appears somewhat surprising, as the visual impression of the failure mode distribution would probably have seen the \emph{Treatment (X.)} process in this role.
\end{enumerate}

Together, the last three figures close the descriptive and semi-quantitative loop of the \emph{No-PSQA} scenario: Fig.~\ref{fig:no_psqa_radar_of_PSs_mean_median_baseline} shows \emph{where} in the workflow skewed or symmetric risk distributions occur, while the Figs. \ref{fig:no_psqa_bar_of PSs_(mean)_baseline} and \ref{fig:sum_of_baseline_RPNs_vs_nFM_by_PS} orders those findings by weight(s). In the following subsections we will revisit both visualisations after successively activating each PSQA modality, allowing a direct, process-resolved comparison of their mitigation power.

\section{Modulation of Absolute RPN Changes by PSQA Modalities}
\label{sec:abs_local_risk_changes}

The baseline silhouette established in Section \ref{sec:no_psqa_risk_landscape} serves as the reference for assessing the absolute overall risk impact of each PSQA approach, following the formal notations defined and derived in Appendix~B, particularly in Subsecs.~\ref{appB:subsec:fm_abs_rel_risk_changes} and \ref{appB:subsec:fm_cumulative_changes}.

Figures \ref{fig:psqa_meas_radar}, \ref{fig:psqa_log_radar}, and \ref{fig:psqa_calc_radar}, overlay the baseline RPN distributions (black) with two PSQA-specific risk traces, one for the \textcolor{red}{data-specific PSQA influence} (\textit{data}: [black,\.red]; \emph{value pair}: $\omega^{*,data}_{i,\,l}$) and the second for the \textcolor{blue}{complete verification effects} (\textit{full}: [red,\.blue]; \emph{value pair}: $\omega^{*,full}_{i,\,l}$) (see Eqs.~\eqref{appB:eq:rpn_pair}, \eqref{appB:eq:abs_risk_change_pair}, and \eqref{appB:eq:rel_risk_change_pair}). An inward shift indicates a reduction in risk, while an outward shift indicates an additional hazard. The risk-scoring-phase-dependent polygons (see legends) connect all $\mathrm{RPN}^{*,\mathrm{pc}}_{i,\,l}$ value spots along all FMs.

\subsection{Measurement-based Verification (\textit{Meas})}

\begin{figure}[h!]
  \centering
  \includegraphics[width=1.0\linewidth]{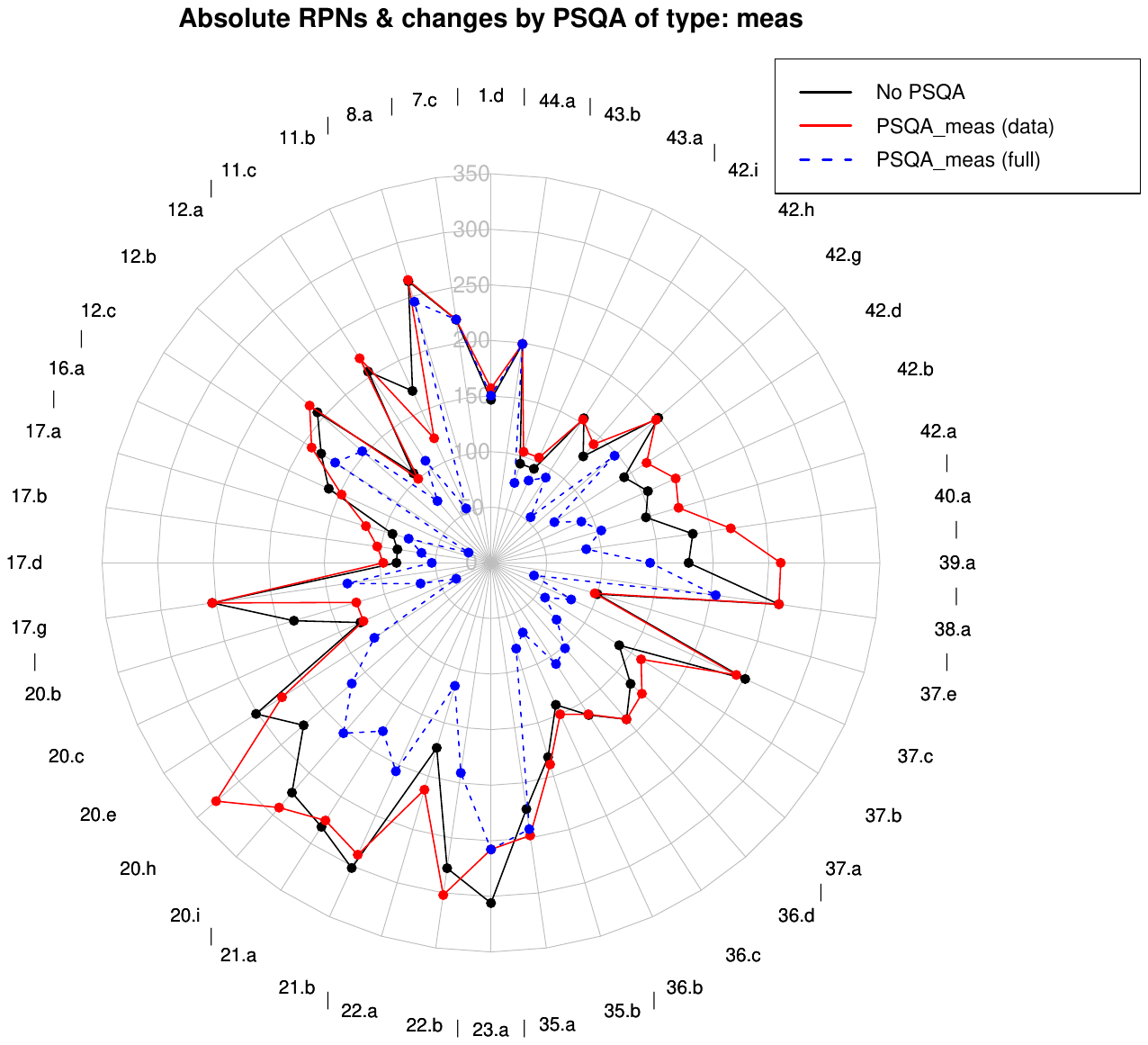}
  \caption{\small \textbf{Baseline RPNs (black) versus risk changes after measurement-based verification (\textit{Meas}).}  
  The blue trace contracts across most axes but to a lesser extent than the log file-based approach, ranking \textit{Meas} in second place visually in terms of aggregate risk mitigation. 
  }
  \label{fig:psqa_meas_radar}
\end{figure}

The blue \textit{Meas-Full} curve in Figure \ref{fig:psqa_meas_radar} reproduces the shape of the black baseline polygon more accurately than any other PSQA modality, suggesting that measurement-based verifications register risk changes that are roughly proportional to the original RPN distribution.
  
The RPN values for PS \#36-40 of \textit{Meas-Full} (blue dots) - reflecting the mitigated risk of correct beam delivery by PSQA\textsubscript{\textit{Meas}}-, decrease continuously, but less significantly than in the case of PSQA\textsubscript{\textit{Log}} (this may reflect the assumption that \textit{Log} includes daily dose recalculation and log file analysis). Interestingly, the RPNs for forward dose calculations (PS\#17) differ only slightly from their counterparts in \textit{Calc-Full} (Fig.~\ref{fig:psqa_calc_radar}), indicating that ISDC does not significantly reduce these treatment planning-related risks in direct comparison to PSQA\textsubscript{\textit{Meas}}. 

Consequently, \textit{Meas} ranks \emph{second} in a visual comparison with regard to overall risk reduction at first glance – substantially better than \textit{Calc}, but significantly behind \textit{Log}.

The red polygon for \textit{Meas-Data} shows an overall inconsistent trend, whereby the general tendency is toward a slight increase in the data-driven risk contribution.

\subsection{Log File–based PSQA (\textit{Log}) — the Strongest Mitigator}
\label{subsec:log_the_strongest_mitigator}

\begin{figure}[H]
  \centering
  \includegraphics[width=1.0\linewidth]{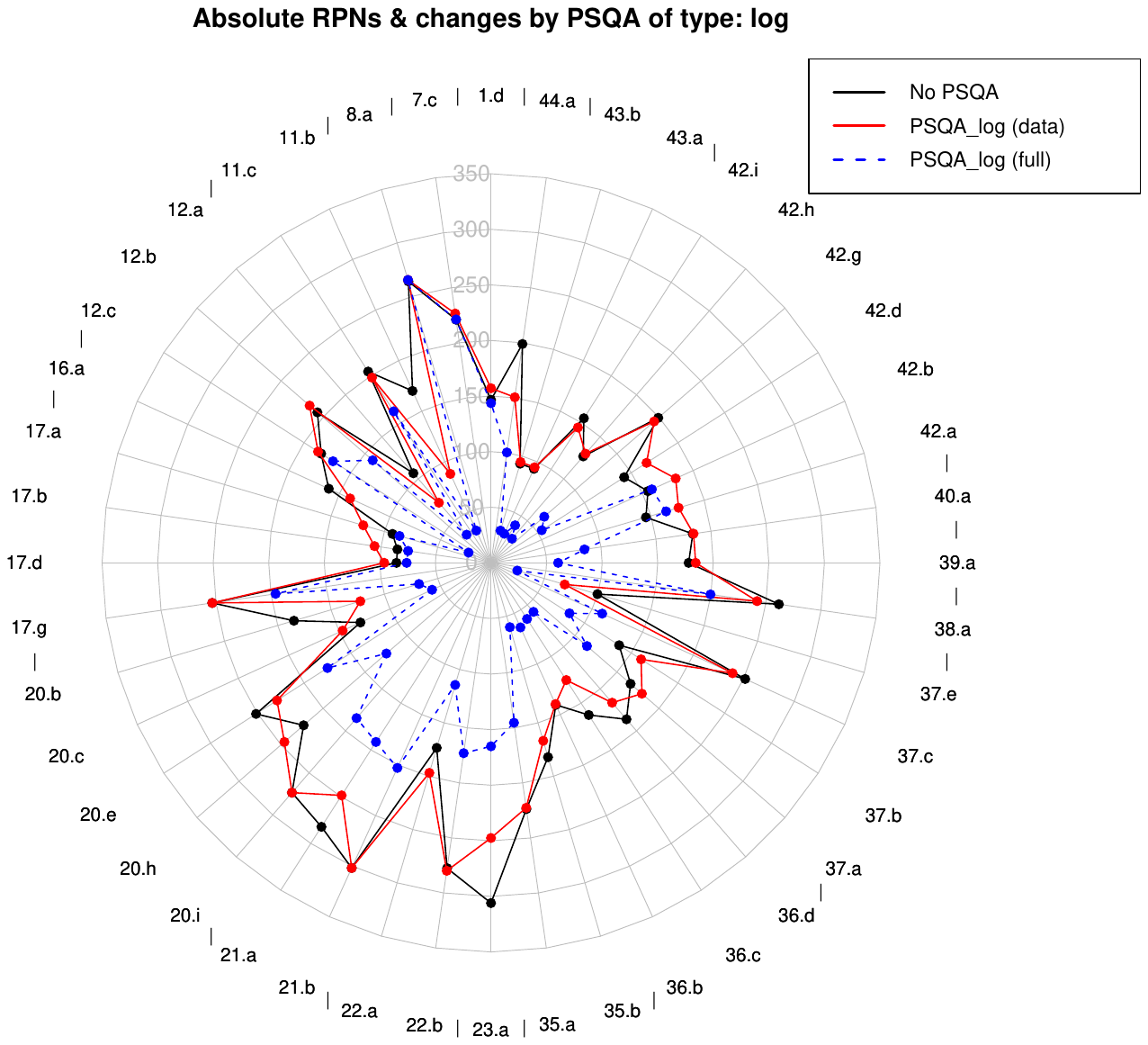}
  \caption{\small \textbf{Baseline RPNs (black) versus risk changes after log file–based PSQA (\textit{Log}).}  
  The blue polygon \textit{Log-Full} shows the strongest contraction of all three PSQA modalities, while the beam delivery and treatment-related axes (PS \#36–\#43) reduce their values toward an RPN range of approximately 50–75, in some cases even below 50. The data-relevant curve marked in red causes only local, minor outward deflections, i.e., \textit{Log-Data} behaves similarly to \textit{NoPSQA} in terms of risk.}
  \label{fig:psqa_log_radar}
\end{figure}

Figure \ref{fig:psqa_log_radar} shows the strongest contraction of the blue polygon: Virtually all axes related to beam delivery (PS \#36–\#43) converge toward the origin\footnote{\label{fn:largest_drel_log_value}It should be explicitly noted that we are \textbf{not} referring to \textbf{all} risk axes in this area. As the attentive reader may have noticed, the RPN values of \textit{Log-Data/Full} within \textbf{PS\#42} for failure modes \textbf{a} and \textbf{b} are \textbf{the only} RPNs (from \textit{Log-Full}) that \textbf{exceed} the corresponding baseline values (red and blue outside black).\\
This risk assessment is based on the assumption that the patient-specific recording of the QA log files could occur under technically faulty conditions which are not part of the TPS calculations or real application conditions, as could happen, e.g., due to an unwanted object in the beam path (loose cable inside the nozzle, etc.). This example illustrates the importance of a thorough pFMEA, in which even the most promising solution package to a problem could still be locally flawed.}, whereby the remaining RPNs are well below 75. 
At the same time, there has also been a notable reduction in the high-risk workflow phase \emph{Plan preparation} (PS \#20–\#23), although the area between RiskIDs $20.e$ and $35.a$ still represents the largest contiguous area with a significant residual risk (despite \textit{Log-Full} measures) with approximately 150 RPN units per RiskID.

The red data track \textit{Log-Data} induces only modest outward spikes, as log file-based data extraction occurs within the treatment-control environment and requires minimal data transfer and file shuffling. Taken together, \textit{Log} visually offers the \emph{largest net reduction} in absolute RPN units across the workflow.

\begin{quote}
\textbf{Caveat:} It should be noted that the advantage of \textit{Log-Full} over \textit{Meas-Full} is partially based on the assumption that log file-based verifications involve daily dose recalculation and performance evaluation based on the machine protocols. If such daily analysis of the log files is not performed, it is probably to be expected that the risk reduction level of PSQA\textsubscript{\textit{Log}} will approach that of PSQA\textsubscript{\textit{Meas}}.
\end{quote}

\subsection{Independent Secondary Dose Calculation (\textit{Calc})}
\begin{figure}[h!]
  \centering
  \includegraphics[width=1.0\linewidth]{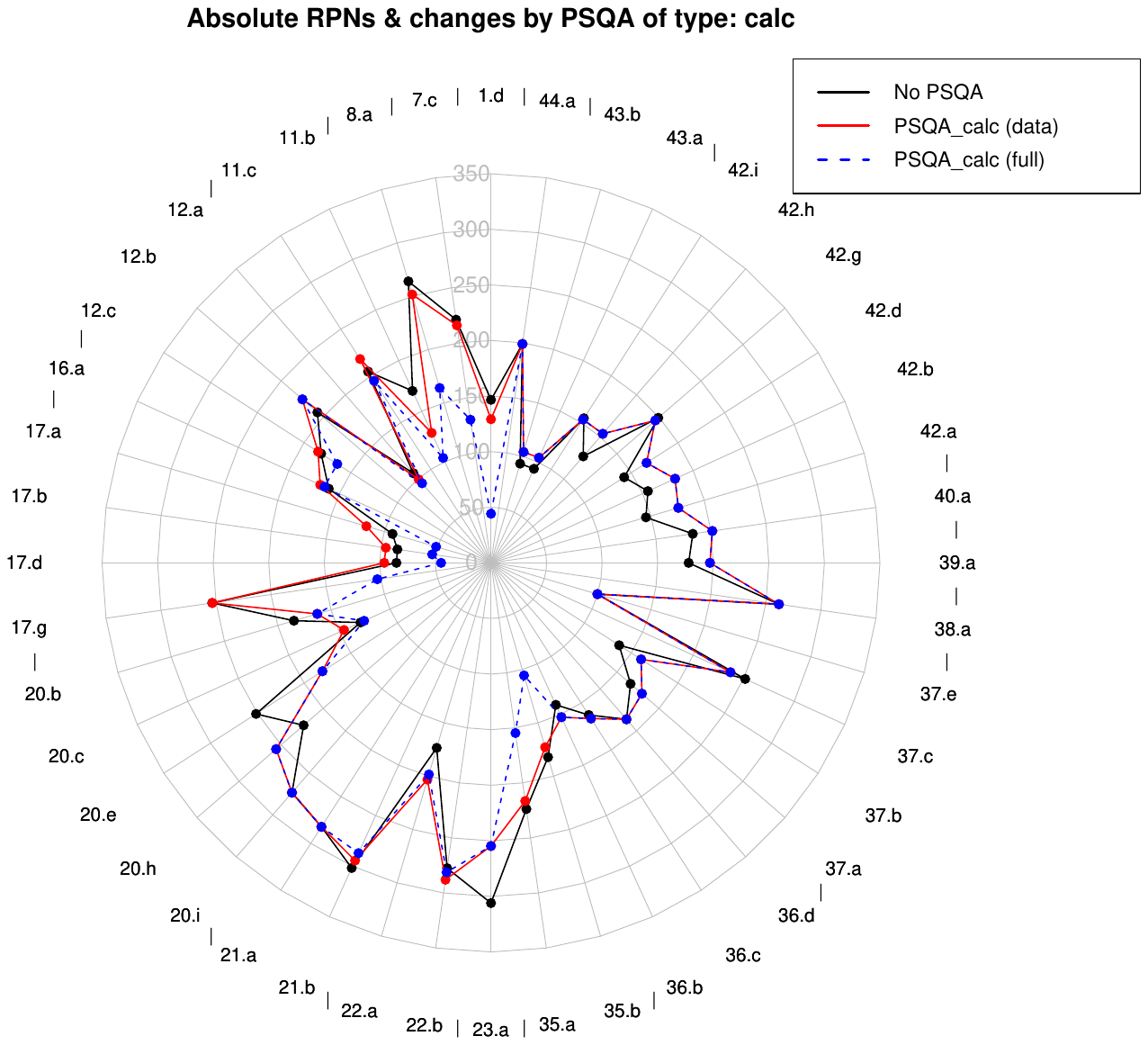}
  \caption{\small \textbf{Baseline RPNs (black) versus risk changes after ISDC.}
  \textcolor{blue}{Blue} denotes full ISDC application; \textcolor{red}{red} captures auxiliary data-handling influence.  Noticeable risk contraction is limited to the initial process steps PS  (forming the processes \emph{Transfer Images} and \emph{Delineation}), PS \#17 (Forward calculation), and PS \#35 (Plan approval), while the high-risk workflow phases of \emph{Plan preparation} and \emph{Treatment} remain essentially unchanged.}
  \label{fig:psqa_calc_radar}
\end{figure}

In Figure \ref{fig:psqa_calc_radar}, the blue \textit{Calc-Full} curve - representing the entire application of PSQA\textsubscript{\textit{Calc}} or ISDC measures - drops mainly on calculation-related axes (PS \#1-8,\#17,\#35), confirming that the independent secondary calculation of the dose distribution significantly reduces the \emph{risk of algorithmic and parameter errors}.

The axes oriented toward beam application (PS \#36–40) remain virtually unchanged, and the red \textit{Calc-Data} track shows a more or less uniform offset toward the outside, indicating that data transfer and database or file transactions required for ISDC can actually increase the baseline risk without full PSQA. Overall, \textit{Calc} achieves the visibly \emph{lowest} total contraction of the blue polygons (here \textit{Calc-Full}) of the three PSQA modalities.

\subsection{Cross-Modality Synthesis}
\label{subsec:cross_modality_synthesis}
A direct comparison of Figures \ref{fig:psqa_meas_radar}–\ref{fig:psqa_calc_radar} yields two robust observations:

\begin{enumerate}
  \item \textbf{Dominance of full verifications.}  
        The blue traces always contribute the lion's share to risk reduction, while the red data traces could potentially lead to an increase in risks throughout the entire workflow, at least locally, making their influence on \textit{Meas} appear to be most pronounced visually.
  \item \textbf{Mitigation hierarchy.}  
        The visual comparison of the blue polygons provides the arrangement
        \begin{equation}
            \label{eq:psqa_risk_mitigation_hierarchy}
            \overline{\textsc{Log}} > \overline{\textsc{Meas}} \gg \overline{\textsc{Calc}}, 
        \end{equation}
        whereby the comparison operator $>$ ($\gg$) indicates a (significantly) larger risk reduction in the sense of a (more) pronounced \emph{polygon contraction} averaged over the global workflow.
        Locally, \textit{Log} reduces the axes that are most relevant to beam application, but also shows significant potential for risk reduction in the axis values for treatment preparation; \textit{Meas} provides a more consistent but slightly less pronounced risk reduction, while \textit{Calc} only has a recognizable positive effect in terms of lower risk on dose calculation-related axes.
\end{enumerate}

These mainly \textbf{qualitative} findings - derived from distributions of \textbf{absolute} risk changes - form the basis for the following sections, which present
\begin{itemize}
    \item a comparative analysis with regard to the distributions of \textbf{relative} risk changes,
    \item a \textbf{semi-quantitative} calculation of key figures for the entire workflow, in which the effects of each PSQA method on the individual risk budgets are analyzed and evaluated.
\end{itemize}

\section{PSQA-Dependent, Unweighted Relative RPN Changes} 
\label{sec:rel_local_risk_changes}

Failure mode-specific relative risk changes were quantified as percentage deviations\footnote{Recall that for each Risk-ID, i.e.\ for each failure mode, relative risk changes are expressed in percent. For the two \textit{NoQA}-anchored stages \textit{Data} and \textit{Cum}, the corresponding quantities are
\[
\delta^{\ast,(d)}_{\mathrm{rel},i,l}
=
\left(
\frac{\mathrm{RPN}^{\ast,\mathrm{data}}_{i,l}}{\mathrm{RPN}^{\mathrm{noQA}}_{i,l}}-1
\right)\times 100
\qquad\text{and}\qquad
\delta^{\ast,(c)}_{\mathrm{rel},i,l}
=
\left(
\frac{\mathrm{RPN}^{\ast,\mathrm{full}}_{i,l}}{\mathrm{RPN}^{\mathrm{noQA}}_{i,l}}-1
\right)\times 100.
\]
Negative values indicate risk mitigation, positive values indicate risk increase. In Subsec.~\ref{subsec:results_d2c_relative}, we additionally use the signed bridge quantity
\[
\delta^{\ast,(d\to c)}_{\mathrm{rel},i,l}
:=
\delta^{\ast,(c)}_{\mathrm{rel},i,l}-\delta^{\ast,(d)}_{\mathrm{rel},i,l},
\]
which measures the extra \textit{Full}-stage displacement from \textit{Data} to \textit{Cum} on the same common \textit{NoQA} scale.}
between the baseline (\textit{NoQA}) and the three independent PSQA approaches (\textit{Meas}, \textit{Log}, \textit{Calc}). In the present ordering of this section, we first analyze the two quantities that are directly comparable on a common reference scale, namely the \textit{Data}-stage and cumulative relative changes (\textit{Cum}), and only thereafter the additional signed transition from \textit{Data} to \textit{Cum}. 

This sequencing is deliberate: both \textit{Data} and \textit{Cum} are normalized to the same \textit{NoQA} baseline and can therefore be displayed and interpreted on identical radar-plot axes, whereas the isolated \textit{Full}-stage quantity remains normalized to the method-specific \textit{Data}-stage state and is therefore less suitable as a primary cross-modality visualization.

Accordingly, Subsec.~\ref{subsec:pc_data_on_relative_RPN_scale} starts with the \textit{Data}-stage polygon tracks, followed in Subsec.~\ref{subsec:pc_full_on_relative_RPN_scale} by the cumulative end-state polygon tracks, both shown for the 44 validated Risk-IDs on the same radial scale from $-100\,\%$ (center) to $+50\,\%$ (outer ring), with dummy pipes separating higher-level process steps. Subsec.~\ref{subsec:results_d2c_relative} then uses this bridge-shift logic to visualize how strongly the completion of the \textit{Full}-stage moves each PSQA method from its \textit{Data}-stage position to its final cumulative state, thereby providing a mathematically and graphically coherent link between the preceding radar-plot analyses\footnote{Recall that the formal \textit{Full}-stage quantity
\[
\delta^{\ast,(f)}_{\mathrm{rel},i,l}
=
\left(
\frac{\mathrm{RPN}^{\ast,\mathrm{full}}_{i,l}}{\mathrm{RPN}^{\ast,\mathrm{data}}_{i,l}}-1
\right)\times 100
\]
is still part of the underlying stage model and remains necessary for the algebraic decomposition of cumulative effects. However, since it is referenced to the already modified \textit{Data}-stage state, it is now introduced only indirectly through the bridge quantity
\[
\delta^{\ast,(d\to c)}_{\mathrm{rel},i,l}
=
\delta^{\ast,(c)}_{\mathrm{rel},i,l}-\delta^{\ast,(d)}_{\mathrm{rel},i,l}
=
\delta^{\ast,(f)}_{\mathrm{rel},i,l}\bigl(1+\delta^{\ast,(d)}_{\mathrm{rel},i,l}\bigr),
\]
which is visualized in Subsec.~\ref{subsec:results_d2c_relative} by dumbbell plots and corresponding histograms. In this way, the additional effect of completing the \textit{Full}-stage is represented on the same common \textit{NoQA} scale that already governs the \textit{Data}- and \textit{Cum}-track analyses.}.

\subsection{Results by Process Category \textit{Data} on Relative RPN Changes}
\label{subsec:pc_data_on_relative_RPN_scale}

\begin{figure}[h]
\begin{center}
\includegraphics[width=1.1\linewidth,trim=3.5cm 1cm 0cm 0cm,clip]{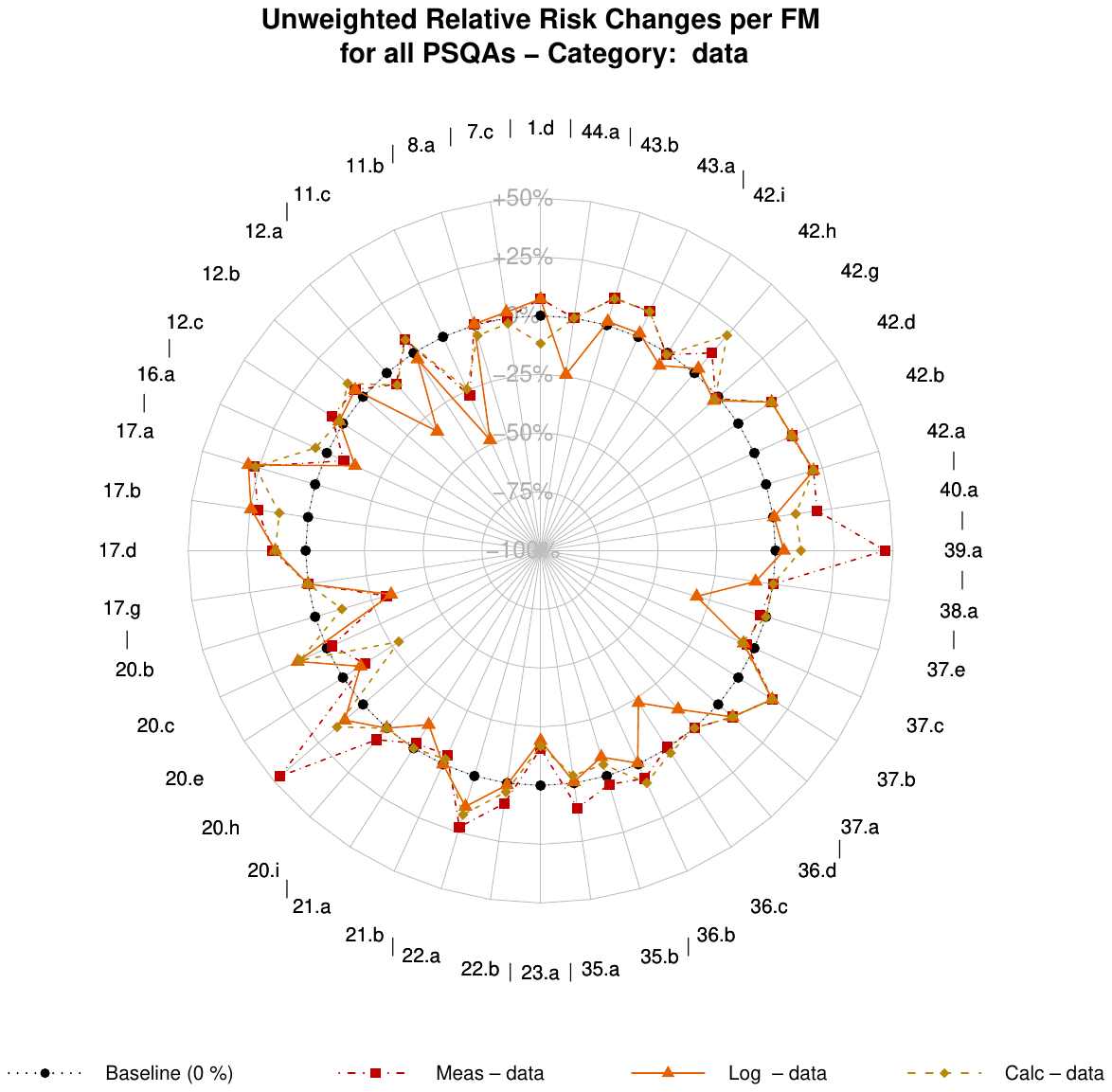}
\definecolor{psqaLog}{HTML}{E66100}
\definecolor{psqaMeas}{HTML}{C00000}
\definecolor{psqaCalc}{HTML}{B8860B}

\caption{Unweighted relative risk changes per FM (= Risk-ID) for \textit{Data}-only PSQA components; legend:
\textit{Baseline} \protect\legendlinemark{dotted}{black}{circle},\;
\textit{Meas-data} \protect\legendlinemark{dash pattern=on 3pt off 2pt on 0.8pt off 2pt}{psqaMeas}{rectangle},\;
\textit{Log-data} \protect\legendlinemark{}{psqaLog}{regular polygon,regular polygon sides=3,shape border rotate=0},\;
\textit{Calc-data} \protect\legendlinemark{dashed}{psqaCalc}{diamond}.
}
\label{fig:rel_risk_change_data_radar}
\end{center}
\end{figure}

\subsubsection{Five Important Observations from the \textit{Data}-only Radar Plot}

The following five observations represent the most noticeable findings resulting from the analysis of the radar plot shown in Fig.\,\ref{fig:rel_risk_change_data_radar}\footnote{The colors of the various PSQA data polygons were deliberately kept in shades of red in order to visually highlight the connection to the PSQA-specific data polygons from section \ref{sec:abs_local_risk_changes}, which are also colored red. To make it easier to identify the individual PSQA methods — despite a similar line color — they were additionally marked with different geometric symbols.}. As before, these are discussed in connection with the content and structure of the workflow from Table\,\ref{tab:riskid_catalogue} and are linked to the results on absolute risk changes.

\begin{enumerate}[label=\arabic*.,itemsep=-2pt]
  \item \textbf{Largest relative risk decrease.}  
        \textit{Log–Data} provides the strongest unweighted change in risk mitigation: the relative risk for Risk-ID~11.b \footnote{PS $\#11$: \textit{Treatment‐field - Insertion}; FM $b$: \textit{Wrong treatment machine}} drops
        by \SI{-48}{\percent} relative to baseline.  The same Risk-ID shows the second strongest decline in \textit{Meas–} and \textit{Calc–Data} (approximately $-28\%$ and $-24\%$), indicating consistent sensitivity of this process element to the data-dependent category for all PSQA approaches.
  \item \textbf{Largest relative risk increase.}
        \textit{Meas-Data} shows the highest unweighted relative risk increases in Risk-IDs 20.h and 39.a\footnote{The failure modes 20.h (PS \#20: \textit{Plan export/import to OIS}) and 39.a (PS \#39: \textit{Set treatment parameters}) are: \textit{Important treatment parameters (e.g. gantry angle, snout, position, MU scaling) changed or set to not intended values} and \textit{Wrong parameters uploaded}.} with $46.7\,\%$ and $46.6\,\%$ respectively, which are also positive for \textit{Log-Data} and \textit{Calc-Data}, but are significantly lower with a range between $3.5\,\%$ and $14.7\,\%$. In both cases, where $\delta^{Meas,(d)}_{rel}$ increases by almost half of the baseline risk, the failure modes are data or data transfer errors prior to the PSQA measurement.
  \item \textbf{\textit{Calc–Data} is consistently narrow–band.}  
        Its entire polygon lies within $\pm 30\,\%$ of the baseline; most points fall within the  $\pm15\,\%$ corridor, which is reflected in Tab.\,\ref{tab:drel_unweighted_data_stats} by the lowest SD ($11.5\%$) of all polygons in Fig. \ref{fig:rel_risk_change_data_radar}. In practical terms, this means that the ISDC only fully unfolds its algorithmic effect in the process category \textit{Calc-Full}.
  \item \textbf{\textit{Log-Data} tends to dominate in \textit{Plan Preparation} and \textit{Treatment} processes.}
        A key finding from Sections \ref{sec:no_psqa_risk_landscape}-\ref{sec:abs_local_risk_changes} was that the two largest absolute aggregations of baseline risks occur in the processes \emph{Plan Preparation} (PS \#20-23) and \emph{Treatment} (PS \#36-43). It is in these phases that \textit{Log-Data} tends to surpass both \textit{Meas-Data} and \textit{Calc-Data} with regard to unweighted relative risk mitigation. This is most evident in process steps 36.c and 37.e\footnote{The failure modes 36.c (PS \#36: \textit{Send plan from OIS to TCS}) and 37.e (PS \#37: \textit{Translation of plan to TCS internal settings}) are \textit{Incorrect recording/resume of partial delivery} and \textit{Plan is not deliverable (TPS allows values not possible in TDS)}.}, which describe TDS problems whose risks can be most effectively mitigated or absorbed by logging measures.
  \item \textbf{\textit{Meas-Data} exhibits bipolar behaviour.}  
        \textit{Meas-Data} shows the largest positive deviation of $+46.7\,\%$ at Risk-ID~20.h in PS\,\#20 \footnote{\textit{Plan export/import to OIS}; FM h: \textit{Important treatment parameters changed or set to not intended values}}.  Conversely, it delivers $-32\,\%$ at Risk-ID~20.b (FM b: \textit{Plan imported into wrong treatment room}) in the same process step. This reversal of polarity illustrates that in this workflow phase, isolated data-driven preparations for the following PSQA measurements can increase the unweighted relative risk, especially if they occur independently within the same process step, i.e., without its complete execution (\textit{Meas-Full}).
\end{enumerate}

\subsubsection{\textit{Data}-Only Verifications - Numerical Summary and Histogram Distributions}
\label{subsubsec:data_only_verifications}

\begin{table}[h!]
\centering
\caption{Distribution parameters of relative RPN changes $\delta^{*,(d)}_{\mathrm{rel},\,i,\,l}$}
\label{tab:drel_unweighted_data_stats}
\begin{tabular}{lccccccc}
\toprule
\textbf{PSQA} & \textbf{\#FM(\#PS)} & \textbf{Min [\%]} & \textbf{Max [\%]} & \textbf{Mean [\%]} & \textbf{SD [\%]} & \textbf{Median [\%]} & \textbf{IQR [\%]} \\
\midrule
\textit{Meas–Data} & $44(20)$ & $-31.6$ & $+46.7$ & $\bm{+5.8}$ & $\bm{14.8}$ & $\bm{+5.0}$ & $\bm{14.0}$ \\
\textit{Log–Data}  & $44(20)$ & $-48.2$ & $+29.6$ & $-\bm{1.8}$ & $\bm{15.9}$ & $\bm{+0.2}$ & $\bm{15.9}$ \\
\textit{Calc–Data} & $44(20)$ & $-28.2$ & $+26.5$ & $\bm{+3.5}$ & $\bm{11.5}$ & $\bm{+2.1}$ & $\bm{14.3}$ \\
\bottomrule
\end{tabular}
\end{table}

\begin{figure}[h!]
  \centering
  \includegraphics[width=0.9\linewidth]{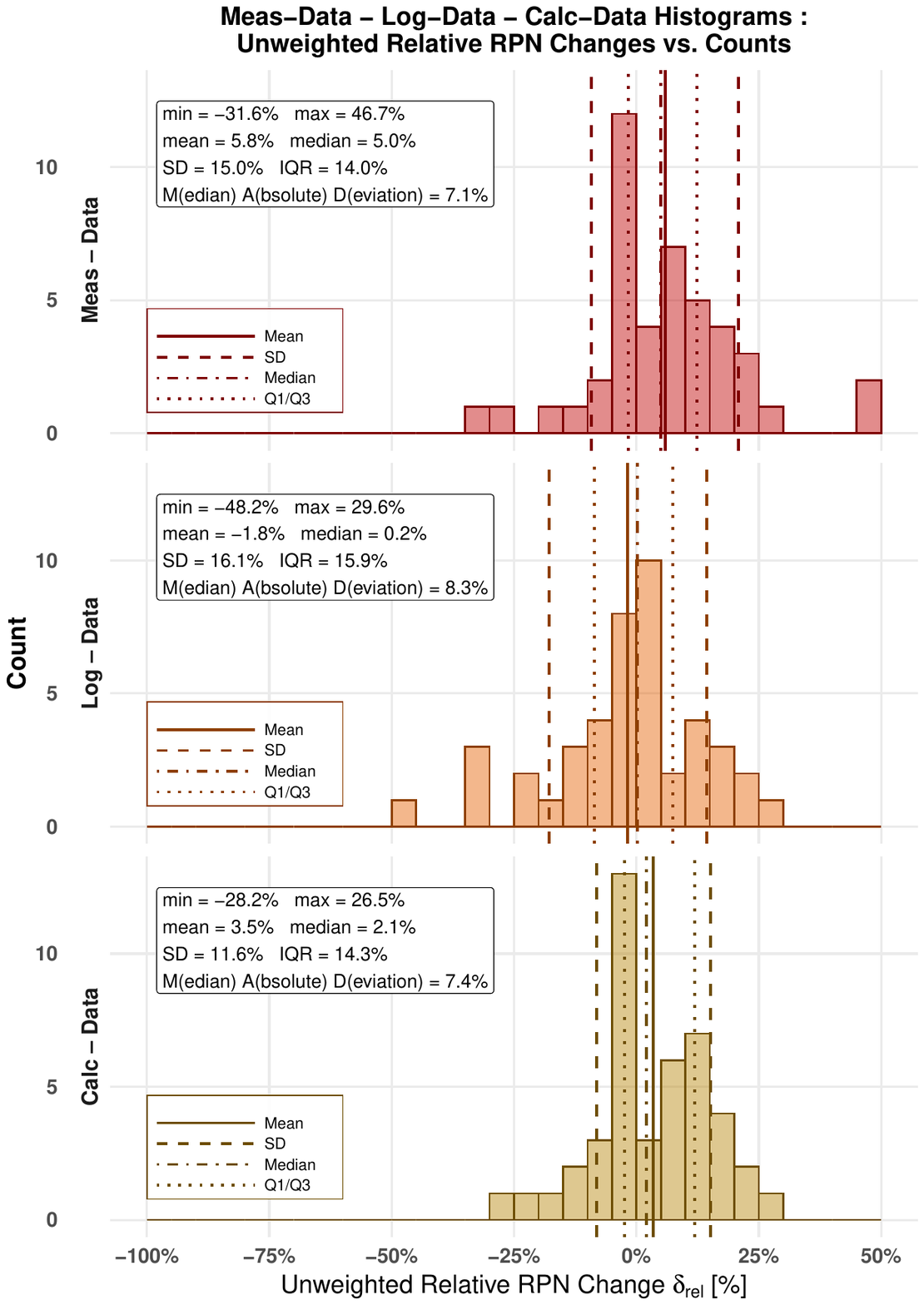}
  \caption{\small Faceted histograms of unweighted relative RPN changes $\delta_{\mathrm{rel}}$ for \textit{Data}-only PSQA components (Meas, Log, Calc) with aligned $x$-axes. Vertical markers indicate mean (solid), $\pm 1$\,SD (dashed), median (dash–dot), and Q1/Q3 (dotted); parameter boxes report \{min, max\}, \{mean, median\}, \{SD, IQR\}, and M(edian) A(bsolute) D(eviation).}
  \label{fig:drel_unweighted_data_hist}
\end{figure}

\paragraph{Numerical aggregation and interpretability.} While the radar plot in \ref{fig:rel_risk_change_data_radar} (only) visualizes the process step-resolved, FM-specific pattern of the relative RPN changes $\delta^{*,(d)}_{\mathrm{rel},\,i,\,l}$ (driven by the PSQA-dependent \textit{Data}-only influence) across the workflow, the following histogram-based aggregation (Fig.~\ref{fig:drel_unweighted_data_hist}) additionally pools all 44 FM-wise values into $5\%$-bins to obtain a semi-quantitative view of central tendency and dispersion. Because this pooling is \textit{without underlying failure mode-process step assignments}, \textit{Data}-only contributions with very different baseline RPNs and clinical leverage are mixed; the resulting distributions therefore exhibit inflated spread and heavier tails by construction, which is consistent with the standard deviations in Table \ref{tab:drel_unweighted_data_stats} (\textit{Meas–Data} $\mathrm{SD}=14.8\%$, \textit{Log–Data} $\mathrm{SD}=15.9\%$, \textit{Calc–Data} $\mathrm{SD}=11.5\%$).

These SDs are informative as descriptive measures of heterogeneity across failure modes, but they are \emph{not} normalized effect sizes and should \emph{not} be used directly to rank PSQA methods or their underlying process categories; for workflow-level inference that reflects the risk budget, the risk-weighted relative metrics introduced in Subsec.~\ref{subsec:psqa_risk_metrics_summary} provide the appropriate lens. These results wil be provided in the next subsections.

\paragraph{Synthesis and rationale.}
The distributional cross-check in Fig.~\ref{fig:drel_unweighted_data_hist}\footnote{The histogram colors mirror the \textit{Data}-only polygon colors used in Fig.~\ref{fig:rel_risk_change_data_radar}, ensuring visual consistency across the Sec. \ref{sec:rel_local_risk_changes}.} confirms that on average the unweighted \textit{Data}-only contributions concentrate near neutrality on the bounded scale $[-100\%,+50\%]$: the medians for \textit{Meas–Data}, \textit{Log–Data}, and \textit{Calc–Data} are close to $0\%$ (approximately $+5.0\%$, $+0.2\%$, and $+2.1\%$, respectively), and the corresponding Inter-Quartile Ranges (IQR) lie in a narrow band of about $14$–$16\%$, with Median Absolute Deviations (MAD)\footnote{\textbf{MAD} (median absolute deviation) is the median of the absolute deviations from the sample median $\mathrm{MAD}=\operatorname{median}_{i}\!\left|x_i-\operatorname{median}_{j}(x_j)\right|$. It is a robust measure of scale with a 50\% breakdown point and the same units as the data; for Gaussian data, \(1.4826\times\mathrm{MAD}\) is a consistent estimator of the standard deviation \(\sigma\).} around $7$–$8\%$ and standard deviations (SD) between roughly $12\%$ and $16\%$. In all three panels the $\pm 1$\,SD intervals straddle $0\%$, and the distributions overlap substantially; the small mean shifts of the different \textit{Data}-only PSQA categories (about $+5.8\%$, $-1.8\%$, and $+3.5\%$), which could potentially indicate a trend in their influence on risks, are therefore modest relative to their dispersion.

\paragraph{Implication for risk-mitigation claims.}
These parameters jointly indicate that any systematic, uni-di\-rectional risk reduction attributable to \textit{Data}-only checks in the context of relative RPN chan\-ges is limited: (i) robust locations are near zero, (ii) typical within-Risk-ID variability (IQR, MAD) is sizable compared with the means, and (iii) classical dispersion keeps the mass across both signs, preventing a consistent separation from the baseline. On a bounded outcome scale, ceiling/floor effects and unweighted pooling across heterogeneous Risk-IDs further attenuate small location shifts. Consequently, \textit{Data}-only elements provide supportive sensitivity but, by themselves, do not deliver a strong or stable mitigation signal across PSQA methods; their primary value is complementary, to be interpreted alongside measurement-, log file-, and calculation-based components rather than as a driver of overall risk reduction.

\subsection{Results on \textit{Cumulative} Relative RPN Changes}
\label{subsec:pc_full_on_relative_RPN_scale}

\begin{figure}[h!]
  \centering
  \includegraphics[width=1.1\linewidth,trim=3.5cm 1cm 0cm 0cm,clip]{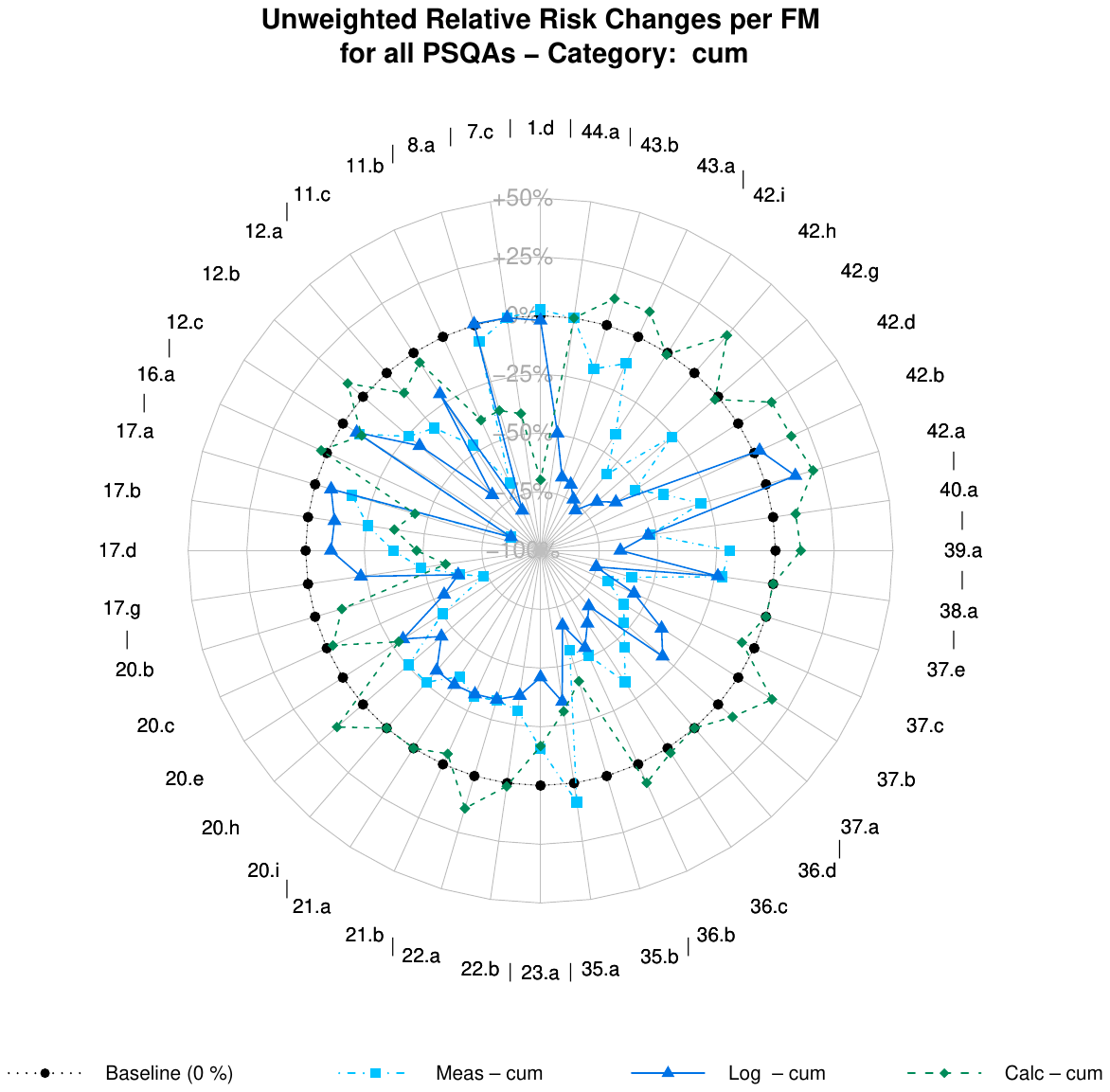}
  \caption{\small Cumulative relative risk changes per FM for \textit{Cum}-track PSQAs:
           Baseline \protect\linestub{dotted,color=black},\;
           \textit{Meas-Cum} \protect\linestub{dash dot,color=cyan},\;
           \textit{Log-Cum}  \protect\linestub{solid,color=azureblue},\;
           \textit{Calc-Cum} \protect\linestub{dashed,color=teal}. A first, rough visual assessment of the risk reduction potential confirms the hierarchy \textit{Log} > \textit{Meas} > \textit{Calc}.}
  \label{fig:rel_risk_change_cum_radar}
\end{figure}

\begin{figure}[h!]
  \centering
  \includegraphics[width=0.95\linewidth]{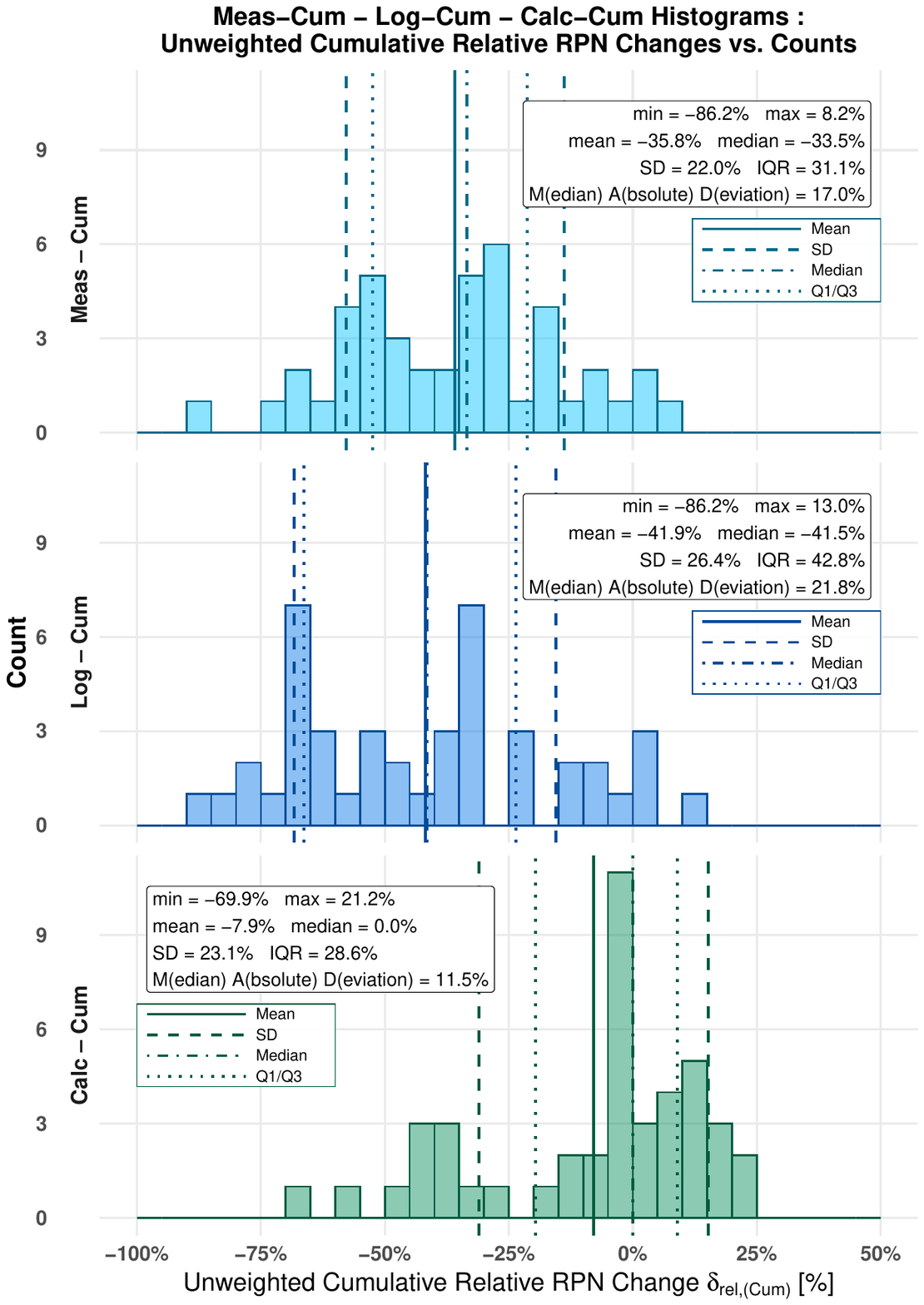}
  \caption{\small Faceted histograms of cumulative relative RPN changes $\delta_{\mathrm{rel,(c)}}$ for \textit{Cum}-track PSQA components (\textit{Meas-Cum}, \textit{Log-Cum}, \textit{Calc-Cum}) with aligned $x$-axes. Vertical markers indicate mean (solid), $\pm 1$\,SD (dashed), median (dash–dot), and Q1/Q3 (dotted); parameter boxes report \{min, max\}, \{mean, median\}, \{SD, IQR\}, and M(edian) A(bsolute) D(eviation).}
  \label{fig:drel_unweighted cum_hist}
\end{figure}

\noindent In this subsection, we extend the analysis from the \textit{Data}-only setting to the \textit{cumulative} PSQA polygon tracks (\textit{Meas–Cum}, \textit{Log–Cum}, \textit{Calc–Cum}) with corresponding $\delta^{*,(c)}_{\mathrm{rel},\,i,\,l}$ changes (see Subsec. \ref{subsec:psqa_risk_metrics_summary}), keeping the same sample of 44 Risk-IDs, axis range ($-100\,\%$ to $+50\,\%$), and baseline at $0\,\%$ (see Fig.~\ref{fig:rel_risk_change_cum_radar}). To ensure strict comparability, we apply identical histogram binning (Fig.~\ref{fig:drel_unweighted cum_hist}), the same color coding between radar plots and histograms, and summary statistics (mean, SD, median, IQR, MAD) as established in Subsec.\ref{subsec:pc_data_on_relative_RPN_scale} (Table~\ref{tab:drel_unweighted_data_stats}). Consequently, we start with a visual analysis of the \textit{Cum}-mode radar plot.

\subsubsection{\textit{Cum}-mode Radar Plot: Five Principal Findings for Cumulative Relative Risk Changes}
\begin{enumerate}[label=\arabic*.]
  \item \textbf{Strongest cumulative relative risk mitigation.}
        \textit{Log–Cum} and \textit{Meas-Cum} deliver the single largest negative $\delta^{*,(c)}_{\mathrm{rel}}$ value at RiskID 16.a \footnote{PS\,\#16: \textit{Optimization}, FM\,$a$b: \textit{Production of undeliverable plan}} with a relative cumulative risk reduction of $-86.2\,\%$, which is confirmed in a similar manner in absolute terms (see Figs. \ref{fig:psqa_log_radar} and \ref{fig:psqa_meas_radar}). Practically, this can be described as a risk exclusion, which is equivalent to a quasi-certain detection of an undeliverable plan. Comparable cumulative relative risk mitigations ($< -75.0\,\%$) are otherwise achieved exclusively through the use of \textit{Log-Cum}, specifically for Risk-IDs 11.b, 37.e and 42.h\footnote{For details of these Risk-IDs, see Sec.\,\ref{sec:res_catalogue}}.
        
  \item \textbf{Largest observed cumulative relative risk increase.} In the \textit{Cum}–track setting, the single largest positive deviation occurs on the \textit{Calc–Cum} polygon, reaching $\delta^{*,(c)}_{\mathrm{rel}}=+21.2\,\%$ for Risk-ID 42.h on a non–calculation axis, i.e., outside PS~\#1–\#8 and PS~\#17 (visually evident in Fig.~\ref{fig:rel_risk_change_cum_radar}; see the maxima in Table~\ref{tab:drel_unweighted_cum_stats}). Distributionally, this outlier coexists with a zero median and a modestly negative mean for \textit{Calc–Cum} ($\text{median}=0.0\,\%$, $\text{mean}=-7.9\,\%$, $\text{SD}=23.1\,\%$, $\text{IQR}=11.5\,\%$), indicating little generic mitigation outside calculation-centric steps and underscoring that the observed risk increase is sparse rather than systematic (Tab.\,\ref{tab:drel_unweighted_cum_stats}).    
      
      By comparison, \textit{Log–Cum} exhibits only small positive excursions (confined to PS~\#42.a–b and bounded by $+13.0\,\%$), while \textit{Meas–Full} peaks at $+8.2\,\%$; thus, both the comparative distributions and the semi-quantitative key figures concur that the largest increase potential lies with \textit{Calc–Cum}, albeit with limited relevance for workflow-wide risk budgets (Tab.\,\ref{tab:drel_unweighted_cum_stats}).  

  \item \textbf{\textit{Meas-Cum}: No outperformer, but a solid, generic risk controller.} $\delta^{Meas,(c)}_{\mathrm{rel}}$ values achieve the best risk reductions for the RiskIDs 20.c, 20.e, 21.a, 37.a to 37.c and 42.a or 42.b. Their risk reduction is inferior compared to \textit{Log-Cum} in the remaining RiskIDs in process steps 42, 43, and 44, which include daily treatment delivery and recording. This is a reflection of the assumption that \textit{Log-Cum} includes daily dose evaluation based on each fraction's treatment log files. Because \textit{Meas-Cum} is performed only once before treatment begins, it is less capable of identifying errors during treatment delivery and recording.
      
        \textit{Cum-Full} does not demonstrate PSQA superiority in any of the workflow phases with the highest risks, but on average offers acceptable to solid risk control for the \textit{Treatment} process (PS \#36 - \#43) and, in some cases, for the \textit{Plan Preparation} phase. The extent to which the high level of effort required for this risk mitigation or control can be considered efficient or justified, which is often used as an argument against the measurement-based PSQA method in general, is not evaluated at this point.

  \item \textbf{\textit{Log-Cum}: The strongest mitigator - still not perfect.} As Fig. \ref{fig:rel_risk_change_data_radar} shows, the vast majority of the $\delta^{Log,(c)}_{\mathrm{rel}}$ values of the \textit{Log-Cum} polygon curve lie in the radial segment of the radar plot between $-50\,\%$ and $-75\,\%$, thus underlining its leading position within the three PSQA approaches in terms of general risk reduction at a cumulative relative level.
      
        Nevertheless, closer inspection reveals that the log file-based method is also prone to errors in various process steps.
        \begin{enumerate}[label=(\alph*)]
            \item As discussed previously, two RiskIDs (42.a and 42.b) show an cumulative relative increase in risk slightly above the baseline value ($\delta^{Log,(c)}_{\mathrm{rel}} > 0$)\footnote{See footnote \ref{fn:largest_drel_log_value} and the discussion in Subsec.\,\ref{subsec:log_the_strongest_mitigator}}, while all other RiskIDs of this method show exclusively negative $\delta^{Log,(c)}_{\mathrm{rel}}$ values, i.e. risk reductions.
            \item For the three initial RiskIDs, 1.d~, 7.c, 8.a, whose risks are preferably mitigated by \textit{Calc-Cum}, the cumulative relative RPN changes for \textit{Log-Cum} – together with the \textit{Meas-Cum} values – are also at baseline level.
            \item In addition to this initial phase of the workflow, in which \textit{Log-Cum} does not provide any risk advantage, the PSQA application of log files for risk reduction performs worst locally in a comparative evaluation of the failure modes of process step PS \#17, where \textit{Calc-Cum} also performs better.
            \item In the range of RiskIDs 20.h to 23.a, which belong to the process \textit{Plan Preparation}, \textit{Log-Cum} again reduces the cumulative relative risk best in comparison, but the $\delta^{Log,(c)}_{rel}$ values do not fall below the $-50\,\%$ threshold, i.e., more than half of the baseline risk still remains. This confirms that this area or phase of the entire workflow appears to be the most risk-prone.
        \end{enumerate}

  \item \textbf{\textit{Calc-Cum} or ISDC: Highly specific but not generic.} The \textit{Calc-Cum} approach shows its greatest potential for unweighted relative risk reduction for the Risk-IDs of PS\,\#17 (\textit{Forward calculation}), whereby it clearly outperforms both \textit{Log-Cum} and \textit{Meas-Cum}. The very first three Risk-IDs of the entire workflow, $1.d$, $7.c$, and $8.a$, which form the initial processes \textit{Transfer Images} and \textit{Anatomy Contouring (Delineation)} are especially interesting, as here too the risk performance of \textit{Calc-Cum} is also significantly better than that of the other two PSQA methods in full mode, without appearing to be obvious a priori.
      
      Otherwise, \textit{Calc-Cum} largely shows an cumulative relative risk level in the remaining process steps of the workflow, which corresponds approximately to the PSQA-free baseline risk or is even scored to be slightly higher (relative risk increase of up to $+25\,\%$). Its suitability for limited, but pronounced risk mitigation in dose calculation-related processes makes \textit{Calc} (in cumulative mode) appear to be a highly specific PSQA method, but it hardly shows the generic potential for general risk reduction.
\end{enumerate}

\subsubsection{Cum-track Verification – Numerical Overview and Histogram Distributions}
\label{subsubsec:cum_track_verifications}

\begin{table}[h!]
\centering
\caption{Distribution parameters of cumulative relative risk changes $\delta^{*,(c)}_{\mathrm{rel}}$}
\label{tab:drel_unweighted_cum_stats}
\begin{tabular}{lccccccc}
\toprule
\textbf{PSQA} & \textbf{\#FM(\#PS)} & \textbf{Min [\%]} & \textbf{Max [\%]} & \textbf{Mean [\%]} & \textbf{SD [\%]} & \textbf{Median [\%]} & \textbf{IQR [\%]} \\
\midrule
\textit{Meas–Cum} & $44(20)$ & $-86.2$ & $+8.2$  & $-\bm{35.8}$ & $\bm{22.0}$ & $-\bm{33.5}$ & $\bm{17.0}$ \\
\textit{Log–Cum}  & $44(20)$ & $-86.2$ & $+13.0$ & $-\bm{41.9}$ & $\bm{26.4}$ & $-\bm{41.5}$ & $\bm{21.8}$ \\
\textit{Calc–Cum} & $44(20)$ & $-69.9$ & $+21.2$ & $-\bm{7.9}$  & $\bm{23.1}$ & $\bm{0.0}$   & $\bm{11.5}$ \\
\bottomrule
\end{tabular}
\end{table}

\paragraph{Numerical aggregation and interpretability.}
While the radar plot visualization in Fig. \ref{fig:rel_risk_change_cum_radar} illustrates the process step-resolved, failure mode-specific representation of the cumulative relative RPN changes under \textit{Cum}-track conditions, which allow an inter-PSQA comparison (see Subsec.~\ref{appB:subsec:fm_cumulative_changes}), the histogram-based aggregation (Fig. \ref{fig:drel_unweighted cum_hist}) summarizes all 44 FM values to provide a semi-quantitative view of the distribution behavior across the entire workflow (analogous to Fig. \ref{fig:drel_unweighted_data_hist}). The resulting values in Table \ref{tab:drel_unweighted_cum_stats} show substantially greater dispersion than in the \textit{Data Only} setting in Table \ref{tab:drel_unweighted_data_stats}: Both \textit{Meas–Cum} and \textit{Log–Cum} exhibit long left tails extending to $-86.2$\%, with standard deviations of 22.0\% and 26.4\%, respectively, while \textit{Calc–Cum} shows a wide bidirectional dispersion ($SD = 23.1\,\%$) around a median of 0\,\%.\\
These wide histogram dispersions arise from the mixing of failure modes with very different base RPNs, a variability that is exacerbated by the fact that unweighted pooling ignores the underlying risk contribution of each mode (within its process step) (see Sec.~\ref{appB:sec:local_risk_metrics}). As a result, the large shifts in mean values (\textit{Meas-Cum}: $-35.8$\%, \textit{Log–Cum}: $-41.9$\%, \textit{Calc–Cum}: $-7.9$\%) must be interpreted in the context of substantial within-method variability and limited overall constraints of the unweighted aggregation framework. In short, the \textit{Cum}-track histograms provide robust evidence of a net risk reduction for \textit{Meas} and \textit{Log}, while at the same time highlighting the limited, occasionally bidirectional influence of \textit{Calc} without risk weighting.

\paragraph{Synthesis and rationale.}
Cross-referencing the \textit{Cum}-track histogram distributions (Fig.~\ref{fig:drel_unweighted cum_hist}) with the structural patterns in the radar plot (Fig.~\ref{fig:rel_risk_change_cum_radar}), three features emerge.
\begin{enumerate}
  \item \textit{Log–Cum} maintains the strongest overall shift toward negative values, consistent with its dominant contraction of the high-risk delivery and recording steps (PS~\#36–\#43). This is quantitatively reflected in its lowest central tendency (mean $-41.9\%$, median $-41.5\%$) and widest IQR ($42.8\%$), which illustrates its broad, consistently negative, i.e., risk-mitigating influence.
  \item \textit{Meas–Cum} delivers a similar but slightly less pronounced risk contraction ($\mathrm{median}=-33.5\%$), with narrower central variability ($\mathrm{IQR}=17.0\%$), confirming its role as a competent but less comprehensive mitigator, particularly in workflow phases that cannot benefit from daily verification. 
  \item \textit{Calc–Cum} demonstrates a highly specific signature: it achieves pronounced reductions only in forward-calculation-related FMs (PS~\#17 and early workflow steps), while residuals across the remaining FMs cluster near baseline, yielding a median of $0\%$ and several small positive outliers (up to $+21.2\%$). 
\end{enumerate}
These functions collectively confirm the qualitative conclusion derived from the pFMEA: \textit{Log} excels in radiation-oriented and data consistency-sensitive phases of the workflow, which—when averaged without weighting—lead to the most significant risk reduction; \textit{Meas} also offers broad-based risk reduction, which is particularly prominent in the initial processes prior to treatment; and \textit{Calc} targets a small group of algorithmic risks with limited generic impact.

\paragraph{Implication for risk-mitigation claims.}
Taken together, the numerical evidence from Table~\ref{tab:drel_unweighted_cum_stats} and the distributions in Fig.~\ref{fig:drel_unweighted cum_hist} supports the qualitative mitigation hierarchy established in Sec. \ref{subsec:cross_modality_synthesis}: 
\begin{equation*}
            \overline{\textsc{Log}}_{\mathrm{Cum}} > \overline{\textsc{Meas}}_{\mathrm{Cum}} \gg \overline{\textsc{Calc}}_{\mathrm{Cum}}, 
\end{equation*}
However, the \textit{Cum}-track aggregates also highlight important caveats for risk-mitigation claims. Because the FM-wise contributions are unweighted, the relative prevalence of large baseline RPNs is obscured, and strong negative shifts in high-impact process steps may be statistically diluted by benign steps with low risk budgets. Therefore, the unweighted metrics in Table~\ref{tab:drel_unweighted_cum_stats} — despite their descriptive value — should not be used to assert high-level performance without the complementary risk-weighted indicators defined in Secs.~\ref{appB:sec:local_risk_metrics} and \ref{appB:sec:global_risk_metrics}. 

Within this appropriate interpretive frame, the \textit{Cum}-track summary robustly confirms that \textit{Log–Cum} provides the deepest and most consistent mitigation across the workflow, \textit{Meas–Cum} replicates much of this benefit but without per-fraction responsiveness, and \textit{Calc–Cum} contributes meaningful but localized reductions with limited influence on the global risk budget.

\subsection{Incremental \textit{Full}-Stage Effects as Signed Relative Risk Shifts from \textit{Data} to \textit{Cum}}
\label{subsec:results_d2c_relative}

In this subsection, we no longer visualize the raw \textit{Full}-stage quantity $\delta^{\ast,(f)}_{\mathrm{rel},i,l}$ as a standalone radar plot. Instead, in direct continuation of Subsecs.~\ref{subsec:pc_data_on_relative_RPN_scale} and \ref{subsec:pc_full_on_relative_RPN_scale}, we use the signed, \textit{NoQA}-anchored bridge quantity
\begin{equation}
\delta^{\ast,(d\to c)}_{\mathrm{rel},i,l}
:=
\delta^{\ast,(c)}_{\mathrm{rel},i,l}
-
\delta^{\ast,(d)}_{\mathrm{rel},i,l}
=
\delta^{\ast,(f)}_{\mathrm{rel},i,l}\bigl(1+\delta^{\ast,(d)}_{\mathrm{rel},i,l}\bigr)
=
\frac{\mathrm{RPN}^{\ast,\mathrm{full}}_{i,l}-\mathrm{RPN}^{\ast,\mathrm{data}}_{i,l}}{\mathrm{RPN}^{\mathrm{noQA}}_{i,l}},
\end{equation}
for each validated Risk-ID $(i,l)$ and PSQA modality $^{\ast} \in \{\mathrm{Meas},\mathrm{Log},\mathrm{Calc}\}$. This quantity represents the signed relative displacement from the \textit{Data}-stage point to the cumulative end state, measured on the same common \textit{NoQA} baseline that already underlies the \textit{Cum}-track analysis. 

Unlike $\delta^{\ast,(f)}_{\mathrm{rel},i,l}$, which remains normalized to the method-specific \textit{Data}-stage state and is therefore primarily a \emph{within-modality} quantity, $\delta^{\ast,(d\to c)}_{\mathrm{rel},i,l}$ is directly compatible with the geometric interpretation of the dumbbell plots: it is exactly the signed segment between the open \textit{Data} marker and the filled \textit{Cum} marker. Negative values indicate additional mitigation induced by completing the \textit{Full}-stage, positive values indicate additional inflation, and zero indicates that the \textit{Full}-stage leaves the \textit{Data}-stage risk estimate unchanged.

\begin{figure}[h!]
\centering
\includegraphics[width=\textwidth]{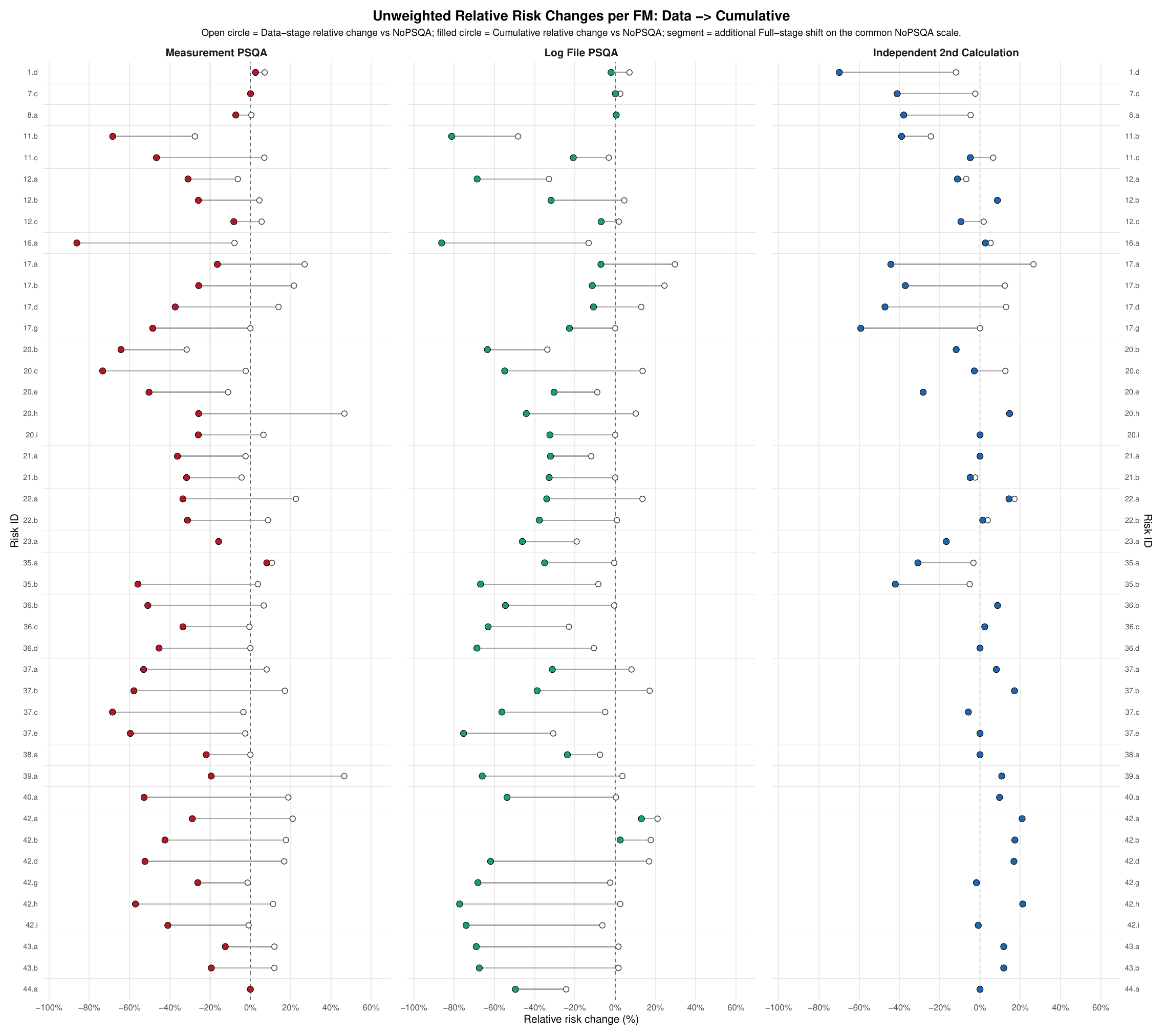}
\caption{\small Unweighted relative risk changes per FM from \textit{Data} to \textit{Cum}. Open circles denote $\delta^{\ast,(d)}_{\mathrm{rel},i,l}$ and filled circles denote $\delta^{\ast,(c)}_{\mathrm{rel},i,l}$; the connecting segment therefore encodes the signed bridge quantity $\delta^{\ast,(d\to c)}_{\mathrm{rel},i,l} = \delta^{\ast,(c)}_{\mathrm{rel},i,l} - \delta^{\ast,(d)}_{\mathrm{rel},i,l}$. Negative segments indicate additional mitigation caused by completing the \textit{Full}-stage, while zero-length segments indicate no further shift beyond the \textit{Data}-stage. Each dumbbell plot displays—parallel to the x-axis—20 light gray horizontal lines that correspond to the “pipe” symbol in the outermost ring of the radar plots. They represent the known segmentation of the 44 Risk-IDs or FMs along the y-axis into the 20 higher-level process steps.}
\label{fig:unweighted_drel_d2c_dumbbell}
\end{figure}

Figure~\ref{fig:unweighted_drel_d2c_dumbbell} displays these unweighted bridge shifts for all 44 validated Risk-IDs and all three PSQA modalities. Figure~\ref{fig:hist_unweighted_drel_d2c} complements this view by
pooling the same FM-wise values into faceted histograms with aligned x-axes and a bin width of $5\%$, thereby providing a distributional cross-check that is mathematically and graphically consistent with the
preceding dumbbell representation.

\subsubsection{\textit{Data}-to-\textit{Cum} Bridge Plot: Five Principal Findings}
The following five observations capture the most relevant patterns emerging from the bridge plot in Fig.~\ref{fig:unweighted_drel_d2c_dumbbell}. As in the previous subsections, they are discussed in conjunction
with the workflow structure in Table~\ref{tab:riskid_catalogue} and the absolute/cumulative findings of Sec.~\ref{sec:abs_local_risk_changes}.

\begin{enumerate}
    \item \textbf{The bridge shifts are systematically non-positive.}
    A first defining feature of Fig.~\ref{fig:unweighted_drel_d2c_dumbbell} is the absence of rightward displacement from \textit{Data} to \textit{Cum}, which is reflected by a maximum value of $0.0\%$ for \textit{Meas–(d$\rightarrow$c)}, \textit{Log–(d$\rightarrow$c)}, and \textit{Calc–(d$\rightarrow$c)} in Table~\ref{tab:drel_signed_rel_risk_shifts_stats}. In the present pFMEA scoring logic, the completion of the \textit{Full}-stage never pushes the \textit{Data}-stage point outward on the common \textit{NoQA} scale. Consequently, $\delta^{\ast,(d\to c)}_{\mathrm{rel},i,l} \le 0$ for all displayed modes, so the bridge quantity acts either as additional mitigation or as a null increment, but not as an additional inflation layer.

    \item \textbf{The largest signed bridge shifts \(\delta^{\ast,(d\to c)}_{\mathrm{rel},i,l}\) occur in sharply modality-specific high-sensitivity FMs.}
    For \textit{Meas}, the deepest \textit{Data}$\to$\textit{Cum} shifts are centered on PS \#16 and the plan-transfer/ translation interface, led by Risk-ID 16.a (production of an undeliverable plan, \(-78.3\%\)), followed by 37.b (\(-75.0\%\)), 20.h (\(-72.4\%\)), 40.a (\(-71.6\%\)), and 20.c (\(-71.1\%\)); this indicates broad incremental mitigation across optimization, plan export/import, translation, and accessory-related checks. For \textit{Log}, the strongest bridge displacements migrate to the late delivery and recording region, with maxima at 42.h (machine delivery failure, \(-79.8\%\)) and 42.d (beam-delivery error, \(-78.7\%\)), followed by 16.a (\(-73.0\%\)), 43.a (\(-70.6\%\)), and 39.a (\(-69.6\%\)), confirming that the \textit{Full}-stage log-based layer is most powerful once failures materialize at the machine-delivery, parameter-upload, and OIS-recording level. By contrast, \textit{Calc} remains highly selective: its dominant bridge shifts are concentrated in the forward-calculation cluster — 17.a (\(-70.8\%\)), 17.d (\(-60.2\%\)), 17.g (\(-59.2\%\)), 1.d (\(-58.0\%\)), and 17.b (\(-49.5\%\)) — showing that the incremental ISDC benefit is strongest for algorithmic, calibration, and dose-modeling errors rather than for generic delivery-state risks.
    
    \item \textbf{\textit{Log}-$\bm{(d\to c)}$ dominates the treatment and recording/documentation part of the workflow.}
    In the late workflow region, especially across PS \#36--44, the \textit{Log} panel displays the longest and most persistent leftward segments. This is most visible around Risk-IDs such as 36.d, 37.e, 42.h, 42.i, 43.a, and 43.b, where the final transition from \textit{Data} to \textit{Cum} is strongest. Thus, once the preparatory data layer has been established, completion of the log file-based verification pathway contributes the largest additional inward displacement in those steps that depend directly on delivery-state and recording-state information.

    \item \textbf{\textit{Meas}-$\bm{(d\to c)}$ provides broad compensatory mitigation across preparation, translation, and setup.}
    The \textit{Meas} panel exhibits a more widely distributed set of leftward segments, particularly in plan preparation phase and plan-to-delivery translation. In practical terms, this means that the completion of
    measurement-based verification compensates a substantial fraction of the local \textit{Data}-stage inflation introduced by preparatory PSQA steps, especially in transfer- and interface-sensitive parts of the workflow.

    \item \textbf{\textit{Calc}-$\bm{(d\to c)}$ remains selective rather than generic.}
    The ISDC panel is characterized by many short or vanishing segments, as well as a smaller subset of distinctly negative shifts. Its bridging effect is therefore real but locally limited, with its most significant contributions concentrated on early failure modes that are sensitive to transmission, contouring, and forward calculation. It is particularly striking that in the range of PS \#36-–44, there is no risk reduction whatsoever in the transition from \textit{Data} to \textit{Cum}, i.e., the data-driven process category of \textit{Calc} contributes almost exclusively to a (slight) increase in risk relative to the \textit{NoQA} baseline level in the late workflow. This pattern confirms that ISDC offers added value, but not in the comprehensive, workflow-wide manner observed with \textit{Meas} and \textit{Log}.
\end{enumerate}

\subsubsection{Bridge-Shift Verification -- Numerical Overview and Histogram Distributions}

\begin{table}[h!]
\centering
\caption{Distribution parameters of signed relative risk shifts $\delta^{*,(d\to c)}_{\mathrm{rel}}$ as incremental \textit{Full}-stage effects}
\label{tab:drel_signed_rel_risk_shifts_stats}
\begin{tabular}{lccccccc}
\toprule
\textbf{PSQA} & \textbf{\#FM(\#PS)} & \textbf{Min [\%]} & \textbf{Max [\%]} & \textbf{Mean [\%]} & \textbf{SD [\%]} & \textbf{Median [\%]} & \textbf{IQR [\%]} \\
\midrule
\textit{Meas–(d$\rightarrow$c)} & $44(20)$ & $-78.3$ & $0.0$  & $-\bm{41.7}$ & $\bm{22.6}$ & $-\bm{42.1}$ & $\bm{33.0}$ \\
\textit{Log–(d$\rightarrow$c)}  & $44(20)$ & $-79.8$ & $0.0$ & $-\bm{40.1}$ & $\bm{21.8}$ & $-\bm{36.5}$ & $\bm{32.9}$ \\
\textit{Calc–(d$\rightarrow$c)} & $44(20)$ & $-70.8$ & $0.0$ & $-\bm{11.4}$  & $\bm{20.3}$ & $\bm{0.0}$   & $\bm{12.2}$ \\
\bottomrule
\end{tabular}
\end{table}

\begin{figure}[h!]
\centering
\includegraphics[width=\textwidth]{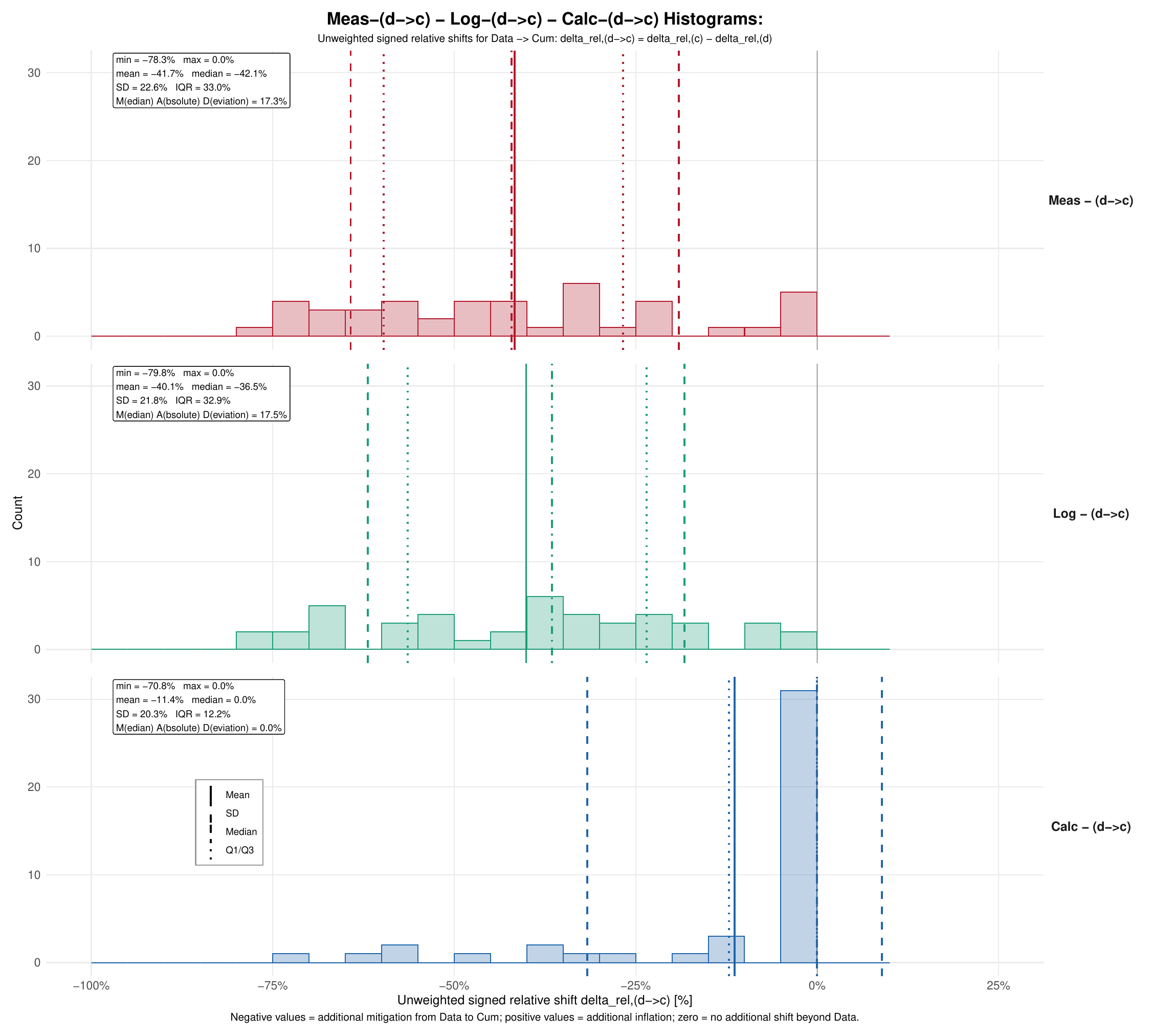}
\caption{\small Faceted histograms of the signed bridge quantity
$\delta^{\ast,(d\to c)}_{\mathrm{rel},i,l}$ for the three PSQA modalities (Meas, Log, Calc), defined by
$\delta^{\ast,(d\to c)}_{\mathrm{rel},i,l} = \delta^{\ast,(c)}_{\mathrm{rel},i,l} - \delta^{\ast,(d)}_{\mathrm{rel},i,l} = \delta^{\ast,(f)}_{\mathrm{rel},i,l}\bigl(1+\delta^{\ast,(d)}_{\mathrm{rel},i,l}\bigr)$.
Thus, the histogram summarizes the same signed Data$\to$Cum displacement that is encoded by the segments in Fig.~\ref{fig:unweighted_drel_d2c_dumbbell}. Histograms are shown with aligned x-axes and a bin width of
$5\%$. Vertical markers indicate mean (solid), $\pm 1$ SD (dashed), median (dash--dot), and Q1/Q3 (dotted); parameter boxes report $\{\min,\max\}$, $\{\mathrm{mean},\mathrm{median}\}$, $\{\mathrm{SD},\mathrm{IQR}\}$, and M(edian) A(bsolute) D(eviation).}
\label{fig:hist_unweighted_drel_d2c}
\end{figure}

At the distributional level, the pooled histograms in Fig.~\ref{fig:hist_unweighted_drel_d2c} summarize the same quantity that is geometrically visible in Fig.~\ref{fig:unweighted_drel_d2c_dumbbell} while Table \ref{tab:drel_signed_rel_risk_shifts_stats} lists its distribution parameters. Each bin counts the signed \textit{NoQA}-anchored bridge shift from \textit{Data} to \textit{Cum} and the histogram can be read directly as a distributional summary of the displayed segment lengths.

Numerically, this means that the central tendencies of the three panels describe how strongly the \textit{Full}-stage, after \textit{Data} has already acted, pulls the corresponding FM-wise risk estimates further inward on the common \textit{NoQA} scale. Negative locations therefore indicate net additional mitigation beyond the \textit{Data}-stage, values at zero indicate that the \textit{Full}-stage leaves the \textit{Data} estimate unchanged, and the spread reflects how unevenly this additional mitigation is distributed over the 44 Risk-IDs.

\subsubsection{Synthesis and Rationale}
Cross-referencing the axis-resolved bridge structure in Fig.~\ref{fig:unweighted_drel_d2c_dumbbell} with the pooled distributions in Fig.~\ref{fig:hist_unweighted_drel_d2c} reveals three robust features.

First, both \textit{Meas}-$(d\to c)$ and \textit{Log}-$(d\to c)$ are concentrated on clearly negative values, confirming that the \textit{Full}-stage generally provides substantial additional mitigation beyond the nearly neutral or locally bipolar \textit{Data}-stage behavior seen in Subsec.~\ref{subsec:pc_data_on_relative_RPN_scale}. Second, Log-$(d\to c)$ exhibits the broadest and deepest negative bridge structure in the late treatment and recording phases, which is fully consistent with the dominant \textit{Cum}-track performance already established in Subsec.~\ref{subsec:pc_full_on_relative_RPN_scale}. Third, \textit{Calc}-$(d\to c)$ remains much more zero-concentrated and selective, showing that its incremental \textit{Full}-stage benefit is real but confined to a smaller subset of algorithmic and model-sensitive failure modes.

Most importantly, the bridge quantity now restores graphical and numerical coherence within this subsection: the dumbbell plot and the histogram describe the same unweighted FM-wise variable, and both are anchored
to the same \textit{NoQA} reference geometry.

\subsubsection{Implication for Risk-Mitigation Claims}
Taken together, Figs.~\ref{fig:unweighted_drel_d2c_dumbbell} and \ref{fig:hist_unweighted_drel_d2c} support a more precise statement about the role of the \textit{Full}-stage. In the current representation, the \textit{Full}-stage is no longer introduced as an isolated alternative baseline quantity; instead, it is interpreted as the additional signed displacement required to move a PSQA modality from its \textit{Data}-stage state to its final \textit{Cum}-stage outcome on a common \textit{NoQA} scale.

Within this interpretive frame, \textit{Log} provides the strongest incremental bridge in delivery- and re\-cording-related workflow regions (primarily PS~\#39 and PS~\#42--43, and more generally across the late treatment block PS~\#36--44), \textit{Meas} provides the broadest compensatory bridge across preparation and translation (especially PS~\#20--23 and PS~\#37, with additional prominent contributions in PS~\#16 and PS~\#40), and \textit{Calc} contributes a narrower bridge concentrated in calculation-sensitive modes (above all PS~\#17, with an additional strong contribution in PS~\#1). Because the present analysis remains unweighted, these results should still be read as descriptive FM-wise and distributional evidence rather than as definitive workflow-level ranking metrics. The latter continue to require the risk-weighted local and global indicators introduced in Sec.~\ref{sec:operational_framework} and analyzed in the following subsections. Nevertheless, the new bridge quantity $\delta^{\ast,(d\to c)}_{\mathrm{rel},i,l}$ provides the most coherent unweighted visualization of how the \textit{Full}-stage actually completes the PSQA mitigation pathway.

\section{Risk-Weighted Results on the Process-Step Level}
\label{sec:process_step_results}

\subsection{Why Local Process-Step Aggregation Is Needed}
\label{subsec:why_local_aggregation}

The FM-level results in Secs.~\ref{sec:res_catalogue}--\ref{sec:rel_local_risk_changes} identify where individual risk changes occur, but they do not yet show how much total risk burden is shifted at the level of a complete process step. This distinction is essential because several process steps contain more than one validated failure mode, so the visual prominence of a single FM does not necessarily reflect the net local consequence for the workflow. 

For this reason, the FM-wise stage-specific changes are now aggregated into the risk-weighted local metrics $\rho^{*,(d)}_{\mathrm{rel},i}$, $\rho^{*,(c)}_{\mathrm{rel},i}$, and $\rho^{*,(d\to c)}_{\mathrm{rel},i}$, which quantify the relative change of the total local step burden under the \textit{Data}-stage, the cumulative end state, and the additional bridge from \textit{Data} to \textit{Cum}, respectively. The process-step level therefore answers the next natural question after the FM analysis: not where isolated vulnerabilities exist, but where they combine into locally meaningful weak zones or mitigation gains.

\subsection{Local Data-Stage Effects Across Process Steps}
\label{subsec:local_data_effects}

\begin{figure}[h!]
\centering
\includegraphics[width=\textwidth]{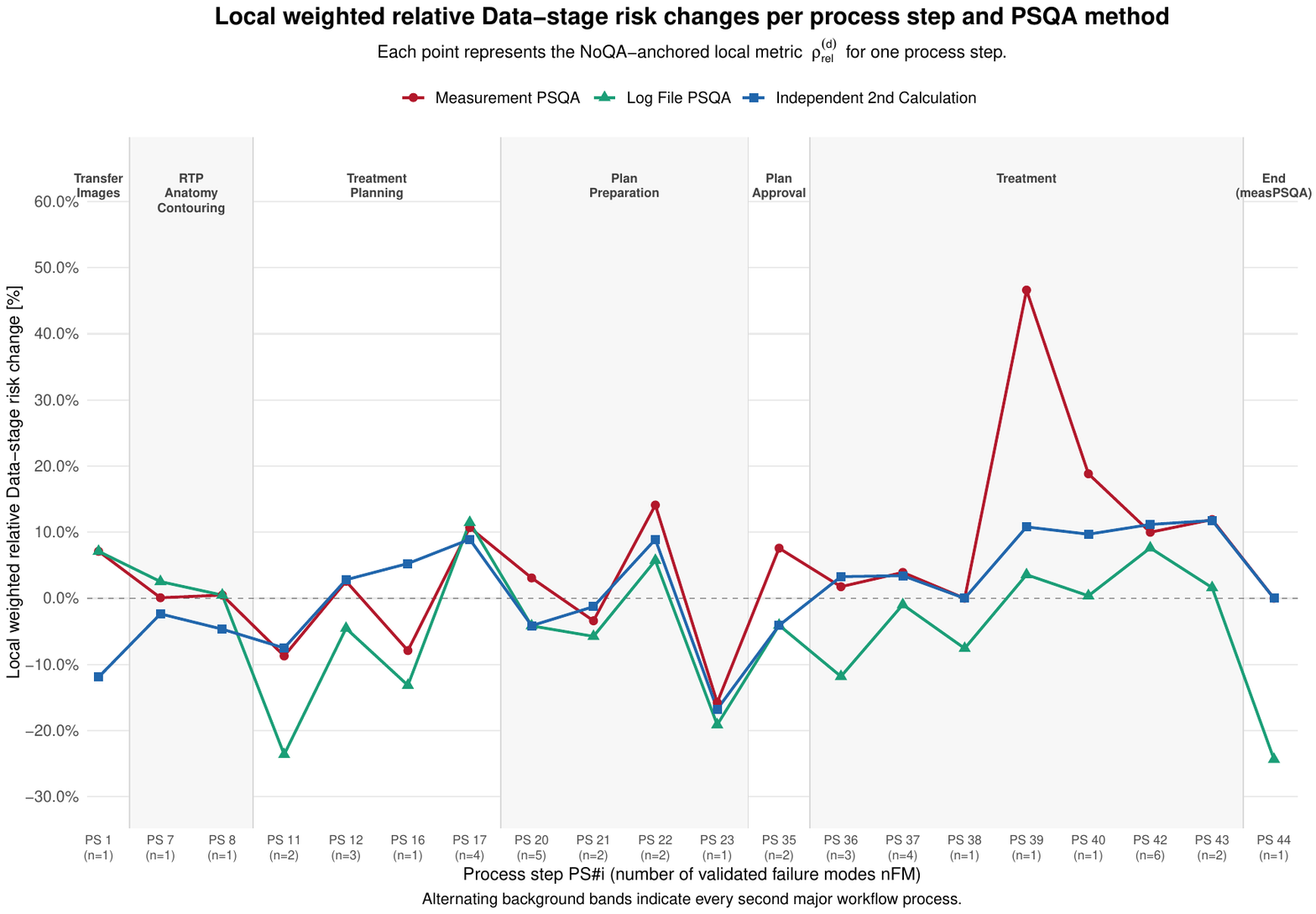}
\caption{\small Local weighted relative \textit{Data}-stage risk changes per process step and PSQA method. Each point represents \(\rho^{*,(d)}_{\mathrm{rel},i}\) for one process step \(i\). Labels below the x-axis show the absolute process-step index together with the number of validated failure modes assigned to that step.}
\label{fig:local_rho_rel_data_per_ps}
\end{figure}

Figure~\ref{fig:local_rho_rel_data_per_ps} shows the locally weighted relative risk changes \(\rho^{*,(d)}_{\mathrm{rel},i}\) during the \textit{Data} phase for all 20 validated process steps and all three PSQA modalities. Since these values are normalized to the total \textit{NoQA} burden of each process step, they quantify the extent to which the total local risk mass of a step is already affected before the final verification stage (\textit{Full}) takes effect. The alternating shaded background assigns the process steps to their respective processes. Note that only those processes are considered whose process steps contain failure modes that are influenced by the respective PSQA method (cf. Sec.~\ref{sec:psqa_fmea_procedures}). At the process-step level, it is therefore possible to determine whether a PSQA path behaves as a low-risk preparatory extension of the base workflow or whether it entails a significant local burden before a cumulative risk reduction is achieved.

At the level of individual process steps, the impact of the \textit{Data} phase shows significant fluctuations and is clearly not without its challenges. \emph{Measurement-based PSQA} exhibits the broadest positive local perturbations, with the largest increases at PS~\#39\footnote{Cf.~Sec.~\ref{sec:res_catalogue} with Tab.~\ref{tab:riskid_catalogue} for the following identifications of the process steps.} (set treatment parameters, \(+46.6\%\)), PS~\#40 (treatment accessories, \(+18.8\%\)), PS~\#22 (review and approve treatment data, \(+14.1\%\)), PS~\#43 (data recording and logging to OIS, \(+11.9\%\)), and PS~\#17 (forward calculation, \(+10.7\%\)). By contrast, its strongest negative local \textit{Data}-stage shifts occur at PS~\#23 (treatment scheduling, \(-15.8\%\)), PS~\#11 (treatment fields insertion, \(-8.7\%\)), and PS~\#16 (optimization, \(-7.9\%\)), showing that measurement-related preparation can either add or remove local burden depending on workflow context.

The \emph{log file-based pathway} remains closest to a low-friction \textit{Data}-stage configuration, but it is not uniformly neutral. Its most pronounced negative local values occur at PS~\#44 (final steps, \(-24.3\%\)), PS~\#11 (\(-23.6\%\)), PS~\#23 (\(-19.1\%\)), PS~\#16 (\(-13.2\%\)), and PS~\#36 (send plan from OIS to TCS, \(-11.8\%\)), whereas positive local perturbations are largely confined to PS~\#17 (\(+11.5\%\)) and PS~\#42 (beam delivery, \(+7.6\%\)). 

\emph{Independent secondary dose calculation} (ISDC) shows the most selective pattern: it reduces local \textit{Data}-stage burden most strongly at PS~\#23 (\(-16.8\%\)) and PS~\#1 (plan CT images import,
\(-11.9\%\)), but produces positive shifts in several late treatment steps, most notably PS~\#43 (\(+11.8\%\)), PS~\#42 (\(+11.2\%\)), PS~\#39 (\(+10.8\%\)), PS~\#40 (\(+9.7\%\)), and PS~\#17 (\(+8.9\%\)). 

Taken together, these local \textit{Data}-stage results confirm that preparatory PSQA activity must not be interpreted as uniformly risk-reducing; rather, it acts as a structured workflow risk perturbation whose direction and magnitude are strongly method- and process step-dependent.

\subsection{Local Cumulative Effects Across Process Steps}
\label{subsec:local_cum_effects}

Whereas Subsec.~\ref{subsec:local_data_effects} quantified the preparatory risk perturbation introduced by the \textit{Data}-stage, the cumulative local metric
\begin{equation}
\rho^{*,(c)}_{\mathrm{rel},i}
\label{eq:local_cum_result_metric}
\end{equation}
describes the final net effect of a complete PSQA pathway on process step \(i\), measured relative to the common \textit{noQA} baseline (cf.~Subsec.~\ref{appB:subsec:local_rel_risk_metric}). In this sense, \(\rho^{*,(c)}_{\mathrm{rel},i}\) is the local end-state quantity and therefore the strictest process step-level basis for comparing the three PSQA methods. It answers the practical question of how much the total local risk burden of a specific process step has been reduced, maintained, or increased once the entire chain of effects has taken place.

This distinction is important because the local cumulative metric does not merely visually summarize the behavior of individual failure modes. Rather, it quantifies the final weighted shift of the entire process step after all FM-specific effects assigned to this step have been aggregated according to their baseline contribution. As a result, \(\rho^{*,(c)}_{\mathrm{rel},i}\) identifies the process steps in which the final net mitigation is truly concentrated, independent of whether that mitigation arises from many moderate FM-level changes or from a smaller number of dominant local contributions.

\begin{figure}[htbp]
\centering
\includegraphics[width=0.99\textwidth]{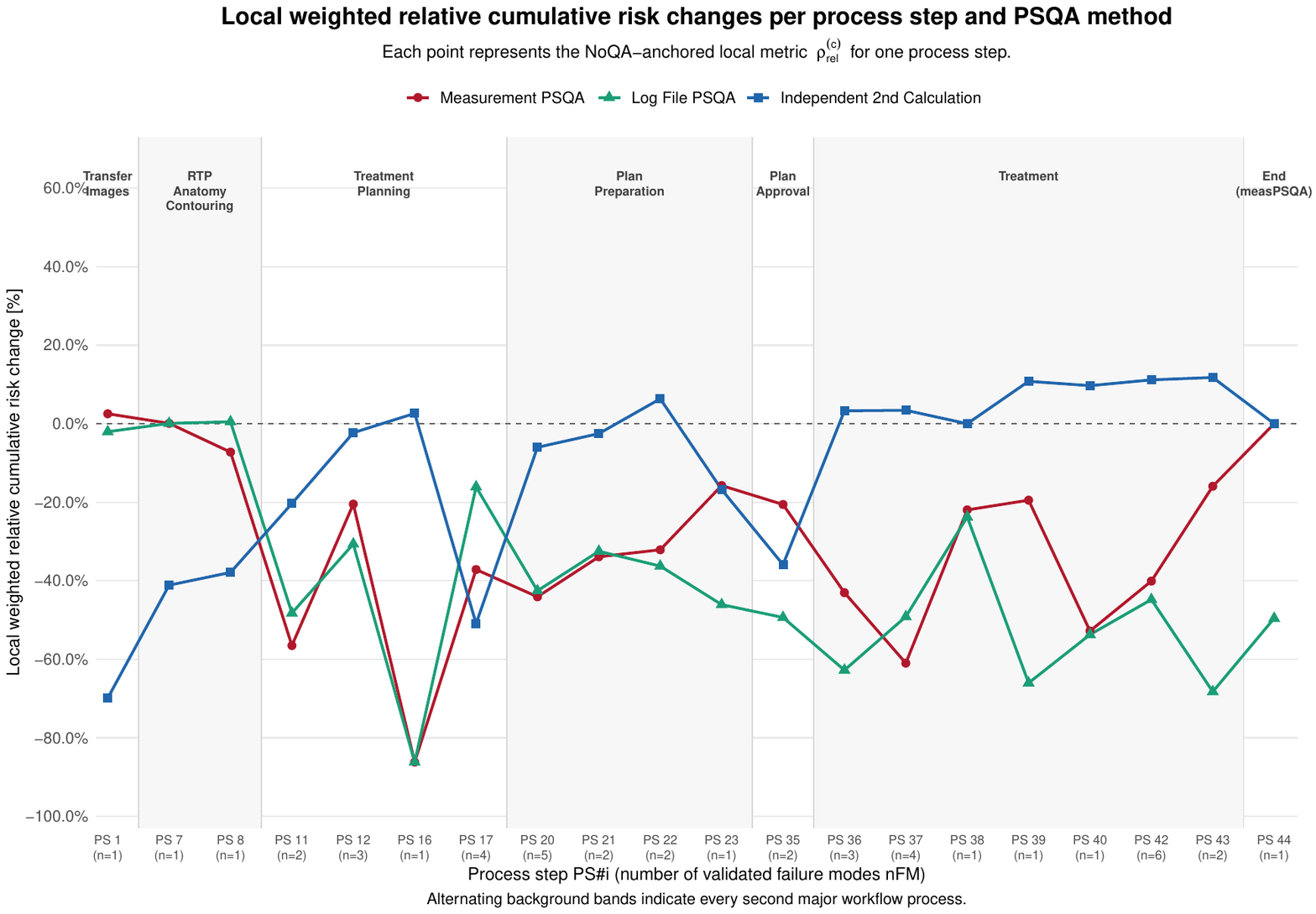}
\caption{\small Local weighted relative cumulative risk changes per process step and PSQA method.}
\label{fig:local_rho_rel_cum_per_ps}
\end{figure}

\paragraph{Interpretive reading.}
At the process-step level, the cumulative picture is already much sharper than the corresponding \textit{Data}-stage view, because the final net effect of the complete PSQA pathway is now visible on the common \textit{noQA} scale. The greatest cumulative risk reductions are relatively concentrated in specific areas and depend on the methods used, rather than being evenly distributed throughout the entire workflow.

For \emph{Measurement PSQA}, the deepest local cumulative reductions are concentrated at PS~\#16 (optimization, \(-86.2\%\)), PS~\#37 (translation of plan to TCS internal settings, \(-61.0\%\)), PS~\#11 (treatment fields insertion, \(-56.5\%\)), PS~\#40 (treatment accessories, \(-52.8\%\)), and PS~\#20 (plan export/import to OIS, \(-44.1\%\)). This pattern confirms that measurement-based verification does not merely compensate local \textit{Data}-stage perturbations, but achieves substantial end-state risk mitigation across a broad set of planning, transfer, and setup-related steps. It is particularly noteworthy that the greatest gains in efficiency are not limited to a single workflow area, but extend across optimization, plan implementation, accessory preparation, and the management of interfaces related to delivery.

For \emph{Log File PSQA}, the cumulative process-step signature is both deeper and more spatially focused in the late workflow. In addition to PS~\#16 (\(-86.2\%\)), the strongest net reductions occur at PS~\#43 (data recording and logging to OIS, \(-68.3\%\)), PS~\#39 (set treatment parameters, \(-66.0\%\)), PS~\#36 (send plan from OIS to TCS, \(-62.8\%\)), PS~\#40 (treatment accessories, \(-53.7\%\)), and PS~\#44 (final steps, \(-49.6\%\)). Thus, once the complete log file-based pathway is in place, its dominant local benefit is concentrated in those process steps that are structurally closest to delivery execution, machine-state transfer, and recording/documentation. This is the local process-step analogue of the FM-level conclusion that \textit{Log} achieves its strongest final mitigation in delivery- and recording-sensitive regions.

For \emph{Independent Secondary Dose Calculation}, the cumulative local pattern remains substantially narrower. Its strongest reductions occur at PS~\#1 (plan CT images import, \(-69.9\%\)), PS~\#17 (forward calculation, \(-50.9\%\)), PS~\#7 (normal tissue delineation, \(-41.1\%\)), PS~\#8 (density overrides, \(-37.9\%\)), and PS~\#35 (final plan approval, \(-35.9\%\)). At the same time, the cumulative metric remains positive in several late treatment steps, most notably PS~\#43 (\(+11.8\%\)), PS~\#42 (\(+11.2\%\)), PS~\#39 (\(+10.8\%\)), and PS~\#40 (\(+9.7\%\)). This confirms that ISDC retains a clearly valuable but highly selective role: it is locally strong where image handling, contouring, calibration, and forward-calculation sensitivity dominate, but it does not provide broad net mitigation in the late delivery and recording part of the workflow.

Taken together, Fig.~\ref{fig:local_rho_rel_cum_per_ps} shows that the final local end-state comparison is far from uniform. Measurement-based PSQA produces the broadest multi-region mitigation pattern, log file-based PSQA produces the strongest and most coherent late-workflow mitigation pattern, and ISDC produces a highly selective early-workflow pattern with limited local reach beyond its core model-sensitive domain. The cumulative process-step view therefore provides the first level at which the final net benefit of each PSQA pathway becomes both mathematically aggregated and still clinically localized.

\subsection{Local Data-to-Cumulative Bridge Shifts}
\label{subsec:local_bridge_effects}

To isolate the additional local contribution generated when the workflow moves from the \textit{Data}-stage state to the final cumulative state, we introduce the process-step bridge metric
\begin{equation}
\rho^{*,(d\to c)}_{\mathrm{rel},i}
:=
\rho^{*,(c)}_{\mathrm{rel},i}
-
\rho^{*,(d)}_{\mathrm{rel},i}.
\label{eq:local_bridge_result_metric}
\end{equation}
Using the formalism developed in Appendix~B (cf.~Sec.~\ref{appB:sec:local_risk_metrics}), this quantity may equivalently be written as
\begin{equation}
\rho^{*,(d\to c)}_{\mathrm{rel},i}
=
\rho^{*,(f)}_{\mathrm{rel},i}\bigl(1+\rho^{*,(d)}_{\mathrm{rel},i}\bigr)
=
\frac{\rho^{*,(f)}_{\mathrm{abs},i}}{G_i^{(d)}}.
\label{eq:local_bridge_result_metric_equiv}
\end{equation}

This is the most informative new quantity at the process-step level because it expresses the additional \textit{Full}-stage contribution on the same common \textit{noQA} scale already used by the \textit{Data} and \textit{Cum} metrics. It is therefore not identical to the classical \textit{Full}-stage metric \(\rho^{*,(f)}_{\mathrm{rel},i}\), which remains normalized to the method-specific \textit{Data}-stage burden. The bridge metric instead answers a different and clinically more transparent question: how much further inward does a complete PSQA pathway move the local process-step burden once the preparatory \textit{Data}-stage state has already been reached?

\paragraph{Interpretive reading.}
The sign of \(\rho^{*,(d\to c)}_{\mathrm{rel},i}\) has direct operational significance. Negative values indicate a local reduction caused by the completion of the \textit{Full}-stage phase; values close to zero suggest that the state of the \textit{Data}-stage phase is already close to the final \textit{Cum} state; and positive values would indicate that the completion of the path entails an additional burden rather than a further reduction. In practice, the bridge metric therefore supports process steps whose local benefits are already initiated during preparation in the data stage phase, in contrast to those that achieve their final reduction only after completion of the verification layer, because an unfavorable risk influence from the \textit{Data}-stage must (first) be compensated for.

\paragraph{Why this quantity matters.}
This local bridge metric is the process-step analogue of the FM-level dumbbell logic developed in Subsec.~\ref{subsec:results_d2c_relative}. On the FM level, the segment between the \textit{Data} and \textit{Cum} markers showed the additional signed shift on a common \textit{noQA} scale; here, \(\rho^{*,(d\to c)}_{\mathrm{rel},i}\) provides exactly the same interpretive layer after the FM-level changes have been aggregated into weighted process-step burdens. For that reason, it is the natural quantity for identifying where the final local mitigation is generated: not merely where a PSQA method ends up, but where the crucial transition from preparatory state to completed end state actually occurs.

\begin{figure}[htbp]
\centering
\includegraphics[width=\textwidth]{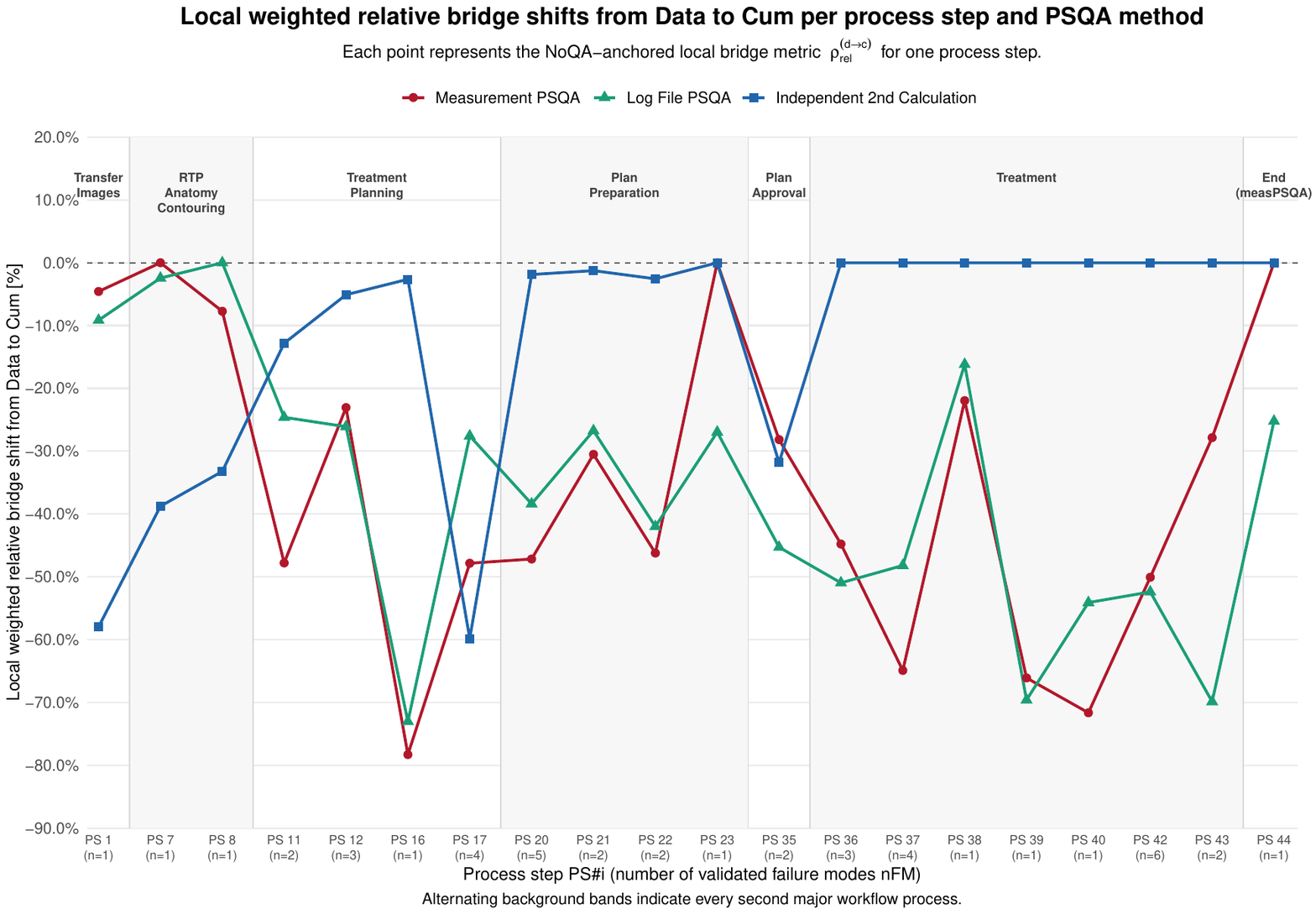}
\caption{\small Local weighted relative bridge shifts from Data to Cum per process step and PSQA method.}
\label{fig:local_rho_rel_d2c_per_ps}
\end{figure}

The bridge representation in Fig.~\ref{fig:local_rho_rel_d2c_per_ps} reveals the mechanism that is largely hidden in the cumulative end-state plot alone. For \emph{Measurement PSQA}, the strongest additional local shifts from \textit{Data} to \textit{Cum} occur at PS~\#16 (optimization, \(-78.3\%\)), PS~\#40 (treatment accessories, \(-71.6\%\)), PS~\#39 (set treatment parameters, \(-66.1\%\)), PS~\#37 (translation to TCS internal settings, \(-64.9\%\)), PS~\#42 (beam delivery, \(-50.1\%\)), and PS~\#17 (forward calculation, \(-47.8\%\)). This shows that the broad cumulative benefit of measurement-based PSQA is not primarily created at one single interface, but through a distributed bridge effect spanning planning, translation, treatment setup, and delivery-adjacent process steps. Conversely, the bridge shift vanishes or becomes negligible at PS~\#7, PS~\#23, and PS~\#44, indicating that the \textit{Data}-stage state in those steps is already close to the final cumulative endpoint.

For \emph{Log File PSQA}, the bridge pattern is even more concentrated and more strongly shifted toward the late workflow. The largest additional local shifts are observed at PS~\#16 (\(-73.0\%\)), PS~\#43 (data recording and logging to OIS, \(-69.9\%\)), PS~\#39 (set treatment parameters, \(-69.6\%\)), PS~\#40 (treatment accessories, \(-54.1\%\)), PS~\#42 (beam delivery, \(-52.4\%\)), PS~\#36 (send plan from OIS to TCS, \(-51.0\%\)), and PS~\#37 (translation to TCS internal settings, \(-48.2\%\)). In other words, the strongest incremental completion of the log-based pathway is created exactly where delivery-state observability, machine-state transfer, and treatment recording are most critical. This is the process-step confirmation of the FM-level conclusion that \textit{Log} dominates the late delivery and documentation block once the pathway is fully completed.

For \emph{Independent Secondary Dose Calculation}, the bridge metric is not only narrower than for \textit{Meas} and \textit{Log}, but structurally more discontinuous. Its strongest additional shifts are concentrated at PS~\#17 (forward calculation, \(-59.8\%\)), PS~\#1 (plan CT images import, \(-58.0\%\)), PS~\#7 (normal tissue delineation, \(-38.8\%\)), PS~\#8 (density overrides, \(-33.2\%\)), and PS~\#35 (final plan approval, \(-31.8\%\)). By contrast, the bridge contribution is essentially absent across the late treatment block from PS~\#36 onward, where the \textit{Calc} pathway remains at or extremely near zero. This is a particularly important mechanistic result: ISDC can generate substantial additional local mitigation, but almost exclusively in early image-, contour-, and calculation-sensitive regions; it contributes essentially no late-stage bridge from \textit{Data} to \textit{Cum} in delivery and recording steps.

Taken together, Fig.~\ref{fig:local_rho_rel_d2c_per_ps} shows that the bridge metric is the clearest local mechanism quantity in the entire process-step hierarchy. It does not merely identify where a method ends up, but where the decisive incremental mitigation is actually produced. Measurement-based PSQA displays the broadest compensatory bridge across preparation, translation, and setup; log file-based PSQA shows the strongest bridge in delivery-, transfer-, and recording-dominated steps; and ISDC shows a narrow but clearly interpretable bridge concentrated in model-sensitive early workflow regions.

\subsection{Process-Step Archetypes and Method-Specific Strengths}
\label{subsec:process_step_archetypes}

The combined reading of the \textit{Data}, \textit{Cum}, and bridge plots suggests that the process-step level is best understood in terms of a small number of recurring local archetypes rather than as a flat list of 20 isolated process steps. Three such archetypes are particularly useful for interpreting the method-specific strengths of the PSQA pathways.

\paragraph{Preparation and translation archetype.}
This archetype encompasses the high-risk planning and transmission areas around PS~\#20–23 and PS~\#36–40, where workflow preparation, implementation in treatment control settings, and the configuration of accessories or parameters dominate the local risk profile. In this area, \emph{Measurement PSQA} exhibits the broadest compensatory effect: Its disruptions during the data phase are often significant, yet its cumulative and bridge diagrams show that completing the entire path subsequently greatly reduces the risk of these steps. The result is a broad, distributed local risk reduction pattern rather than a single, sharply localized risk minimum.

\paragraph{Delivery and recording archetype.}
This archetype is dominated by PS~\#39, PS~\#42, PS~\#43, and PS~\#44, i.e., parameter setting, beam delivery, recording to the OIS, and final treatment-adjacent completion steps. Here, \emph{Log File PSQA} is clearly dominant. Its cumulative process-step values are among the most negative in the late workflow, and its bridge shifts are deepest precisely in this region. This confirms that once machine-state observability and post-delivery record integrity become decisive, log file-based verification provides the strongest local contribution to final risk reduction.

\paragraph{Calculation-sensitive archetype.}
This archetype includes PS~\#1, PS~\#7, PS~\#8, and especially PS~\#17, where CT import, contouring, density override handling, and forward calculation determine the local risk response. In this domain, \emph{Independent Secondary Dose Calculation} is most distinctive. Its strongest cumulative and bridge reductions are concentrated exactly in these steps, while its late-workflow contribution remains weak or vanishing. This makes \textit{Calc} locally powerful but structurally selective: it is not a broad workflow-wide mitigator, but it is highly effective where the dominant uncertainty is algorithmic, calibration-related, or model-dependent.

\paragraph{Overall process-step synthesis.}
These three archetypes provide a compact interpretation of the complete local result triad. Measurement-based PSQA is the broadest local compensator across preparation, translation, and setup. Log file-based PSQA is the strongest local finisher in the treatment and recording block. ISDC is the most selective local specialist, with its strongest contribution confined to early image- and calculation-sensitive steps. The process-step level therefore sharpens the FM-level results into a clinically interpretable map of where each method achieves its most meaningful local benefit and why those strengths differ structurally across the workflow.

\section{Risk-Weighted Results on the Workflow Level}
\label{sec:workflow_results}

\subsection{Why Workflow-Level Aggregation Is Needed}
\label{subsec:why_workflow_aggregation}

The process-step results in Sec.~\ref{sec:process_step_results} provide the first risk-weighted aggregation beyond the FM level. They identify which local workflow locations are most strongly affected by the \textit{Data} stage, the \textit{Cum} end state, and the \textit{Data-to-Cum} bridge. However, even this process-step representation remains too granular to answer the final system-level comparison question. Twenty local quantities still describe a structured map of the workflow, not yet a single workflow-wide risk signature.

The workflow-level or global analysis therefore asks a different question: after all local process-step burdens have been weighted by their baseline contribution and aggregated, which PSQA method changes the total workflow risk burden most effectively? This question cannot be answered by visually inspecting individual FMs or by averaging process-step percentages. A process step that carries a large baseline burden must contribute more to the workflow-level result than a step with minimal baseline burden. In simple terms, a step that carries ten times more baseline burden should not count the same as a step that carries almost none.

For this reason, the global metrics introduced in Appendix~B (cf.~Sec.~\ref{appB:sec:global_risk_metrics}) aggregate the local process-step quantities using burden-based workflow weights. For the complete workflow, this yields the global \textit{Data}-stage, cumulative, and \textit{Data-to-Cum} bridge quantities
\begin{equation}
\Rho^{*,(d)}_{\mathrm{rel},\mathcal{W\!F}},
\qquad
\Rho^{*,(c)}_{\mathrm{rel},\mathcal{W\!F}},
\qquad
\Rho^{*,(d\to c)}_{\mathrm{rel},\mathcal{W\!F}}.
\end{equation}
These quantities retain the same interpretive logic as their process-step counterparts, but they now answer the system-level question. The \textit{Data}-stage metric describes the preparatory workflow perturbation, the \textit{Cum} metric describes the final NoQA-anchored end state, and the bridge metric describes the additional \textit{NoQA}-anchored shift generated when moving from \textit{Data} to \textit{Cum}.

Thus, the workflow-level analysis completes the hierarchy
\begin{equation}
\delta \;\longrightarrow\; \rho \;\longrightarrow\; \Rho,
\end{equation}
moving from individual failure modes to process steps and finally to the full workflow. This is the point at which the risk analysis changes from a local map into a global PSQA comparison.

\subsection{Global Data-Stage and Cumulative Effects}
\label{subsec:global_data_cum_effects}

The first workflow-level comparison concerns the relationship between the preparatory \textit{Data}-stage effect and the final cumulative effect. These two quantities are directly comparable because both are anchored to the same common \textit{NoQA} workflow burden:
\begin{equation}
\Rho^{*,(d)}_{\mathrm{rel},\mathcal{W\!F}}
=
\frac{\Rho^{*,(d)}_{\mathrm{abs},\mathcal{W\!F}}}{G^{(d)}_{\mathcal{W\!F}}},
\qquad
\Rho^{*,(c)}_{\mathrm{rel},\mathcal{W\!F}}
=
\frac{\Rho^{*,(c)}_{\mathrm{abs},\mathcal{W\!F}}}{G^{(d)}_{\mathcal{W\!F}}}.
\end{equation}
The \textit{Data}-stage value therefore describes how much the introduction of the PSQA-specific preparatory workflow changes the total baseline burden before the final verification layer has acted. The cumulative value describes the final net effect after the complete pathway has been applied.

At the workflow level, the \textit{Data}-stage effects remain relatively small compared with the \textit{Cum}-end-state effects, but their signs are highly informative. \emph{Measurement-based PSQA} shows a positive \textit{Data}-stage contribution of \(+4.5\%\), indicating that its preparatory workflow introduces measurable additional risk before the verification layer compensates for it. \emph{Log file-based PSQA} shows a slightly negative \textit{Data}-stage contribution of \(-2.9\%\), consistent with a comparatively low-risk preparatory layer that already reduces part of the workflow burden. \emph{Independent Secondary Dose Calculation} remains close to neutral, with a small positive \textit{Data}-stage perturbation of \(+1.6\%\). Thus, the \textit{Data} stage alone does not explain the final ranking of PSQA methods; it mainly describes the initial workflow friction or early preparatory relief.

The cumulative workflow metrics provide the strict global comparison. They show the final \textit{NoQA}-anchored effect of the complete PSQA pathways and therefore represent the appropriate workflow-level endpoint for inter-method comparison. In this final state, \textit{Log} achieves the strongest workflow-wide reduction with \(-40.3\%\), \textit{Meas} follows with a still substantial reduction of \(-34.7\%\), and \textit{Calc} (or ISDC) provides a smaller but meaningful global reduction of \(-8.5\%\). This ordering is consistent with the process-step results: \textit{Log} benefits most from its strong late-workflow delivery and recording contributions, \textit{Meas} benefits from broad compensatory mitigation across planning, preparation, translation, setup, and delivery-adjacent steps, and \textit{Calc} remains concentrated in early image-, contour-, and calculation-sensitive regions whose contribution is important but less workflow-wide.

This comparison also clarifies why the cumulative metric must be distinguished from the classical \textit{Full}-stage metric. The \textit{Full}-stage relative quantity is normalized to the method-specific \textit{Data}-stage risk burden and is therefore best interpreted as a within-method incremental quantity. By contrast, the cumulative quantity is normalized to the common \textit{NoQA} workflow burden and is therefore the appropriate global endpoint for comparing PSQA methods.

\subsection{Global \textit{Data-to-Cum} Bridge Shifts and Decomposition of Net Benefit}
\label{subsec:global_bridge_effects}

The workflow-level bridge quantity isolates the additional \textit{NoQA}-anchored shift that occurs when the system moves from the \textit{Data}-stage state to the cumulative end state:
\begin{equation}
\Rho^{*,(d\to c)}_{\mathrm{rel},\mathcal{W\!F}}
:=
\Rho^{*,(c)}_{\mathrm{rel},\mathcal{W\!F}}
-
\Rho^{*,(d)}_{\mathrm{rel},\mathcal{W\!F}}.
\end{equation}
Equivalently, using the closed-form relations from Appendix~B (cf.~Subsec.~\ref{appB:subsec:global_abs_rel_relation}),
\begin{equation}
\Rho^{*,(d\to c)}_{\mathrm{rel},\mathcal{W\!F}}
=
\Rho^{*,(f)}_{\mathrm{rel},\mathcal{W\!F}}
\left(1+\Rho^{*,(d)}_{\mathrm{rel},\mathcal{W\!F}}\right)
=
\frac{\Rho^{*,(f)}_{\mathrm{abs},\mathcal{W\!F}}}{G^{(d)}_{\mathcal{W\!F}}}.
\end{equation}
This quantity is not a new scoring stage. Rather, it is a derived bridge metric that expresses the additional \textit{Full}-stage contribution on the same common \textit{NoQA} scale used by the \textit{Data} and cumulative metrics.

This bridge representation gives a particularly transparent decomposition of the final net workflow benefit:
\begin{equation}
\Rho^{*,(c)}_{\mathrm{rel},\mathcal{W\!F}}
=
\Rho^{*,(d)}_{\mathrm{rel},\mathcal{W\!F}}
+
\Rho^{*,(d\to c)}_{\mathrm{rel},\mathcal{W\!F}}.
\end{equation}
This additive decomposition should not be confused with the multiplicative stage relation developed in Appendix~B (cf.~Subsec.~\ref{appB:subsec:global_bridge_metrics}). The multiplicative identity remains the stage-consistent relationship between \textit{Data} and \textit{Full}. The equation above is instead an interpretive decomposition on the common \textit{NoQA} scale: it separates the final cumulative effect into the contribution already present after the Data stage and the additional shift generated when moving from \textit{Data} to \textit{Cum}.

\begin{figure}[h!]
\centering
\includegraphics[width=\textwidth]{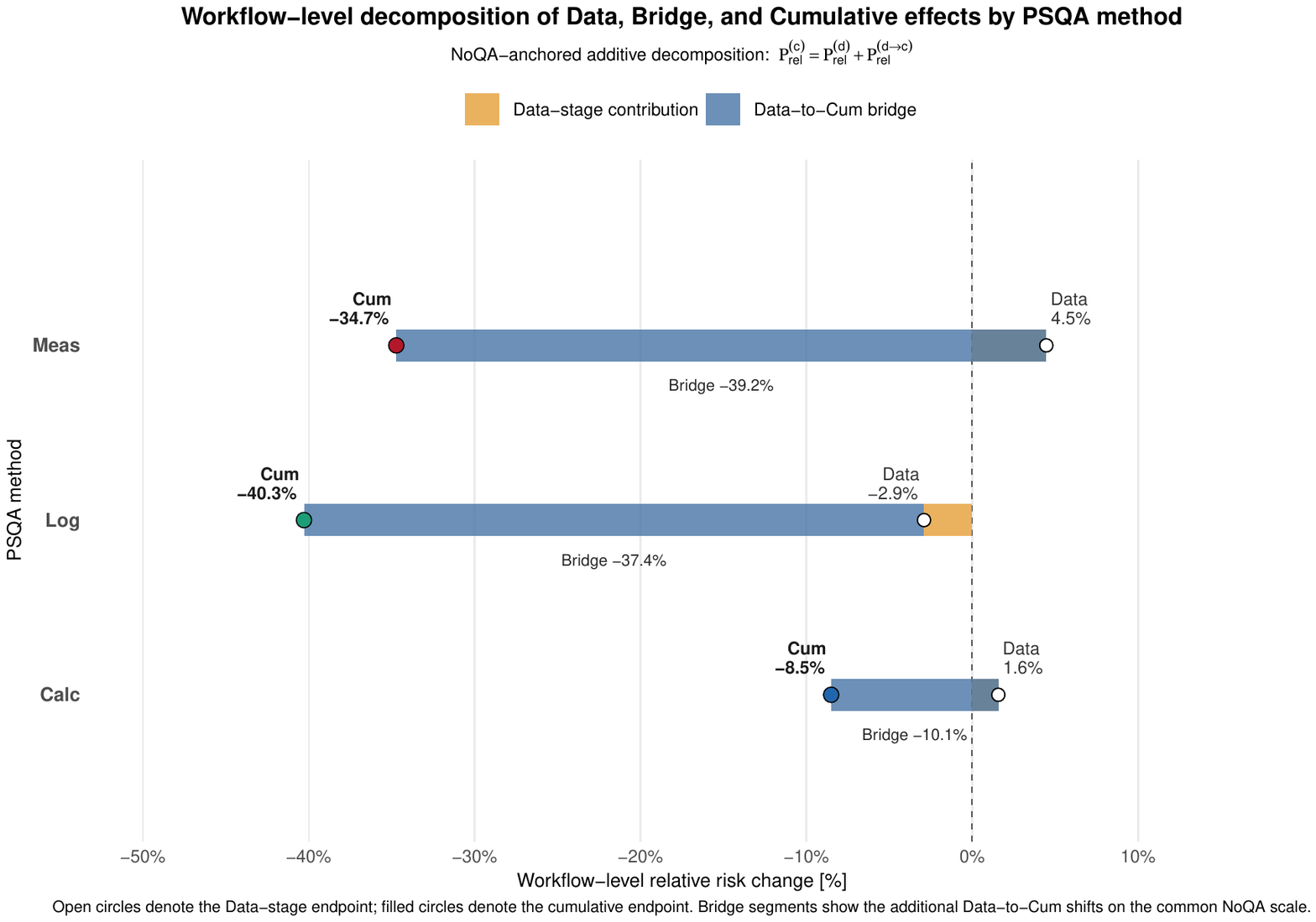}
\caption{\small Workflow-level decomposition of Data-stage, Data-to-Cumulative bridge, and cumulative relative effects by PSQA method.}
\label{fig:global_workflow_decomposition_data_bridge_cum}
\end{figure}

Figure~\ref{fig:global_workflow_decomposition_data_bridge_cum} makes this decomposition explicit. \emph{Measurement-based PSQA} starts with a positive \textit{Data}-stage perturbation of \(+4.5\%\), but then produces a large additional bridge shift of \(-39.2\%\), resulting in a cumulative workflow-level effect of \(-34.7\%\). In practical terms, the method initially adds workflow burden through preparatory measurement-related data activity, but the subsequent completion of the verification pathway more than compensates for this initial risk friction. The \textit{Meas} pathway therefore wins its global effect mainly through a broad compensatory bridge.

\emph{Log file-based PSQA} behaves differently. It begins with a mildly favorable \textit{Data}-stage effect of \(-2.9\%\) and then adds a very strong bridge contribution of \(-37.4\%\), reaching the strongest cumulative workflow-level reduction of \(-40.3\%\). This means that \textit{Log} does not win globally because of a large preparatory advantage alone. It wins because a low-risk \textit{Data}-stage is followed by a deep additional \textit{Data-to-Cum} bridge, driven especially by delivery-, transfer-, and recording-related workflow regions.

\emph{Independent Secondary Dose Calculation} shows a third pattern. Its \textit{Data}-stage contribution is slightly positive at \(+1.6\%\), and its bridge contribution reaches only \(-10.1\%\), yielding a cumulative workflow effect of \(-8.5\%\). This does not imply that ISDC is ineffective. Rather, it shows that its strong local benefits are concentrated in a comparatively narrow subset of image-, contour-, and calculation-sensitive regions and do not dominate the total workflow burden. The global decomposition therefore explains why \textit{Calc} can be locally powerful but globally less dominant than \textit{Meas} or \textit{Log}.

The bridge metric therefore answers the decisive workflow-level mechanism question: does a method achieve its final net benefit because its \textit{Data}-stage is already favorable, or because the additional completion of the pathway generates a strong risk mitigation shift from \textit{Data} to \textit{Cum}? In the present workflow, the answer is method-specific: \textit{Log} combines low \textit{Data}-stage friction with the strongest overall bridge; \textit{Meas} requires a strong bridge to compensate a positive \textit{Data}-stage burden; and \textit{Calc} remains a selective bridge contributor whose strongest effects do not cover enough high-burden workflow regions to determine the global ranking.

\subsection{Workflow-Level Comparison by Major Processes}
\label{subsec:workflow_region_comparison}

The complete workflow-level metrics in Subsec.~\ref{subsec:global_bridge_effects} provide the final global ranking, but they intentionally collapse the entire process map into one number per PSQA method. This is necessary for system-level comparison, but it hides an important mechanistic question: \emph{which major workflow processes produce the final global result?} To answer this, the workflow was further partitioned according to the major process groupings used in the workflow maps (cf.~Appendix A), and each region was analyzed using the same risk-weighted re\-lative metrics as the full workflow, i.e.~the general process-step collection $\mathcal{I}$ was identified with a single major process $\mathcal{P}$, such that $\mathcal{I} = \mathcal{P}$ instead of $\mathcal{I} = \mathcal{W\!F}$ (global case; cf.~\ref{appB:subsec:set_representation}).

\begin{figure}[htbp]
\centering
\includegraphics[width=\textwidth,keepaspectratio]{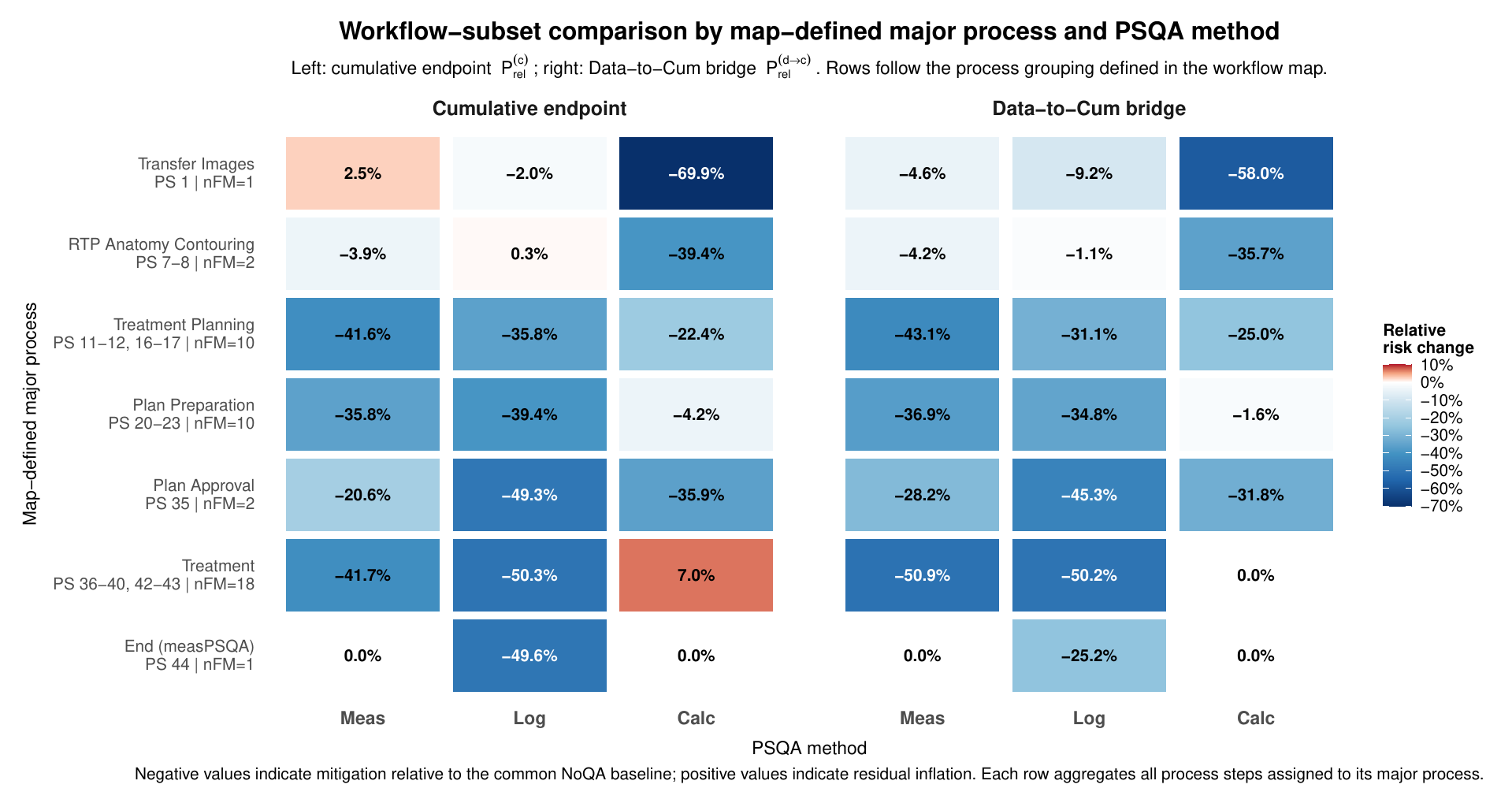}
\caption{\small Workflow-subset comparison by major processes and PSQA method. The left panel shows the cumulative endpoint \( \Rho^{*,(c)}_{\mathrm{rel},\mathcal{I}} \), which represents the total aggregated risk change per process applying a given PSQA approach. The right panel shows the \textit{Data-to-Cum} bridge \( \Rho^{*,(d\to c)}_{\mathrm{rel},\mathcal{I}} \) that illustrates the risk-reduction potential of each PSQA method without taking into account the inherent impact that the selected method has on data integrity within the process under consideration. Rows correspond to map-defined major workflow processes.}
\label{fig:workflow_region_heatmap_cum_bridge}
\end{figure}

Figure~\ref{fig:workflow_region_heatmap_cum_bridge} should be understood as a regional map showing where each PSQA method achieves its ultimate strength at the process level. The left panel shows the cumulative endpoint, i.e., the final \textit{NoQA}-anchored net effect. The right panel shows the corresponding \textit{Data-to-Cum} bridge, i.e., the additional \textit{NoQA}-anchored shift generated after the \textit{Data}-stage. The two panels together distinguish \emph{where a process ends} from \emph{where its incremental completion effect is generated}.

The strongest early-workflow result is the \textit{Calc} effect in 'Transfer Images' process. In that region, ISDC reaches a cumulative reduction of \(-69.9\%\) and a bridge contribution of \(-58.0\%\), whereas \textit{Meas} remains mildly positive at \(+2.5\%\) and \textit{Log} is nearly neutral at \(-2.0\%\). This is a clean example of method specificity: the dominant risk mechanism in this region is image- and calibration-sensitive, so \textit{Calc} provides the strongest regional contribution. A similar, though less extreme, pattern appears in 'RTP Anatomy Contouring' process, where \textit{Calc} reaches \(-39.4\%\) cumulatively and \(-35.7\%\) as bridge, while \textit{Meas} and \textit{Log} remain near neutral.

'Treatment Planning' and 'Plan Preparation' form the first broad multi-method mitigation block. In the 'Treatment Planning' process, all three methods reduce regional risks, with cumulative values of \(-41.6\%\) for \textit{Meas}, \(-35.8\%\) for \textit{Log}, and \(-22.4\%\) for \textit{Calc}. The corresponding bridge values are \(-43.1\%\), \(-31.1\%\), and \(-25.0\%\), respectively. This shows that the final regional risk mitigation is mainly created by the \textit{Data-to-Cum} completion step, especially for \textit{Meas}. In 'Plan Preparation', \textit{Meas} and \textit{Log} remain closely matched at the cumulative endpoint (\(-35.8\%\) and \(-39.4\%\)), while \textit{Calc} is much weaker at \(-4.2\%\). The bridge panel explains why: \textit{Meas} and \textit{Log} retain large additional inward shifts (\(-36.9\%\) and \(-34.8\%\)), whereas \textit{Calc} contributes only \(-1.6\%\).

'Plan Approval' shows another distinct pattern. \textit{Log} achieves the strongest cumulative risk reduction in this region (\(-49.3\%\)), followed by \textit{Calc} (\(-35.9\%\)) and \textit{Meas} (\(-20.6\%\)). The bridge values mirror this ordering: \textit{Log} reaches \(-45.3\%\), \textit{Calc} \(-31.8\%\), and \textit{Meas} \(-28.2\%\). Thus, 'Plan Approval' is not purely a calculation-sensitive region nor purely a delivery-sensitive region; it is a convergence point where \textit{Log} and \textit{Calc} both provide strong additional mitigation, while \textit{Meas} still contributes but less dominantly.

The 'Treatment' process is the decisive late-workflow discriminator. \textit{Log} achieves the strongest cumulative reduction at \(-50.3\%\), closely followed by \textit{Meas} at \(-41.7\%\), whereas \textit{Calc} shows a residual positive cumulative value of \(+7.0\%\). The bridge panel is even more revealing: \textit{Meas} and \textit{Log} both produce very strong additional local-to-regional completion effects (\(-50.9\%\) and \(-50.2\%\)), while \textit{Calc} contributes \(0.0\%\). This is the clearest regional explanation for the global ranking. In the largest and most delivery-relevant workflow block, \textit{Calc} essentially does not add a \textit{Data-to-Cum} bridge, while \textit{Meas} and \textit{Log} both do; \textit{Log} ultimately dominates because its cumulative late-workflow risk profile is slightly deeper and more coherent.

Finally, the 'End' process is specific to measurement-related finalization and log-based post-delivery information. \textit{Log} reaches a cumulative reduction of \(-49.6\%\) with a bridge of \(-25.2\%\), while \textit{Meas} and \textit{Calc} remain at \(0.0\%\). This does not mean that the other methods are globally irrelevant; rather, it shows that this specific terminal region is structurally visible to \textit{Log} and not meaningfully modified by \textit{Calc} or \textit{Meas} in the same way.

Taken together, the heatmap shows that the global ranking does not arise from a uniform advantage of one PSQA method everywhere. Instead, the workflow is regionally heterogeneous. \textit{Calc} is strongest in early image-, contour-, and calculation-sensitive process regions, \textit{Log} dominates the late delivery, treatment, recording, and finalization processes, and \textit{Meas} provides broad compensatory mitigation across planning, preparation, translation, and treatment. The global superiority of \textit{Log} is therefore not a generic property of all regions, but the result of strong effects in high-burden late-workflow processes combined with a low-risk \textit{Data}-stage. The broad strength of \textit{Meas} comes from its distributed bridge across several mid- and late-workflow regions. The more limited global effect of \textit{Calc} follows directly from its regional selectivity.

\subsection{Workflow-Level Synthesis and PSQA Strategy Implications}
\label{subsec:workflow_synthesis}

The complete workflow-level analysis leads to a simple but powerful interpretation. No single PSQA method is uniformly best across all regions of the workflow. Instead, each method has a characteristic risk signature.

\emph{Log File PSQA} provides the strongest workflow-wide net mitigation. Its global cumulative effect is largest because it combines a mildly favorable \textit{Data}-stage contribution with a deep \textit{Data-to-Cum} bridge, especially in 'Treatment', recording, parameter transfer, and finalization-related workflow parts. Its strength is therefore not merely that it observes delivered treatment data, but that it does so precisely in the regions of the workflow where delivery-state and recording-state information carry high-risk relevance.

\emph{Measurement PSQA} provides the broadest compensatory pattern. It begins with the highest \textit{Data}-stage burden, reflecting the additional workflow complexity introduced by measurement preparation and execution. However, its subsequent bridge is deep enough to convert this initial risk friction into a strong cumulative reduction. \textit{Meas} therefore acts as a broad physical and procedural compensator across planning, preparation, translation, setup, and treatment-related steps.

\emph{Independent Secondary Dose Calculation} is the most selective strategy. It is locally and regionally powerful in the parts of the workflow where CT handling, contouring, density overrides, algorithm selection, and forward calculation dominate the risk mechanism. However, its bridge contribution largely disappears in late delivery and recording process regions, which explains why its global cumulative effect remains much smaller than that of \textit{Meas} and \textit{Log}. Therefore, ISDC is not globally dominant, but it is strategically indispensable.

The workflow-level result therefore argues against a winner-takes-all interpretation of PSQA. \textit{Log} appears strongest as a workflow-wide backbone for delivery- and recording-sensitive risk reduction; \textit{Meas} remains essential as a broad physical and interface-sensitive verification layer; \textit{Calc} provides a targeted, structurally independent safeguard against image-, model-, and calculation-related failure modes. The most defensible PSQA strategy is therefore not replacement of one method by another, but a hybrid architecture in which each method is assigned to the workflow regions where its risk signature is strongest.

In this sense, the hierarchy developed throughout Secs.~\ref{sec:res_catalogue}--\ref{sec:workflow_results} closes the argument of the Results section. The FM-level analysis identifies individual vulnerabilities, the process-step analysis shows where those vulnerabilities combine into local weak zones, and the workflow-level analysis reveals the net safety signature of the entire PSQA pathway. Only after all three levels are considered together does the comparative role of \textit{Meas}, \textit{Log}, and \textit{Calc} become fully interpretable.

\section{Root-Cause Structure of the PSQA-Relevant Failure Modes}
\label{sec:root_cause_structure}

As a final result-level perspective, the 44 validated PSQA-relevant failure modes were reviewed with respect to their documented causes, effects, contextual notes, and PSQA-specific detectability comments collected during the pFMEA. This analysis is intended as a structured result that links the pFMEA outcome back to the socio-technical premise established in the Prologue and the broad workflow-oriented definition of PSQA introduced in Sec.~\ref{sec:general_psqa_definition}.

The most prominent finding is that the root-cause structure is dominated by ``human error'', i.e. procedural, communicative, and human-system interaction components. 20 of the 44 validated FMs fall into this human/procedural category. A further 8 FMs explicitly name human or procedural mechanisms such as manual data transfer, miscommunication, insufficient review, or plan-selection complexity. In addition, 5 FMs describe mixed human-system or human-technical causes, such as user interaction with TCS, network, or human-driven equipment malfunction. Taken together, at least 33 of the 44 validated FMs, corresponding to 75.0\%, are therefore human-, procedure-, communication-, or human-system-interaction-related.

This result provides an important empirical closure to the conceptual framing of this PSQA risk assessment report. The Prologue argues that radiotherapy quality assurance cannot be reduced to technical quality control alone, because modern particle therapy workflows are complex socio-technical systems involving human procedures, software behavior, data integrity, machine function, and communication processes. The present pFMEA results reproduce this structure at the level of individual PSQA-relevant failure modes: most relevant hazards arise not from isolated machine failure, but from the interaction of personnel, data, software, treatment planning systems, OIS/TCS transfer pathways, and delivery documentation.

The FM-related effects confirm that these root causes are clinically meaningful rather than merely administrative. Across the validated FM set, wrong dose, wrong dose distribution, or wrong dose location is the dominant consequence class, appearing in 28 of 44 FMs. Absolute dose or fractionation-related effects are present in 6 FMs, suboptimal plan or treatment effects in 7 FMs, and patient/staff inconvenience or treatment delay in 6 FMs. Thus, the predominance of human and process-related causes does not imply low clinical relevance. On the contrary, the root-cause pattern shows how procedural and information-flow failures can propagate into dosimetric or treatment-delivery consequences.

The collected PSQA FM notes further clarify why no single PSQA method can be considered universally sufficient. \emph{Measurement-based PSQA} is frequently identified as useful where deliverability, treatment-room compatibility, accessories, or physical dose-delivery behavior may reveal an error. \emph{Log file-based PSQA} is particularly relevant for per-fraction delivery, machine-state, recording, and interruption-related mechanisms. \emph{Independent Secondary Dose Calculation} is most naturally linked to CT/HU handling, dose-calculation model sensitivity, forward-calculation errors, and certain transfer-related discrepancies. These qualitative notes are consistent with the quantitative results of Secs.~\ref{sec:rel_local_risk_changes}--\ref{sec:workflow_results}: \textit{Meas} provides broad physical and interface-sensitive coverage, \textit{Log} is strongest in delivery and recording-related workflow regions, and \textit{Calc} is selective but valuable in image-, contour-, and calculation-sensitive parts of the workflow.

Overall, the root-cause distribution reinforces the central result of the pFMEA: PSQA is not merely a technical verification task performed after treatment planning, but a workflow-embedded risk-control strategy. The dominant failure causes are primarily located in human procedures, information transfer, system interaction, and process execution, while the dominant effects remain clinically meaningful dose-, plan-, and delivery-related consequences. This finding supports the fundamental premise that patient-specific QA in particle therapy must be interpreted as a risk-based, socio-technical quality assurance process rather than as a narrow extension of machine-centered quality control. 

\ManualNumberedChapter{4}{DISCUSSION AND IMPLEMENTATION}{ch:discussion_and_implementation}
\label{ch:log_implement}

\renewcommand{\sectionmark}[1]{\markright{\thesection.\ #1}}
Having quantified how measurement-based PSQA, log file-based PSQA, and independent secondary dose calculation reshape the same PBS proton therapy workflow at the failure-mode, process-step, and workflow levels, we now turn from risk description to clinical interpretation. The results have shown that PSQA is not a single technical endpoint, but a set of method-specific risk-control layers whose effects depend on where in the workflow they act, which failure modes they influence, and whether their benefit appears already at the data manipulation stage or only after completion of the full verification pathway.

The structure of this chapter reflects the history of the report. The original PTCOG Treatment Efficiency Subcommittee initiative was motivated by a focused practical question: whether and how treatment log records could be used for patient-specific QA in particle therapy (cf.~Sec.~\ref{sec:development_of_psqa_report}). As the work progressed, however, it became clear that this question could not be answered responsibly by considering log files in isolation. A meaningful evaluation of log file-based PSQA required a common reference workflow, a comparison with measurement-based PSQA and independent secondary dose calculation, and a formal way to distinguish the preparatory data-related workflow burden from the final verification benefit. This historical evolution is the reason why the present report combines practical implementation guidance with the broader PSQA-pFMEA formalism developed in the preceding sections.

Accordingly, the discussion begins by situating the proposed framework within the context of the current literature on risk management, learning from incidents, and data-driven PSQA. This first step highlights the report's shared features with current developments in the field of radiation therapy safety and identifies where it makes its own contribution: the introduction of a baseline-anchored, stage-specific, and risk-weighted method for comparing PSQA strategies. The chapter then returns to the areas of application that motivated the original work: data transmission integrity, requirements for treatment systems and log files, the role of independent secondary dose calculation, and practical recommendations for centers considering a transition to log-file-based or hybrid PSQA.

In this sense, the following discussion is not a separate add-on to the mathematical results, but rather it serves as their operational translation. It moves from the question ``How does each PSQA method change risk?'' to the more practical question ``How should a clinic use that knowledge when designing, validating, and evolving its PSQA program?''

\section{Contemporary Risk Management and PSQA Context}
\label{sec:discussion_relation_to_rm_psqa_context}

This risk assessment report lies at the intersection of three converging trends in radiation therapy: process-based risk management, data-driven learning from incidents, and increasingly automated or phantom-free patient-specific quality assurance. The relevant current literature does not point to a single dominant solution. Rather, it reveals a field in transition. The classic prospective (p)FMEA is being challenged by feedback loops in incident-based learning; traditional measurement-based patient-specific quality assurance (PSQA) is being challenged by alternatives such as log files, independent secondary dose calculations, in vivo measurements, and detector-integrated methods; and purely technical PSQA validation is being challenged by the need to understand which failure modes are actually controlled by which method. A key contribution of this report is to provide a formal framework within which these developments can be compared using a common risk language.

A particularly important reference point is the analysis by O'Daniel et al., based on the AAPM--ESTRO TG-360, which examines the errors that PSQA systems must detect \cite{odaniel2025which_failures_psqa}. This study collected reports of errors in radiation therapy from various sources and synthesized them into a set of failure modes relevant to treatment verification. Its central message is highly consistent with the premise of this report: PSQA should not be evaluated solely by the performance of a generic gamma analysis or by whether a single technical test is passed, but by whether the workflow detects failure modes that are significant for treatment quality and patient safety. O'Daniel et al.\ therefore identify and prioritize \emph{what} PSQA should detect (cf.~Sec.~\ref{sec:root_cause_structure}); the present framework asks \emph{how} different PSQA strategies alter the risk state of the same workflow, \emph{where} this change occurs, and \emph{how} the consequences of these failure modes propagate from process steps to larger workflow regions.

\begin{figure}[h!]
\centering
\begin{tikzpicture}[
  x=1cm,y=1cm,
  >=Latex,
  every node/.style={font=\small},
  line cap=round,
  line join=round
]
\tikzset{
  stagebox/.style={
    draw=black!70,
    thick,
    rounded corners=2pt,
    minimum width=3.25cm,
    minimum height=1.15cm,
    align=center,
    inner sep=4pt
  },
  notebox/.style={
    draw=black!55,
    rounded corners=2pt,
    fill=gray!6,
    align=center,
    inner sep=5pt,
    text width=6.15cm
  }
}

\node[stagebox, fill=gray!15]   (noqa) at (0,0)
  {\textbf{\textit{NoQA} baseline}};

\node[stagebox, fill=orange!20] (data) at (5.1,0)
  {\textbf{\textit{Data}-stage state}};

\node[stagebox, fill=green!18, minimum width=3.45cm] (cum) at (10.6,0)
  {\textbf{Final PSQA state}\\[-1pt]\footnotesize (= cumulative endpoint)};

\draw[-Latex, very thick, orange!85!black]
  (1.65,0) -- (3.45,0)
  node[midway, above=9pt, align=center, text=orange!85!black]
  {\textbf{\textit{Data}-stage metric}\\[-1pt]
   preparatory\\[-1pt]
   data measures\\[-1pt]
   {\footnotesize $\bm{M^{*,(d)}}$}};

\draw[-Latex, very thick, blue!75!black]
  (6.75,0) -- (8.85,0)
  node[midway, above=9pt, align=center, text=blue!75!black]
  {\textbf{\textit{Full}-stage metric}\\[-1pt]
   method-specific incremental verification layer\\[-1pt]
   {\footnotesize (normalized against the \textit{Data}-stage state)}\\[-1pt]
   {\footnotesize $\bm{M^{*,(f)}}$}};

\draw[-Latex, very thick, green!50!black]
  (0,-2.15) -- (10.6,-2.15)
  node[midway, above=7pt, align=center, text=green!40!black]
  {\textbf{Cumulative metric}\\[-1pt]
   final endpoint strictly aligned with the common \textit{NoQA} baseline\\[-1pt]
   {\footnotesize $\bm{M^{*,(c)}}$}};

\draw[densely dashed, gray!70] (5.1,-0.6) -- (5.1,-3.55);
\draw[densely dashed, gray!70] (10.6,-0.7) -- (10.6,-3.55);

\draw[<->, very thick, purple!80!black]
  (5.1,-4.0) -- (10.6,-4.0)
  node[midway, above=7pt, align=center, text=purple!80!black]
  {\textbf{Bridge from \textit{Data} to \textit{Cum}}\\[-1pt]
   additional \textit{Full}-stage contribution on the same \textit{NoQA} scale\\[-1pt]
   {\footnotesize $\bm{M^{*,(d\to c)} = \bm{M}^{*,(c)} - \bm{M}^{*,(d)}}$}};

\node[notebox, anchor=north west] at (-1.65,-4.6)
  {\textbf{Generic notation:} $\bm{M}$ denotes the corresponding FM-, process-step-, or workflow-level metric (e.g.\ $\bm{\delta_{\mathrm{rel}}}$, $\bm{\rho_{\mathrm{rel}}}$, or $\bm{\Rho_{\mathrm{rel}}}$).};

\node[notebox, text width=5.85cm, anchor=north east] at (12.35,-4.6)
  {\textbf{Explanation:} The framework outlines a step-by-step approach to risk control rather than merely listing criteria for identifying risks.};

\end{tikzpicture}
\caption{\small Conceptual relation between stage-specific and baseline-anchored PSQA metrics. The \textit{Data}-stage captures the perturbation introduced by preparatory data measures, the \textit{Full}-stage captures the method-specific incremental verification layer relative to the \textit{Data}-stage state, the cumulative metric reports the final endpoint relative to \textit{NoQA}, and the bridge from \textit{Data} to \textit{Cum} re-expresses the added \textit{Full}-stage contribution on the common \textit{NoQA} scale.}
\label{fig:concept_stage_logic_psqa}
\end{figure}
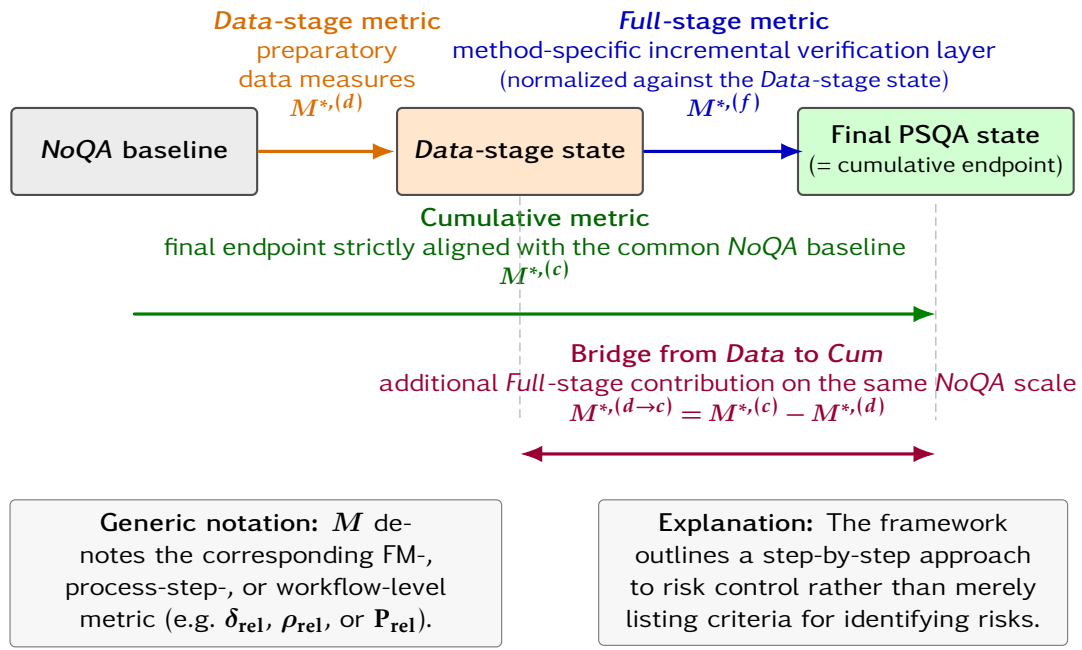

This distinction is important because a catalog of clinically relevant failure modes is necessary for comparing PSQA methods, but it is not sufficient. If a failure mode is significant, one must still know whether a particular PSQA strategy influences it through preparatory data measures, through final verification, through both, or not at all. This is precisely the purpose of the stage formalism developed here (see Fig.~\ref{fig:concept_stage_logic_psqa}). The \textit{Data} stage captures the perturbation introduced by preparatory data measures, the \textit{Full} stage captures the method-specific incremental verification layer, the cumulative metric defines the endpoint strictly aligned with \textit{NoQA} and thereby enables an \emph{inter}-PSQA comparison, and the bridge from data to cumulative expresses the additional contribution of the full stage on the same baseline scale. The result is a description of workflow-level risk control, not merely a list of detectability criteria.

The SEAFARER Head and Neck benchmark provides an empirical counterpart to this conceptual argument. In a multicenter photon PSQA study with 89 submissions from 44 centers, May et al.\ found substantial variation in sensitivity and specificity among clinically used PSQA systems; importantly, every commonly used device achieved perfect sensitivity and specificity in at least one submission, while the same device class could perform substantially differently under other local protocols \cite{May2026SEAFARER_HN}. This finding supports the interpretation that PSQA performance is determined by the complete clinical implementation, including device configuration, analysis settings, tolerances, workflow application, and user decisions, rather than by detector type alone. It also reinforces the need to evaluate PSQA as an implemented workflow with measurable failure-detection performance, not as an abstract device category.

The proton-therapy literature then adds the modality-specific counterpart to this argument. Wolter et al.\ investigate a one-time, phantom-free, log-file-based PSQA in PBS proton therapy and demonstrate that the irradiation system's log files enable reliable dose reconstruction, provided that relevant deviations in spot parameters are reproducible throughout the course of treatment and dosimetrically negligible \cite{wolter2025phantomless_psqa_proton}. Their work directly supports the feasibility of log-file-based PSQA as a clinically attractive alternative to time-consuming phantom measurements.

A subsequent direct comparison by Wolter et al.~extends this evidence from the reproducibility of beam delivery to the sensitivity of error detection and operational effort: in a clinically implemented PBS plan for the head-and-neck region with 29 input plan parameter errors, a phantom-free workflow combining log-file-based PSQA and automated physical checks detected 90\% of the scenarios, compared to 52\% for the local phantom-based workflow under clinically realistic measurement conditions, while the estimated manual workload was reduced by about one-third \cite{Wolter2026SensitivityEfficiency}.

Although both studies validate and quantify the reproducibility and dosimetric reliability of a log-based workflow, they do not determine which parts of the proton therapy workflow are best protected by log analysis, which remain unprotected, and how log-based verification compares structurally to measurement-based or secondary calculation strategies. However, in the terminology of both reports, Wolter et al. provide strong evidence and numbers for the technical credibility and reliability of a PSQA method, while the pFMEA formalism provides the risk architecture map against which this method can be comprehensively compared with others.

The same distinction applies to the detector-integrated PBS-PSQA approach described by Bateman et al.\ \cite{bateman2025pbs_psqa_integrated_detector}. Their integrated detector system combines independent measurements of beam range/energy, spot position/size, and beam intensity with a Monte Carlo-based 3D dose reconstruction. This is a technically powerful development, as it addresses a limitation of purely machine-generated log files: the lack of independent measurements of critical beam parameters. In the context of this white paper, such a system does not simply belong in the old category of conventional phantom measurement. It points to a more modern form of measurement-based PSQA, in which the physical verification of beam parameters and dose reconstruction are more automated, better aligned with irradiation, and potentially more compatible with efficient clinical workflows. This reinforces, rather than weakens, the conclusion that measurement-based PSQA should be interpreted broadly as a physical and interface-sensitive verification layer, rather than narrowly as a static phantom procedure.

Extensive multimodal PSQA results point in the same direction. The retrospective approach described by Decabooter and Nijsten for conventional radiotherapy combines secondary dose calculations, log-file-based verification, Electronic Portal Dosimetry (EPD), pre-treatment quality assurance, and in vivo quality assurance using a very extensive clinical dataset \cite{decabooter2026data_driven_psqa}. The most important finding of this work is not only that large-scale PSQA analyses are possible, but that different PSQA methods reveal different error mechanisms. Independent secondary dose calculations can reveal algorithmic or model-related discrepancies; log file- and delivery-based reviews can reveal issues with transmission, delivery, or device status; and in vivo methods can reveal interfractional or patient-specific deviations that remain invisible in purely pre-treatment workflows. Although this preprint should be interpreted with appropriate caution until the peer review process is complete, its core conceptual message is highly consistent with the present report: PSQA methods are complementary; they are not interchangeable.

The incident-learning literature — whose robust published evidence base originates primarily from conventional EBRT — clarifies how a prospective pFMEA can evolve into a learning risk-management instrument: initially expert-driven, subsequently constrained by observed events, and ultimately embedded in institutional feedback \cite{Paradis2021IncidentLearningFMEA,kornek2024incident_driven_fmea_feedback}. This feedback-oriented view prepares the methodological positioning of the present framework. Before photon-derived PSQA concepts can be translated into PBS proton therapy, the framework itself must first be understood as a proton-specific risk-coordinate system rather than as a direct import of photon QA doctrine (cf. Sec.~\ref{sec:intro_psqahistory}).

\subsection{Incident Learning and Institutional RM Feedback}
\label{subsec:discussion_incident_learning_rm_feedback}

The present framework adds a missing prospective layer to this empirical multimodality argument. A retrospective multimodal PSQA database can show which errors were detected by which method. The pFMEA formalism developed here can predict, structure, and compare why those detections should occur in the first place. It therefore functions as a conceptual translation layer between clinical PSQA performance data and risk-informed QA governance. It does not replace empirical validation; it organizes the question that empirical validation must answer.

The contemporary risk-management literature also supports this interpretation. Sölkner et al.\ demonstrate a data-driven radiation-oncology risk-management system built around real-time incident reporting, workflow-resolved cause and detection points, and targeted strengthening of safety barriers \cite{soelkner2025data_driven_risk_management}. Their work is especially relevant because it treats safety not as a static checklist, but as a process problem: errors originate at one workflow location, travel through one or more barriers, and are detected or missed at another. The reported reduction in the average distance between cause and detection is an important practical safety endpoint. In the language of the present report, their study measures how far an error travels before being caught; the PSQA-pFMEA formalism asks how a specific PSQA architecture is expected to change that journey.

Li et al.\ extend this logic to multi-site incident learning \cite{li2025radiation_incident_learning_multisite}. Their hybrid method combines statistical incident analysis, root-cause classification, word-cloud visualization, and FMEA-based proactive modeling across multiple clinical sites. The key insight is that risk management must be efficient, comparable, and learnable across heterogeneous clinical environments. This is directly relevant for PSQA in proton therapy, where workflows, delivery systems, TPS/OIS/TCS interfaces, and local QA traditions vary substantially between institutions. A method that only works as a local expert exercise is not enough. A useful framework must be structured enough to support comparison, yet flexible enough to be adapted to institutional workflows. The present report aims precisely at that balance: it fixes the mathematical structure of comparison while allowing the scores and failure-mode set to remain institution-specific.

Kornek et al.\ address another essential limitation of classical FMEA: the fact that prospective expert scoring may be incomplete or insufficiently anchored to observed error rates \cite{kornek2024incident_driven_fmea_feedback}. By linking an incident reporting system to an FMEA database, they show how additional failure modes can be identified after clinical use and how occurrence estimates can, at least in principle, be tested against reported events. This is a crucial point for interpreting the present work. The PSQA-pFMEA formalism is not intended to be a final, immutable truth table. It is a structured prospective model. Its long-term value increases if it can be recalibrated with incident-learning data, tool-performance data, and institution-specific experience.

\subsection{Methodological Position of the Present Framework}
\label{subsec:discussion_methodological_position}

Taken together, these studies define a clear boundary around the contribution of this risk assessment report. The reviewed literature already provides important pieces of the puzzle: high-priority PSQA failure modes, log-file reproducibility for phantomless proton PSQA, independent detector-based PBS verification, large-scale multimodal PSQA evidence, data-driven risk-management systems, multi-site incident learning, and incident-driven FMEA feedback loops. What it does not yet provide is a unified, baseline-anchored, mathematically explicit framework for comparing how different PSQA methods transform the same proton-therapy workflow.

This is the methodological gap addressed here. The common \textit{NoQA} baseline makes the three PSQA pathways comparable. The \textit{Data/Full/Cum/bridge} decomposition separates preparatory workflow burden from verification benefit. The hierarchical aggregation
\[
\delta \;\longrightarrow\; \rho \;\longrightarrow\; \Rho
\]
prevents a visually prominent single failure mode from being mistaken for a dominant workflow-level risk, and prevents low-burden process steps from receiving the same weight as high-burden steps. The cumulative metric provides the strict inter-method endpoint, while the bridge metric explains how the final endpoint is reached. In this respect, the proposed methodology is not merely another FMEA table. It is a risk-coordinate system for PSQA.

This risk-coordinate system also changes how the roles of the three PSQA methods should be interpreted. Measurement-based PSQA should not be viewed only as conventional phantom measurement. It includes the broader class of physical, detector-based, and interface-sensitive verification strategies, including emerging integrated detector and 3D-reconstruction approaches. Log file-based PSQA should not be viewed merely as a convenient automation of delivery review. It is a delivery-state and recording-state observability layer whose strength is concentrated where machine delivery, parameter transfer, and treatment documentation dominate the risk mechanism. Independent secondary calculation should not be viewed as globally weaker simply because its workflow-level cumulative effect is smaller. It is a selective and structurally independent safeguard for image-, contour-, model-, and calculation-sensitive failure modes.

The practical implication is that the present report does not support a winner-takes-all PSQA doctrine. It supports a risk-informed hybrid architecture. Log file-based PSQA may serve as a workflow-wide backbone for delivery and recording risk. Measurement-based or detector-integrated PSQA remains essential where independent physical verification, deliverability, accessory configuration, and beam-parameter confidence are required. Independent secondary calculation remains essential where the dominant risks arise from dose calculation, CT/HU handling, model configuration, or algorithmic assumptions. The optimal PSQA program is therefore not the one with the largest single technical validation metric, but the one whose combined methods cover the relevant risk architecture of the workflow.

The formalism in Appendix~\ref{ch:math_framework} is also practically implementable. Its inputs are not exotic: a process map, a validated set of failure modes, stage-specific \(O\), \(S\), and \(D\) scores, and a spreadsheet-based aggregation structure are sufficient to reproduce the analysis. The framework can therefore be implemented as an expert-scored institutional pFMEA, then updated as local incident data, log-file performance, measurement outcomes, or secondary-calculation discrepancies become available. In a mature implementation, the pFMEA table would not be a static document but a living risk model: prospective at first, then progressively constrained by observed safety events and QA-performance data.

At the same time, the framework must be interpreted within its intended semi-quantitative scope. RPN-derived values are not calibrated probabilities of harm, and the numerical outputs should not be overread as absolute risk estimates. Their strength lies in comparative consistency: the same scoring rubric is applied to the same workflow under alternative PSQA states, and the resulting changes are aggregated with explicit baseline-burden weights. This is why the proposed methodology can support rational PSQA governance even though it does not claim to be a fully probabilistic risk model.

The broader conclusion is therefore positive but deliberately critical. Contemporary PSQA and RM literature has made major progress in tool validation, phantomless verification, multimodal detection, incident learning, and data-driven safety-barrier design. The present report does not replace those developments. It gives them a common comparative structure. It shows how to ask, for each PSQA method, not only whether it works, but where it works, which failure modes it affects, how its \textit{Data}-stage burden is compensated, how its \textit{Full}-stage contribution appears on a common baseline, and how all of this changes the total workflow risk burden. That is the central methodological advance of this work.

With this methodological position established, the discussion now returns to the implementation questions that originally motivated this report, but only after making one further boundary condition explicit. Much of the contemporary PSQA literature, including the recent review by Decabooter et al., is photon-centered and is therefore highly informative, but not directly prescriptive for PBS proton therapy \cite{decabooter2026psqa}. The relevant question is not whether photon PSQA concepts are useful for proton therapy, but which concepts can be translated as general risk-control principles and which must be reformulated because the underlying delivery physics, observability, data structures, and workflow risks are proton-specific. 

The following subsection defines this translation boundary and thereby prepares the operational sections that follow: data transfer integrity addresses the reliability of the information pathway before delivery; the treatment delivery system section addresses the machine- and log-file layer where delivered treatment becomes observable; the section on independent secondary dose calculation addresses the complementary algorithmic and model-based verification layer; and the final adoption recommendations translate these elements into a clinically usable pathway for centers considering a transition toward log file-based or hybrid PSQA.

\subsection{Transferability and Boundaries of Photon PSQA Concepts for PBS Proton Therapy}
\label{subsec:photon_to_proton_transferability}

The contemporary photon PSQA literature provides an important conceptual reference for the evolution of patient-specific verification in radiation therapy. In particular, the review by Decabooter et al.\ frames PSQA as a final patient-specific safety layer within a broader quality-assurance ecosystem and describes the transition from routine pretreatment phantom measurements toward risk-adapted, data-driven, image-based, log file-supported, in vivo, and clinically interpretable PSQA \cite{decabooter2026psqa}. This evolution is directly relevant for the present report because it reflects the same general movement away from a single pass/fail verification paradigm and toward a layered risk-control architecture.

However, the transfer from photon PSQA to PBS proton therapy must be made at the level of principles rather than at the level of devices, metrics, or action thresholds \cite{May2026SEAFARER_HN}. The central photon-derived lesson is not that measurement-based PSQA is obsolete. Rather, the central lesson is that PSQA effort should be allocated according to the failure modes that are clinically relevant, insufficiently controlled by upstream QA layers, and insufficiently observable by alternative verification methods. This interpretation is consistent with the hybrid architecture proposed in the present report: PSQA methods are complementary risk-control layers, not interchangeable tests.

Direct technical transfer is limited because the dominant physical, technological, and workflow determinants differ substantially between photon therapy and PBS proton therapy. Photon workflows are often shaped by linac delivery, MLC behavior, gantry and dose-rate modulation, photon beam modeling, EPID-based transit dosimetry, CBCT- or MR-based adaptive workflows, and gamma-based detector comparisons. PBS proton therapy instead depends on spot position, spot size, monitor units per spot, energy-layer sequencing, scanning dynamics, range shifters, apertures, nozzle geometry, CT-to-stopping-power conversion, robustness assumptions, range uncertainty, motion/interplay effects, and the structure of DICOM RT Ion Plan and treatment-record data. These differences alter both the relevant failure modes and the observability of those failure modes by measurement-based PSQA, log file-based PSQA, and independent secondary dose calculation.

Several photon-derived principles remain transferable. These include risk-based selection of PSQA depth, class-solution commissioning, known-error validation of PSQA sensitivity and specificity, structure-based dose evaluation, automated consistency checks, treatment-course monitoring, and the progressive integration of log files, independent calculation, imaging, and in vivo information. For PBS proton therapy, however, these principles must be reformulated around proton-specific risk mechanisms. Known-error validation should include, for example, spot-position shifts, energy-layer errors, range-shifter errors, HU-to-stopping-power errors, aperture misassignment, gantry/couch parameter errors, partial-delivery recording errors, and log-file corruption. Class solutions must be narrower and more physics-aware than in many photon settings, because range sensitivity, robustness, beam arrangement, motion/interplay, beam modifiers, and vendor-specific delivery-system behavior can substantially modify the risk profile.

Other photon-specific concepts are not directly transferable. EPID transit dosimetry is a prominent example: photon exit fluence can support EPID-based in vivo dose reconstruction, whereas proton beams require different observables, such as prompt-gamma signals, PET activation, ionoacoustic approaches, range probes, detector-integrated verification, or indirect reconstruction from images and delivery records. Likewise, MLC-centric failure-mode rankings, photon-specific gamma action levels, and photon-derived assumptions about dose-calculation heterogeneity correction cannot be imported directly into PBS proton therapy. In proton therapy, the clinically relevant translation must be expressed in terms of range, stopping-power conversion, spot delivery, beam-modifier state, machine-state observability, patient-anatomy dependence, and dose reconstruction on the appropriate anatomical and delivery-state information.

The distinction between transferable photon-derived principles and non-transferable photon-specific implementations is summarized in Table~\ref{tab:photon_psqa_transferability}, which provides a compact crosswalk between major photon PSQA concepts, the corresponding risk-control principles that may inform PBS proton therapy, and the proton-specific implementation boundaries that must be respected before these concepts are used in local PSQA governance.

\begin{center}
\small
\begin{longtable}{p{0.25\textwidth}p{0.34\textwidth}p{0.34\textwidth}}
\caption{\small Photon PSQA concepts: transferability to PBS proton therapy.}\\
\toprule
\textbf{Photon PSQA concept} & \textbf{Transferable principle} & \textbf{PBS proton-specific boundary}\\
\midrule
\endfirsthead

\multicolumn{3}{@{}l}{\textit{Table \thetable{} – continued}}\\
\toprule
\textbf{Photon PSQA concept} & \textbf{Transferable principle} & \textbf{PBS proton-specific boundary}\\
\midrule
\endhead
  
\midrule
\multicolumn{3}{r@{}}{\textit{Continued on next page}}\\
\endfoot

\bottomrule
\endlastfoot
  
Risk-based PSQA & QA effort should match clinically relevant, insufficiently controlled failure modes. & Proton risk ranking must include range, stopping power, spot delivery, beam modifiers, robustness, and motion/interplay.\\
Class solutions\footnote{In this context, a class solution should be understood as a locally commissioned and prospectively bounded treatment class for which planning, delivery, data-transfer, dose-calculation, and PSQA performance have been validated sufficiently to justify a risk-adapted verification pathway, while cases outside the validated class remain subject to escalated PSQA.} & Stable treatment classes can justify reduced routine measurement after commissioning. & Proton class solutions must be narrower and range/robustness-aware\footnote{For PBS proton therapy, class solutions must be defined more narrowly than in many photon settings because range sensitivity, robustness assumptions, beam arrangement, motion/interplay, beam modifiers, delivery-system behavior, and treatment-record observability can substantially alter both the risk profile and the detectability of clinically relevant failures.}.\\
Gamma limitations & Generic pass rates may not predict clinical dose error. & Proton action levels must reflect range and structure-based consequences, not only dose-distance agreement.\\
Secondary calculation & Independent recalculation can reveal model and calculation errors. & True independence requires separate algorithmic and workflow pathways; Monte Carlo is especially important.\\
Log-file based PSQA & Delivery records can support efficient verification and dose reconstruction. & Proton logs must be spot-, energy-, beam-modifier-, and timing-resolved and must be independently QA-validated.\\
EPID in vivo dosimetry & Treatment-course verification is superior to pretreatment-only verification. & Photon EPID transit dosimetry is not directly transferable; proton in vivo verification requires different observables.\\
Adaptive workflow QA & Online adaptation requires fast, automated, risk-based QA. & Adaptive proton therapy has stronger anatomy/range sensitivity and requires proton-specific risk controls.\\
AI and automation & Automation can improve scalability and anomaly detection. & Automation must be validated against proton-specific failure modes and local data pathways.\\

\label{tab:photon_psqa_transferability}
\end{longtable}
\end{center}

Table~\ref{tab:photon_psqa_transferability} should therefore be read as an implementation boundary rather than as a simple literature summary. It identifies which photon-derived concepts may inform PBS proton therapy at the level of risk-control principles, and which elements require proton-specific reformulation before they can be used for local PSQA governance. This distinction is essential for the remainder of the implementation discussion. Data-transfer integrity, treatment delivery system requirements, log-file based QA, independent secondary dose calculation, and adoption of log file-based or hybrid PSQA cannot be specified by importing photon-based devices, metrics, or action levels. They must be defined around proton-specific observability: range sensitivity, spot- and energy-layer-resolved delivery, beam-modifier state, treatment-record fidelity, anatomy-dependent dose reconstruction, and the degree to which each PSQA layer provides independent and clinically interpretable information.

The first implementation consequence of this translation boundary is data-transfer integrity. If photon-derived concepts are translated into PBS proton therapy at the level of risk-control principles, then the reliability of the information pathway from imaging and treatment planning through OIS, TCS, treatment records, log files, and dose reconstruction becomes a central safety condition. The following section therefore turns to the data-transfer layer as the first practical domain in which proton-specific PSQA implementation must be specified.

\section{Data Transfer Integrity}
\label{sec:data_transfer}

The first implementation domain is data transfer integrity, because every PSQA pathway depends on the correct propagation of treatment information through the TPS $\longrightarrow$ OIS $\longrightarrow$ TCS chain. In the terminology of the risk formalism, this is the practical substrate of the \textit{Data}-stage: before any final verification layer can mitigate risk, the workflow must preserve the identity, consistency, and clinical meaning of the plan data being transferred and transformed.

Data transfer integrity during the treatment workflow is a major concern for the safe and correct treatment delivery for the individual patient. The integrity of data transfer must be verified among several connected systems, including treatment planning, oncology information systems, treatment delivery systems, and treatment records. An important element of measurement-based PSQA is the verification of data transfer integrity through patient-specific end-to-end testing. Although only a limited number of voxels are measured, this process still gives a reasonable assurance of data transfer integrity. When moving away from measurement-based PSQA for every patient, the integrity of data transfer must be addressed in another way. Solutions to this problem are proposed in the AAPM TG 201 report \cite{AAPM_201} on Quality management (QM) in external beam radiotherapy. 

Following the recommendation from AAPM TG 201, the overall aim to assure the correct treatment delivery relies on an understanding of data structures, the type of information systems and interfaces, and the configurations in the clinic. Every clinic is advised to implement an individual QM program. The responsibility includes strategies to handle standard format data, such as DICOM, as well as vendor specific data and manual documentation. 

\begin{figure}[h!]
    \centering
    \includegraphics[width=1\linewidth]{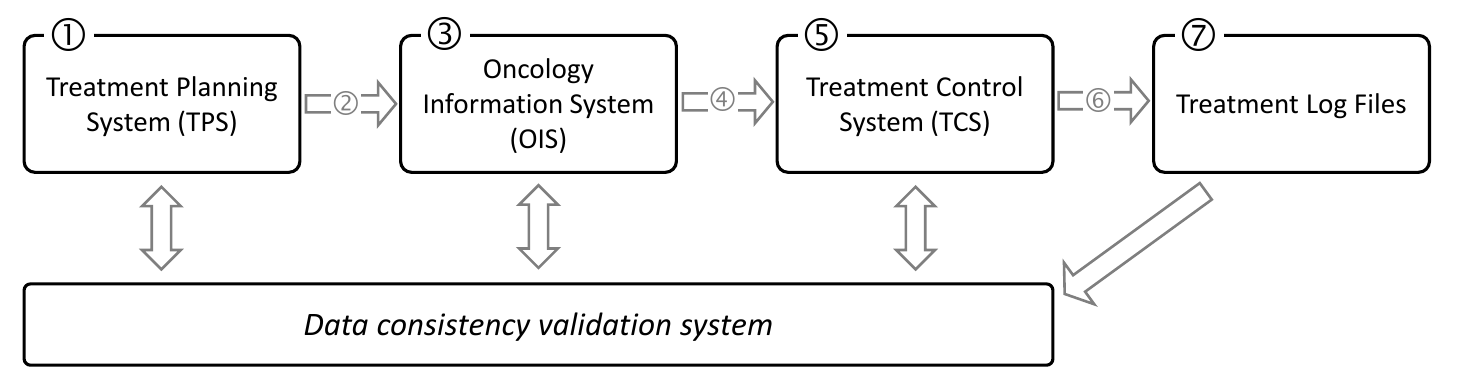}
    \caption{\small Schematic of a data consistency validation system. The Treatment Planning System (TPS) produces a plan (1), The plan is transferred (2) to the Oncology Information System (OIS) for verification and approval (3). The Treatment Control System (TCS) retrieves the treatment plan from the OIS (4) to deliver the treatment (5). During the delivery, the TCS produces a log in a treatment record (6). This log (7) provides a record of the treatment parameters. An ideal data consistency validation system verifies the data integrity after each step in the process, so that the plan delivered corresponds to the intended and approved plan. }
    \label{fig:data-transfer}
\end{figure}

While verifying data transfer is straightforward conceptually, it may be complicated by differing levels of data transparency in various systems, as well as different storage and transmission formats. One way to simplify data integrity management would be to implement an external system for verifying data consistency, as shown in Fig.~\ref{fig:data-transfer}.
 
For a common clinical workflow, with different vendors, the data verification depends on the level of data transparency, the ability to use checksums, the use of standard data, and the structure of the non-standard data. The DICOM format is considered to offer enough flexibility to contain all the required information for treatment delivery logging. However, conformance to the DICOM standard varies considerably between vendors and systems, making the format less reliable in practice. The working group highlighted the importance of conformance to data standard formats as a future work for the community and the vendors. This work is being pursued by the Integrating the Healthcare Enterprise – Radiation Oncology, or IHE-RO organization. Vendors are encouraged to follow the Technical Framework Supplement “Treatment Delivery Record Content for Ion” (TDRC-ION) \cite{TDRC-ION_full}, which had just entered trial implementation at the time of the writing of this report.

Once the integrity of the treatment data pathway has been considered, the next practical question is whether the delivered treatment can be observed with sufficient fidelity. This shifts the discussion from information transfer to treatment execution, machine-state representation, and log-file reliability. In the staged risk framework, this is the point where log file-based PSQA becomes more than a convenience tool: it becomes a potential observability layer for the delivered treatment itself. 
\section{Treatment Delivery System}
\label{sec:TDS}

\subsection{Treatment Delivery Accuracy}
\label{subsec:delivery_accuracy}
As discussed in section \ref{subsec:TDS_QA}, a move away from measurement-based PSQA for each patient should prompt a reconsideration of the frequency and thoroughness of machine QA.  If a robust measurement-based PSQA program is currently used to supplement elements of the machine QA program (whether explicitly or implicitly), it may be appropriate to increase the frequency of some machine QA tests or perform them in a more comprehensive way.  The AAPM TG 224 report \cite{arjomandy2019taskgroup224} on comprehensive proton therapy machine quality assurance should be consulted during this process, and it is recommended that risk analysis methods should be employed to consider the impact of this change on quality assurance of the treatment delivery system.

\subsection{Quality Assurance of Log Files}
\label{subsec:log_QA}
If treatment log files are to be used for PSQA, it is important to consider how the accuracy of the log file data will be verified.  This should be done through measurements as part of the regular machine QA program. For example, the recorded treatment position and MU could be compared to treatment plans, as well as to film, scintillating screen, or 2D detector array measurements \cite{li2013use}. Most standard QA equipment lacks the high acquisition speeds required to validate the timing of individual spot delivery or the dynamics of continuous scanning, but it is possible to indirectly validate these parameters by evaluating the delivery time of an entire field.  Additionally, new QA tools, such as multi-strip ionization chamber arrays, can measure spot delivery at very high speeds, enabling more comprehensive verification of the timing aspects of treatment log files. 

\subsection{Log File Requirements and Recommendations}
\label{subsec:log_requirements}

\begin{center}
\small
\begin{table}[h!]
    \caption{\small Required minimum semantic information in treatment log files or treatment records for patient-specific QA and delivered-dose reconstruction in PBS proton therapy.}
    \label{tab:log_required}
    \begin{tabular}{p{0.25\textwidth}p{0.68\textwidth}}
    \hline
    \multicolumn{2}{c}{\textbf{Plan and record identifiers}}\\
    \hline
    General &
    Patient ID or anonymized patient identifier\\
    & Plan name/ID and referenced RT Ion Plan UID\\
    & Treatment record UID or treatment session/fraction identifier\\
    & Irradiation date and time\\
    \hline
    \multicolumn{2}{c}{\textbf{Field parameters}}\\
    \hline
    General &
    Field or beam name/ID and referenced beam number\\
    \hline
    Machine configuration &
    Treatment room or delivery device identifier\\
    & Gantry angle\\
    & Patient support/couch angle and table-top position\\
    & Snout/nozzle position\\
    & Range-shifter ID, type, water-equivalent thickness, position, or status\\
    & Aperture ID, dynamic-aperture file, or beam-modifying-device status\\
    & X/Y spot projection ratio to isocenter, or an equivalent coordinate transformation\\
    \hline
    Beam delivery &
    Total delivered monitor units or delivered meterset\\
    & Total number of delivered spot positions\\
    & Treatment termination or interruption status and details\\
    \hline
    \multicolumn{2}{c}{\textbf{Spot parameters, per delivered spot}}\\
    \hline
    Beam delivery &
    Delivered spot position at isocenter or machine isocentric plane, with X/Y coordinates\\
    & Delivered monitor units or spot meterset from the primary dose monitor\\
    & Planned monitor units or planned spot meterset\\
    & Delivered spot size at isocenter, preferably as X/Y FWHM\\
    & Spot number, delivered order, reordered flag, or prescribed spot index\\
    & Delivered or nominal beam energy / energy layer\\
    & Dynamic range-shifter position or status\\
    \hline
    \end{tabular}
\end{table}
\end{center}

Treatment log files are not always made available to the customer, or are available only in vendor-specific machine-data formats that are not immediately suitable for log file-based PSQA. In contemporary clinical systems, however, at least part of the treatment delivery state is commonly reported to the OIS or treatment management system to generate a treatment record. This is often implemented using the format described in the Digital Imaging and Communications in Medicine (DICOM) standard \cite{nem_dicom_full}. The DICOM standard includes radiotherapy extensions for ion therapy, in particular the~\texttt{RT Ion Beams Treatment Record IOD}~ and the corresponding ion-beam session-record content. While DICOM is not the only possible mechanism for recording or transmitting treatment-delivery information, it is widely used and is capable of representing much of the information required for treatment verification, intended-versus-delivered comparison, and dose reconstruction.

Vendor conformance to the current DICOM RT Ion Beams Treatment Record definition, and preferably to the IHE-RO Technical Framework Supplement \emph{Treatment Delivery Record Content for Ion} (TDRC-ION), should therefore be regarded as an important implementation objective \cite{TDRC-ION_full}. For PBS proton therapy workflows that intend to use log file-based PSQA or treatment-record-based dose reconstruction, the semantic information categories listed in Table~\ref{tab:log_required} should be available as a minimum requirement.

This is a requirement for the observability of the workflow and not merely a requirement for the availability of files. If clinically relevant variables regarding device status — such as information on the radiation field or air gap — are missing from treatment records or proprietary log files, the resulting gap in observability should be explicitly addressed through automated data transfer, consistency, or physical checks, rather than assuming that it is controlled solely by dose reconstruction based on log files \cite{Wolter2026SensitivityEfficiency}.

The required information should be accessible to automated workflows, traceable to the corresponding approved plan and treatment fraction, and archived according to the institutionally required retention period or the user's specified retention policy. Vendors are strongly encouraged to support export of a standardized treatment record even when additional proprietary log files remain available, because standardized records reduce implementation burden and improve interoperability across independent QA and dose-reconstruction systems.

\begin{center}
\small
\begin{table}[h!]
    \caption{\small Additional useful log file or treatment-record information for advanced verification, machine QA, time-resolved delivery analysis, and adaptive or 4D proton therapy workflows.}
    \label{tab:log_bonus}
    \begin{tabular}{p{0.25\textwidth}p{0.68\textwidth}}
    \hline
    \multicolumn{2}{c}{\textbf{Plan and machine parameters}}\\
    \hline
    Machine configuration &
    Monitor chamber calibration factor or conversion factor, e.g., charge/MU or pulses/MU\\
    & Room temperature and pressure, if needed for monitor calibration\\
    \hline
    \multicolumn{2}{c}{\textbf{Field-level delivery parameters}}\\
    \hline
    Beam delivery &
    Total planned number of spot positions\\
    & Spot scanning mode or spot scanning speed\\
    & Field delivery duration and cumulative delivery time\\
    & Warning, override, interruption, resume, and error logs\\
    & Treatment termination reason and machine-specific termination codes\\
    \hline
    \multicolumn{2}{c}{\textbf{Spot-level and time-resolved delivery parameters}}\\
    \hline
    Beam delivery &
    Scan-spot time offset or delivered spot timestamp relative to treatment start\\
    & Delivered monitor units or meterset from a secondary dose monitor\\
    & Delivered MU rate, delivered meterset rate, or beam-current surrogate\\
    & Time-resolved scanning magnet signals or spot trajectory samples\\
    & Per-spot gantry or patient-support angle, if variable during delivery\\
    & Prescribed spot indices, especially when delivered spot order differs from planned order\\
    \hline
    \end{tabular}
\end{table}
\end{center}

Additional valuable log file data elements should be considered, as described in Table~\ref{tab:log_bonus}. These elements are not always required for basic intended-versus-delivered comparison, but they are important for more advanced treatment validation, machine QA, time-resolved delivery analysis, motion/interplay evaluation, adaptive workflows, and future 4D optimization or dose-accumulation strategies.

Treatment log records can strengthen the observability of delivery, but they do not by themselves provide an independent recalculation of the intended dose model. A complementary verification layer is therefore required for risks rooted in calculation assumptions, CT/HU handling, beam modeling, or algorithmic implementation. This motivates the subsequent discussion of independent secondary dose calculation as a distinct PSQA strategy rather than merely an auxiliary technical check.

\section{Independent Secondary Dose Calculations}
\label{sec:ISDC_discussion}
The independent secondary dose calculation, or independent dose or MU calculation, is a valuable risk reduction tool that is complementary to measurement-based or log file-based PSQA. Because it mitigates some risks that are not addressed by either form of PSQA, an independent secondary dose calculation should be a part of any patient-specific verification program.

AAPM task group reports have provided recommendations on the implementation and uses of independent calculation-based dose or MU verification for non-IMRT and IMRT radiotherapy \cite{zhu2021report,stern2011verification}. 
In fact, the physics and beam modeling for proton PBS beams are simpler compared with X-ray IMRT. First, the energy-dependent lateral spot profiles and integral depth-dose curves measured for the TPS commissioning can be used to generate the beam data set for the independent secondary dose calculation program. The point dose or 3D dose distribution in either water or patient CT voxels can then be calculated using the PBS beam parameters in the DICOM RT ion plan, including proton energies, spot positions and monitor units per spot. The results can be reported as a point dose difference for each field, 2D planar doses at multiple depths, or 3D dose comparison with dose-volume histograms.

A simple pencil-beam convolution dose algorithm for PBS proton therapy could be programmed as the in-house independent secondary dose calculation in homogeneous water with the achievable point dose difference within 5\%.  A 3D dose calculation in the patient CT scan is recommended, because it enables a more comprehensive assessment of dose calculation accuracy, including dose-volume histogram analysis and geometric evaluation of the dose to targets and critical structures.  The Monte Carlo method provides more accurate dose calculations in 3D heterogeneous CT voxels due to its more physically-accurate handling of proton interactions with different materials and material interfaces. Open-source codes such as matRad or MCSquare can be used for independent secondary dose calculation.\cite{deng2020technical,souris2016fast,wieser2017development} 

A different dose calculation algorithm in the same TPS could be used for secondary dose calculations. For example, if the Monte Carlo dose engine AcurosPT in the Eclipse TPS is used as the primary dose calculation engine, the pencil-beam convolution superposition (PCS) algorithm could be used as a secondary dose calculation tool.  However, this would not qualify for our definition of an \textit{independent} secondary dose calculation, because the two algorithms are housed within the same software application.  

A fully independent secondary dose calculation would be a structurally-independent piece of software, ideally produced by another entity than the vendor providing the treatment planning system. In general, such an ISDC should be commissioned as a clinical dose-calculation system rather than treated as a simple numerical cross-check since algorithmic limitations, statistical uncertainty, dose scoring, CT calibration, beam modeling, comparison metrics, and clinically meaningful dose-volume interpretation require explicit validation.

The use of an independent software system for secondary dose calculations provides a more robust check against failure modes that may be specific to a given vendor's software, data handling, or interfaces.  However, the introduction of a separate system may increase risks due to data transfer and plan manipulation.  Decisions about what system to use for independent secondary dose calculation should be made after a thorough risk analysis to determine which system will be most effective at reducing risk as a whole throughout the entire workflow.

The preceding sections separate the practical implementation problem into three domains: integrity of transferred data, observability of treatment delivery, and independent verification of dose calculation. In clinical reality, however, these domains must be combined into a coherent QA strategy. The following recommendations therefore translate the risk-analysis results and the implementation considerations into practical guidance for centers considering whether and how to expand the role of log file-based PSQA.

\section[Risk-Informed Adoption of Log File-Based PSQA]{Recommendations for Risk-Informed Adoption of Log File-based PSQA with\-in a Hybrid PSQA Architecture}
\label{sec:log_adoption_recommend}

Transitioning from routine measurement-based PSQA to a log file-supported or log file-dominant PSQA workflow should be considered only as a risk-managed change in the overall QA architecture, not as a simple technical substitution. Potential advantages include reduced measurement burden, improved delivery-state observability, and detection of some treatment-delivery errors that are difficult to identify with conventional measurement-based PSQA. However, these advantages are clinically meaningful only if the log data, treatment records, data-transfer pathways, machine-QA dependencies, and independent dose-verification layers are explicitly validated for the local workflow \cite{huq2016taskgroup100,arjomandy2019taskgroup224}.

The generalized risk analysis presented in this report can be used as a structured starting point, but it cannot replace a customized risk analysis tailored to the equipment, procedures, software environment, staffing model, automation level, and clinical practice of a given particle therapy center. Adequate attention should be given to verification of correct data transfer from the treatment planning system through the oncology information system, treatment control system, delivery system, treatment record, and any downstream QA or dose-reconstruction software. The frequency and comprehensiveness of the machine-QA program should also be re-evaluated whenever routine patient-specific measurements are reduced, because some risk controls may have been implicitly provided by the measurement-based PSQA pathway \cite{arjomandy2019taskgroup224}.

\subsection{Minimum Conditions before Reducing Routine Measurement-based PSQA}

A reduction of routine measurement-based PSQA should not be based on historical pass rates alone. Before measurement frequency is reduced, the following minimum conditions should be documented locally \cite{huq2016taskgroup100,siochi2021report}:

\begin{enumerate}
    \item The TPS-to-OIS-to-TCS-to-delivery-record data pathway has been verified for semantic completeness, traceability, and failure handling \cite{siochi2021report,nem_dicom_full}.
    \item The log file or treatment-record content has been validated against independent measurements and routine machine-QA data \cite{arjomandy2019taskgroup224,li2013use}.
    \item The log file-based workflow has been tested with clinically meaningful known-error scenarios, including spot-position, spot-MU, energy-layer, range-shifter, aperture, gantry/couch/nozzle-state, data-transfer, treatment-record, and log-file errors \cite{Wolter2026SensitivityEfficiency,Lehmann2022SEAFARER}.
    \item An independent secondary dose calculation is clinically commissioned and structurally independent of the primary treatment-planning calculation whenever feasible \cite{zhu2021report}.
    \item At least one primary or secondary dose-calculation layer is capable of accurate three-dimensional dose calculation in patient geometry, preferably using Monte Carlo \cite{aitkenhead2020automated,souris2016fast,wieser2017development}.
    \item The local pFMEA has been updated to reflect the institution's equipment, software versions, staffing model, automation level, treatment sites, and case mix.
    \item Escalation criteria, documentation requirements, and periodic revalidation intervals are defined before clinical transition.
\end{enumerate}

These conditions should be interpreted as safeguards against a common misinterpretation: a log file-based workflow is not automatically safer because it is more automated, more efficient, or closer to the nominal treatment delivery state. It becomes a clinically meaningful PSQA layer only if the information on which it relies is complete, independently verified, and embedded in a controlled institutional QA process.

\subsection{Validation of Error-Detection Performance}

A log file-based PSQA system must undergo explicit error-detection validation before it is used to reduce or replace any routine patient-specific measurements. This requirement should be interpreted not merely as a period of operational familiarization, but as a formal test of whether the proposed workflow detects clinically relevant deviations. Purposeful treatment-delivery edits or known-error scenarios should be embedded in otherwise clinically plausible plans, and the resulting dosimetric or delivery-state impact should be characterized independently. The institution should then evaluate whether its routine PSQA pathway detects these errors without prior knowledge of which cases contain deviations \cite{Lehmann2022SEAFARER}.

The Wolter et al. benchmark provides a proton therapy-specific example of this validation logic, because it separates routine phantom-measurement geometry from ideal error-informed measurement placement and thereby shows that detectability is a property of the complete PSQA workflow, not of the detector class alone \cite{Wolter2026SensitivityEfficiency}.

For PBS proton therapy, such validation should include proton-specific scenarios that are meaningful for the local workflow. These should include, where applicable, spot-position errors, spot-size errors, spot-MU errors, energy-layer errors, range-shifter or aperture errors, gantry/couch/nozzle-state errors, treatment-record errors, data-transfer errors, delivery-interruption or resumed-delivery errors, dose-calculation errors, and log-file generation or interpretation errors. The objective is not only to compare pass-rate behavior between measurement-based, log file-based, and calculation-based PSQA, but to estimate sensitivity, specificity, false-negative risk, failure-mode coverage, and clinical interpretability.

Concurrent operation of the old and new PSQA pathways should therefore be complemented, whenever feasible, by known-error injection and by retrospective testing of plans with known measurement failures, delivery anomalies, treatment interruptions, or calculation discrepancies. A high historical pass rate, including a high gamma passing rate, is insufficient evidence of safety if the proposed PSQA pathway has not been shown to detect errors that matter clinically in the local PBS-PT workflow \cite{nelms2011per,carrasco2012dvh,kry2014institutional,nelms2013evaluating,miften2018tolerance}.

This requirement is reinforced by the SEAFARER head-and-neck study, in which nearly half of the participating centres passed at least one plan classified as a should-fail case despite using established clinical PSQA systems. The relevant validation endpoint is therefore not the nominal availability of a PSQA device or a historically acceptable pass-rate distribution, but the demonstrated sensitivity and specificity of the local PSQA workflow against clinically meaningful known-error scenarios \cite{May2026SEAFARER_HN}.

\subsection{Quality Assurance of Log Files and Treatment Records}

A weakness of log file-based PSQA is its reliance on the treatment machine or treatment-management infrastructure to report its own performance. A robust program for verification and quality assurance of the log files themselves is therefore a prerequisite for using log file-based PSQA as a primary clinical risk-control layer. This program should include independent verification that recorded spot positions, spot metersets, energy layers, field identifiers, range-shifter status, aperture or beam-modifying-device status, interruption states, and delivered metersets are complete, correctly associated with the approved plan and fraction, and consistent with independent machine-QA or measurement data.

Treatment log records can strengthen the observability of delivery, but they do not by themselves provide an independent recalculation of the intended dose model. They also cannot detect all failures upstream of the recorded delivery state. For this reason, log file-based PSQA should be interpreted as a delivery-state and recording-state observability layer, not as a complete replacement for independent dose verification, machine QA, commissioning, or physical measurements. When the log file is generated downstream of some system components but upstream of others, this location in the information chain should be explicitly documented, because it determines which failure modes the log file can and cannot observe.

Log files and treatment records should be available in documented, accessible, and preferably standardized formats. Vendor conformance with relevant DICOM RT Ion treatment-record objects and, where applicable, IHE-RO treatment-delivery-record profiles should be considered an important implementation objective. At minimum, the treatment record should provide enough information to support intended-versus-delivered comparison, dose reconstruction, traceability to the approved plan and delivered fraction, and automated review. Proprietary logs may remain valuable, especially for high-frequency or time-resolved delivery analysis, but they should not be the only accessible representation of clinically relevant treatment-delivery information \cite{arjomandy2019taskgroup224,li2013use,toscano2019impact}.

\subsection{Role of Measurement-based PSQA during and after Transition}

Measurement-based PSQA should not be considered obsolete when a log file-based workflow is introduced. Even when a log file-based and automated workflow shows superior sensitivity in a local benchmark, the transition should remain conditional on parallel validation and on the preservation of measurement-based checks for non-loggable or insufficiently validated failure modes \cite{Wolter2026SensitivityEfficiency}.

Measurements remain important for failure modes that are not observable in log data, for independent physical verification, for commissioning and validation of new techniques, for treatment sites or delivery configurations with limited prior experience, and for workflows involving patient-specific accessories or beam-modifying devices. Measurements may also be required to validate the log file content itself, especially when the log-file-based pathway becomes a primary or dominant clinical risk-control layer \cite{mackin2014spot,zhu2015towards}.

After implementing log file-based PSQA, regular end-to-end tests for selected patients or representative treatment classes should be continued, even if the process is not completed for every patient. The frequency of such tests should be determined through local risk analysis. It is advisable to begin a transition with simple and standardized classes of treatments with a long record of successful measurement-based PSQA. New treatment sites, modified treatment or planning approaches, new treatment accessories, unusual beam arrangements, motion-sensitive sites, adaptive workflows, and sites with a history of PSQA failures should be prioritized for continuing measurement-based or detector-based verification.

\subsection{Independent Secondary Dose Calculation and Monte Carlo Verification}

An independent secondary dose calculation should be performed for all treatment plans, in addition to either measurement-based or log file-based PSQA. Its role is not limited to detecting numerical dose differences; it provides a structurally independent check against errors in dose-calculation assumptions, CT/HU calibration or interpretation, material assignment, beam modeling, plan export, DICOM interpretation, image-to-dose translation, and algorithmic implementation \cite{zhu2021report}.

For contemporary PBS-PT workflows, independent dose verification should ideally include an independently commissioned three-dimensional dose calculation in patient geometry, with clinically interpretable dose-volume and structure-based review. At least one of the primary or secondary dose-calculation layers should preferably use an accurate Monte Carlo algorithm whenever technically and clinically feasible. If the primary and secondary dose calculations are performed by different algorithms within the same treatment-planning system, this may provide useful algorithmic comparison, but it should not be assumed to provide the same level of structural independence as an independently commissioned external calculation system \cite{aitkenhead2020automated,souris2016fast,wieser2017development}.

The commissioning of an independent secondary dose-calculation system should be treated as the commissioning of a clinical dose-calculation pathway, not as the installation of a simple numerical cross-check. Algorithmic limitations, statistical uncertainty, dose-to-water or dose-to-medium conventions, CT calibration, heterogeneity handling, range-shifter and aperture modeling, beam model parameters, grid size, output normalization, comparison metrics, and dose-volume interpretation should be explicitly validated. Decisions about which secondary calculation system to use should be based on a local risk analysis of the entire workflow, including the possibility that the introduction of a separate system may itself add data-transfer, configuration, or interpretation risks.

\subsection{Per-Fraction and Delivered-Dose Verification}

The infrastructure required for log file-based PSQA can also support ongoing periodic or per-fraction delivery verification. Treatment log files may be acquired for one or more fractions during patient treatment, and the delivered dose may be reconstructed or recalculated from these logs to verify treatment-delivery accuracy and consistency. In the present risk analysis, routine or daily log file-based dose evaluation was an important contributor to the favorable risk-score profile of log file-based PSQA. If this per-fraction or repeated delivered-dose evaluation is not performed, the overall risk-control benefit of log file-based PSQA may be closer to that of measurement-based PSQA than to the full log-file-supported architecture modeled here \cite{meijers2019log,matter2018alternatives,winterhalter2019log,ates2023development,scandurra2016assessing}.

Because log files can contain large quantities of treatment-delivery information, active and automated monitoring is recommended. Ideally, the monitoring system should flag deviations from expected delivery behavior, detect missing or inconsistent records, identify abnormal interruptions or resumed-delivery sequences, and notify the therapy team for further evaluation. Such monitoring should be integrated with defined action levels, escalation pathways, documentation requirements, and incident-learning feedback.

\subsection{Adaptive and (Online) Adaptive Proton Therapy}

Centers planning for adaptive or online adaptive proton therapy should evaluate whether the proposed PSQA architecture remains valid under time-constrained conditions. In such workflows, conventional phantom-based pre-treatment measurement may be infeasible, and PSQA must shift toward fast independent dose calculation, automated data-integrity checks, adapted-plan plausibility checks, delivery-system consistency checks, and post-delivery or near-real-time verification. The local pFMEA should therefore explicitly include adaptive workflow steps before measurement frequency is reduced in clinical scenarios likely to require adaptation \cite{paganetti2021apt,gambetta2025online_apt_closed_loop,albertini2024daily_online_apt,bobic2024experimental_oapt_workflows}.

Adaptive workflows also change the meaning of redundancy. Repeating the same model-dependent calculation may add less safety than combining independent information streams, such as image-based anatomical assessment, independent dose calculation, treatment log analysis, physical or detector-based verification where feasible, in-vivo range or dose verification, structured human review, and incident-learning feedback. A log file-based or calculation-based PSQA architecture intended for adaptive proton therapy should therefore be evaluated not only for speed, but also for independence, failure-mode coverage, interpretability, and ability to support timely clinical action.

\subsection{Recommended Implementation Pathway}

A practical transition from routine measurement-based PSQA toward a log file-supported hybrid PSQA architecture should proceed in stages \cite{huq2016taskgroup100,soelkner2025data_driven_risk_management,li2025radiation_incident_learning_multisite}:

\begin{enumerate}
    \item \textbf{Baseline characterization:} document the current measurement-based, calculation-based, machine-QA, data-transfer, and review workflows, including known failure modes and historical PSQA outcomes.
    \item \textbf{Local pFMEA update:} adapt the generic PSQA-pFMEA to the local equipment, software, workflow, treatment sites, staffing model, and automation level.
    \item \textbf{Parallel operation:} operate measurement-based, log file-based, and independent calculation-based pathways in parallel for a representative transition period.
    \item \textbf{Known-error validation:} test the proposed PSQA pathway against clinically meaningful injected or retrospective error scenarios, and evaluate sensitivity, specificity, and failure-mode coverage rather than pass-rate behavior alone \cite{May2026SEAFARER_HN}.
    \item \textbf{Logfile and data-pathway QA:} validate the completeness, accuracy, accessibility, and traceability of log files, treatment records, and upstream data-transfer pathways.
    \item \textbf{Decision review:} evaluate whether the new hybrid architecture provides independent, timely, clinically interpretable, and non-overlapping risk-control information.
    \item \textbf{Controlled reduction of measurements:} reduce routine measurement frequency only for treatment classes where the local evidence supports this change, while retaining measurements for new, complex, high-risk, or insufficiently validated workflows.
    \item \textbf{Continuous feedback:} update the pFMEA and QA program using log-file performance, measurement results, secondary-calculation discrepancies, delivery anomalies, incident-learning data, and clinical workflow experience.
\end{enumerate}

The final decision should therefore be local, documented, and reversible. Log file-based PSQA can become a central component of a mature proton therapy PSQA program, but only as part of a hybrid architecture that preserves independent physical verification where needed, maintains structurally independent dose calculation, validates the data foundations on which log interpretation depends, and updates its risk model as clinical practice evolves.

\ManualNumberedChapter{5}{CONCLUSION AND OUTLOOK}{ch:conclusions_and_outlook}

\renewcommand{\sectionmark}[1]{\markright{\thesection.\ #1}}
\section{From Verification to Risk-Informed PSQA}
\label{sec:from_verification_to_risk_informed_psqa}

The work on this report began with a focused practical question: can treatment log records be used safely and effectively for patient-specific quality assurance in pencil beam scanning proton therapy? The analysis ultimately led to a broader conclusion: the central PSQA question is not whether one verification method can replace another, but how complementary verification layers should be combined to control the dominant risks of a defined clinical workflow. Log file-based PSQA is therefore not merely a more efficient substitute for measurement-based verification. It is one component of a larger risk-control architecture in which measurement-based PSQA, log file-based PSQA, and independent secondary dose calculation each reshape different parts of the same clinical workflow.

To make this comparison meaningful, we developed a process-driven PSQA-pFMEA formalism anchored to a common no-PSQA baseline. Forty-four validated PSQA-relevant failure modes were mapped onto 20 process steps and evaluated across three PSQA pathways. The staged \textit{Data}/\textit{Full}/\textit{Cum} framework, together with the derived \textit{Data-to-Cum} bridge metric, allowed us to separate preparatory workflow burden from final verification benefit and to propagate risk changes from individual failure modes to process steps, workflow regions, and the full treatment workflow. This hierarchy is the central methodological contribution of the report: within a common expert-scored and baseline-anchored model, PSQA is not treated as a single pass/fail event, but as a structured, workflow-embedded risk-control process whose results should be interpreted as semi-quantitative comparative risk-score signatures rather than absolute clinical risk estimates.

Within the present model, all investigated PSQA pathways reduce the baseline risk-score burden of the generic PBS-PT workflow, but the resulting method comparison should not be read as a universal or probability-calibrated hierarchy. Log file-based PSQA shows the strongest overall cumulative workflow-level risk-score reduction, particularly because it combines low preparatory friction with strong mitigation in delivery-, re\-cording-, and machine-state-related workflow regions. Measurement-based PSQA remains a powerful and broadly compensatory method, especially for physical delivery, interface, accessory, setup, commissioning-sensitive, and non-loggable risks. Independent secondary dose calculation produces a smaller global effect in the present workflow-level aggregation, but remains essential for image-, contour-, model-, and calculation-sensitive failure modes that are not adequately covered by either measurement-based or log file-based PSQA alone.

The practical message is therefore not that one PSQA method should simply replace another, nor that device class alone determines clinical PSQA performance \cite{May2026SEAFARER_HN}. Rather, the safest and most efficient strategy is a baseline-anchored, stage-resolved, risk-informed hybrid PSQA architecture. Pre-treatment log file-based PSQA can reduce workload and equipment dependence and can test treatment fields under delivery conditions close to the actual patient treatment \cite{Wolter2026SensitivityEfficiency}. When combined with routine or daily log file-based dose evaluation, it can provide a very strong risk-control layer. Nevertheless, log data cannot reveal all clinically relevant failure modes. Measurement-based PSQA remains indispensable where independent physical verification is required, and phantom-based measurements may be necessary whenever new techniques, treatment sites, delivery configurations, treatment accessories, or commissioning states are introduced.

Independent secondary dose calculation should also remain part of a mature PSQA program. Its value is not that it dominates the entire workflow, but that it provides a structurally independent check in precisely those regions where calculation assumptions, CT/HU handling, dose modeling, image-to-dose translation, and algorithmic implementation determine the risk profile. Whenever technically and clinically feasible, this layer should include an independently commissioned three-dimensional calculation in patient geometry, preferably using an accurate Monte Carlo algorithm in either the primary or secondary calculation pathway, so that PSQA interpretation can move beyond numerical agreement alone toward structure-based and dose-volume-relevant review.

The safe adoption of log file-based PSQA also requires explicit quality assurance of the log data themselves and of all upstream data-transfer pathways on which their interpretation depends. Vendors and clinical users share responsibility here. Log files must be complete, transparent, accessible, and sufficiently documented for PSQA use; data transfer between TPS, OIS, TCS, delivery system, and record-and-verify components must be traceable and verifiable. Without such controls, log file-based PSQA risks becoming a technically elegant method built on insufficiently protected data foundations.

Finally, the framework presented here should be viewed as a structured starting point for local implementation and governance, not as a universal prescription, generic acceptance standard, or substitute for center-specific risk analysis. Each proton therapy center should adapt the pFMEA to its own workflow, equipment, software environment, staffing model, automation level, clinical case mix, and institutional QA maturity. The long-term outlook is a living PSQA risk model: prospective at first, then progressively refined by log-file performance, measurement results, secondary-calculation discrepancies, incident-learning data, adaptive-workflow experience, and local clinical outcomes. In that form, PSQA can evolve from a routine verification obligation into an adaptive, evidence-informed safety strategy for modern particle therapy. 
\section{Outlook: APT, PSQA, and the Challenge of Complexity}
\label{sec:outlook_apt_complex}

Adaptive proton therapy (APT) represents a natural but demanding extension of the workflow logic developed in this risk assessment report. It should also be regarded as a stress test for any proposed PSQA architecture, because the available time for measurement, recalculation, review, approval, and corrective action is compressed while the dependence on image quality, contour propagation, adaptive optimization, data integrity, and delivery verification increases. The motivation is simple, but its consequences are profound: because proton dose distributions are highly sensitive to anatomical, density, and range-related changes, the treatment plan that was optimal at simulation may no longer be optimal at the time of delivery \cite{paganetti2021apt,gambetta2025online_apt_closed_loop}. 

Offline APT addresses this problem between fractions, online APT aims to adapt the plan while the patient remains on the treatment couch, and near real-time APT extends the concept further by attempting to close the loop between treatment verification, adaptation decision, plan modification, QA, and delivery. The first clinical daily online adaptive proton therapy workflows and multi-institutional experimental validations demonstrate that such concepts are moving from feasibility toward implementation, but they also show that the adaptive workflow is no longer a simple extension of conventional planning \cite{albertini2024daily_online_apt,bobic2024experimental_oapt_workflows}.

This shift changes the role of PSQA. In conventional workflows, PSQA can often be performed before the treatment course or before delivery with sufficient time for measurement, calculation, review, and repetition if needed. In online adaptive proton therapy, this luxury largely disappears. Phantom-based pre-treatment PSQA is generally incompatible with the time constraints of online adaptation, and log-file information may become available only after delivery unless an independent pre-delivery verification strategy is available. As a result, APT requires fast, automated, and layered QA: independent dose calculation, data-integrity checks, sanity checks of contours and plan parameters, validation of adapted-plan plausibility, verification of delivery-system consistency, and ultimately online or near-real-time treatment verification \cite{gambetta2025online_apt_closed_loop}. PSQA therefore becomes less a single test and more a sequence of control points embedded in a time-critical adaptive workflow.

This is precisely where the risk-assessment structure developed in this report becomes relevant beyond its original static setting. The Data/Full/Cum framework already separates preparatory workflow burden from verification benefit, and the Data-to-Cum bridge quantifies the additional mitigation produced when the pathway is completed. In APT, these concepts would not disappear; they would become more dynamic. Each adaptive fraction can be interpreted as a time-constrained traversal of a shortened PSQA workflow: imaging and data generation modify the workflow state, contour propagation and adaptation decisions introduce new data and decision risks, adapted plan optimization and approval compress planning and review into a narrow time window, and pre-delivery or in-vivo verification provides the bridge toward a safe cumulative delivery state.

The same logic also defines the empirical validation endpoint for future extensions of the present framework. The RPN-derived quantities reported in this study should not be interpreted as direct measurements of delivered dose or clinical outcome; rather, they identify where risk-control effort is expected to have the greatest workflow-level effect. In future clinical implementations, these risk signatures should be tested against reconstructed delivered dose, anatomy-of-the-day dose recalculation, accumulated dose where appropriate, dose-volume metrics, structure-based deviations, incident-learning outcomes, and clinically relevant failure-detection performance.

Such validation would close the loop between prospective pFMEA prediction and observed clinical PSQA performance. The pFMEA indicates which workflow regions and failure modes should be controlled by a given PSQA architecture, whereas delivered-dose and structure-based endpoints test whether that architecture actually protects clinically meaningful target coverage and organ-at-risk sparing. This is particularly important in adaptive proton therapy, where the clinically relevant endpoint is not agreement with a static pretreatment phantom plan, but confidence that the delivered or adapted dose remains acceptable on the anatomical and delivery-state information available at the time of treatment.

Thereby, APT should be regarded not merely as a more complicated version of proton therapy, but as a candidate complex adaptive system. In the context of radiotherapy, complexity should not be understood simply as the presence of many components. A system is complex when clinically relevant outcomes and safety states emerge from nonlinear, context-dependent interactions among people, technologies, data flows, organizational constraints, automation, and feedback loops, rather than from isolated components with stable one-to-one cause-effect behavior \cite{bakhtiari2026complex_adaptive_ro}. This distinction matters. A conventional proton workflow may already be complicated; an adaptive proton workflow becomes complex when the patient state, imaging system, contouring engine, optimizer, dose calculation, QA checks, human approval, delivery system, and treatment verification all interact within the same short clinical decision cycle.

Systems-theory-based evaluations of pre-treatment PSQA provide an important conceptual warning \cite{wong2024systems_theory_psqa}. In a modern, tightly integrated treatment environment, pre-treatment PSQA may add limited safety value if it provides feedback that overlaps with already well-controlled commissioning, machine QA, and integrated data-management safeguards. That conclusion is valuable, but it should not be transferred uncritically to proton therapy or APT. Proton therapy adds range sensitivity, heterogeneous vendor ecosystems, spot-delivery dynamics, robustness requirements, and stronger dependence on anatomical state. APT adds still more: compressed decision-making, automated contouring, rapid plan adaptation, online QA, and potential treatment-verification feedback. In this setting, the key question is not whether PSQA in general is redundant, but whether a given PSQA layer provides independent, timely, and non-overlapping information at the point where adaptive decisions are made.

This complexity has two implications for future risk assessment. First, prospective pFMEA remains necessary but is not sufficient. APT workflows require systematic anticipation of failure modes, but they may also generate emergent risks that are difficult to enumerate before clinical operation. Second, redundancy must be diverse rather than duplicative. Repeating the same model-dependent calculation is less valuable than combining independent information streams: image-based anatomical assessment, independent dose calculation, treatment log analysis, physical or detector-based verification, in-vivo range or dose verification, structured human review, and incident-learning feedback. In this sense, the hybrid PSQA architecture recommended in this report becomes even more important in APT.

The future application of the present formalism to APT should therefore follow two directions. The first is structural: the same failure-mode to process-step to workflow hierarchy can be used to map adaptive workflow risks, including imaging, contour propagation, adaptation-need decisions, plan re-optimization, online QA, delivery interruption, resumed delivery, and post-delivery verification. The second is dynamic: the pFMEA should evolve from a static expert table into a living risk model that is recalibrated by log files, online QA outcomes, incident learning, adaptation frequency, failed or overridden adaptation decisions, delivery-verification data, and local clinical experience. The natural evolution is therefore

\[
  \boxed{
  \mathrm{static\ PSQA\mbox{-}pFMEA}
  \quad \longrightarrow \quad
  \mathrm{feedback\mbox{-}updated\ adaptive\ risk\ model}.
  }
\]

Such a model would preserve the transparency of pFMEA while allowing observed workflow behavior to refine the assumed occurrence, detectability, and mitigation patterns over time. It would also provide a practical bridge between prospective risk analysis and continuous quality improvement: the pFMEA would define what should be controlled, whereas operational feedback would test whether the controls actually perform as intended.

Viewed in this way, APT is not a threat to the formalism proposed here; it is its most demanding future test case. If PSQA in standard PBS proton therapy asks which method best reduces the risk of a planned treatment workflow, adaptive proton therapy asks a harder question: how can risk be controlled when the workflow itself changes during treatment? The answer is unlikely to be a single QA method. It will require a baseline-anchored, stage-aware, feedback-driven, and hybrid risk architecture. The PSQA-pFMEA framework developed here provides a first practical language for building such an architecture: transparent enough to support local clinical governance, structured enough to compare complementary PSQA layers, and explicitly conditional enough to avoid mistaking model-derived risk-score reductions for universal clinical risk estimates. 

\renewcommand{\sectionmark}[1]{\markright{\thesection.\ #1}}

\printglossary[nonumberlist]

\newpage
\setcounter{page}{105}
\addcontentsline{toc}{chapter}{Bibliography}
\bibliographystyle{unsrt}
\bibliography{./references.bib}
\nocite{*}


\clearpage
\appendix
\pagestyle{appendixfancy}
\fancyfoot[CE]{}

\begingroup
  \SetAppendixHyperPrefix{appA}
  \SetAppendixNumbering{A}

  \clearpage
  \newgeometry{
    a4paper,
    top=1.5cm,
    bottom=1.5cm,
    left=1.5cm,
    right=1.5cm,
    twoside=false
  }

  \ManualAppendixChapter{A}{TREATMENT WORKFLOWS FOR PSQA SCENARIOS}{ch:appendix_workflows}

  \BeginAppendixStarredSectionTOCAnchors
  \section*{A.0 List of Figures}
\label{sec:appA_list_of_figures}
\addcontentsline{toc}{section}{A.0 ~~ List of Figures}

\begin{description}
    \setlength\itemsep{2pt} 
    \setlength\parskip{0pt} 
    \item[\hspace{0.42cm}Figure A1: Generic PBS-PT workflow model with all possible PSQA tracks] ~\\ {\small Generic workflow model for a representative PBS-PT treatment with all possible PSQA tracks [process VIII: 2\textsuperscript{nd} dose calculation – left (\textcolor{yellow}{yellow}) | measurement-based -- mid (\textcolor{red}{red}] | logfile-based -- right (\textcolor{green}{green})]. Visible are all processes, sub-processes, and process steps as functional, self-explaining units. The small number on the right side of a process step indicates the number of related failure modes (hidden). The timeline runs top-down in the order of the process step numbers (1.) to (44.). Iterative feedback loops due to adaptive planning are not explicitly shown \ref{fig:PSQA_all}.}
    \item[\hspace{0.42cm}Figure A2: Generic PBS-PT Workflow Model with all Failure Modes (no PSQA)] ~\\ {\small Generic workflow model for a representative PBS-PT treatment without PSQA [process VIII: ignored]. Visible are all relevant processes, sub-processes, and process steps as functional, self-explaining units. The magenta-filled capsules report the process step-associated failure modes. The black-framed rhombus on the right of a failure mode represents the initially rated RPN [NoQA] as required data. The timeline runs top-down in the order of the process step numbers (1.) to (44.). Iterative feedback loops due to adaptive planning are not explicitly shown \ref{fig:PSQA_noQA}.}
    \item[\hspace{0.42cm}Figure A3: Generic PBS-PT Workflow Model with ISDC-based (2\textsubscript{nd} Calc) PSQA] ~\\ {\small Generic workflow model for a representative PBS-PT treatment with 2nd  dose calculation PSQA [process VIII]. To enhance readability of the graph, the [V. Treatment Planning] and [X. Treatment] processes are shown without their process steps. Visible are the remaining, processes, sub-processes, and process steps as functional, self-explaining units. The magenta-filled capsules report the process step-associated failure modes. The yellow-framed rhombi on the right of a failure mode represent the newly rated RPN [calc] replacing the formerly rated RPN [noQA] (black frame).  A small arrow indicates the transfer. The timeline runs top-down in the order of the process step numbers (1.)  to (44.).  Iterative feedback loops due to adaptive planning are not explicitly shown \ref{fig:PSQA_calc}.}
    \item[\hspace{0.42cm}Figure A4: Generic PBS-PT Workflow Model with Measurement-based PSQA] ~\\ {\small Generic workflow model for a representative PBS-PT treatment with measurement-based PSQA [process VIII]. To enhance readability of the graph, the [V. Treatment Planning] and [X. Treatment] processes are shown without their process steps. Visible are the remaining, processes, sub\nobreakdash-processes, and process steps as functional, self\nobreakdash-explaining units input variables. The magenta-filled capsules report the process step-associated failure modes. The red-framed rhombi on the right of a failure mode represent the newly rated RPN [meas] replacing the formerly rated RPN [noQA] (black frame). A small arrow indicates the transfer. The timeline runs top-down in the order of the process step numbers (1.) to (44.). Iterative feedback loops due to adaptive planning are not explicitly shown. \ref{fig:PSQA_meas}.}
    \item[\hspace{0.42cm}Figure A5: Generic PBS-PT Workflow Model with Log File-based PSQA] ~\\ {\small Generic workflow model for a representative PBS-PT treatment with log file-based PSQA [process VIII]. To enhance readability of the graph, the [V. Treatment Planning] and [X. Treatment] processes are shown without their process steps. Visible are the remaining, processes, sub-processes, and process steps as functional, self-explaining units. The magenta-filled capsules report the process step-associated failure modes. The green-framed rhombi on the right of a failure mode represent the newly rated RPN [log] replacing the formerly rated RPN [noQA] (black frame). A small arrow indicates the transfer. The timeline runs top-down in the order of the process step numbers (1.) to (44.). Iterative feedback loops due to adaptive planning are not explicitly shown \ref{fig:PSQA_log}.}
\end{description}
\pagebreak

\vspace{-1cm}
\section*{A.1 Generic PBS-PT workflow model with all possible PSQA tracks}
\label{sec:appA_wf_psqa_all}
\addcontentsline{toc}{section}{A.1 ~~ Generic PBS-PT workflow model with all possible PSQA tracks}
\samepage
\vspace{-0.5cm}
\begin{figure}[H]
    \centering
    \includegraphics[width=1.05\textwidth,height=1.0\textheight,keepaspectratio]{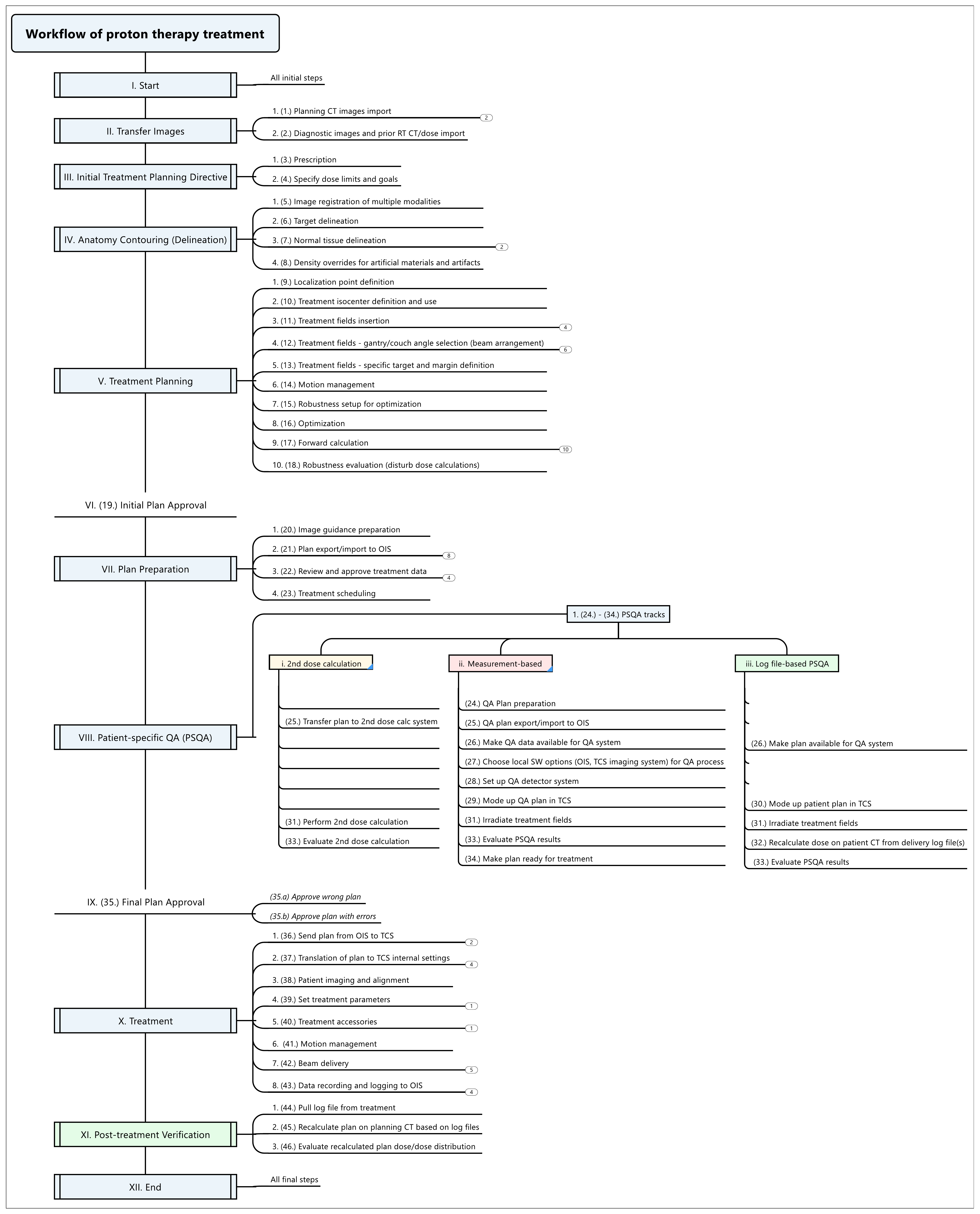}
    \caption{\footnotesize Generic workflow model for a representative PBS-PT treatment with all possible PSQA tracks [process VIII: 2\textsuperscript{nd} dose calculation – left (\textcolor{yellow}{yellow}) | measurement-based -- mid (\textcolor{red}{red}) | logfile-based -- right (\textcolor{green}{green})]. Visible are all processes, sub-processes, and process steps as functional, self-explaining units. The small number on the right side of a process step indicates the number of related failure modes (hidden). The timeline runs top-down in the order of the process step numbers (1.) to (44.). Iterative feedback loops due to adaptive planning are not explicitly shown.}
    \label{fig:PSQA_all}
\end{figure}

\section*{A.2 Generic PBS-PT Workflow Model with all Failure Modes (no PSQA)}
\label{sec:appA_wf_psqa_noqa}
\addcontentsline{toc}{section}{A.2 ~~ Generic PBS-PT Workflow Model with all Failure Modes (no PSQA)}
\samepage
\vspace{-0.5cm}
\begin{figure}[H]
    \centering
    \hspace*{-0.05\textwidth} 
    \includegraphics[width=1.\textwidth,height=0.9\textheight,keepaspectratio]{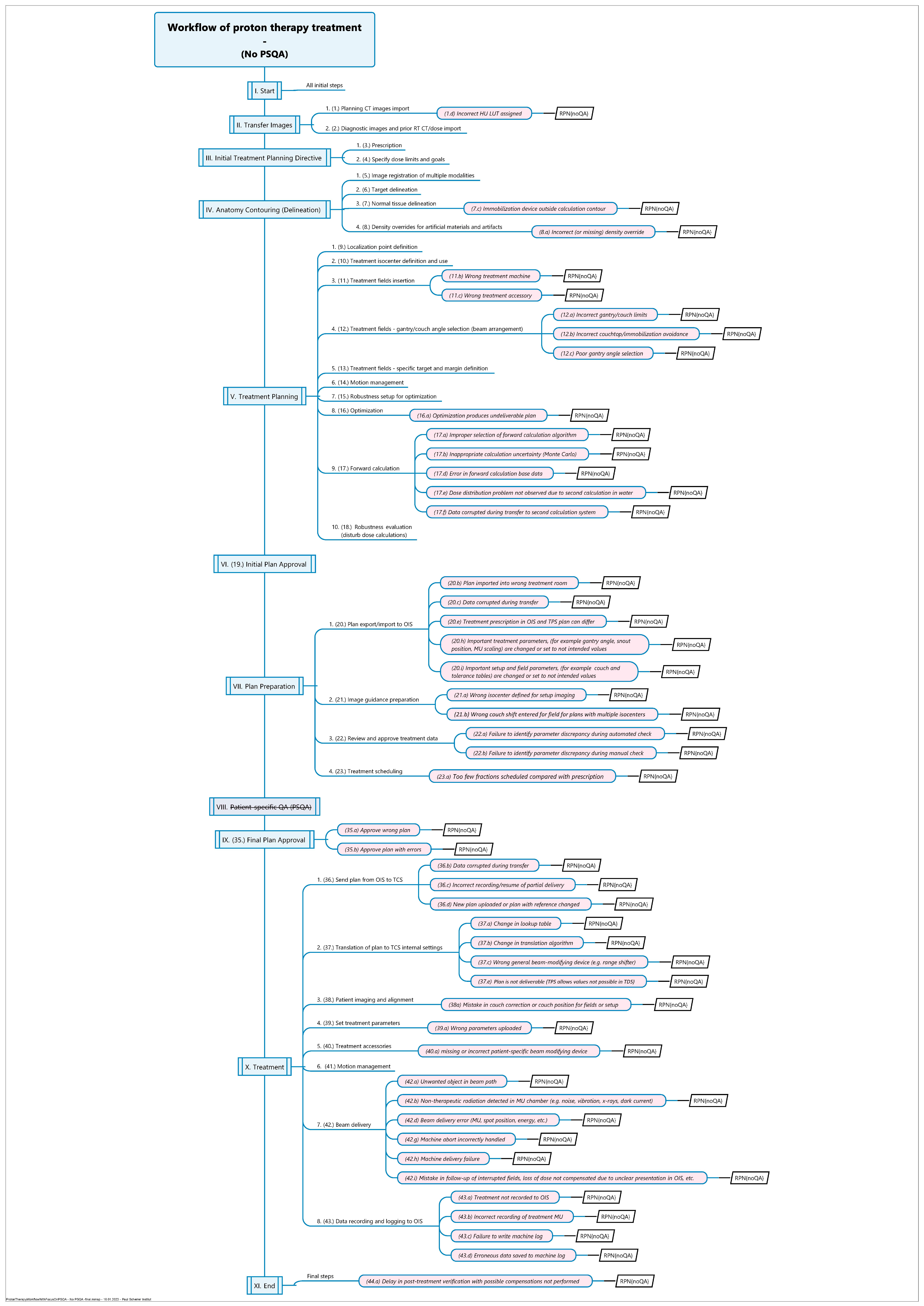}
    \caption{\footnotesize Generic workflow model for a representative PBS-PT treatment without PSQA [process VIII: ignored] but all possible failure modes listed. Visible are all relevant processes, sub-processes, and process steps as functional, self-explaining units. The magenta-filled capsules report the process step-associated failure modes. The black-framed rhombus on the right of a failure mode represents the initially rated RPN [NoQA] as required data. The timeline runs top-down in the order of the process step numbers (1.) to (44.). Iterative feedback loops due to adaptive planning are not explicitly shown.}
    \label{fig:PSQA_noQA}
\end{figure}

\section*{A.3 Generic PBS-PT Workflow Model with ISDC-based (2\textsuperscript{nd} Calc) PSQA}
\label{sec:appA_wf_psqa_calc}
\addcontentsline{toc}{section}{A.3 ~~ Generic PBS-PT Workflow Model with ISDC-based (2nd Calc) PSQA}
\samepage
\vspace{-0.5cm}
\begin{figure}[H]
    \centering
    \hspace*{-0.025\textwidth} 
    \includegraphics[width=1.05\textwidth,height=0.875\textheight,trim={0.0cm 12.0cm 0.0cm 12.0cm},clip]{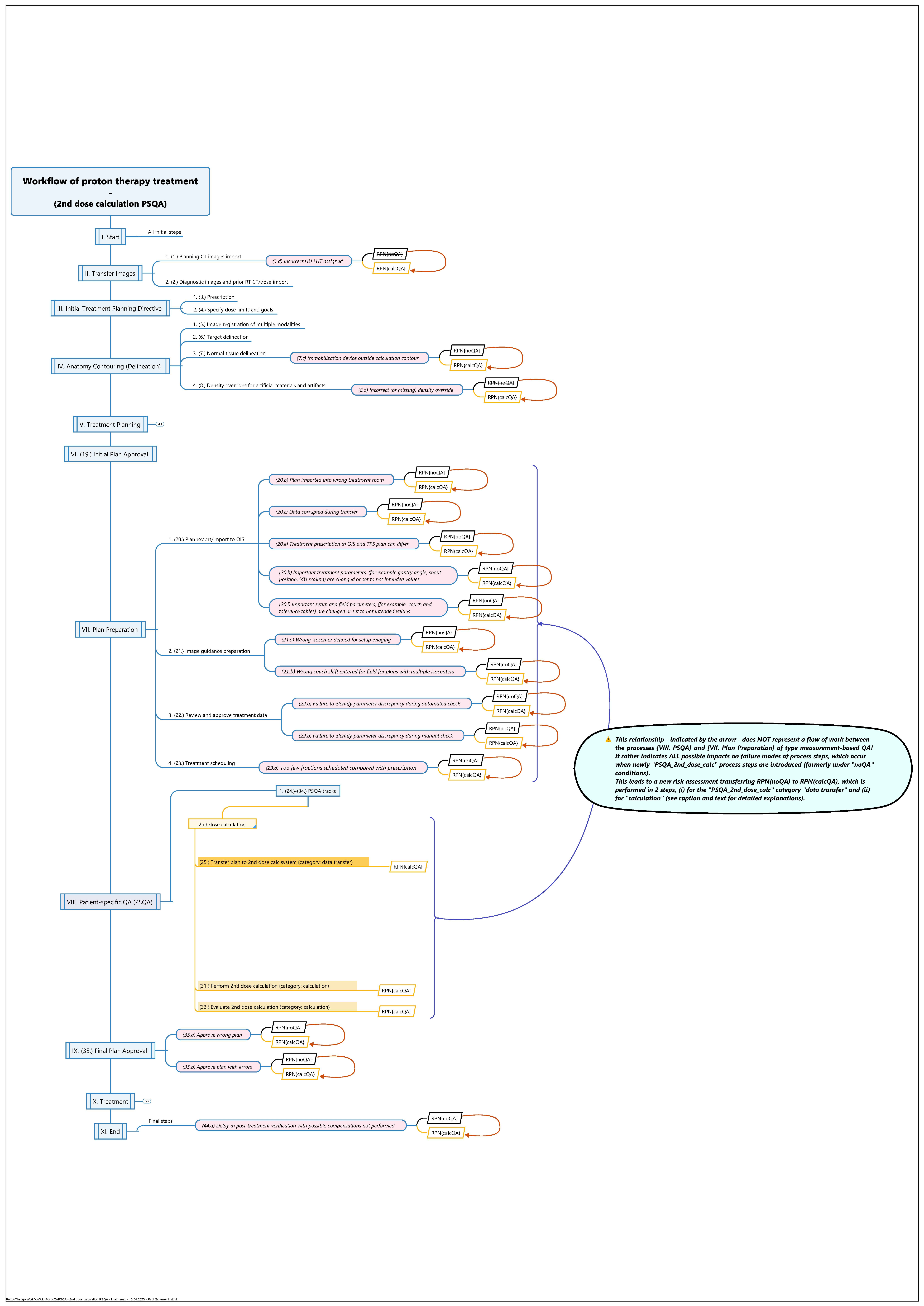}
    \caption{\footnotesize Generic workflow model for a representative PBS-PT treatment with 2\textsuperscript{nd} dose calculation PSQA [process VIII]. To enhance readability of the graph, the [V. Treatment Planning] and [X. Treatment] processes are shown without their process steps. Visible are the remaining, processes, sub-processes, and process steps as functional, self-explaining units. The magenta-filled capsules report the process step-associated failure modes. The yellow-framed rhombi on the right of a failure mode represent the newly rated RPN [calc] replacing the formerly rated RPN [noQA] (black frame). A small arrow indicates the transfer. The timeline runs top-down in the order of the process step numbers (1.) to (44.). Iterative feedback loops due to adaptive planning are not explicitly shown.}
    \label{fig:PSQA_calc}
\end{figure} 

\section*{A.4 Generic PBS-PT Workflow Model with Measurement-based PSQA}
\label{sec:appA_wf_psqa_meas}
\addcontentsline{toc}{section}{A.4 ~~ Generic PBS-PT Workflow Model with Measurement-based PSQA}
\samepage
\vspace{-0.5cm}
\begin{figure}[H]
    \centering
    \hspace*{-0.025\textwidth} 
    \includegraphics[width=1.05\textwidth,height=0.875\textheight,trim={0.0cm 12.0cm 0.0cm 12.0cm},clip]{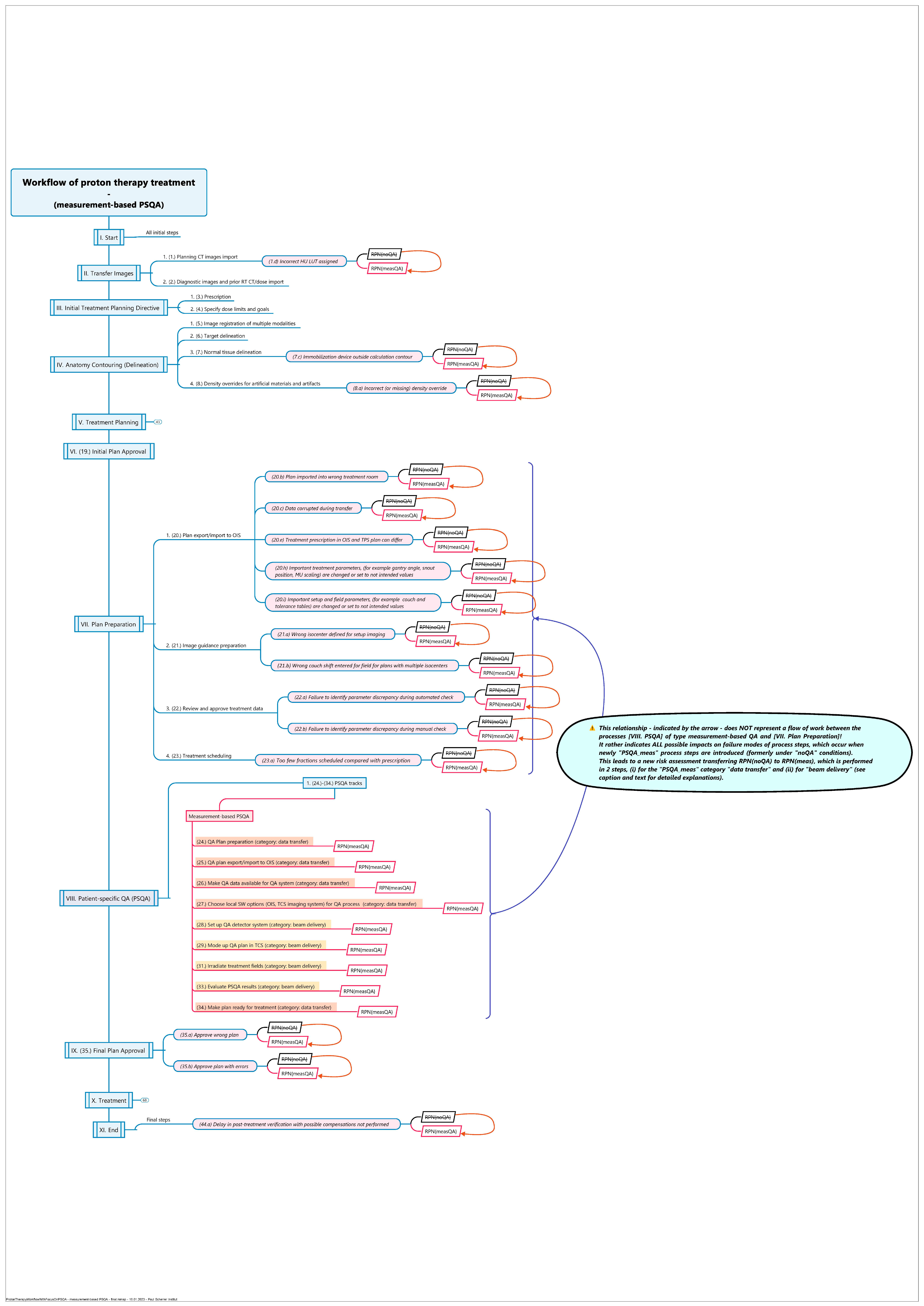}
    \caption{\footnotesize Generic workflow model for a representative PBS-PT treatment with measurement-based PSQA [process VIII]. To enhance readability of the graph, the [V. Treatment Planning] and [X. Treatment] processes are shown without their process steps. Visible are the remaining, processes, sub-processes, and process steps as functional, self-explaining units input variables. The magenta-filled capsules report the process step-associated failure modes. The red-framed rhombi on the right of a failure mode represent the newly rated RPN [meas] replacing the formerly rated RPN [noQA] (black frame). A small arrow indicates the transfer.  The timeline runs top-down in the order of the process step numbers (1.) to (44.). Iterative feedback loops due to adaptive planning are not explicitly shown.}
    \label{fig:PSQA_meas}
\end{figure}

\section*{A.5 Generic PBS-PT Workflow Model with Log File-based PSQA}
\label{sec:appA_wf_psqa_log}
\addcontentsline{toc}{section}{A.5 ~~ Generic PBS-PT Workflow Model with Log File-based PSQA}
\samepage
\vspace{-0.5cm}
\begin{figure}[H]
    \centering
    \hspace*{-0.025\textwidth} 
    \includegraphics[width=1.05\textwidth,height=0.875\textheight,trim={0.0cm 15.0cm 0.0cm 15.0cm},clip]{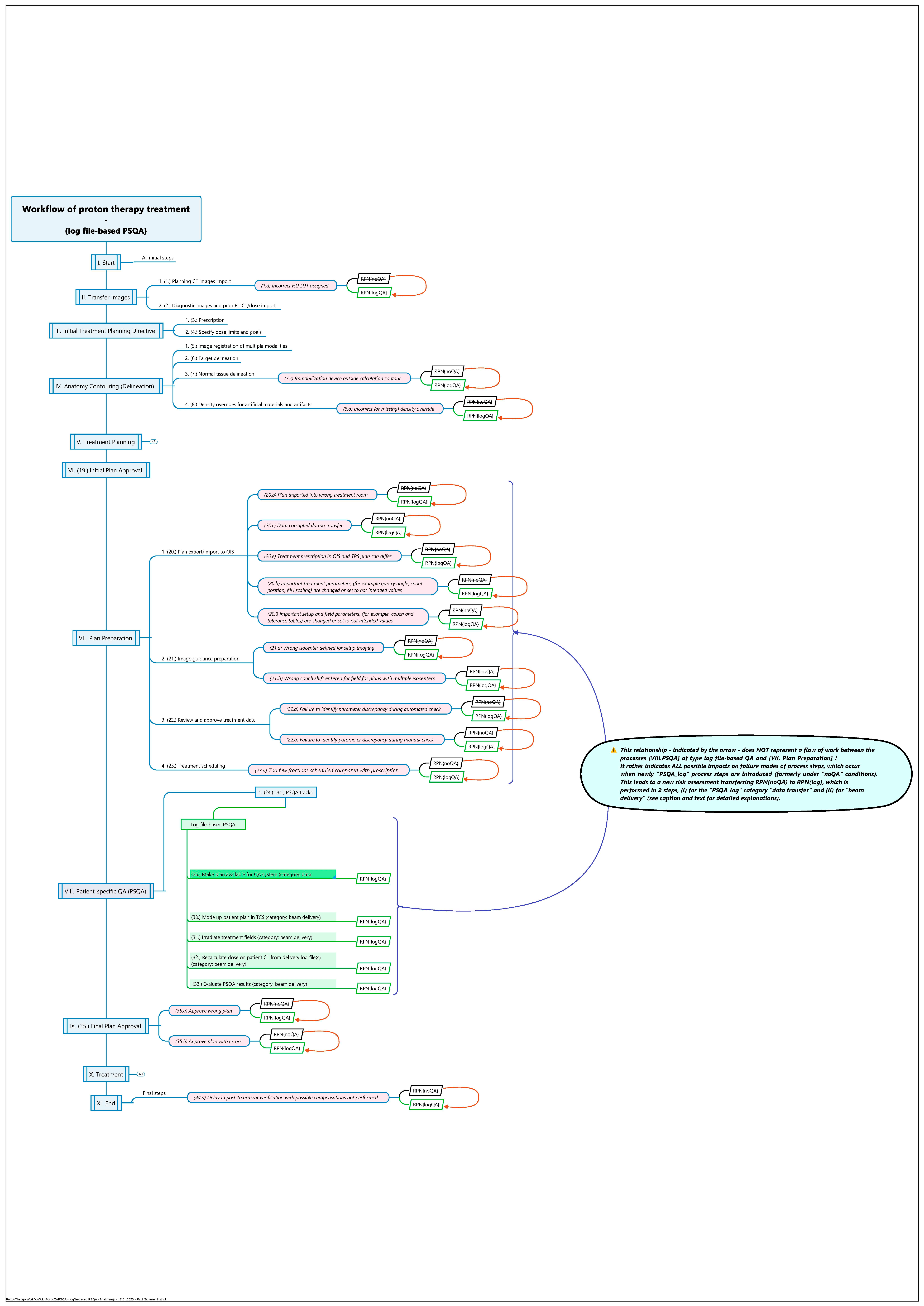}
    \caption{\footnotesize Generic workflow model for a representative PBS-PT treatment with log file-based PSQA [process VIII]. To enhance readability of the graph, the [V. Treatment Planning] and [X. Treatment] processes are shown without their process steps. Visible are the remaining, processes, sub-processes, and process steps as functional, self-explaining units. The magenta-filled capsules report the process step-associated failure modes. The green-framed rhombi on the right of a failure mode represent the newly rated RPN [log] replacing the formerly rated RPN [noQA] (black frame). A small arrow indicates the transfer.  The timeline runs top-down in the order of the process step numbers (1.) to (44.). Iterative feedback loops due to adaptive planning are not explicitly shown.}
    \label{fig:PSQA_log}
\end{figure}

  \clearpage
  \restoregeometry
\endgroup

\begingroup
  \BeginTextAppendixLayout

  \SetAppendixHyperPrefix{appB}
  \SetAppendixNumbering{B}

  \renewcommand{\sectionmark}[1]{}
  \markright{}

  \ManualAppendixChapter{B}{MATHEMATICAL FRAMEWORK FOR A COMPARATIVE RISK ASSESSMENT}{ch:math_framework}

  \BeginAppendixLabelPrefix{appB}

\section{Introduction}

This appendix presents a unified mathematical framework for comparing patient-specific quality assurance (PSQA) strategies within particle therapy workflows. The framework builds on classical process-driven Failure Modes and Effects Analysis (pFMEA) using the Risk Priority Number (RPN) methodology, but it is used here in a deliberately restricted sense: not as an absolute metrology of risk, but as an operational, semi-quantitative scoring system that enables baseline-consistent comparisons among different PSQA methods (measurement-based, log file-based, and independent secondary dose calculation). The central structural idea is a staged intervention logic,
\[
\mathrm{NoQA}\rightarrow \mathrm{data}\rightarrow \mathrm{full}\rightarrow \mathrm{cum},
\]
where the \textit{Data}-stage captures preparatory workflow modifications, the \textit{Full}-stage captures the incremental verification layer acting on the already modified \textit{Data}-stage configuration, and the cumulative end state captures the final effect relative to the common \textit{NoQA} baseline.

\section{Foundational Principles}
\label{sec:foundational_principles}

Risk management in radiotherapy requires that alternative quality assurance strategies be compared in a structured, quantitative, and operationally consistent way.\footnote{Throughout this appendix, the term \emph{quantitative} is used only in the limited sense of a number-based formal comparison. As discussed in the main text and Appendix~C, the present framework remains semi-quantitative and measurement-theoretically conditional, because the underlying pFMEA scores are expert-assigned ordinal quantities rather than directly measured physical risks.} In this framework we assume that:\\

\begin{enumerate}
    \item The pFMEA-based RPN approach can serve as an operational basis for \emph{comparative, semi-quantitative} risk assessment within a fixed scoring rubric.
    \item Failure modes can be identified for every relevant process step in the treatment workflow.
    \item PSQA methods modify the key parameters (Occurrence, Severity, Detectability) in ways that can be represented through structured expert assessment.
    \item Differential risk reduction across PSQA modalities provides meaningful guidance for clinical implementation when evaluated relative to a common baseline workflow.
\end{enumerate}
~\\
Although risk scoring is partly subjective and failure modes may be interdependent, the present process-driven formalism provides a rigorous basis for comparing how alternative PSQA workflows modify the same underlying treatment process.

\section{Risk Priority Number Calculation}
\label{sec:RPN_calculation}

Following established pFMEA methodology (adapted for radiotherapy as in AAPM TG-100), each failure mode (FM) is characterized by three parameters:\\
\begin{itemize}
    \item \textbf{Occurrence (O)}: the probability or frequency of the failure,
    \item \textbf{Severity (S)}: the impact or clinical consequence if the failure occurs,
    \item \textbf{Detectability (D)}: the likelihood that the failure will remain undetected.
\end{itemize}
~\\
Each parameter is scored on a scale from 1 to 10 according to predefined criteria. Table~\ref{tab:tg100_fmea} reproduces the qualitative descriptions and corresponding numerical ranks used in this study in the form envisaged in the AAPM TG-100 report in order to standardize the evaluation across different errors~\cite{huq2016taskgroup100}.\footnote{The terminology and detailed meaning of the TG-100 categories (e.g.\ ``Limited toxicity or tumor underdose'' or ``Wrong dose, dose distribution, location, or volume'') are described in full in~\cite{huq2016taskgroup100} and are therefore not repeated here.}\\

The \textbf{baseline} Risk Priority Number (RPN) is then computed as
\begin{subequations}
  \label{eq:RPN_pair}
  \begin{align}
      \mathrm{RPN}^{\mathrm{noQA}}_{\mathrm{FM}} &= O^{\mathrm{noQA}}_{\mathrm{FM}} \times S^{\mathrm{noQA}}_{\mathrm{FM}} \times D^{\mathrm{noQA}}_{\mathrm{FM}}\,,
      \label{eq:RPN_noQA}
      \\
      \mathrm{RPN}^{\mathrm{PSQA},\mathrm{pc}}_{\mathrm{FM}} &= O^{\mathrm{PSQA},\mathrm{pc}}_{\mathrm{FM}} \times S^{\mathrm{PSQA},\mathrm{pc}}_{\mathrm{FM}} \times D^{\mathrm{PSQA},\mathrm{pc}}_{\mathrm{FM}}\,,
      \label{eq:RPN_pc}
  \end{align}
\end{subequations}
where $\mathrm{PSQA} \in \{\textit{meas},\textit{log},\textit{calc}\}$ and $\mathrm{pc} \in \{\textit{data},\textit{full}\}$ (for the complete notation, see Sec.~\ref{sec:notation_system}).\\

The TG-100 scales for \(O\) and \(D\) are strongly non-linear. Although an exponential representation would in principle be possible, the standard multiplicative form in Eqs.~\eqref{eq:RPN_pair} is retained for continuity and interpretive consistency with the existing pFMEA framework.

\begin{table}[tbp]
\centering
\footnotesize
\renewcommand{\arraystretch}{1.5}
\setlength{\tabcolsep}{5pt}

\begin{tabular}{|m{1.0cm}|m{2.0cm}|m{1.7cm}|m{3.0cm}|m{2.7cm}|m{3.5cm}|}
\hline \hline
\multirow{2}{*}{\centering \textbf{Rank}}
& \multicolumn{2}{m{3.7cm}|}{\centering \textbf{Occurrence (O)}}
& \multicolumn{2}{m{5.7cm}|}{\centering \textbf{Severity (S)}}
& \multicolumn{1}{m{3.5cm}|}{\centering \textbf{Detectability (D)}} \\ \cline{2-6}
& \multicolumn{1}{m{2.0cm}|}{\centering \textbf{Qualitative}}
& \multicolumn{1}{m{1.7cm}|}{\centering \textbf{Frequency in \%}}
& \multicolumn{1}{m{3.0cm}|}{\centering \textbf{Qualitative}}
& \multicolumn{1}{m{2.7cm}|}{\centering \textbf{Categorization}}
& \multicolumn{1}{m{3.5cm}|}{\centering \textbf{Estimated Probability of failure going undetected in \%}} \\ \hline

\centering 1
& \multirow{2}{*}{\parbox{2cm}{\centering Failure\hspace{1cm} unlikely}}
& \centering 0.01
& \centering No effects
& \centering -
& \parbox{3.5cm}{\centering 0.01} \\ \cline{1-1} \cline{3-6}
\centering 2
&
& \centering 0.02
& \multirow{2}{*}{\parbox{2.7cm}{\centering Inconvenience}}
& \multirow{2}{*}{\parbox{2.7cm}{\centering Inconvenience}}
& \parbox{3.5cm}{\centering 0.2} \\ \cline{1-3} \cline{6-6}

\renewcommand{\arraystretch}{5}
\centering 3
& \multirow{3}{*}{\parbox{2cm}{\centering Relatively\hspace{1cm} few~failures}}
& \centering 0.05
&
&
& \parbox{3.5cm}{\centering 0.5} \\ \cline{1-1} \cline{3-6}
\centering 4
&
& \centering 0.1
& \parbox{2.7cm}{\centering Minor dosimetric error}
& \parbox{2.7cm}{\centering Suboptimal plan or treatment}
& \parbox{3.5cm}{\centering 1.0} \\ \cline{1-1} \cline{3-6}
\centering 5
&
& \centering <0.2
& \multirow{2}{*}{\parbox{2.7cm}{\centering Limited toxicity or tumor underdose}}
& \multirow{4}{*}{\parbox{2.7cm}{\centering Wrong dose, dose distribution, location, or volume}}
& \parbox{3.5cm}{\centering 2.0} \\ \cline{1-3} \cline{6-6}
\renewcommand{\arraystretch}{1.5}

\centering 6
& \multirow{2}{*}{\parbox{2cm}{\centering Occasional\hspace{1cm} failures}}
& \centering <0.5
&
&
& \parbox{3.5cm}{\centering 5.0} \\ \cline{1-1} \cline{3-4} \cline{6-6}
\centering 7
&
& \centering <1
& \multirow{2}{*}{\parbox{2.7cm}{\centering Potentially serious toxicity or tumor underdose}}
&
& \parbox{3.5cm}{\centering 10} \\ \cline{1-3} \cline{6-6}

\centering 8
& \multirow{2}{*}{\parbox{2cm}{\centering Repeated\hspace{1cm} failures}}
& \centering <2
&
&
& \parbox{3.5cm}{\centering 15} \\ \cline{1-1} \cline{3-6}
\centering 9
&
& \centering <5
& \centering Possible very serious toxicity or tumor underdose
& \multirow{2}{*}{\parbox{2.7cm}{\centering Very wrong dose, dose distribution, location, or volume}}
& \parbox{3.5cm}{\centering 20} \\ \cline{1-4} \cline{6-6}

\centering 10
& \centering Failures inevitable
& \centering >5
& \centering Catastrophic
&
& \parbox{3.5cm}{\centering >20} \\ \hline \hline

\end{tabular}
\caption{\small Scales and descriptions of the \(O\), \(S\), and \(D\) values used in the AAPM TG-100 FMEA.}
\label{tab:tg100_fmea}
\end{table}

\FloatBarrier
\Needspace{10\baselineskip}
\section{Notation System for a Process-Driven Risk Assessment}
\label{sec:notation_system}

Symbolically and in terms of workflow modeling (Sec.~\ref{subsec:workflow_granularity}), a PBS-PT workflow \(\mathcal{WF}\) is represented as a hierarchical process map: high-level clinical processes \(\mathcal{P}\) (Roman numerals) are decomposed into process steps (PS), and each PS carries one or more associated failure modes (FM). In the workflow maps of Appendix A, the PS labels encode both (a) a relative counter \(\kappa\) within the surrounding process and (b) an absolute index \(i\) in the full workflow, written as ``\(\kappa.\,(i)\)''; the corresponding FMs are depicted as branch twigs and marked as ``\((i.\alpha)\)'' by appending an alphabetic tag to the absolute PS index (Fig.~\ref{fig:procVII_excerpt}).\footnote{Recall that, according to our discussion about workflow granularity in Sec.~\ref{subsec:workflow_granularity}, a workflow consists of a sequence of functional high-level processes. In Appendix~A, the figures \ref{fig:PSQA_calc}, \ref{fig:PSQA_meas}, and \ref{fig:PSQA_log} illustrate the three complete PSQA-dependent workflows. Each consists of eleven processes (numbered from I to XI with PSQA as process VIII), symbolized by double-lined rectangles. On the right-hand side, the associated process steps are labeled with their absolute workflow indices (numbered from \((1)\) to \((44)\)) and modeled as branch structures whose individual twigs represent the assigned failure modes (marked by an additional alphabetic tag). Figure~\ref{fig:procVII_excerpt} reproduces the same graphical encoding for Process~VII (\textit{Plan Preparation}) as a standalone example.} \vspace{0.1cm}
\FloatBarrier

\begin{figure}[h!]
  \centering
  \includegraphics[
    page=1,
    width=\linewidth,
    clip,
    trim=9.7mm 215.7mm 38.0mm 10.1mm
  ]{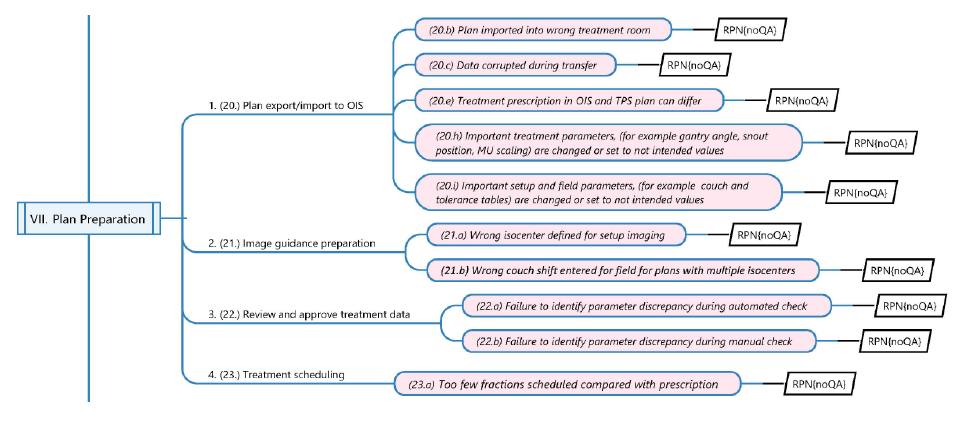}
  \caption{\small
  Standalone excerpt reproducing the graphical encoding of the PBS-PT workflow map for \textbf{Process~VII (\textit{Plan Preparation})}. Process steps are labeled by their relative index within the process and their absolute index in the overall workflow (in parentheses), while failure modes inherit the absolute PS index and add an alphabetic tag. The rightmost rhombuses indicate that each FM is initially scored by a \textit{noQA} baseline risk priority number (here: \(\mathrm{RPN}^{\mathrm{noQA}}\)).}
  \label{fig:procVII_excerpt}
\end{figure}

The entire workflow $\mathcal{W\!F}$ is represented by the set of process steps $i$:
\begin{equation}
  \mathcal{W\!F} \triangleq \{ i \mid i = 1,2,\ldots,N \}\,.
  \label{eq:workflow}
\end{equation}

For each process step \(i\) (PS \#$i$), the associated set of failure modes $\mathrm{FMI}_i$ (with $\mathrm{FMI}$ abbreviating \emph{Failure Mode Indexation}) consists of the corresponding FM indices \(l\):
\begin{equation}
  \mathrm{FMI}_i \triangleq \{(i,l) \mid l = 1,2,\ldots,L_i \} \equiv \{(i.\alpha) \mid \alpha = a,b,\ldots,\alpha_i\}\,.
  \label{eq:failure_modes}
\end{equation}

Any meaningful collection of subsequent process steps is represented by a subset $\mathcal{I}\subseteq\mathcal{W\!F}$; in particular, an isolated process corresponds to a specific subset $\mathcal{P}=\mathcal{I}$, while the complete workflow corresponds to $\mathcal{W\!F}=\mathcal{I}$ (cf. Sec.~\ref{subsec:workflow_granularity}).

In the mathematical framework developed below, the intermediate process layer is deliberately not parameterized. Instead, the formalism is written directly on the two-level index structure PS \(\to\) FM and then completed by a third aggregation layer acting on arbitrary collections of process steps. This yields a transparent hierarchy that remains fully compatible with the process map:\\

\begin{enumerate}
    \item each individual failure mode is quantified by \(\mathrm{RPN}_{i,l}\) and its stage-wise risk changes,
    \item each process step \(i\) aggregates its failure modes into local indicators,
    \item any set of process steps \(\mathcal{I}\subseteq\mathcal{WF}\) can be summarized by global indicators.
\end{enumerate}
~\\
To prevent ambiguity between descriptive figure labels and the symbols used in the equations, we adopt the \emph{absolute} process-step index \(i\) (the number in parentheses) as the primary mathematical identifier, whereas the leading within-process counter remains purely mnemonic. Likewise, the alphabetic FM tags in the figures provide human-readable labels and correspond one-to-one to the numerical FM index \(l\) used below.
~\\

Within our generic PBS-PT workflows, we use four main building blocks:\\

\begin{enumerate}[start=0,noitemsep]
    \item \textit{noQA} for the core baseline,
    \item \textit{meas} for measurement-based verification,
    \item \textit{log} for log file-based verification,
    \item \textit{calc} for independent secondary dose calculation.
\end{enumerate}
~\\
We further distinguish the two-stage sequential pFMEA process categories
\begin{equation}
  \mathrm{pc} \in \{\mathrm{data}, \mathrm{full}\}\,,
  \label{eq:process_category}
\end{equation}
and use the corresponding elementary RPN notation
\begin{subequations}
  \label{eq:RPN_pair_notation}
  \begin{equation}
    \mathrm{RPN}^{\mathrm{noQA}}_{i,\,l} = \mathrm{RPN}^{\mathrm{noQA}}_{\mathrm{PS},\mathrm{FM}},
    \label{eq:RPN_noQA_notation}
  \end{equation}
  \begin{equation}
    \mathrm{RPN}^{*,\mathrm{pc}}_{i,\,l} = \mathrm{RPN}^{\mathrm{PSQA},\mathrm{pc}}_{\mathrm{PS},\mathrm{FM}}.
    \label{eq:RPN_psqa_notation}
  \end{equation}
\end{subequations}

\Needspace{10\baselineskip}
\section{Failure-Mode Level: Mathematical Formalism for Risk Changes}
\label{sec:fm_risk_changes}

As described in Subsec.~\ref{subsec:psqa_classification}, each complete determination of a process-step-dependent, failure-specific risk change is carried out in two phases. For each PSQA method \(^{\ast}\), the risk modification under the process category \textit{Data} is first estimated, resulting in \(\mathrm{RPN}^{*,\mathrm{data}}_{i,l}\).\\

This phase is evaluated relative to the common baseline element \(\mathrm{RPN}^{\mathrm{noQA}}_{i,l}\). In other words, each individual \textit{Data}-stage risk change answers the question of what has changed in risk due to the PSQA-dependent, data-related activity compared to the core workflow risk without PSQA.\\

Assuming that the \textit{Data}-driven measures have modified the local risk conditions, the second and final scoring phase (\textit{full}) yields \(\mathrm{RPN}^{*,\mathrm{full}}_{i,l}\), which --- because it operates on an already changed risk state --- must be interpreted relative to the \textit{Data}-stage baseline \(\mathrm{RPN}^{*,\mathrm{data}}_{i,l}\).

\subsection{Absolute and Relative Risk Changes}
\label{subsec:fm_abs_rel_risk_changes}

For each failure mode, define the two primary stage pairs
\begin{subequations}
\label{eq:rpn_pair}
\begin{align}
  \omega^{*,\mathrm{data}}_{i,l} &= \Bigl[ \mathrm{RPN}^{\mathrm{noQA}}_{i,l},\, \mathrm{RPN}^{*,\mathrm{data}}_{i,l} \Bigr]\,,
  \label{eq:rpn_pair_noQA4data}
  \\
  \omega^{*,\mathrm{full}}_{i,l} &= \Bigl[ \mathrm{RPN}^{*,\mathrm{data}}_{i,l},\, \mathrm{RPN}^{*,\mathrm{full}}_{i,l} \Bigr]\,.
  \label{eq:rpn_pair_data4full}
\end{align}
\end{subequations}

The \textbf{absolute} risk changes are
\begin{subequations}
\label{eq:abs_risk_change_pair}
\begin{align}
    \Delta^{*,(d)}_{\mathrm{abs},i,l}
    &\triangleq
    \Delta_{\mathrm{abs}}\bigl[\omega^{*,\mathrm{data}}_{i,l}\bigr]
    :=
    \mathrm{RPN}^{*,\mathrm{data}}_{i,l}-\mathrm{RPN}^{\mathrm{noQA}}_{i,l}\,,
    \label{eq:abs_risk_change_noQA4data}
    \\
    \Delta^{*,(f)}_{\mathrm{abs},i,l}
    &\triangleq
    \Delta_{\mathrm{abs}}\bigl[\omega^{*,\mathrm{full}}_{i,l}\bigr]
    :=
    \mathrm{RPN}^{*,\mathrm{full}}_{i,l}-\mathrm{RPN}^{*,\mathrm{data}}_{i,l}\,,
    \label{eq:abs_risk_change_data4full}
\end{align}
\end{subequations}
and the corresponding \textbf{relative} changes are
\begin{subequations}
\label{eq:rel_risk_change_pair}
\begin{align}
    \delta^{*,(d)}_{\mathrm{rel},i,l}
    &\triangleq
    \delta_{\mathrm{rel}}\bigl[\omega^{*,\mathrm{data}}_{i,l}\bigr]
    :=
    \frac{\Delta^{*,(d)}_{\mathrm{abs},i,l}}{\mathrm{RPN}^{\mathrm{noQA}}_{i,l}}
    =
    \frac{\mathrm{RPN}^{*,\mathrm{data}}_{i,l}}{\mathrm{RPN}^{\mathrm{noQA}}_{i,l}} - 1\,,
    \label{eq:rel_risk_change_noQA4data}
    \\
    \delta^{*,(f)}_{\mathrm{rel},i,l}
    &\triangleq
    \delta_{\mathrm{rel}}\bigl[\omega^{*,\mathrm{full}}_{i,l}\bigr]
    :=
    \frac{\Delta^{*,(f)}_{\mathrm{abs},i,l}}{\mathrm{RPN}^{*,\mathrm{data}}_{i,l}}
    =
    \frac{\mathrm{RPN}^{*,\mathrm{full}}_{i,l}}{\mathrm{RPN}^{*,\mathrm{data}}_{i,l}} - 1\,,
    \label{eq:rel_risk_change_data4full}
\end{align}
\end{subequations}
where \(\delta^{*,(d)}_{\mathrm{rel},i,l}\) captures the \textit{Data}-only change relative to the common \textit{NoQA} baseline, while \(\delta^{*,(f)}_{\mathrm{rel},i,l}\) captures the incremental \textit{Full}-stage change relative to the already modified \textit{Data}-stage baseline.

\subsection{Full-stage Sign Convention}
\label{subsec:fm_full_sign_convention}

By definition,
\begin{equation*}
  \delta^{*,(f)}_{\mathrm{rel},i,l}
  =
  \frac{\mathrm{RPN}^{*,\mathrm{full}}_{i,l}}{\mathrm{RPN}^{*,\mathrm{data}}_{i,l}} - 1,
\end{equation*}
so mitigation at the \textit{Full}-stage is equivalent to
\begin{equation}
  \delta^{*,(f)}_{\mathrm{rel},i,l}\le 0
  \;\Longleftrightarrow\;
  \mathrm{RPN}^{*,\mathrm{full}}_{i,l}\le \mathrm{RPN}^{*,{data}}_{i,l}\,.
  \label{eq:condition_delta_rel_full_le_0}
\end{equation}

Hence, the condition \(\mathrm{RPN}^{*,\mathrm{full}}_{i,l}\le \mathrm{RPN}^{\mathrm{noQA}}_{i,l}\) does \emph{not} by itself imply \(\delta^{*,(f)}_{\mathrm{rel},i,l}\le 0\), because the overall \textit{NoQA-to-Full} change is governed by the cumulative relation derived below.
In practical terms, certain data-driven activities --- such as additional exports, format conversions, or manual transfers --- may increase risk, in which case
\begin{equation}
  \mathrm{RPN}^{*,\mathrm{data}}_{i,l} > \mathrm{RPN}^{\mathrm{noQA}}_{i,l}
  \quad\Longleftrightarrow\quad
  \delta^{*,(d)}_{\mathrm{rel},i,l} > 0.
  \label{eq:case_delta_rel_data_gt_0}
\end{equation}

This explains why a strongly negative \textit{Full}-stage quantity may still be required merely to compensate an unfavorable \textit{Data}-stage perturbation.
Typical examples observed in this study are discussed in Subsec.~\ref{subsec:pc_data_on_relative_RPN_scale}. For ease of interpretation, all relative quantities are primarily discussed within the interval \((-1,0]\), where \(0\%\) denotes no change and \(-100\%\) denotes complete elimination of the corresponding burden. Positive values are explicitly interpreted as risk inflation.\\

\subsection{Cumulative Absolute and Relative Changes (\textit{Full} vs.\ \textit{NoQA})}
\label{subsec:fm_cumulative_changes}
Because the relative changes in the \textit{Data} and \textit{Full} phases are defined with respect to their own baselines, the corresponding ratios are
\[
\frac{\mathrm{RPN}^{*,\mathrm{data}}_{i,l}}{\mathrm{RPN}^{\mathrm{noQA}}_{i,l}}
=
1+\delta^{*,(d)}_{\mathrm{rel},i,l}
\qquad\text{and}\qquad
\frac{\mathrm{RPN}^{*,\mathrm{full}}_{i,l}}{\mathrm{RPN}^{*,\mathrm{data}}_{i,l}}
=
1+\delta^{*,(f)}_{\mathrm{rel},i,l}.
\]

At this point it is crucial to recognize that these are both method-specific quantities. They describe what happens \emph{within} a given PSQA pathway. Since the purpose of the overall framework is to enable \emph{strict inter-PSQA comparison}, the final \textit{Full}-stage state must additionally be related back to the same common reference \(\mathrm{RPN}^{noQA}_{i,l}\).
We therefore define the cumulative pair and its associated changes:
\begin{subequations}
\label{eq:cum_risk_change_triple}
  \begin{equation}
    \omega^{*,(c)}_{i,l} = \Bigl[ \mathrm{RPN}^{\mathrm{noQA}}_{i,l},\, \mathrm{RPN}^{*,\mathrm{full}}_{i,l} \Bigr]\,,
    \label{eq:rpn_pair_noQA4full}
  \end{equation}
  \begin{equation}
      \Delta^{*,(c)}_{\mathrm{abs},i,l}
      \triangleq
      \Delta_{\mathrm{abs}}\bigl[\omega^{*,(c)}_{i,l}\bigr]
      :=
      \mathrm{RPN}^{*,\mathrm{full}}_{i,l}-\mathrm{RPN}^{\mathrm{noQA}}_{i,l}\,,
      \label{eq:cumabs_risk_change}
  \end{equation}
  \begin{equation}
      \delta^{*,(c)}_{\mathrm{rel},i,l}
      \triangleq
      \delta_{\mathrm{rel}}\bigl[\omega^{*,(c)}_{i,l}\bigr]
      :=
      \frac{\Delta^{*,(c)}_{\mathrm{abs},i,l}}{\mathrm{RPN}^{\mathrm{noQA}}_{i,l}}
      =
      \frac{\mathrm{RPN}^{*,\mathrm{full}}_{i,l}}{\mathrm{RPN}^{\mathrm{noQA}}_{i,l}}-1\,.
      \label{eq:cumrel_risk_change_data4full}
  \end{equation}
\end{subequations}

Consequently, \(\delta^{*,(c)}_{\mathrm{rel},i,l}\) factorizes as
\begin{equation}
\begin{aligned}
\delta^{*,(c)}_{\mathrm{rel},i,l}
&=
\frac{\mathrm{RPN}^{*,\mathrm{full}}_{i,l}}{\mathrm{RPN}^{\mathrm{noQA}}_{i,l}}-1 \\
&=
\bigl(1+\delta^{*,(d)}_{\mathrm{rel},i,l}\bigr)\bigl(1+\delta^{*,(f)}_{\mathrm{rel},i,l}\bigr)-1 \\
&=
\delta^{*,(d)}_{\mathrm{rel},i,l}
+\delta^{*,(f)}_{\mathrm{rel},i,l}
+\delta^{*,(d)}_{\mathrm{rel},i,l}\,\delta^{*,(f)}_{\mathrm{rel},i,l}.
\end{aligned}
\label{eq:rel_risk_change_cum_noQA4full}
\end{equation}

\paragraph{Remark on strict inter-PSQA comparability in the \textit{Cum} track.}
For between-method comparisons at the final end state (e.g.\ \textit{Meas} vs.\ \textit{Log} vs.\ \textit{Calc}), the stage-consistent incremental quantity \(\delta^{*,(f)}_{\mathrm{rel},i,l}\) is anchored to the method-dependent baseline \(\mathrm{RPN}^{*,\mathrm{data}}_{i,l}\) and therefore must be interpreted as an \emph{incremental} effect within a given method.\footnote{Concrete counterexample (same \textit{Full} outcome, different \(\delta^{*,(f)}_{\mathrm{rel},i,l}\)): fix a failure mode \((i,l)\) with \(\mathrm{RPN}^{\mathrm{noQA}}_{i,l}=100\) and suppose two PSQA methods \(A\) and \(B\) yield the same \textit{Full}-track outcome \(\mathrm{RPN}^{A,\mathrm{full}}_{i,l}=\mathrm{RPN}^{B,\mathrm{full}}_{i,l}=25\), but differ at the \textit{Data}-stage, \(\mathrm{RPN}^{A,\mathrm{data}}_{i,l}=50\) and \(\mathrm{RPN}^{B,\mathrm{data}}_{i,l}=80\). Then \(\delta^{A,(f)}_{\mathrm{rel},i,l}=25/50-1=-0.50\), whereas \(\delta^{B,(f)}_{\mathrm{rel},i,l}=25/80-1\approx -0.6875\). A ranking based on \(\delta^{*,(f)}_{\mathrm{rel},i,l}\) would therefore incorrectly declare method \(B\) ``better'' at \textit{Full}, despite identical final outcomes. In contrast, the cumulative changes coincide: \(\delta^{A,(c)}_{\mathrm{rel},i,l}=\delta^{B,(c)}_{\mathrm{rel},i,l}=25/100-1=-0.75\).} In contrast, \(\delta^{*,(c)}_{\mathrm{rel},i,l}\) is anchored to the method-independent NoQA baseline and is therefore the natural FM-level basis for strict inter-PSQA comparison of the net \textit{Full}-track effect.

\subsection{Derived Data-to-Cumulative Bridge Quantity}
\label{subsec:fm_bridge_quantity}

To compare the additional \textit{Full}-stage contribution on the same \textit{NoQA}-anchored scale used by the \textit{Data} and cumulative quantities, define the bridge quantity
\begin{equation}
\delta^{*,(d\to c)}_{\mathrm{rel},i,l}
:=
\delta^{*,(c)}_{\mathrm{rel},i,l}
-
\delta^{*,(d)}_{\mathrm{rel},i,l}.
\label{eq:bridge_rel_change_fm_def}
\end{equation}

Using Eq.~\eqref{eq:rel_risk_change_cum_noQA4full}, this becomes
\begin{equation}
\delta^{*,(d\to c)}_{\mathrm{rel},i,l}
=
\delta^{*,(f)}_{\mathrm{rel},i,l}\bigl(1+\delta^{*,(d)}_{\mathrm{rel},i,l}\bigr),
\label{eq:bridge_rel_change_fm_product}
\end{equation}
and directly in RPN form,
\begin{equation}
\delta^{*,(d\to c)}_{\mathrm{rel},i,l}
=
\frac{\mathrm{RPN}^{*,\mathrm{full}}_{i,l}-\mathrm{RPN}^{*,\mathrm{data}}_{i,l}}{\mathrm{RPN}^{\mathrm{noQA}}_{i,l}}.
\label{eq:bridge_rel_change_fm_rpn}
\end{equation}

This bridge quantity is not a new scoring stage. Rather, it is a derived comparison layer that measures how much additional relative shift occurs when moving from the \textit{Data}-stage point to the final cumulative point, but on the same \textit{NoQA}-anchored scale already used by \textit{Data} and \textit{Cum}. It is therefore the natural quantity for \textit{Data-to-Cum} dumbbell plots and bridge-shift histograms.

\Needspace{10\baselineskip}
\section{Process-Step Level: Mathematical Formalism for Local Risk Metrics}
\label{sec:local_risk_metrics}

Aggregating the FM-level risk changes over all failure modes of a process step \(i\) defines local (step-level) risk metrics. These metrics provide bounded and numerically stable summaries that can be rolled up to global, workflow-level indicators without semantic distortion (see Sec.~\ref{sec:global_risk_metrics}).

\subsection{Local Absolute Risk Metrics}
\label{sub:local_abs_risk_metric}

The aggregation of the three absolute risk changes in Eqs.~\eqref{eq:abs_risk_change_pair} and \eqref{eq:cumabs_risk_change} over all failure modes of process step \(i\) defines the local absolute metrics
\begin{subequations}
\label{eq:local_abs_metric_triple}
  \begin{equation}
    \rho^{*,(d)}_{\mathrm{abs},i}
    :=
    \sum_{(i,l)\in \mathrm{FMI}_i}\Delta^{*,(d)}_{\mathrm{abs},i,l}\,,
    \label{eq:local_abs_metric_data}
  \end{equation}
  \begin{equation}
    \rho^{*,(f)}_{\mathrm{abs},i}
    :=
    \sum_{(i,l)\in \mathrm{FMI}_i}\Delta^{*,(f)}_{\mathrm{abs},i,l}\,,
    \label{eq:local_abs_metric_full}
  \end{equation}
  \begin{equation}
    \rho^{*,(c)}_{\mathrm{abs},i}
    :=
    \sum_{(i,l)\in \mathrm{FMI}_i}\Delta^{*,(c)}_{\mathrm{abs},i,l}\,.
    \label{eq:local_abs_metric_cum}
  \end{equation}
\end{subequations}

\subsection{Local Relative Risk Metrics}
\label{subsec:local_rel_risk_metric}

Suitable local relative indicators should:\\

\begin{enumerate}[label=(\roman*)]
  \item scale each FM contribution proportionally to its baseline relevance,
  \item remain bounded and interpretable,
  \item avoid pathologies due to vanishing denominators,
  \item preserve hierarchical additivity for later workflow-level aggregation.
\end{enumerate}
~\\
Define the stage-specific total step burdens
\begin{equation}
  G^{(d)}_i := \sum_{(i,l)\in \mathrm{FMI}_i} \mathrm{RPN}^{\mathrm{noQA}}_{i,l},
  \qquad
  G^{*,(f)}_i := \sum_{(i,l)\in \mathrm{FMI}_i} \mathrm{RPN}^{*,\mathrm{data}}_{i,l}.
  \label{eq:local_total_step_burdens}
\end{equation}
Here \(G^{*,(f)}_i\) should be read as the \emph{Data-stage burden underlying the Full-stage normalization}; the superscript indicates the stage being normalized, not the numerical stage from which the burden is taken.

The corresponding FM weights are
\begin{subequations}
\label{eq:local_weight_triple}
\begin{align}
g^{(d)}_{i,l}
&=
\frac{\mathrm{RPN}^{\mathrm{noQA}}_{i,l}}{G^{(d)}_i},
\label{eq:local_weight_data}
\\[0.3em]
g^{*,(f)}_{i,l}
&=
\frac{\mathrm{RPN}^{*,\mathrm{data}}_{i,l}}{G^{*,(f)}_i},
\label{eq:local_weight_full}
\\[0.3em]
g^{(c)}_{i,l}
&=
\frac{\mathrm{RPN}^{\mathrm{noQA}}_{i,l}}{G^{(d)}_i}.
\label{eq:local_weight_cum}
\end{align}
\end{subequations}
Hence \(g^{(d)}_{i,l}=g^{(c)}_{i,l}\) for all methods, because both \textit{Data} and \textit{Cum} are anchored to the same \textit{NoQA} baseline.

The local weighted relative metrics are
\begin{subequations}
\label{eq:local_rel_metric_weighted_triple}
  \begin{align}
    \rho^{*,(d)}_{\mathrm{rel},i}
    &=
    \sum_{(i,l)\in \mathrm{FMI}_i} g^{(d)}_{i,l}\,\delta^{*,(d)}_{\mathrm{rel},i,l},
    \label{eq:local_rel_metric_weighted_data}
    \\[0.3em]
    \rho^{*,(f)}_{\mathrm{rel},i}
    &=
    \sum_{(i,l)\in \mathrm{FMI}_i} g^{*,(f)}_{i,l}\,\delta^{*,(f)}_{\mathrm{rel},i,l},
    \label{eq:local_rel_metric_weighted_full}
    \\[0.3em]
    \rho^{*,(c)}_{\mathrm{rel},i}
    &=
    \sum_{(i,l)\in \mathrm{FMI}_i} g^{(d)}_{i,l}\,\delta^{*,(c)}_{\mathrm{rel},i,l}.
    \label{eq:local_rel_metric_weighted_cum}
  \end{align}
\end{subequations}

\subsection{Closed-form Relation between Local Absolute and Local Relative Process Step Metrics}
\label{subsubsec:local_abs_rel_relation}

Substituting the FM-level relative-change definitions into the weighted local metrics of Eq.~\eqref{eq:local_rel_metric_weighted_triple}, using the weights in Eq.~\eqref{eq:local_weight_triple}, the local burdens in Eq.~\eqref{eq:local_total_step_burdens}, and the absolute local sums in Eq.~\eqref{eq:local_abs_metric_triple}, yields
\begin{subequations}
\label{eq:rho_rel_in_terms_of_rho_abs}
\begin{align}
  \rho^{*,(d)}_{\mathrm{rel},i}
  &=
  \frac{\rho^{*,(d)}_{\mathrm{abs},i}}{G^{(d)}_i},
  \label{eq:rho_rel_vs_abs_data}
  \\[0.3em]
  \rho^{*,(f)}_{\mathrm{rel},i}
  &=
  \frac{\rho^{*,(f)}_{\mathrm{abs},i}}{G^{*,(f)}_i},
  \label{eq:rho_rel_vs_abs_full}
  \\[0.3em]
  \rho^{*,(c)}_{\mathrm{rel},i}
  &=
  \frac{\rho^{*,(c)}_{\mathrm{abs},i}}{G^{(d)}_i}.
  \label{eq:rho_rel_vs_abs_cum}
\end{align}
\end{subequations}

Since all stage-specific burdens are strictly positive under the scoring scheme, Eqs.~\eqref{eq:rho_rel_vs_abs_data}--\eqref{eq:rho_rel_vs_abs_cum} can be rearranged to recover the corresponding absolute metrics:
\begin{equation}
  \rho^{*,(d)}_{\mathrm{abs},i}=G^{(d)}_i\,\rho^{*,(d)}_{\mathrm{rel},i},\qquad
  \rho^{*,(f)}_{\mathrm{abs},i}=G^{*,(f)}_i\,\rho^{*,(f)}_{\mathrm{rel},i},\qquad
  \rho^{*,(c)}_{\mathrm{abs},i}=G^{(d)}_i\,\rho^{*,(c)}_{\mathrm{rel},i}.
  \label{eq:rho_abs_from_rho_rel}
\end{equation}

Moreover,
\begin{equation}
1+\rho^{*,(c)}_{\mathrm{rel},i}
=
\bigl(1+\rho^{*,(d)}_{\mathrm{rel},i}\bigr)\bigl(1+\rho^{*,(f)}_{\mathrm{rel},i}\bigr),
\label{eq:local_rel_product}
\end{equation}
and thus
\begin{equation}
\rho^{*,(c)}_{\mathrm{rel},i}
=
\rho^{*,(d)}_{\mathrm{rel},i}
+
\rho^{*,(f)}_{\mathrm{rel},i}
+
\rho^{*,(d)}_{\mathrm{rel},i}\,\rho^{*,(f)}_{\mathrm{rel},i}.
\label{eq:local_rel_change_cum_factorization}
\end{equation}

\subsection{Local Data-to-Cumulative Bridge Metrics}
\label{subsec:local_bridge_metrics}

In complete analogy to the FM-level bridge quantity, define
\begin{equation}
\rho^{*,(d\to c)}_{\mathrm{rel},i}
:=
\rho^{*,(c)}_{\mathrm{rel},i}
-
\rho^{*,(d)}_{\mathrm{rel},i}.
\label{eq:local_bridge_def}
\end{equation}

Using Eq.~\eqref{eq:local_rel_change_cum_factorization}, one obtains
\begin{equation}
\rho^{*,(d\to c)}_{\mathrm{rel},i}
=
\rho^{*,(f)}_{\mathrm{rel},i}\bigl(1+\rho^{*,(d)}_{\mathrm{rel},i}\bigr),
\label{eq:local_bridge_product}
\end{equation}
and with Eqs.~\eqref{eq:rho_rel_vs_abs_full} and \eqref{eq:local_total_step_burdens},
\begin{equation}
\rho^{*,(d\to c)}_{\mathrm{rel},i}
=
\frac{\rho^{*,(f)}_{\mathrm{abs},i}}{G^{(d)}_i}.
\label{eq:local_bridge_abs_closed}
\end{equation}

Thus, the local bridge metric is the \textit{NoQA}-anchored process-step-level counterpart of the FM bridge quantity. It answers a different question from \(\rho^{*,(f)}_{\mathrm{rel},i}\):
\begin{itemize}
  \item \(\rho^{*,(f)}_{\mathrm{rel},i}\) measures the \textit{Full}-stage increment relative to the \textit{Data}-stage burden,
  \item \(\rho^{*,(d\to c)}_{\mathrm{rel},i}\) measures the same increment on the common \textit{NoQA} scale.
\end{itemize}

This distinction is especially useful for process steps with \(L_i>1\), where heterogeneous FM-level bridge shifts are compressed into a single weighted local summary.

\subsection{Meaning and Implications}
\label{subsec:local_meaning}

Eqs.~\eqref{eq:rho_rel_vs_abs_data}--\eqref{eq:rho_rel_vs_abs_cum} and \eqref{eq:rho_abs_from_rho_rel} show that the selected risk-weighted local relatives are not arbitrary averages of FM-level relatives; rather, they are exactly the fractional changes of the total step-level burden relative to the stage-appropriate baseline. In particular,
\begin{subequations}
\label{eq:ratio_of_totals_local_relatives}
\begin{align}
  1+\rho^{*,(d)}_{\mathrm{rel},i}
  &=
  \frac{\sum_{(i,l)\in \mathrm{FMI}_i}\mathrm{RPN}^{*,\mathrm{data}}_{i,l}}
       {\sum_{(i,l)\in \mathrm{FMI}_i}\mathrm{RPN}^{\mathrm{noQA}}_{i,l}},
  \label{eq:ratio_of_totals_local_rel_data}
  \\[0.3em]
  1+\rho^{*,(f)}_{\mathrm{rel},i}
  &=
  \frac{\sum_{(i,l)\in \mathrm{FMI}_i}\mathrm{RPN}^{*,\mathrm{full}}_{i,l}}
       {\sum_{(i,l)\in \mathrm{FMI}_i}\mathrm{RPN}^{*,\mathrm{data}}_{i,l}},
  \label{eq:ratio_of_totals_local_rel_full}
  \\[0.3em]
  1+\rho^{*,(c)}_{\mathrm{rel},i}
  &=
  \frac{\sum_{(i,l)\in \mathrm{FMI}_i}\mathrm{RPN}^{*,\mathrm{full}}_{i,l}}
       {\sum_{(i,l)\in \mathrm{FMI}_i}\mathrm{RPN}^{\mathrm{noQA}}_{i,l}}.
  \label{eq:ratio_of_totals_local_rel_cum}
\end{align}
\end{subequations}

Consequently:\\

\begin{itemize}[label=$\rightarrow$,leftmargin=8.4mm]
  \item \(\rho^{*,(d)}_{\mathrm{rel},i}\) is the \textit{NoQA}-anchored relative change induced by the \textit{Data}-stage,
  \item \(\rho^{*,(f)}_{\mathrm{rel},i}\) is the stage-consistent incremental change from \textit{Data} to \textit{Full},
  \item \(\rho^{*,(c)}_{\mathrm{rel},i}\) is the cumulative \textit{NoQA}-anchored relative change from \textit{NoQA} to \textit{Full},
  \item \(\rho^{*,(d\to c)}_{\mathrm{rel},i}\) is the additional \textit{NoQA}-anchored bridge shift needed to move from the \textit{Data}-stage point to the cumulative end state.
\end{itemize}

\section{Process and Workflow-Level: Mathematical Formalism for Global Risk Metrics}
\label{sec:global_risk_metrics}

Having defined stage-consistent, risk-weighted local indicators, we now introduce workflow-level aggregates as global indicators. A global indicator compresses the collection of step-level results for a chosen part of the workflow into a single system-level quantity that can be compared across stages and --- when the baseline anchoring is appropriate --- across PSQA methods.

\subsection{Set-theoretical Representation}
\label{subsec:set_representation}
\paragraph{Process-step collections.}
In general, let
\begin{equation}
  \mathcal{I}\subseteq\mathcal{WF}
  \label{eq:I_as_process_subset_of_WF}
\end{equation}
denote any non-empty subset of subsequent process steps that forms a continuous process step-collection. If \(\mathcal{I}\) builds an isolated process, it may be referred to as \(\mathcal{P}\); in its fully extended form, \(\mathcal{I}\) can coincide with the entire workflow.

\paragraph{The three PSQA workflows.}
As the set-theoretical risk equivalent of the three PSQA workflows, define
\begin{subequations}
  \label{eq:omega_sets}
  \begin{align}
    \Omega^{meas}
    &= \bigl\{\,\omega^{\mathrm{meas},\mathrm{pc}}_{i,l}\mid i\in\mathcal{WF}^{\mathrm{meas}},\ l\in \mathrm{FMI}_i,\ \mathrm{pc}\in\{\mathrm{data},\mathrm{full}\}\bigr\},
    \label{eq:omega_meas}\\
    \Omega^{log}
    &= \bigl\{\,\omega^{\mathrm{log},\mathrm{pc}}_{i,l}\mid i\in\mathcal{WF}^{\mathrm{log}},\ l\in \mathrm{FMI}_i,\ \mathrm{pc}\in\{\mathrm{data},\mathrm{full}\}\bigr\},
    \label{eq:omega_log}\\
    \Omega^{calc}
    &= \bigl\{\,\omega^{\mathrm{calc},\mathrm{pc}}_{i,l}\mid i\in\mathcal{WF}^{\mathrm{calc}},\ l\in \mathrm{FMI}_i,\ \mathrm{pc}\in\{\mathrm{data},\mathrm{full}\}\bigr\}.
    \label{eq:omega_calc}
  \end{align}
\end{subequations}
The cumulative \((c)\) quantities are derived \textit{NoQA-to-Full} comparison layers, not additional scoring phases.

\paragraph{Process-restricted analysis sets.}
For any isolated process \(\mathcal{P}\subseteq\mathcal{WF}\), define
\begin{equation}
  \Omega^{*,\mathrm{pc}}_{\mathcal{P}}
  =
  \bigl\{\, \omega^{*,\mathrm{pc}}_{i,l}\mid i\in\mathcal{P},\ l\in \mathrm{FMI}_i \,\bigr\},
  \qquad
  \left|\Omega^{*,\mathrm{pc}}_{\mathcal{P}}\right|
  =
  \sum_{i\in\mathcal{P}} L_i.
  \label{eq:Omega_P_pc}
\end{equation}

\subsection{Global Absolute Risk Metrics}
\label{subsec:global_abs_metrics}

For each PSQA method \(^{\ast}\in\{\mathrm{meas},\mathrm{log},\mathrm{calc}\}\) and any non-empty comparison set \(\mathcal{I}\subseteq\mathcal{WF}\), define
\begin{subequations}
  \label{eq:global_abs_metric_triple}
  \begin{align}
    \Rho^{*,(d)}_{\mathrm{abs},\mathcal{I}}
    &:=
    \sum_{i\in\mathcal{I}}\rho^{*,(d)}_{\mathrm{abs},i},
    \label{eq:global_abs_metric_data}\\
    \Rho^{*,(f)}_{\mathrm{abs},\mathcal{I}}
    &:=
    \sum_{i\in\mathcal{I}}\rho^{*,(f)}_{\mathrm{abs},i},
    \label{eq:global_abs_metric_full}\\
    \Rho^{*,(c)}_{\mathrm{abs},\mathcal{I}}
    &:=
    \sum_{i\in\mathcal{I}}\rho^{*,(c)}_{\mathrm{abs},i}.
    \label{eq:global_abs_metric_cum}
  \end{align}
\end{subequations}
Hence,
\begin{equation}
\Rho^{*,(c)}_{\mathrm{abs},\mathcal{I}}
=
\Rho^{*,(d)}_{\mathrm{abs},\mathcal{I}}
+
\Rho^{*,(f)}_{\mathrm{abs},\mathcal{I}}.
\label{eq:global_abs_telescoping}
\end{equation}

\subsection{Global Relative Risk Metrics}
\label{subsec:global_rel_metrics}

Define the stage-specific step burdens
\begin{equation}
  G^{(d)}_i := \sum_{(i,l)\in \mathrm{FMI}_i} \mathrm{RPN}^{\mathrm{noQA}}_{i,l},
  \qquad
  G^{*,(f)}_i := \sum_{(i,l)\in \mathrm{FMI}_i} \mathrm{RPN}^{*,\mathrm{data}}_{i,l},
  \label{eq:global_step_burdens_recall}
\end{equation}
and the induced normalized step weights
\begin{equation}
  \gamma^{(d)}_{\mathcal{I},i}
  :=
  \frac{G^{(d)}_i}{\sum_{j\in\mathcal{I}} G^{(d)}_j},
  \qquad
  \gamma^{*,(f)}_{\mathcal{I},i}
  :=
  \frac{G^{*,(f)}_i}{\sum_{j\in\mathcal{I}} G^{*,(f)}_j},
  \qquad i\in\mathcal{I}.
  \label{eq:global_rel_metric_weight_coeffs}
\end{equation}

Then the corresponding global relative metrics are
\begin{subequations}
  \label{eq:global_rel_metric_weighted_triple}
  \begin{align}
    \Rho^{*,(d)}_{\mathrm{rel},\mathcal{I}}
    &:=
    \sum_{i\in\mathcal{I}} \gamma^{(d)}_{\mathcal{I},i}\,\rho^{*,(d)}_{\mathrm{rel},i},
    \label{eq:global_rel_metric_weighted_data}\\
    \Rho^{*,(f)}_{\mathrm{rel},\mathcal{I}}
    &:=
    \sum_{i\in\mathcal{I}} \gamma^{*,(f)}_{\mathcal{I},i}\,\rho^{*,(f)}_{\mathrm{rel},i},
    \label{eq:global_rel_metric_weighted_full}\\
    \Rho^{*,(c)}_{\mathrm{rel},\mathcal{I}}
    &:=
    \sum_{i\in\mathcal{I}} \gamma^{(d)}_{\mathcal{I},i}\,\rho^{*,(c)}_{\mathrm{rel},i}.
    \label{eq:global_rel_metric_weighted_cum}
  \end{align}
\end{subequations}

Thus, \(\Rho^{*,(c)}_{\mathrm{rel},\mathcal{I}}\) remains explicitly anchored to the common \textit{NoQA} burden and is therefore the natural basis for strict inter-method comparison of the final cumulative effect.

\subsection{Closed-form Relation between Global Absolute and Global Relative Metrics}
\label{subsec:global_abs_rel_relation}

Define the total stage-specific burdens on \(\mathcal{I}\):
\begin{equation}
  G^{(d)}_{\mathcal{I}}
  :=
  \sum_{i\in\mathcal{I}} G^{(d)}_i
  =
  \sum_{i\in\mathcal{I}}\sum_{(i,l)\in \mathrm{FMI}_i} \mathrm{RPN}^{\mathrm{noQA}}_{i,l},
  \qquad
  G^{*,(f)}_{\mathcal{I}}
  :=
  \sum_{i\in\mathcal{I}} G^{*,(f)}_i
  =
  \sum_{i\in\mathcal{I}}\sum_{(i,l)\in \mathrm{FMI}_i} \mathrm{RPN}^{*,\mathrm{data}}_{i,l}.
  \label{eq:global_total_burdens}
\end{equation}

Substituting the local closed-form relations into the global weighted definitions yields
\begin{subequations}
\label{eq:Rho_rel_in_terms_of_Rho_abs}
\begin{align}
  \Rho^{*,(d)}_{\mathrm{rel},\mathcal{I}}
  &=
  \frac{\Rho^{*,(d)}_{\mathrm{abs},\mathcal{I}}}{G^{(d)}_{\mathcal{I}}},
  \label{eq:Rho_rel_vs_abs_data}
  \\[0.3em]
  \Rho^{*,(f)}_{\mathrm{rel},\mathcal{I}}
  &=
  \frac{\Rho^{*,(f)}_{\mathrm{abs},\mathcal{I}}}{G^{*,(f)}_{\mathcal{I}}},
  \label{eq:Rho_rel_vs_abs_full}
  \\[0.3em]
  \Rho^{*,(c)}_{\mathrm{rel},\mathcal{I}}
  &=
  \frac{\Rho^{*,(c)}_{\mathrm{abs},\mathcal{I}}}{G^{(d)}_{\mathcal{I}}}.
  \label{eq:Rho_rel_vs_abs_cum}
\end{align}
\end{subequations}

Equivalently,
\begin{equation}
  \label{appB:eq:Rho_abs_from_Rho_rel}
  \Rho^{*,(d)}_{\mathrm{abs},\mathcal{I}}=G^{(d)}_{\mathcal{I}}\,\Rho^{*,(d)}_{\mathrm{rel},\mathcal{I}},\qquad
  \Rho^{*,(f)}_{\mathrm{abs},\mathcal{I}}=G^{*,(f)}_{\mathcal{I}}\,\Rho^{*,(f)}_{\mathrm{rel},\mathcal{I}},\qquad
  \Rho^{*,(c)}_{\mathrm{abs},\mathcal{I}}=G^{(d)}_{\mathcal{I}}\,\Rho^{*,(c)}_{\mathrm{rel},\mathcal{I}}.
\end{equation}

Moreover,
\begin{equation}
1+\Rho^{*,(c)}_{\mathrm{rel},\mathcal{I}}
=
\bigl(1+\Rho^{*,(d)}_{\mathrm{rel},\mathcal{I}}\bigr)\bigl(1+\Rho^{*,(f)}_{\mathrm{rel},\mathcal{I}}\bigr),
\label{eq:global_telescoping_cum}
\end{equation}
hence
\begin{equation}
\Rho^{*,(c)}_{\mathrm{rel},\mathcal{I}}
=
\Rho^{*,(d)}_{\mathrm{rel},\mathcal{I}}
+
\Rho^{*,(f)}_{\mathrm{rel},\mathcal{I}}
+
\Rho^{*,(d)}_{\mathrm{rel},\mathcal{I}}\Rho^{*,(f)}_{\mathrm{rel},\mathcal{I}}.
\label{eq:global_rel_change_cum_factorization}
\end{equation}

\subsection{Global \textit{Data-to-Cum} Bridge Metrics}
\label{subsec:global_bridge_metrics}

Define the workflow-level bridge metric by
\begin{equation}
\Rho^{*,(d\to c)}_{\mathrm{rel},\mathcal{I}}
:=
\Rho^{*,(c)}_{\mathrm{rel},\mathcal{I}}
-
\Rho^{*,(d)}_{\mathrm{rel},\mathcal{I}}.
\label{eq:global_bridge_def}
\end{equation}

Using Eq.~\eqref{eq:global_rel_change_cum_factorization}, this becomes
\begin{equation}
\Rho^{*,(d\to c)}_{\mathrm{rel},\mathcal{I}}
=
\Rho^{*,(f)}_{\mathrm{rel},\mathcal{I}}
\bigl(1+\Rho^{*,(d)}_{\mathrm{rel},\mathcal{I}}\bigr),
\label{eq:global_bridge_product}
\end{equation}
and equivalently,
\begin{equation}
\Rho^{*,(d\to c)}_{\mathrm{rel},\mathcal{I}}
=
\frac{\Rho^{*,(f)}_{\mathrm{abs},\mathcal{I}}}{G^{(d)}_{\mathcal{I}}}.
\label{eq:global_bridge_abs_closed}
\end{equation}

Thus, the global bridge metric is the workflow-level analogue of the FM- and process-step-level bridge quantities. It measures the additional \textit{Full}-stage contribution on the same \textit{NoQA}-anchored scale used by the \textit{Data} and cumulative quantities, and is therefore especially useful whenever the bridge from \textit{Data} to \textit{Cum} is to be compared across methods or workflow segments.

\subsection{Interpretation and Conclusion}
\label{subsec:global_meaning}

Eqs.~\eqref{eq:Rho_rel_in_terms_of_Rho_abs}--\eqref{eq:Rho_abs_from_Rho_rel} show that the global relatives are not arbitrary process-step averages: they correspond exactly to the fractional change of the total burden in \(\mathcal{I}\) relative to the corresponding stage-specific baseline.

In particular,
\begin{subequations}
  \label{eq:ratios_of_totals_for_global_relatives}
  \begin{equation}
    1+\Rho^{*,(d)}_{\mathrm{rel},\mathcal{I}}
    =
    \frac{\sum_{i\in\mathcal{I}}\sum_{(i,l)\in \mathrm{FMI}_i}\mathrm{RPN}^{*,\mathrm{data}}_{i,l}}
         {\sum_{i\in\mathcal{I}}\sum_{(i,l)\in \mathrm{FMI}_i}\mathrm{RPN}^{\mathrm{noQA}}_{i,l}}\,,
    \label{eq:ratio_of_totals_for_global_rel_data}
  \end{equation}
  \begin{equation}
    1+\Rho^{*,(f)}_{\mathrm{rel},\mathcal{I}}
    =
    \frac{\sum_{i\in\mathcal{I}}\sum_{(i,l)\in \mathrm{FMI}_i}\mathrm{RPN}^{*,\mathrm{full}}_{i,l}}
         {\sum_{i\in\mathcal{I}}\sum_{(i,l)\in \mathrm{FMI}_i}\mathrm{RPN}^{*,\mathrm{data}}_{i,l}}\,,
    \label{eq:ratio_of_totals_for_global_rel_full}
  \end{equation}
  \begin{equation}
    1+\Rho^{*,(c)}_{\mathrm{rel},\mathcal{I}}
    =
    \frac{\sum_{i\in\mathcal{I}}\sum_{(i,l)\in \mathrm{FMI}_i}\mathrm{RPN}^{*,\mathrm{full}}_{i,l}}
         {\sum_{i\in\mathcal{I}}\sum_{(i,l)\in \mathrm{FMI}_i}\mathrm{RPN}^{\mathrm{noQA}}_{i,l}}\,.
    \label{eq:ratio_of_totals_for_global_rel_cum}
  \end{equation}
\end{subequations}

Accordingly:\\

\begin{itemize}[label=$\rightarrow$,leftmargin=8.4mm]
  \item \(\Rho^{*,(d)}_{\mathrm{rel},\mathcal{I}}\) is the \textit{NoQA}-anchored relative workflow shift induced by the \textit{Data}-stage,
  \item \(\Rho^{*,(f)}_{\mathrm{rel},\mathcal{I}}\) is the stage-consistent \textit{Full}-stage increment relative to the \textit{Data}-stage burden,
  \item \(\Rho^{*,(c)}_{\mathrm{rel},\mathcal{I}}\) is the cumulative \textit{NoQA}-anchored final workflow shift,
  \item \(\Rho^{*,(d\to c)}_{\mathrm{rel},\mathcal{I}}\) is the additional \textit{NoQA}-anchored bridge shift required to move from \textit{Data} to \textit{Cum}.
\end{itemize}
~\\
As at the lower levels, only \(\Rho^{*,(c)}_{\mathrm{rel},\mathcal{I}}\) is the strict basis for inter-PSQA comparison of final net outcomes. The classical \textit{Full}-stage quantity \(\Rho^{*,(f)}_{\mathrm{rel},\mathcal{I}}\) should be retained because it measures the stage-consistent marginal effect relative to the \textit{Data}-stage burden. The newly introduced bridge quantity \(\Rho^{*,(d\to c)}_{\mathrm{rel},\mathcal{I}}\) complements this by recasting the same incremental \textit{Full}-stage contribution on the common \textit{NoQA} scale already used by the \textit{Data} and cumulative metrics.\\

This completes the hierarchical bridge structure of the formalism:
\[
\delta^{*,(d\to c)}_{\mathrm{rel},i,l}
\quad\longrightarrow\quad
\rho^{*,(d\to c)}_{\mathrm{rel},i}
\quad\longrightarrow\quad
\Rho^{*,(d\to c)}_{\mathrm{rel},\mathcal{I}},
\]
thereby making the \textit{Data-to-Cum} bridge logic available consistently at the failure-mode, process-step, and workflow levels. 

  \EndTextAppendixLayout
\endgroup

\begingroup
  \BeginTextAppendixLayout

  \SetAppendixHyperPrefix{appC}
  \SetAppendixNumbering{C}

  \renewcommand{\sectionmark}[1]{}
  \markright{}

  \ManualAppendixChapter{C}{CRITICAL ANALYSIS OF METHODOLOGICAL IMPLICATIONS}{ch:methodological_implications}

  \BeginAppendixLabelPrefix{appC}

\section{Background}
\label{sec:critical_analysis}

Risk analyses in radiotherapy are increasingly institutionalized by regulatory and professional guidance and are commonly  operationalized using process‐based FMEA (pFMEA) with RPN scoring. The central metho\-dological tension is well known: the RPN is easy to compute and communicate, yet it is mathematically problematic because it multiplies ordinal scores and invites overinterpretation as a quantitative risk measure.\\

In this document, a new mathematical framework is built on a structured, generic PBS proton therapy workflow with a baseline condition without PSQA (\textit{noQA}) and staged PSQA transformations ($\mathrm{data},\, \mathrm{full},\,\mathrm{cum}$). The framework defines absolute and relative risk changes at the failure‐mode level and aggregates these into local (process‐step) and global (workflow‐level) metrics using risk‐weighted normalization. Because these operations use ratios and weighted averages of RPN‐based quantities, the question arises whether the framework remains valid in light of published criticisms \cite{BUCHGEISTER2021343} of the RPN approach to risk assessment.

\section{Limitations of Risk Analysis Using an RPN-Based pFMEA}
\label{sec:RPN_limitations}
Several weaknesses of RPN-based pFMEA have been identified, which could cast doubt on the validity of the analysis methods used in this document. These weaknesses and concerns will be briefly described, and their bearing on the methodology will be evaluated.

\subsection{Arithmetic of Ordinal Scales}
\label{subsec:ordinal_arithmetic}
A foundational critique of the RPN score generally is in its use of a multiplication operation on three ordinally-scaled ratings.  Fundamentally, if the numbers do not correspond to well-defined intervals, multiplication and ratio statements (“twice the risk”) lack formal validity.  Consequently,
computing means across raters is not justified.

\subsection{Non-Uniqueness and Distorted Prioritization}
\label{subsec:non-uniqueness}
A single RPN value can arise from many $(S, O, D)$ combinations. Multiplicative aggregation of these values can over-weight intermediate ratings, such that a larger RPN does not necessarily imply a “higher risk” in a  decision‐relevant sense, particularly when severity is high but occurrence is low (or vice versa).  This undermines the RPN’s name: it is not reliably suited for prioritization.

\subsection{Uncertainty Amplification and Poor Reproducibility}
\label{subsec:uncertainty_amplification}
The subjectivity of the pFMEA approach leads to RPN variability even among experienced assessors, with one study reporting an order of magnitude RPN score variability \cite{BUCHGEISTER2021343}.  This underlying uncertainty in $S$, $O$, and $D$ ratings compounds under multiplication, threatening the reproducibility and stability of rankings.

\subsection{Ambiguity Between Occurrence and Detectability}
\label{subsec:OD_ambiguity}
There can be some practical difficulty in separating “occurrence” and “detectability” ($O$ vs $D$), due to questions about when an error is considered to be "detected in time".  This problem can lead to inconsistent scoring logic across institutions and within teams.

\subsection{Acceptability Thresholds and Inter-Institutional Comparability}
\label{subsec:acceptability_comparability}
Because different risk profiles can share the same RPN, and because RPN does not map to an absolute risk probability, it is problematic to determine fixed acceptability thresholds for RPN that can be compared between institutions.

\Needspace{8\baselineskip}
\section{Addressing RPN Limitations in the Current Formalism}
\label{sec:addressing_RPN_limitations}
The formalism developed here retains the TG-100-style multiplicative RPN for pragmatic continuity, but it does so under an explicitly restricted interpretation. This positioning is consistent with a substantial body of FMEA literature showing both the usefulness and the limitations of classical RPN-based prioritization, including well-known concerns about non-unique rankings, distorted prioritization, and the proliferation of alternative prioritization schemes~\cite{chang2009generalrpn,ciani2019criticalrpn,liu2013fmea_review}.\\

The framework therefore does not assume that multiplication, averaging, or ratio formation on ordinal scores is fully valid in a strict measurement-theoretic sense. Instead, it treats the RPN as a semi-quantitative surrogate score and uses it to construct a comparative, intervention-oriented description of how PSQA modifies a common workflow baseline.\\

A central methodological departure from conventional one-pass pFMEA is the introduction of a baseline workflow state (\textit{noQA}) together with staged PSQA transformations (\textit{Data}, \textit{Full}, and \textit{Cum}) applied to the same generic process model. This recasts the analysis from a static ranking exercise into a structured intervention comparison. Within that setting, the burden-weighted local and global metrics derived in Appendix~B are especially important because they recast relative changes as normalized changes in total baseline burden rather than as unstable averages that can be dominated by low-burden failure modes.\\

This narrowed interpretation is also consistent with probabilistic and generalized reformulations of the RPN. Probabilistic priority numbers were introduced by Sant'Anna, and Kim and Zuo proposed a more general mathematical framework for the RPN itself~\cite{parrachosantanna2012ppn,kimzuo2018general}. Recent work by Mahmoudvand et al. extends this line further by modeling the distribution of RPN values from ordered categorical inputs in a Bayesian-multinomial framework~\cite{mahmoudvand2025rpn}. The implication for the present study is not that the RPN thereby becomes a calibrated physical risk measure, but rather that RPN-derived quantities may still be used coherently when their scope is stated explicitly and their interpretation is kept conditional.

\subsection{The Central Plausibility Claim}
\label{subsec:plausibility_claim}
The formalism is most defensible under the following restricted interpretation:
\begin{quote}
    \vspace{0.1cm}
    \begin{mdframed}[linewidth=1pt]
    \emph{The RPN is used as a monotone surrogate score for a latent risk construct, and the framework is applied primarily to compare PSQA interventions and PSQA stages within a common workflow model and a common scoring rubric.}
    \end{mdframed}
\end{quote}

Under this interpretation, the framework does not require the RPN to be a probability, nor does it require ratios of RPN values to carry a literal physical meaning. It requires only that: \vspace{0.15cm}
\begin{enumerate}
    \item scoring is applied consistently across baseline and PSQA-transformed states;
    \item baseline anchoring and burden weighting stabilize comparisons by reference to the same underlying risk mass;
    \item conclusions are drawn from pattern-level effects (e.g., process-step profiles and workflow-level aggregates), not from isolated single-FM rankings.
\end{enumerate}

\subsection{Conceptual Importance of Baseline Anchoring}
\label{subsec:baseline_anchoring}
A major criticism of conventional RPN use is that fixed thresholds and raw cross-institutional comparisons invite a false sense of absolute meaning. The present framework avoids that pitfall by asking a different question:
\begin{quote}
\textit{How does a given PSQA modality change the baseline risk-score structure of this workflow?}
\end{quote}
rather than:
\begin{quote}
\textit{Is this absolute RPN value acceptable in general?}
\end{quote}
This shift is not merely rhetorical. It changes the analysis from a universal acceptability judgment to a within-workflow intervention study. In particular, the cumulative track is anchored to the shared \textit{noQA} baseline and therefore provides the strict basis for inter-modality comparison, whereas the \textit{Full}-stage metrics remain inherently within-modality quantities because they are normalized to modality-specific \textit{Data}-stage states.

\subsection{Mathematical Significance of Risk-Weighted Normalization}
\label{subsec:significance_risk_weighted_normalization}
A further strength of the formalism lies in the choice of baseline-burden weights. In simplified form, suppressing stage indices for readability (cf.~Eqs.~\eqref{appB:eq:local_weight_triple} and \eqref{appB:eq:local_rel_metric_weighted_triple}), the local relative metric may be written as
\begin{equation}
\rho_{\mathrm{rel},i}
=
\sum_{l=1}^{L_i} g_{i,l}\,\delta_{\mathrm{rel},i,l},
\qquad
g_{i,l}
=
\frac{\mathrm{RPN}^{(\mathrm{baseline})}_{i,l}}
{\sum_{l'=1}^{L_i}\mathrm{RPN}^{(\mathrm{baseline})}_{i,l'}}.
\end{equation}
Likewise, for any comparison set of process steps $\mathcal{I}$,
\begin{equation}
\Rho_{\mathrm{rel},\mathcal{I}}
=
\sum_{i\in\mathcal{I}} \gamma_{\,\mathcal{I},i}\,\rho_{\mathrm{rel},i},
\qquad
\gamma_{\,\mathcal{I},i}
=
\frac{\sum_{l}\mathrm{RPN}^{(\mathrm{baseline})}_{i,l}}
{\sum_{j\in\mathcal{I}}\sum_{l'}\mathrm{RPN}^{(\mathrm{baseline})}_{j,l'}}.
\end{equation}

These expressions ensure that low-burden failure modes cannot dominate the local or global result solely because they exhibit large relative fluctuations. Conversely, changes affecting high-burden failure modes contribute proportionally to the aggregate metric. The weighted relatives are therefore not arbitrary averages of relative score changes; they are exactly the corresponding absolute burden changes normalized by the stage-appropriate baseline burden. This is the precise mathematical sense in which the formalism adds structure beyond conventional pFMEA usage.

\subsection{What the Formalism Does and Does Not Claim}
\label{subsec:what_formalism_claims}
The strongest scientifically defensible claim of the present framework is therefore limited but precise. The formalism does not establish the RPN as a quantitative or probabilistic measure of real-world risk. What it does establish is an algebraically exact, baseline-anchored, and stage-consistent transformation and aggregation scheme for comparing PSQA interventions once a common scoring rubric has been fixed.\vspace{0.1cm}

In this sense, the framework is best characterized as \emph{algebraically rigorous but measurement-theoretically conditional}. The rigor resides in the exact identities of Appendix~\ref{ch:math_framework} and in the burden-preserving normalization from failure modes to process steps and workflow segments. The conditionality resides in the fact that the underlying $O$, $S$, and $D$ values remain ordered expert scores. The resulting metrics should therefore be interpreted as semi-quantitative comparative effect-size indices rather than as absolute risks.

\Needspace{8\baselineskip}
\section{Summary and Recommendations}
\label{sec:conclusions_recommendations}
Risk‐weighted normalization is a substantive methodological improvement over standard pFMEA approaches. It operationalizes a transparent normalization by baseline risk contribution, which is consistent with the idea that changes in high‐burden failure modes matter more than changes in negligible ones.\\

Uncertainty remains the dominant practical vulnerability. High inter‐rater variability can still propagate into derived metrics; however, aggregation and baseline weighting provide partial stabilization. Robustness procedures during the scoring process, such as anonymous scoring and moderator facilitation, are essential tools to decrease scoring uncertainty.\\

The framework is largely agnostic to the specific “risk score” function. While multiplication of $O$, $S$, and $D$ were used herein, conceptually, the same staged and weighted structure could incorporate additive scores or other categorization methods, provided that consistent baseline anchoring and weighting can be defined.

\Needspace{8\baselineskip}
\section{Conclusions}

Published critiques and comparative studies of RPN remain directly relevant: ordinal inputs do not by themselves justify ratio statements, raw RPN orderings can be unstable or non-unique, fixed RPN thresholds are not generally defensible, and cross-institutional comparison of unqualified RPN values is methodologically fragile. The present framework does not overturn those critiques, nor does it attempt to promote the RPN to a calibrated probabilistic risk measure. Instead, it narrows the claim and changes the use case. By introducing a common \textit{NoQA} baseline, stage-specific PSQA transformations, and burden-weighted aggregation, the framework repurposes RPN-derived scores into comparative indicators of intervention effect within a defined workflow model.\\

Recent probabilistic and generalized reformulations of the RPN are fully compatible with this limited interpretation. Probabilistic priority numbers, general mathematical RPN models, and recent Bayesian-multinomial modeling of the RPN distribution itself all show that richer formal treatment of RPN-derived quantities is possible without implying that the RPN thereby becomes an absolute measure of real-world risk \cite{parrachosantanna2012ppn,kimzuo2018general,mahmoudvand2025rpn}.\\

Under this restricted interpretation, the formalism is best described as \emph{algebraically rigorous but measurement-theoretically conditional}. Its algebraic rigor follows from the exact identities developed in Appendix~\ref{ch:math_framework}, which guarantee stage consistency and burden-preserving aggregation from the failure-mode level to process steps and workflow segments. Its conditional character reflects the fact that the underlying scores remain ordered expert judgments.\vspace{0.1cm}

Accordingly, the formalism can be regarded as methodologically justified provided that four conditions remain explicit: first, the same scoring rubric must be applied consistently across baseline and PSQA-transformed states; second, strict inter-modality comparisons should be based primarily on cumulative metrics anchored to the common \textit{NoQA} baseline; third, conclusions should be drawn from pattern-level burden changes rather than isolated single-RPN rankings; and fourth, uncertainty should be treated as approximate unless supplemented by richer categorical, Monte Carlo, bootstrap, or Bayesian analyses.\\

Within these limits, the framework provides a coherent and practically useful architecture for comparative PSQA decision support and remains sufficiently general to accommodate future non-RPN prioritization schemes without sacrificing its baseline-anchored hierarchical logic. 

  \EndTextAppendixLayout
\endgroup

\end{document}